\documentclass{aa}  

\usepackage[switch]{lineno}
\usepackage{graphicx}
\usepackage{natbib}
\usepackage{txfonts}
\usepackage[draft]{hyperref}

\usepackage[usenames]{color}

\newcommand{\bcep}{$\beta$~Cep }
\newcommand{\dsct}{$\delta$~Sct }
\newcommand{\gdor}{$\gamma$~Dor }
\newcommand{\Kepler}{\textit{Kepler} }

\usepackage{amssymb}
\usepackage{mathrsfs}

\begin{document} 


   \title{On the photometric detection of Internal Gravity Waves \\ in upper main-sequence stars}
   \titlerunning{On the photometric detection of IGWs in CoRoT targets}
   
   \subtitle{I. Methodology and application to CoRoT targets}

   \author{D. M. Bowman \inst{1} 
          \and
          C.~Aerts \inst{1,2}
          \and
          C.~Johnston \inst{1}
          \and
          M.~G.~Pedersen\inst{1}
          \and
          T.~M.~Rogers \inst{3,4}
          \and
          P.~V.~F.~Edelmann \inst{3}
          \and
          S.~Sim{\' o}n-D{\' i}az\inst{5,6}
          \and
          T.~Van~Reeth \inst{7,8}      
          \and
          B.~Buysschaert \inst{1,9}
          \and
          A.~Tkachenko \inst{1}   
          \and
          S.~A.~Triana \inst{10}
          }

    \institute{Instituut voor Sterrenkunde, KU Leuven, Celestijnenlaan 200D, 3001 Leuven, Belgium \\
              \email{dominic.bowman@kuleuven.be} 
         \and
        		Department of Astrophysics, IMAPP, Radboud University Nijmegen, NL-6500 GL Nijmegen, The Netherlands
         \and
             	School of Mathematics, Statistics and Physics, Newcastle University, Newcastle-upon-Tyne NE1 7RU, UK
         \and
           	Planetary Science Institute, Tucson, AZ 85721, USA
	\and
		Instituto de Astrof{\' i}sicade Canarias, E-38200 La Laguna, Tenerife, Spain
	\and	
		Departamento de Astrof{\' i}sica, Universidad de La Laguna, E-38205 La Laguna, Tenerife, Spain
	\and
		Sydney Institute for Astronomy (SIfA), School of Physics, The University of Sydney, NSW 2006, Australia
	\and
		Stellar Astrophysics Centre, Department of Physics and Astronomy, Aarhus University, Ny Munkegade 120, DK-8000 Aarhus C, Denmark
	\and
		LESIA, Observatoire de Paris, PSL Research University, CNRS, Sorbonne Universit{\' e}s, UPMC Univ. Paris 06, Univ. Paris Diderot, Sorbonne Paris Cit{\' e}, 5 place Jules Janssen, F-92195 Meudon, France
	\and
		Royal Observatory of Belgium, Ringlaan 3, B-1180 Brussels, Belgium
            }
            
   \date{Received Month, Day 2017; accepted Month Day, 2017}

 
  \abstract
   {Main sequence stars with a convective core are predicted to stochastically excite Internal Gravity Waves (IGWs), which effectively transport angular momentum throughout the stellar interior and explain the observed near-uniform interior rotation rates of intermediate-mass stars. However, there are few detections of IGWs, and fewer still made using photometry, with more detections needed to constrain numerical simulations.}
   {We aim to formalise the detection and characterisation of IGWs in photometric observations of stars born with convective cores ($M \gtrsim 1.5$~M$_{\rm \odot}$) and parameterise the low-frequency power excess caused by IGWs.}
   {Using the most recent CoRoT light curves for a sample of O, B, A and F stars, we parameterise the morphology of the flux contribution of IGWs in Fourier space using an MCMC numerical scheme within a Bayesian framework. We compare this to predictions from IGW numerical simulations and investigate how the observed morphology changes as a function of stellar parameters.}
   {We demonstrate that a common morphology for the low-frequency power excess is observed in early-type stars observed by CoRoT. Our study shows that a background frequency-dependent source of astrophysical signal is common, which we interpret as IGWs. We provide constraints on the amplitudes of IGWs and the shape of their detected frequency spectrum across a range of mass, which is the first ensemble study of stochastic variability in such a diverse sample of stars.}
   {The evidence of a low-frequency power excess across a wide mass range supports the interpretation of IGWs in photometry of O, B, A and F stars. We also discuss the prospects of observing hundreds of massive stars with the Transiting Exoplanet Survey Satellite (TESS) in the near future.}

   \keywords{asteroseismology -- stars: early-type -- stars: oscillations -- stars: evolution -- stars: rotation -- stars: individual: HD~46150, HD~46223, HD~46966 -- techniques: photometric}

   \maketitle


\section{Introduction}
\label{section: intro}

Understanding the physics at work within massive stars is a significant goal for astronomy as these stars are dominant in stellar and galactic evolution theory \citep{Maeder2000a}. Specifically, the radial rotation profile, interior mixing and angular momentum transport mechanisms are poorly constrained, yet they dramatically influence stellar structure and evolution theory \citep{Zahn1992, Heger2000a, Maeder_rotation_BOOK, Meynet_rotation_BOOK, Aerts2018c*}. Only detailed observational constraints of stellar interiors provide the ability to mitigate uncertainties in current stellar models (e.g. \citealt{Aerts2017b, Ouazzani2018a*}).

Asteroseismology is a vastly successful method for directly probing the interior structure and physical processes occurring within a star. This methodology uses surface variability caused by stellar oscillations generated within a star's interior to probe the physics at work at different depths, with a detailed monograph provided by \citet{ASTERO_BOOK}. Stellar oscillations are governed by the physics of driving and damping. Some driving mechanisms produce standing waves, i.e. coherent pulsation modes, with characteristic frequencies and long mode lifetimes. Pressure (p) modes are sound waves, for which pressure is the dominant restoring force and are most sensitive to the surface layers within a star \citep{ASTERO_BOOK}. Whereas gravity (g) modes have buoyancy as the dominant restoring force and are most sensitive to the near-core region. For practically all stars, rotation and specifically the Coriolis force also act as a restoring force. When the Coriolis force is the dominant restoring force, one also gets inertial modes (also known as r~modes). In the case of gravito-inertial pulsation modes, both buoyancy and the Coriolis force are important. 

The research field of asteroseismology has greatly and rapidly expanded in the last decade because of high-precision and high duty-cycle photometric data sets from space missions such as CoRoT \citep{Auvergne2009}, \Kepler \citep{Borucki2010} and K2 \citep{Howell2014}. These telescopes began a space photometry revolution and provided the first truly high-quality asteroseismic data sets, which have photometric precisions that are hundreds of times better than achievable from the ground. Furthermore, these data sets were amongst the first continuous data sets long enough to provide the high frequency resolution required for detailed asteroseismic studies of different types of pulsating stars. Of course, ground-based multi-site campaigns can offer long-term monitoring of pulsating stars, but typically have lower duty cycles, much higher noise levels and are subject to aliasing problems (see e.g. \citealt{Breger2000d, Breger2002e, Breger2004a, Kurtz2005b, Bowman2015a}), which can make identification of individual pulsation modes difficult.

The 4-yr \Kepler mission data have proven particularly useful for studying the interior properties of intermediate-mass stars \citep{Aerts2017b}. To date, approximately 70 main-sequence B, A and F stars have been measured to have near-uniform radial rotation rates \citep{Degroote2010a, Kurtz2014, Papics2014, Saio2015b, VanReeth2015b, VanReeth2018a, Triana2015, Murphy2016a, Ouazzani2017a, Ouazzani2018a*, Papics2017a, Zwintz2017a}. However, more observational studies of intermediate- and especially high-mass stars are needed to address the large shortcomings in the theory of angular momentum transport when comparing observations of main sequence and evolved stars \citep{Tayar2013, Cantiello2014, Eggenberger2017a, Aerts2017b, Ouazzani2018a*, Aerts2018c*}. To produce a near-uniform rotation rate within a star during its evolution, a strong angular momentum transport mechanism must be at work, the exact nature of which is not currently understood. A possible explanation of near-rigid rotation in stars on the upper main sequence is transport by Internal Gravity Waves (IGWs). These travelling waves (i.e. damped modes) are stochastically driven at the interface of a convective region and a stably stratified zone, possibly by plumes of material perturbing the interface, such that IGWs propagate and dissipate within radiative regions. It has been shown numerically and theoretically that IGWs are efficient at transporting angular momentum and chemical mixing within stars of various masses and evolutionary stages (see, e.g. \citealt{Press1981a, Schatzman1993b, Montalban1994, Montalban1996, Talon1997b, Talon2003b, Talon2004a, Talon2008b, Pantillon2007, Charbonnel2005a, Charbonnel2013b, Shiode2013, Lecoanet2013, Fuller2014a, Fuller2015c, Rogers2013b, Rogers2015, Rogers2017c}), but also in laboratory experiments (e.g. \citealt{Plumb1978}).

However, the few inferred detections of IGWs have so far been limited to massive stars, and were made from a qualitative comparison of the observed frequency spectrum with predictions from state-of-the-art simulations of IGWs \citep{Rogers2013b, Rogers2015, Aerts2015c, Johnston2017a, Aerts2017a, Aerts2018a, Simon-Diaz2018a}. In this paper, we perform a detailed search for observational evidence of IGWs across a wide range in mass -- the first dedicated study of its kind -- to provide observational constraints of IGWs to numerical simulations. In Section~\ref{section: red noise observations}, we provide an overview of the various causes of low-frequency variability in intermediate- and high-mass stars, and we interpret the predicted frequency spectra from numerical simulations of IGWs in Section~\ref{section: IGW simulations}. In Section~\ref{section: method} we discuss the method of characterising observations of IGWs and in Sections~\ref{section: CoRoT results} and \ref{section: discussion} the results from our ensemble study of stars are discussed. Finally, we discuss the prospects of detecting IGWs in massive stars observed by the Transiting Exoplanet Survey Satellite (TESS; \citealt{Ricker2015}) in Section~\ref{section: future}, and draw conclusions in Section~\ref{section: conclusions}.


\section{The sources of low-frequency variability in early-type stars}
\label{section: red noise observations}

Stochastic variability with periods of order days appears to be almost ubiquitous in massive stars, with photometric variability\footnote{Note that the conversion between flux variations expressed in $\mu$mag and ppm is $2.5\log_{10}(e) = 1.08574$.} of order a few hundred $\mu$mag and spectroscopic variability of at least tens of km\,s$^{-1}$ typically observed in O stars (see e.g. \citealt{Balona1992c, Walker2005a, Rauw2008, Buysschaert2015, Buysschaert2017a}). In spectroscopy, large values of macroturbulent broadening are usually needed when fitting line profiles in O and B stars, with macroturbulence typically correlated with mass on the main sequence (see, e.g. \citealt{Simon-Diaz2010b, Simon-Diaz2014a, Grassitelli2015a, Simon-Diaz2017a}). Since macroturbulence is a non-rotational form of broadening, it has been interpreted to be caused by stellar pulsations as these would explain the observed large-scale turbulent motions in stellar atmospheres of massive stars \citep{Fullerton1996, Aerts2009b, Simon-Diaz2010b, Simon-Diaz2011h}. 

Photometry from the MOST satellite of the rapidly-rotating ($v\,\sin\,i \simeq 400$~km\,s$^{-1}$) O9.5\,V star $\zeta$~Oph showed low-frequency ($\nu \lesssim 10$~d$^{-1}$; 116~$\mu$Hz)\footnote{Note that the conversion between frequency expressed in d$^{-1}$ and $\mu$Hz is $\frac{625}{54} = 11.57407$.} variability that was attributed to \bcep pulsation modes \citep{Walker2005a}. Later, \citet{Howarth2014a} study the variable amplitudes and stochastic variability of the pulsation modes in $\zeta$~Oph by combining spectroscopy and multiple sources of space photometry that span several years. Another example of a star observed by MOST with low-frequency variability is the Wolf-Rayet star HD~165763, which was interpreted to be caused by its strong wind \citep{Moffat2008c}. These studies discuss the possibility of radial and non-radial pulsation modes as the cause of the low-frequency variability, yet none of them were able to identify individual pulsation modes because of their non-coherent nature. Further examples of non-coherent variability in massive stars include the three O~stars, HD~46150, HD~46223 and HD~46966, which were found to each have a significant low-frequency excess in their frequency spectra using CoRoT photometry \citep{Blomme2011b}. Similarly, \citet{Buysschaert2015} used K2 space photometry to study five main-sequence O stars and found stochastic low-frequency variability in at least two of them. \citet{Aerts2017a} were able to extract the photometric variability of the O~supergiant HD~188209 and also discovered a significant low-frequency power excess. More recently, \citet{Aerts2018a} investigated the variability of the blue supergiant HD~91316 ($\rho$~Leo), which revealed rotational modulation, but additional multiperiodic low-frequency variability indicative of IGWs.

The observed low-frequency power excess, commonly referred to as red noise, in the photometry of massive stars has been determined to be astrophysical. However, there are multiple physical mechanisms that may occur concurrently, which produce stochastic low-frequency variability:
\begin{itemize}
\item sub-surface convection and/or granulation;
\item stellar pulsations (coherent and/or damped modes);
\item modulation from an inhomogeneous and aspherical wind.
\end{itemize}

Each of these phenomena produce a different frequency spectrum because of the different physical timescales associated with the variability. Our aim is to undertake a systematic investigation of stars across the Hertzsprung--Russell (HR) diagram and unravel the different possible observational signatures of their variability, for the purpose of constraining the amplitudes and frequencies of IGWs in upper main sequence stars. Below we briefly discuss the different sources of possible variability.


	\subsection{Granulation and surface convection}
	
	One of the earliest discussions of how red noise can be caused by granulation was made by \citet{Schwarzschild1975}. This stochastic and non-periodic form of variability is commonly observed in stars with large surface convection zones such as solar-type and red giant stars \citep{Michel2008a, Chaplin2013c, Kallinger2014, Hekker2017a}. \citet{Harvey1985} was the first to fit the background in the Fourier spectrum of the Sun with a Lorentzian function including a characteristic granulation frequency.
	
	Red noise in the photometry of intermediate-mass main-sequence A stars has also been investigated. The two \dsct stars HD~50844 and HD~174936 observed by CoRoT were claimed to contain hundreds of statistically significant peaks using iterative pre-whitening that assumed only white noise \citep{GH2009, Poretti2009}. However, \citet{Kallinger2010c} characterised the red noise in these stars using a Lorentzian-like model and demonstrate that these stars have characteristic frequencies of order a few hundred $\mu$Hz and amplitudes of order a few tens to a few hundred $\mu$mag, which are consistent with solar granulation scaled to higher mass stars. Other examples of an observed red noise background attributed to granulation include the A5~III star $\alpha$~Ophiuchi (HD~159561) studied using MOST photometry by \citet{Monnier2010}. Granulation has been claimed to be a possible explanation for the astrophysical red noise in post-main sequence late-A and early-F stars because of their non-negligible surface convection zones \citep{Kallinger2010c, Monnier2010, PG2015d}. However, no similar \dsct stars have been found amongst the thousands of A stars observed by the \Kepler Space Telescope (see, e.g. \citealt{Balona2014b, Bowman2018a}).

	
	\subsection{Coherent and damped pulsation modes}
	
	Coherent and/or damped pulsation modes have been independently suggested as explanations for macroturbulence in spectroscopy and for red noise in photometry, yet discrepancies between observations and theoretical models of pulsation excitation remain to this day. Specifically, pulsation modes predicted to be stable by theoretical models have been detected in some stars and vice versa (e.g. \citealt{Dziembowski2008, Handler2009b, Briquet2011, Aerts2011}).
		
	The excitation and detectability of convectively-driven g~modes in high-mass stars was studied by \citet{Shiode2013}, who predicted spectroscopic variability of order mm\,s$^{-1}$ and photometric variability of order tens of $\mu$mag. For more evolved stars that are near the terminal-age main-sequence (TAMS) or have evolved off the main sequence, the amplitudes were predicted to be of order a few hundred $\mu$mag and tens of cm\,s$^{-1}$ \citep{Shiode2013}. Importantly, \citet{Shiode2013} also predict that the amplitudes correlate with stellar mass. However, the theoretical predictions by \citet{Shiode2013} of pulsation mode lifetimes were of the order of years to Myr, which are more representative of coherent pulsation modes as opposed to damped modes or IGWs. Furthermore, these predictions were inconsistent with the observed lifetimes of order hours and days observed by \citet{Blomme2011b} and \citet{Aerts2015c}. 
	
	\citet{Cantiello2009b} and \citet{Shiode2013} demonstrated that a sub-surface convection layer in high-mass stars is able to stochastically excite IGWs of high-degree ($\ell \gtrsim 30$). These small-scale IGWs would produce an undetectably small velocity field and photometric amplitudes at the stellar surface because of geometric cancellation effects. A large-scale velocity field was detected in spectroscopy and stochastic variability of order $0.1$~mmag in photometry for the high-mass primary of V380~Cyg by \cite{Tkachenko2014a}, which was interpreted to be caused by IGWs. The higher amplitudes in more massive stars and/or more evolved stars increases the likelihood of detecting IGWs. This may be in part why detections of IGWs in the literature have so far been in main-sequence O stars and blue supergiants \citep{Aerts2015c, Aerts2017a, Aerts2018a, Simon-Diaz2017a, Simon-Diaz2018a}.
	
	Pure inertial waves, and the corresponding subset known as Rossby waves, have recently been discovered in the \Kepler photometry of intermediate-mass stars \citep{VanReeth2016a, Saio2018a}. In their pioneering work, \citet{Saio2018a} find that r~modes (i.e. global Rossby waves) appear as groups of closely-spaced peaks in a frequency spectrum. Furthermore, in experiments that investigate inertial waves in rotating spheres of fluid \citep{Zimmerman2011, Zimmerman2014, Triana_PhD, Rieutord2012b}, these waves have been observed to produce a low-frequency power excess (see Figure~6.15 from \citealt{Triana_PhD}). For a detailed discussion of the various types of waves that can occur within stars, we refer the reader to \citet{Mathis2011b} and \citet{Mathis2014a} and references therein.


	\subsection{Stellar winds}
	
	In addition to sub-surface convection, granulation and IGWs, red noise in stars with masses above $M \gtrsim 15$~M$_{\rm \odot}$, can also be caused by a clumpy, aspherical and inhomogeneous stellar wind. The exact cause of the wind clumping in massive stars is not known \citep{Puls2008c}, but a radiative driving mechanism was originally proposed by \citet{Owocki1984}, with the onset of a clumpy wind occurring in the stellar photosphere \citep{Puls2006}. Wind clumping is a possible explanation for photometric observations of red noise in massive stars (e.g. \citealt{Aerts2018a, Rami2018a, Krticka2018d}), which also may synonymously explain the large macroturbulence observed in spectroscopy of massive stars (e.g. \citealt{Simon-Diaz2017a}). 
	
	In their study of red noise in the CoRoT photometry of three O stars, \citet{Blomme2011b} comment on a tentative dichotomy of low-frequency photometric variability in massive stars. Specifically, early-O stars typically show red noise and late-O stars are more likely to have \bcep pulsations, with the transition occurring at a spectral type of O8 \citep{Blomme2011b}. A similar transition has also been noted in spectroscopic studies of variability on the upper main sequence \citep{Simon-Diaz2017a}. The various photometric and spectroscopic studies of stochastic variability in early-type stars provide strong motivation to study how the morphology of red noise changes as a function of mass on the main sequence. Furthermore, by characterising the photometric red noise in intermediate- and high-mass stars, one is able to search for the observational signatures of IGWs and provide much-needed constraints of their amplitudes and frequencies to numerical simulations.


\section{Interpreting simulations of Internal Gravity Waves}
\label{section: IGW simulations}
	
Any star with a convective region is capable of generating IGWs, with changing properties imposed by the radial rotation profile \citep{Pantillon2007, Rogers2013b, Rogers2015, Mathis2014a} or strength of an internal magnetic field \citep{Rogers2010, Rogers2011a, Mathis2011b, Augustson2016a, Lecoanet2017a}. The amplitudes of IGWs are expected to scale with stellar mass, as high-mass main-sequence stars have a larger convective core compared to intermediate-mass stars \citep{Samadi2010c, Shiode2013, Rogers2015}. Also, the frequencies of IGWs are predicted to inversely scale with stellar mass on the main sequence, since high-mass main-sequence stars have larger radii compared to intermediate-mass stars \citep{Samadi2010c, Shiode2013}. 

The first 3D simulations of core convection for an intermediate-mass star, specifically a 2-M$_{\rm \odot}$ A~star, were performed by \citet{Browning2004a}, but were restricted to only the inner 30~per~cent in stellar radius and provided no predictions of the amplitudes of IGWs, or of the occurrence of wave breaking close to the stellar surface. Other studies of wave excitation due to core convection exist (e.g. \citealt{Samadi2010c, Shiode2013}), yet are typically limited to 1D, a parameterised treatment of convection and neglect rotation. The predictions from \citet{Samadi2010c} and \citet{Shiode2013} were not consistent with each other in terms of the amplitudes of IGWs, which is likely because these 1D models were for non-rotating stars; it was shown by \citet{Mathis2014a} that rotation and the Coriolis force play a key role in the determination of IGW amplitudes.


	\subsection{2D numerical simulations in a 3-M$_{\rm \odot}$ star}
	
	The most interesting aspect of IGWs for stellar structure and evolution is their ability to transport angular momentum within a star's interior, which was demonstrated using 2D numerical simulations of a zero-age main-sequence (ZAMS) 3-M$_{\rm \odot}$ star by \citet{Rogers2013b}. These 2D simulations have limitations in their dimensionality and degree of turbulence, but produce predictions of an IGW frequency spectrum that can be confronted with observations. As noted by \citet{Rogers2013b}, these IGW simulations have enhanced thermal diffusivity for numerical reasons. The consequences of these limitations strongly damp variability with frequencies below $\nu \lesssim 1$~d$^{-1}$. Furthermore, the absolute values of IGW amplitudes in a predicted spectrum are uncalibrated and possibly more likely resemble a higher mass main-sequence star \citep{Aerts2015c}. The amplitudes of IGWs are dependent on the mass of the star, specifically the mass of the convective core, but the frequencies of IGWs are only weakly dependent of the stellar mass \citep{Shiode2013, Aerts2015c}. For instance, the scaling of the global frequency spectrum is estimated to be approximately $\sim 0.75$ for a ZAMS 3-M$_{\rm \odot}$ star and a 30-M$_{\rm \odot}$ star \citep{Shiode2013, Aerts2015c}. However, without further IGW simulations for different masses and evolutionary stages, this scaling is only qualitative and approximate.

	\begin{figure}
	\centering
	\includegraphics[width=0.99\columnwidth]{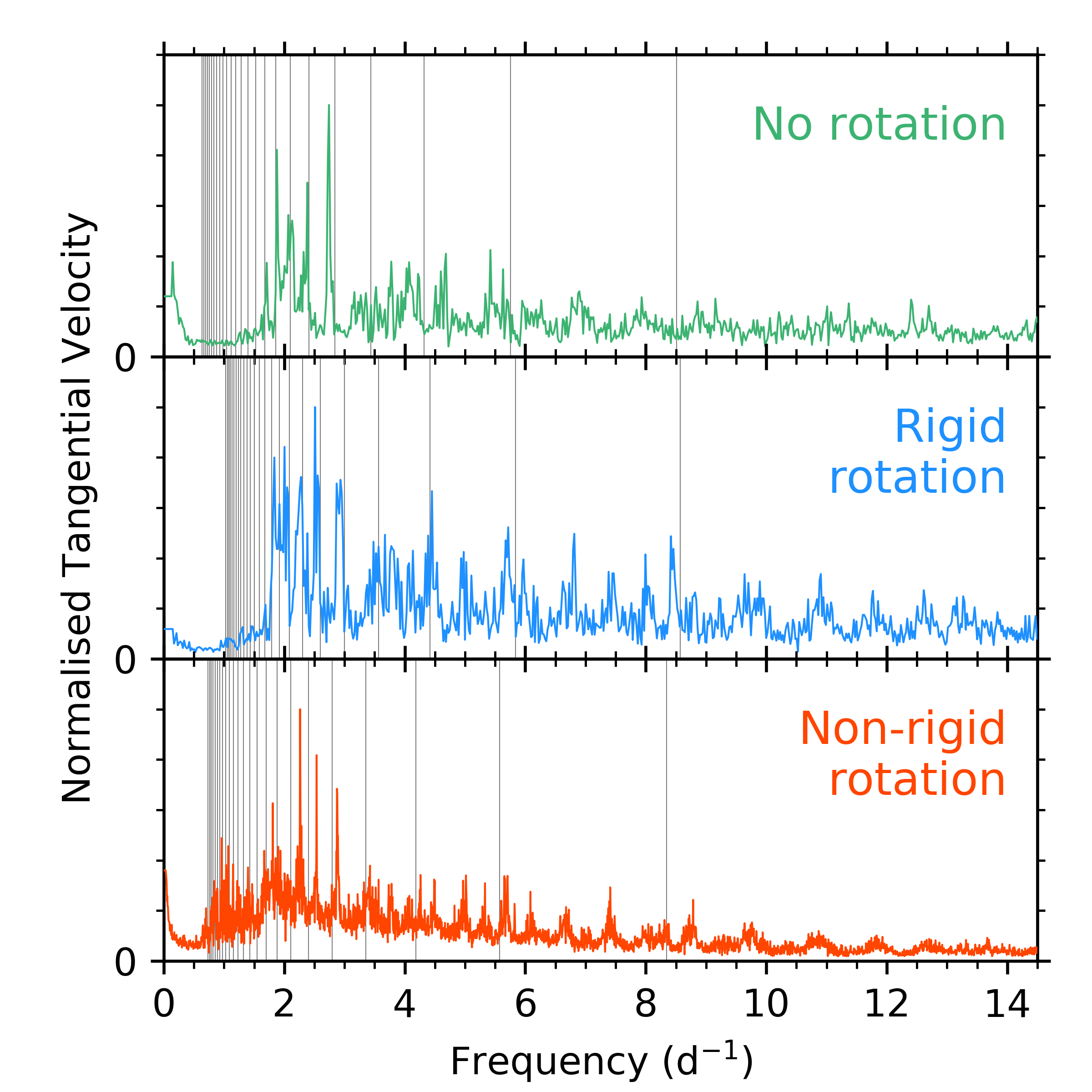}
	\includegraphics[width=0.99\columnwidth]{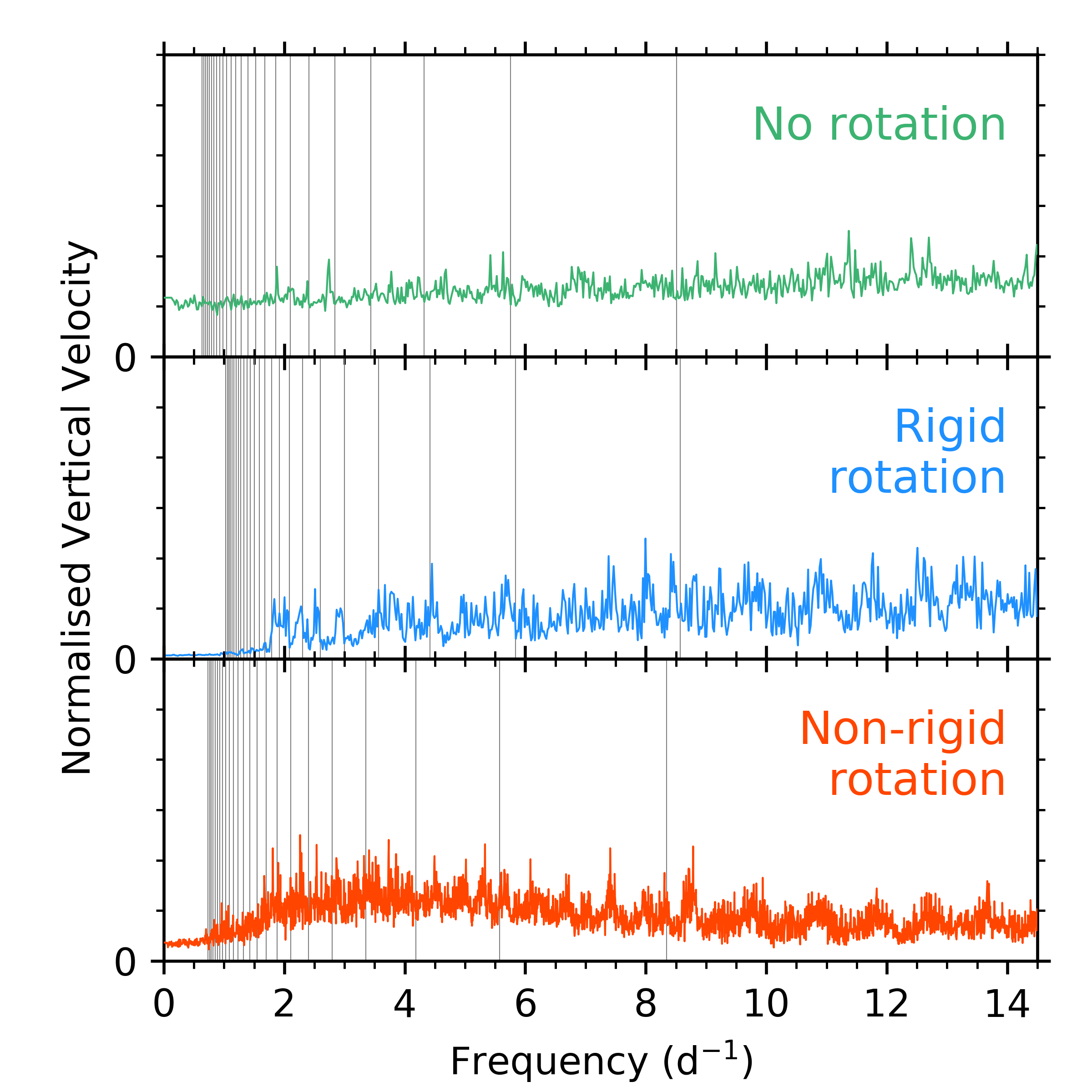}
	\caption{The top and bottom panels show the tangential and vertical velocity spectra predicted by the 2D IGW simulations from \citet{Rogers2013b}, for three scenarios for the radial rotation profile: no rotation; rigid rotation with a rotation frequency of 1.1~d$^{-1}$ (12.7~$\mu$Hz); and non-rigid rotation (${\Omega_{\rm core}}/{\Omega_{\rm env}} = 1.5$) with a surface rotational frequency of 0.275~d$^{-1}$ (8.7~$\mu$Hz), which are shown in green, blue and red, respectively. Note that the amplitudes of these spectra have been normalised to the dominant peak in each spectrum and have arbitrary units. The vertical grey lines show the theoretical pulsation mode frequencies for $(\ell,m) = (1,0)$ in a 3-M$_{\rm \odot}$ ZAMS star for radial orders in the range $n \in [-25,-1]$ calculated by \texttt{GYRE} \citep{Townsend2013b} and each rotation scenario.}
	\label{figure: IGW simulations}
	\end{figure}
		
	The combined effect of hundreds of IGWs at the surface of a star is predicted to produce a low-frequency power excess in the spectra of velocity and flux variations \citep{Rogers2013b, Rogers2015}, with the surface tangential and vertical velocities from the 2D IGW simulations by \citet{Rogers2013b} shown in the top and bottom panels of Fig.~\ref{figure: IGW simulations}, respectively. For any gravity wave, we expect the amplitudes of vertical velocities to be much smaller than the tangential velocities \citep{ASTERO_BOOK}, but the ordinate axis for the spectra in Fig.~\ref{figure: IGW simulations} are essentially in arbitrary units. In Fig.~\ref{figure: IGW simulations}, three scenarios for the radial rotation profile are shown, which are labelled as no rotation; rigid rotation with a rotation frequency of 1.1~d$^{-1}$ (12.7~$\mu$Hz); and non-rigid rotation (${\Omega_{\rm core}}/{\Omega_{\rm env}} = 1.5$) with a surface rotational frequency of 0.275~d$^{-1}$ (8.7~$\mu$Hz), which are shown in green, blue and red, respectively. As can be seen in Fig.~\ref{figure: IGW simulations}, the non-rigid rotation typically shifts the IGW spectrum to lower frequencies, but the overall morphology of the IGW tangential velocity spectrum remains similar for all cases. Thus the dominant feature of IGWs in the frequency spectrum of a star should be a stochastic low-frequency power excess.


	\subsection{IGW frequency comparison with \texttt{GYRE}}
	\label{subsection: gyre}

	To further understand the IGW spectra from \citet{Rogers2013b}, we calculate a non-rotating 1D stellar structure model of a ZAMS 3-M$_{\rm \odot}$ star using \texttt{MESA} (v9793; \citealt{Paxton2011, Paxton2013, Paxton2015, Paxton2018a}), and calculate its zonal dipole g-mode frequencies using the adiabatic module of \texttt{GYRE} (v5.0; \citealt{Townsend2013b, Townsend2018a}). The IGW frequency spectra shown in Fig.~\ref{figure: IGW simulations} have known rotation profiles from \citet{Rogers2013b}, hence we employ the traditional approximation for rotation (TAR) within \texttt{GYRE} when calculating numerical g-mode pulsation frequencies, with uniform rotation for the rigidly rotating model, and differential rotation as derived by \citet{Mathis2009d} and recently demonstrated by \citet{VanReeth2018a} for the non-rigid model. The frequencies corresponding to $(\ell,m) = (1,0)$ for $n \in [-25,-1]$ are overplotted on the IGW spectra as vertical grey lines in Fig.~\ref{figure: IGW simulations} for comparison. For a given radial order $n$, higher angular degrees $\ell$ have higher frequencies, which is why the IGW spectra in Fig.~\ref{figure: IGW simulations} appear structured, since these spectra contain many values of $n$, $\ell$ and $m$ and thus contain hundreds of individual IGWs.
	
	For all three scenarios of rotation in the tangential velocity spectrum in the top panel of Fig.~\ref{figure: IGW simulations}, a series of dominant peaks are present between $1 \lesssim \nu \lesssim 4$~d$^{-1}$ ($10 \lesssim \nu \lesssim 50$~$\mu$Hz). Precise frequencies cannot be extracted from the IGW simulations using classical pre-whitening techniques because of the stochastic nature of the driving mechanism which produces broad peaks. However, if one assumes that the dominant peaks seen in Fig.~\ref{figure: IGW simulations} correspond to zonal dipole modes, $(\ell,m) = (1,0)$, then one can see that they appear close to the pulsation mode frequencies calculated using \texttt{GYRE}, as such they are approximately equal to consecutive radial order modes in the range of $n \in [-9, -5]$ with a characteristic period spacing between $\Delta\,P \simeq 4000 - 5000$~s. This is consistent with the expected asymptotic period spacing for a ZAMS 3-M$_{\rm \odot}$ star \citep{Papics2017a, Pedersen2018a}. A similar interpretation is obtained if one makes a comparison to zonal quadrupole, retrograde dipole or prograde dipole g-mode frequencies using \texttt{GYRE}. Therefore, we speculate that the dominant peaks in the 2D IGW spectra from \citet{Rogers2013b} correspond to g-mode pulsation frequencies that are stochastically excited through resonance, yet we cannot claim a unique identification of individual mode geometries for a given frequency. This interpretation is supported by the study of the Be star HD~51452, which was concluded to exhibit gravito-inertial modes that were stochastically excited by a convective region and produced broad peaks with variable amplitudes in the observed frequency spectrum \citep{Neiner2012d}.
	
	A detailed quantitative comparison of individual pulsation mode frequencies using \texttt{GYRE} and the theoretical IGW spectra from \citet{Rogers2013b} must await 3D IGW numerical simulations. Yet, there are two important conclusions from the comparison of the 2D IGW spectra and \texttt{GYRE}, which are relevant for the characterisation of photometric variability due to an entire spectrum of IGWs, containing both travelling and standing wave components:
	\begin{enumerate}
	\item The dominant broad peaks in the IGW spectra are compatible with standing g-mode pulsation eigenfrequencies, which we interpret to be stochastically-excited;
	\item The background low-frequency power excess is caused by a spectrum of IGWs of various scales, which can be decomposed in multiple spherical harmonics of various $\ell$ and $m$ values.
	\end{enumerate}
	
	It is the goal of this study to interpret the observational signatures of this background low-frequency power excess caused by a spectrum of IGWs in photometry. Specifically, we aim to measure the characteristic frequency and slope of the low-frequency power excess for stars across the HR diagram, as these observables provide useful constraints to the theory and numerical simulations of IGWs and angular momentum transport \citep{Rogers2013b, Rogers2015}.


\section{Parameterising photometric red noise in massive stars}
\label{section: method}

\begin{table*}
\caption{The properties of the stars included in this work, including the $V$-band magnitude, spectral type, and approximate fundamental parameters including the effective temperature, surface gravity, and projected surface rotational velocity. We also include the run and length of CoRoT data for each star, and a reference for the stellar parameters.} 
\begin{center}


\begin{tabular}{l r r r r r r r r r}
\hline \hline
\multicolumn{1}{c}{Name} & \multicolumn{1}{c}{$V$~mag} & \multicolumn{1}{c}{Sp.\,Type} & \multicolumn{1}{c}{$T_{\rm eff}$} & \multicolumn{1}{c}{$\log\,g$} & \multicolumn{1}{c}{$v\,\sin\,i$} & \multicolumn{2}{c}{CoRoT data} & \multicolumn{1}{c}{Reference} \\

\multicolumn{1}{c}{} & \multicolumn{1}{c}{} & \multicolumn{1}{c}{} & \multicolumn{1}{c}{(K)} & \multicolumn{1}{c}{(cm\,s$^{-2}$)} & \multicolumn{1}{c}{(km\,s$^{-1}$)} & \multicolumn{1}{c}{run} & \multicolumn{1}{c}{$\Delta\,T$~(d)} & \multicolumn{1}{c}{} \\
\hline

\multicolumn{8}{l}{O stars from \citet{Blomme2011b} inferred to have IGWs by \citet{Aerts2015c}:} \vspace{0.1cm} \\
HD~46150	&	$6.73$	&	O5\,V((f))z		&	$42\,000$		&	$4.0$	&	$100$	&	SRa02	&	34.33	&	\citet{Martins2015d} \\
HD~46223 	&	$7.28$	&	O4\,V((f))		&	$43\,000$		&	$4.0$	&	$100$	&	SRa02	&	34.33	&	\citet{Martins2015d} \\
HD~46966 	&	$6.87$	&	O8.5\,IV		&	$35\,000$		&	$3.8$	&	$50$		&	SRa02	&	34.32	&	\citet{Martins2015d} \\

\hline
\multicolumn{8}{l}{Additional CoRoT OBAF stars:} \vspace{0.1cm} \\
HD~45418	&	$6.47$	&	B5\,V	&	$16\,750$	&	$4.2$	&	$237$	&	SRa04	&	54.14	&	\citet{Mugnes2015} \\
HD~45517	&	$7.58$	&	A0\,V	&	$10\,000\tablefootmark{$\ddagger$}$	&	$$		&	$$		&	SRa04	&	53.77	&	$-$ \\
HD~45546	&	$5.04$	&	B2\,V	&	$19\,000$	&	$4.0$	&	$61$		&	SRa04	&	54.29	&	\citet{Silaj2014a} \\
HD~46149\tablefootmark{$\dagger$} 	&	$7.61$	&	O8\,V	&	$36\,000$		&	$3.7$	&	$30$		&	SRa02	&	34.33	&	\citet{Degroote2010b} \\
HD~46179\tablefootmark{$\dagger$}		&	$6.71$	&	B9\,V	&	$11\,000$		&	$4.0$	&	$152$	&	SRa02	&	34.10 &	\citet{Niemczura2009b} \\
HD~46202 	&	$8.19$	&	O9.2\,V	&	$33\,500$		&	$4.2$	& 	$15$		&	SRa02	&	34.15	&	\citet{Martins2015d} \\
HD~46769 	&	$5.86$	&	B5\,II		&	$13\,000$	&	$2.7$	&	$72$		&	SRa03	&	26.38	&	\citet{Aerts2013} \\
HD~47485 	&	$8.83$	&	A5\,IV/V	&	$7300$	&	$2.9$	&	$40$		&	SRa03	&	26.37	&	\citet{Gebran2016} \\
HD~48784	&	$6.65$	&	F0\,V	&	$6900$	&	$3.5$	&	$110$	&	SRa01	&	25.05	&	\citet{Barcelo2017a} \\
HD~48977	&	$5.92$	&	B2.5\,V	&	$20\,000$	&	$4.2$	&	$25$		&	SRa01	&	25.28	&	\citet{Thoul2013} \\
HD~49677	&	$8.07$	&	B9\,V	&	$9200$	&	$4.0$	&	$10$	&	SRa01	&	25.28	&	\citet{Degroote2011} \\
HD~50747\tablefootmark{$\dagger$}		&	$5.44$	&	A4\,IV	&	$7900$	&	$3.5$	&	$70$		&	IRa01	&	60.76	&	\citet{Dolez2009} \\
HD~50844 	&	$9.09$	&	A2\,II		&	$7400$	&	$3.6$	&	$58$		&	IRa01	&	57.71	&	\citet{Poretti2009} \\
HD~50870\tablefootmark{$\dagger$} 	&	$8.85$	&	A8\,III	&	$7600$	&	$3.9$	&	$38$		&	LRa02	&	114.41	&	\citet{Mantegazza2012} \\
HD~51193 	&	$8.06$	&	B1.5\,IVe	&	$23\,000$	&	$3.6$	&	$220$	&	LRa02	&	114.41	&	\citet{Fremat2006c} \\
HD~51332 	&	$7.01$	&	F0\,V	&	$6700$	&	$4.1$	&	$$		&	LRa02	&	114.41	&	\citet{David2015a} \\
HD~51359 	&	$8.5\tablefootmark{$\ddagger$}$	&	F0\,IV	&	$6800$	&	$$	&	$$		&	LRa02	&	117.41	&	\citet{McDonald2012a} \\
HD~51452 	&	$8.08$	&	B0I\,Ve	&	$29\,500$	&	$3.9$	&	$320$	&	LRa02	&	114.41	&	\citet{Neiner2012d} \\
HD~51722 	&	$7.53$	&	A9\,V	&	$7000$	&	$$	&	$$		&	LRa02	&	117.37	&	\citet{McDonald2012a} \\
HD~51756\tablefootmark{$\dagger$} 	&	$7.18$	&	B0.5\,IV	&	$30\,000$	&	$3.8$	&	$28$		&	LRa02	&	114.41	&	\citet{Papics2011} \\
HD~52130 	&	$9.11$	&	A2\,III/IV	&	$8200\tablefootmark{$\ddagger$}$	&	$$	&	$$		&	LRa02	&	117.41	&	$-$ \\
HD~174532	&	$6.88$	&	A2\,III/IV	&	$7200$	&	$3.6$	&	$32$		&	SRc02	&	26.24	&	\citet{Fox-Machado2010a} \\
HD~174589	&	$6.06$	&	F2\,III	&	$7000$	&	$3.5$	&	$97$		&	SRc02	&	26.17	&	\citet{Fox-Machado2010a} \\
HD~174936 	&	$8.56$	&	A3\,IV	&	$8000$	&	$4.1$	&	$170$	&	SRc01	&	27.19	&	\citet{GH2009} \\
HD~174966	&	$7.69$	&	A7\,III/IV 	&	$7500$	&	$4.2$	&	$126$	&	SRc01	&	27.20	&	\citet{GH2009} \\
HD~174967	&	$9.31$	&	B9.5\,V	&	$11\,000\tablefootmark{$\ddagger$}$	&	$$	&	$$		&	SRc02	&	26.24	&	$-$ \\
HD~174990	&	$8.74$	&	A3\,V	&	$8500\tablefootmark{$\ddagger$}$	&	$$	&	$$		&	SRc02	&	23.23	&	$-$ \\
HD~175272	&	$7.40$	&	F5\,V	&	$6700$	&	$4.1$	&	$23$		&	SRc01	&	27.19	&	\citet{Hekker2014b} \\
HD~175445	&	$7.79$	&	hA5mA2V	&	$8500$	&	$4.0$	&	$$		&	SRc02	&	26.24	&	\citet{Paunzen2002c} \\
HD~175542	&	$9.00$	&	A1\,V	&	$8500\tablefootmark{$\ddagger$}$	&	$$	&	$$	&	SRc01	&	27.19	&	$-$ \\
HD~175640	&	$6.20$	&	B9\,V\,HgMn	&	$12\,000$	&	$4.0$	&	$5$		&	SRc02	&	23.09	&	\citet{Lefever2010} \\
HD~263425	&	$9.41$	&	A0\,V	&	$9500\tablefootmark{$\ddagger$}$	&	$$	&	$$	&	SRa01	&	25.28	&	$-$ \\

\hline \hline
\end{tabular} 
\end{center}
\tablefoot{$\dagger$ indicates a confirmed multiple system for which we list the parameters of the primary, and $\ddagger$ indicates a value was estimated from its spectral type.}
\label{table: stars}
\end{table*}


	\subsection{CoRoT stars of spectral types O, B, A and F}
	\label{subsection: CoRoT stars}
		
	\begin{figure}
	\centering
	\includegraphics[width=0.49\textwidth]{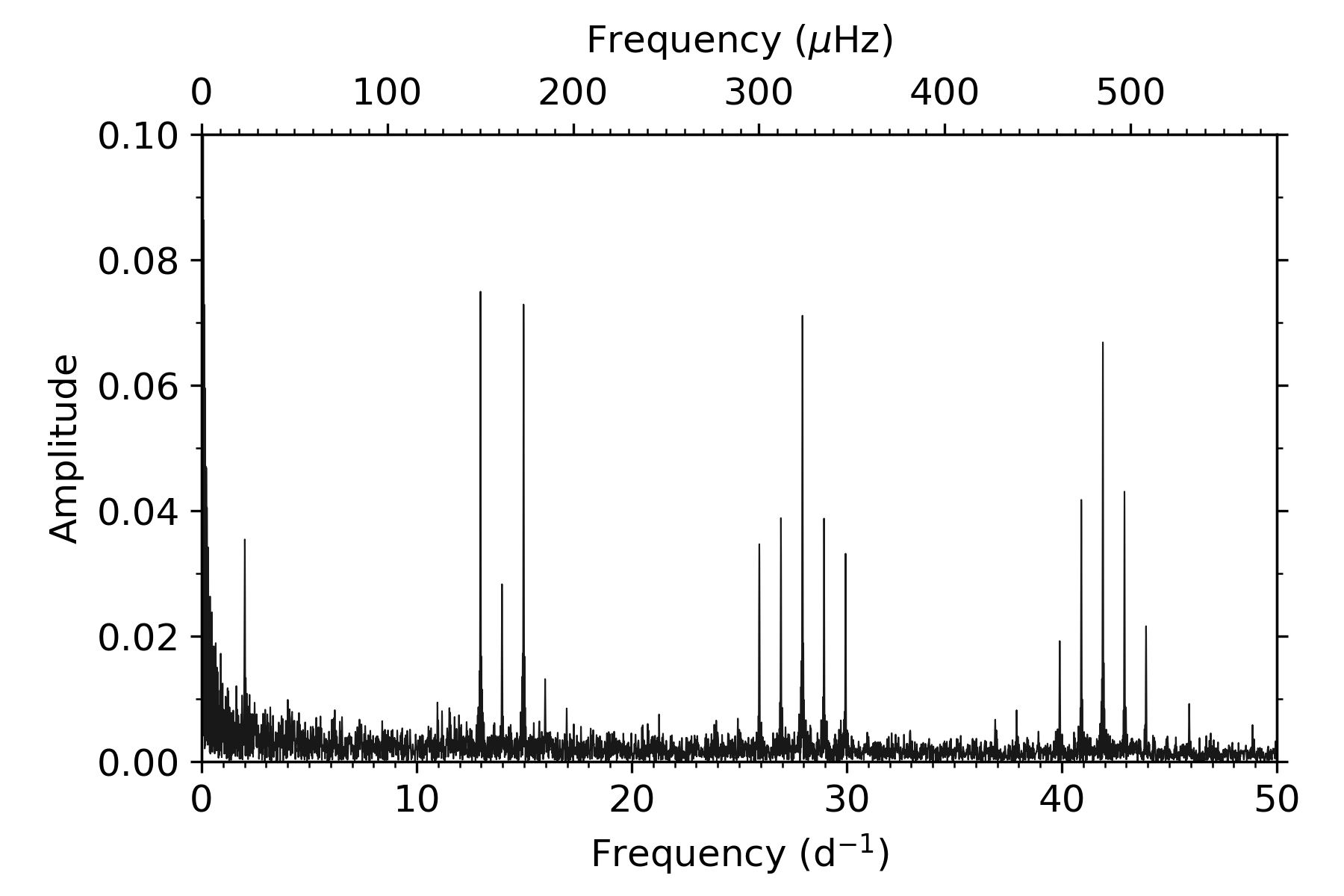}
	\includegraphics[width=0.49\textwidth]{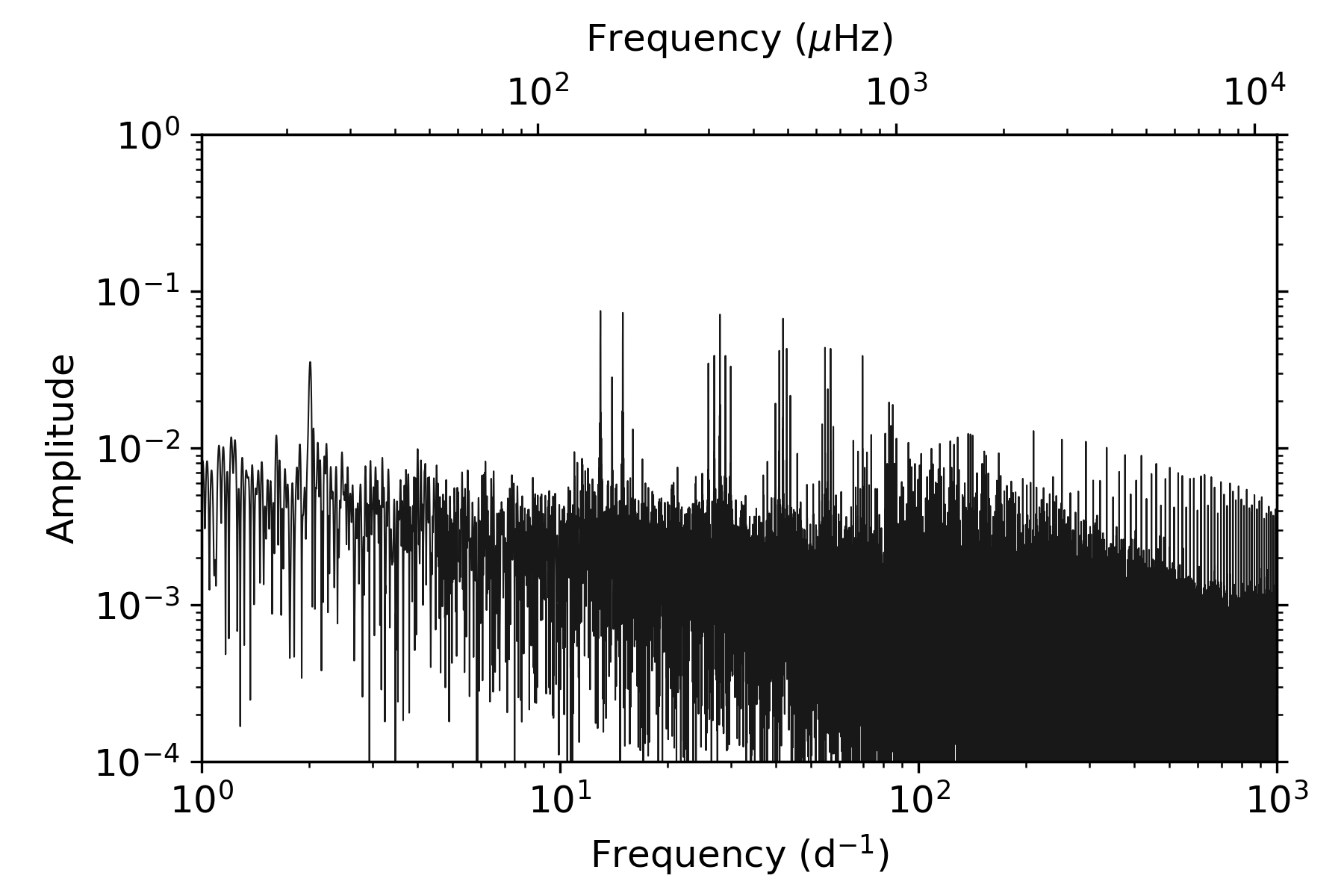}
	\caption{The spectral window function of the CoRoT data for HD~46150 up to 50~d$^{-1}$ (578.7~$\mu$Hz) is plotted on a linear scale in the top panel, and on a logarithmic scale in the bottom panel.}
	\label{figure: CoRoT window}
	\end{figure}
	
	To perform a systematic search for IGWs across a wide range in stellar mass, we use the most recent, high-quality and high-cadence photometry from the CoRoT satellite \citep{Auvergne2009}. We use only a single telescope to avoid introducing biases from the differences in instrument passbands and sampling frequencies of other space missions. The early-type stars included in our study are stars of spectral type O, B, A and F from the CoRoT asteroseismology runs IRa01, LRa02, SRa01, SRa02, SRa03, SRa04, SRc01 and SRc02, with the specific details for each star, including the $V$-band magnitude, spectral type and fundamental parameters given in Table~\ref{table: stars}. 
	
	Note that spectroscopic studies prior to the last decade typically have larger estimates of $v\,\sin\,i$ values for massive O and B stars as only rotational broadening was assumed and macroturbulence was usually ignored. Furthermore, estimates derived from only a single spectrum represent only a snapshot of the true variability, which explains why typically a variance of $20$~km\,s$^{-1}$ is often found in literature values of $v\,\sin\,i$ and $v_{\rm macro}$ for O and B stars --- see Table~1 of \citet{Aerts2014b}. 
	
	The CoRoT light curves range in length between $\sim$25 and $\sim$115~d and have a median cadence of approximately 32~s \citep{Michel2006c}. The CoRoT satellite has an orbital frequency of 13.972~d$^{-1}$ (161.7~$\mu$Hz) and its orbit crosses the South Atlantic Anomaly (SAA) twice each sidereal day, which introduces alias frequencies in a spectrum because of invalid flux measurements and periodically discarded hot pixels. The spectral window of CoRoT data is shown in Fig.~\ref{figure: CoRoT window}, using the 34-d time series of HD~46150.

	
	\subsection{Pre-whitening of high signal-to-noise peaks}
	\label{subsection: prewhitening}
	
	Amongst our sample are stars with coherent pulsation modes, including for example $\delta$~Sct, SPB and \bcep pulsators. The coherent pulsation modes in these stars can be p and g~modes, which are driven by an opacity or heat engine driving mechanism and have periods of order hours and days \citep{ASTERO_BOOK}. The mode lifetimes of coherent pulsation modes are essentially infinite compared to the length of our observations, hence they can be extracted using iterative pre-whitening if time series data of sufficient length is available (see, e.g. \citealt{Degroote2009a, Papics2012a, ASTERO_BOOK, Bowman_BOOK}). 
	
	Since we are interested in characterising the morphology of the background low-frequency power excess caused by a spectrum of IGWs, representing spherical harmonics of multiple $\ell$ and $m$ values, we remove $S/N > 4$ peaks corresponding with the detected standing waves via pre-whitening (see e.g. \citealt{Breger1993b}). The number of peaks extracted depends on the quality and quantity of photometric data available for an individual star. The noise level for each extracted peak is calculated using a window of $1$~d$^{-1}$ centred at the extracted frequency, which in reality is an estimate that includes instrumental noise, white noise and any low-$S/N$ astrophysical signal. Hence our conservative approach to pre-whitening by allowing for a local $S/N$ estimate of a peak in a spectrum ensures we are not overfitting \citep{Degroote2009a, Papics2012a}. Several studies have warned against the over-extraction of hundreds of frequencies using iterative pre-whitening, with this method injecting variability rather than removing it in the cases with variable amplitude and/or frequency peaks or strongly correlated data sets (see, e.g. \citealt{Degroote2009a, Papics2012a, PG2015d, Kurtz2015b, Bowman_BOOK}). We opt for this conservative approach to only pre-whitening high-$S/N$ peaks since we are interested in parameterising the background low-frequency power excess in a spectrum. This pre-whitening methodology produces a residual light curve, from which we calculate a residual power density spectrum in units of ppm$^2$/$\mu$Hz versus $\mu$Hz to allow for an accurate comparison of stochastic signal in stars with different length CoRoT runs.
	
	In the summary figures for these stars, we show the original and residual power density spectra in orange and black, respectively, in Appendix~\ref{section: appendix: MCMC results}. In stars for which pre-whitening was employed, the residual power density spectra represent our observations. The only exceptions were HD~46150 and HD~46223, in which no peaks with $S/N \geq 4$ were detected. Thus we used the original power density spectra as our data for HD~46150 and HD~46233, and the residual power density spectra as our data for all other stars to search for signatures of IGWs.


	\subsection{Red noise fitting with a Bayesian MCMC numerical scheme}
	\label{subsection: MCMC}
	
	Following \citet{Stanishev2002} and \citet{Blomme2011b}, but also numerous studies of solar-like oscillators (see e.g. \citealt{Michel2008a, Chaplin2013c, Kallinger2014, Hekker2017a}), the morphology of a stochastically-excited low-frequency power excess in a power density spectrum can be physically interpreted by a Lorentzian function:
	
	\begin{equation}
	\alpha \left( \nu \right) = \frac{ \alpha_{0} } { 1 + \left( \frac{\nu}{\nu_{\rm char}} \right)^{\gamma}} + P_{\rm W} ~ ,	
	\label{equation: Blomme}
	\end{equation}
	
	\noindent where $\alpha_{0}$ is a scaling factor and represents the amplitude at zero frequency, $\gamma$ is the gradient of the linear part of the profile in a log-log plot, $\nu_{\rm char}$ is the characteristic frequency and $P_{\rm W}$ is a frequency-independent (i.e. white) noise term. The characteristic frequency (i.e. the frequency at which the amplitude of background equals half of $\alpha_0$) is the inverse of the characteristic timescale: $\nu_{\rm char} = (2\pi\tau)^{-1}$. 
	
	Previously, \citet{Blomme2011b} discussed how the difference in the parameters of $\alpha_{0}$, $\tau$ and $\gamma$ may be related to the mass and evolutionary stage of the O stars HD~46150, HD~46223 and HD~46966. Specifically, HD~46966 is more evolved than HD~46223 and HD~46150, such that the fundamental parameters of HD~46966 may define the different morphology of the observed red noise. Furthermore, \citet{Blomme2011b} also clearly demonstrate that this red noise is not instrumental, since different stars required different profiles. In this work, we extend the morphology characterisation of the photometric low-frequency power excess to more stars observed by CoRoT. We are motivated by the use of Eqn.~(\ref{equation: Blomme}) as stochastic signals in the time domain are well-represented by Lorentzian profiles in the Fourier domain. Furthermore, the fitting parameters of $\nu_{\rm char}$ and $\gamma$ provide observables that can be directly compared to predictions from IGW simulations.
	
	In our study, we use a Markov Chain Monte Carlo (MCMC) numerical scheme within a Bayesian framework using the \texttt{python} \texttt{emcee} package \citep{Foreman-Mackey2013}, which is an ensemble, affine-invariant approach to sampling the parameter posterior distributions and provides a rigorous error assessment of the individual model parameters. Previous examples of this approach within asteroseismology include deriving orbital parameters for binary systems \citep{Hambleton_PhD, Schmid2016a, Johnston2017a, Hambleton2018a, Kochukhov2018c}, peak-bagging of solar-like oscillations (e.g. \citealt{Toutain1994b, Appourchaux1998g, Appourchaux2003a, Appourchaux2008a, Appourchaux2008b, Benomar2009a, Gaulme2009, Gruberbauer2009, Handberg2011, Davies2016a}), and extraction of unresolved pulsation mode frequencies in \dsct stars \citep{Bowman_PhD, Bowman_BOOK}. In our usage, the CoRoT residual power density spectra in the frequency range $1 \leq \nu \leq 15\,500$~$\mu$Hz are the data, the model and its parameters are given in Eqn.~(\ref{equation: Blomme}). Since we are using the same model for all stars, we opt to use non-informative (flat) priors for all parameters and 128 parameter chains. At every iteration, each parameter chain is used to construct a model and is subject to a log-likelihood evaluation of:
	
	\begin{equation}
	\ln\mathcal{L} \propto - \frac{1}{2} \sum_i \left( \frac{y_i - M(\Theta_i)}{\sigma_i} \right)^{2} ~ ,
	\label{equation: likelihood}
	\end{equation}
	
	\noindent where $\ln\mathcal{L}$ is the log-likelihood, $y_i$ are the data, $\sigma_i$ are their uncertainties and $M_{\Theta}$ is the model produced with parameters $\Theta$. Since Eqn.~(\ref{equation: likelihood}) uses an unnormalised $\chi^2$ statistic, then $\ln\mathcal{L}$ converges to approximately $-0.5n$, where $n$ is the number of data points. Typically, the first few hundred iterations are burned, but the exact number varies from star to star. Convergence is confirmed using the parameter variance criterion from \citet{Gelman1992}, which is typically achieved after approximately 1000 iterations, and we produced the posteriors for the remaining {$\sim 1000$} iterations. 
	
\begin{figure}
	\centering
	\includegraphics[width=0.49\textwidth]{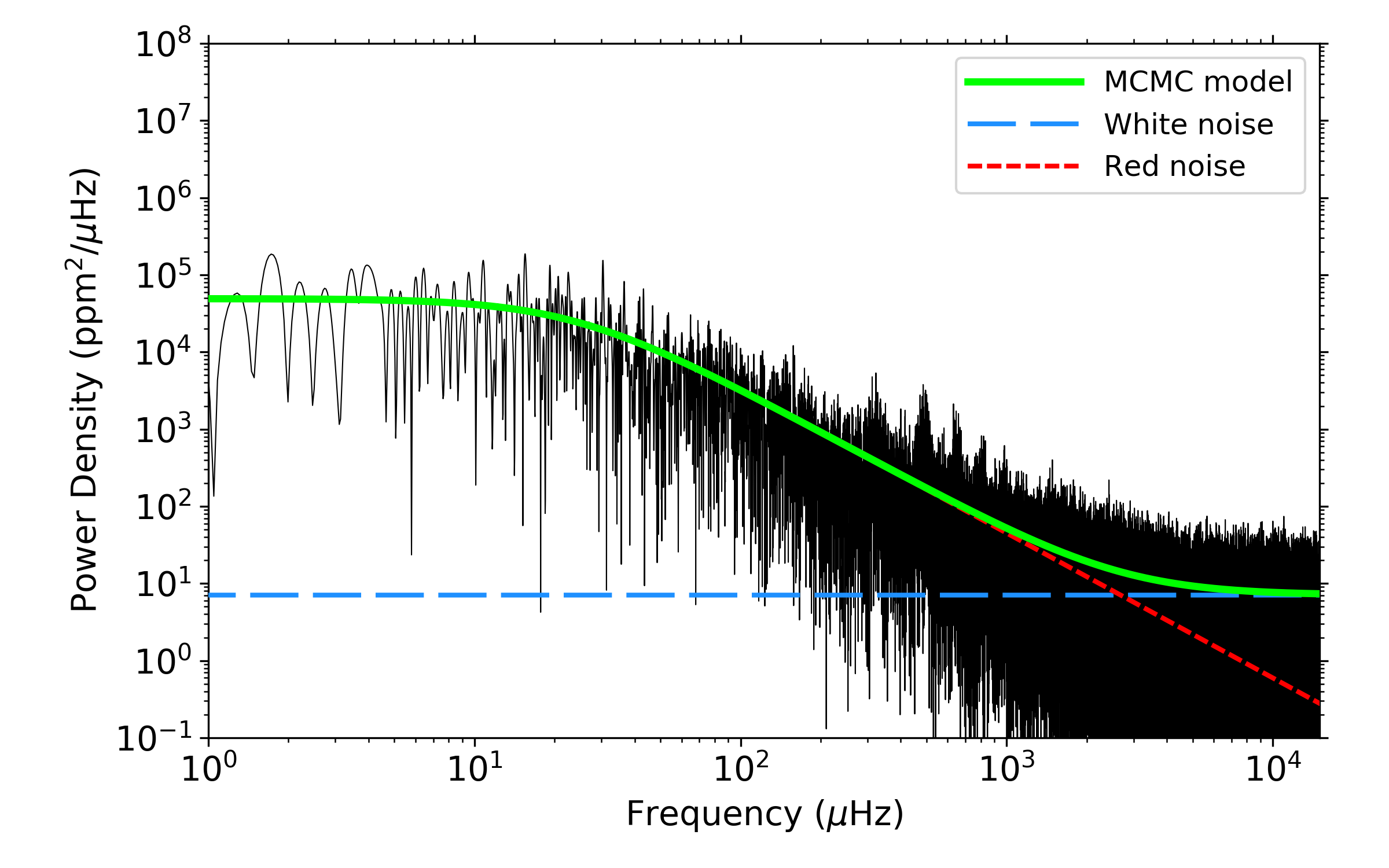}
	\includegraphics[width=0.49\textwidth]{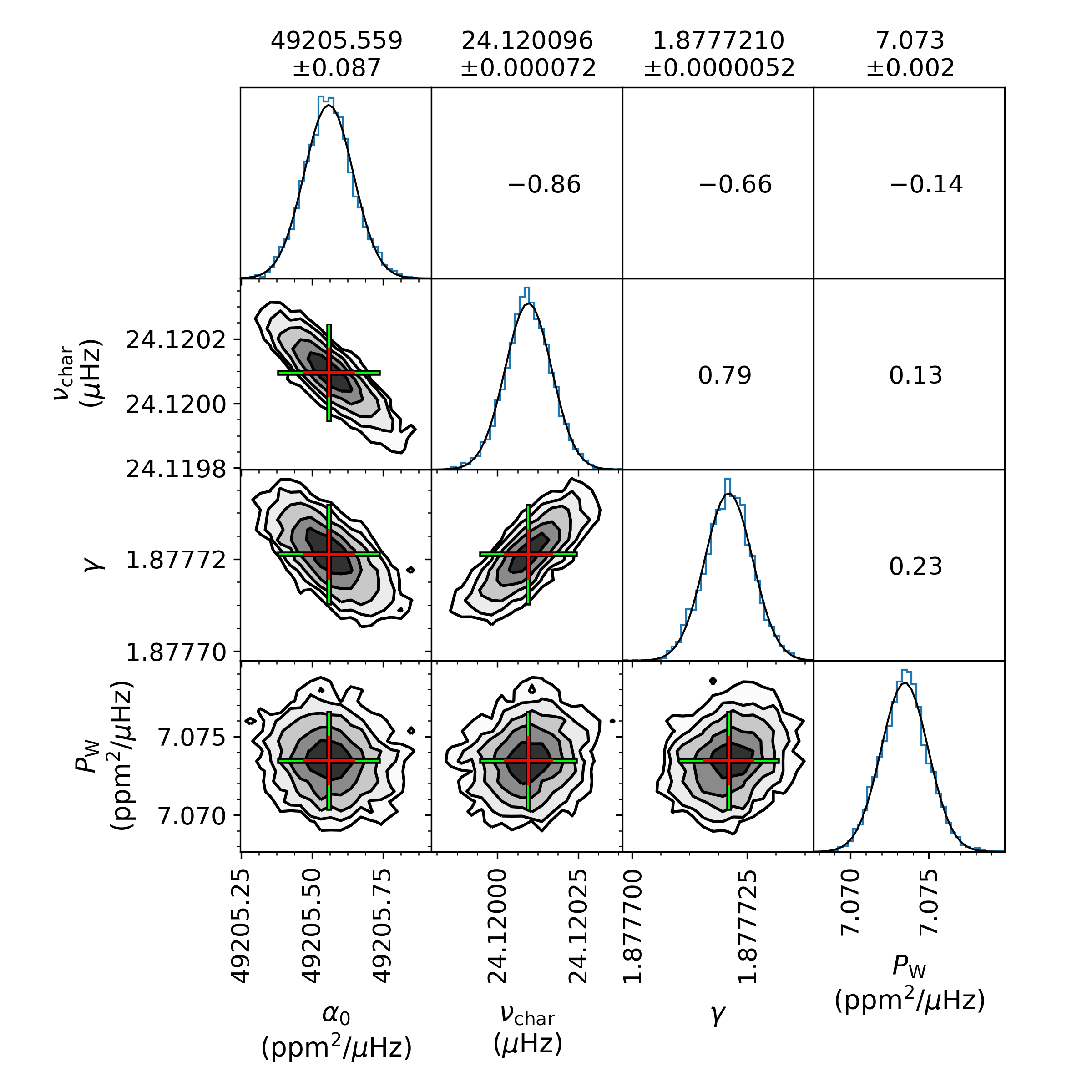}
	\includegraphics[width=0.49\textwidth]{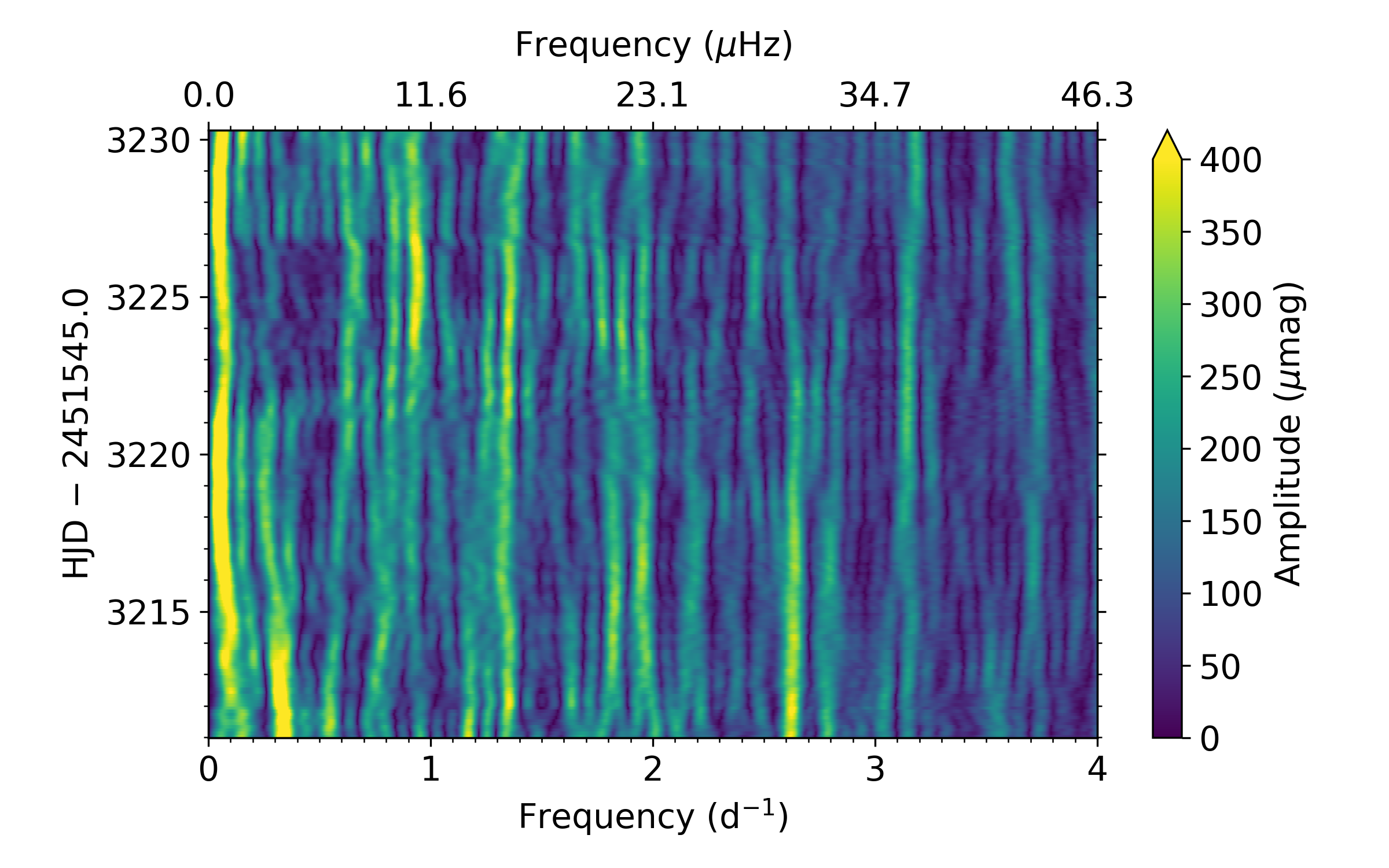}
	\caption{The logarithmic power density spectrum of the CoRoT photometry for the O star HD~46150 is shown in the top panel in black, in which the best-fit model obtained using Eqn.~(\ref{equation: Blomme}) is shown as the solid green line, and its red and white components shown as the red short-dashed and blue long-dashed lines, respectively. The middle panel shows the posterior distributions from the Bayesian MCMC, with best-fit parameters obtained from Gaussian fits (shown in black) to the 1D histograms (shown in blue). The bottom panel shows a sliding Fourier transform in the (linear) frequency range of $0 \leq \nu \leq 4$~d$^{-1}$ using a 15-d bin and a step of 0.1~d.}
	\label{figure: HD46150}
	\end{figure}

	The results of our analysis for the O star HD~46150 are shown in Fig.~\ref{figure: HD46150}, in which the top panel shows the logarithmic power density spectrum and the best-fit of Eqn.~(\ref{equation: Blomme}) shown as a solid green line\footnote{Note that the difference in the model parameters from our study and that of \citet{Blomme2011b} is caused by the difference in units.}. The middle panel of Fig.~\ref{figure: HD46150}, shows the marginalised posteriors of each parameter as 1D histograms and 2D contours with the red and green crosses corresponding to the 68\% and 95\% credible levels, respectively. $1\sigma$ statistical uncertainties of each parameter are determined from a Gaussian fit to each histogram of the converged MCMC parameter chains. The numbers in the top-right sub-plots of the middle panel in Fig.~\ref{figure: HD46150} give the pair-wise correlation between parameters, with $1$ indicating direct correlation and $-1$ indicating direct anti-correlation. The output parameters and their respective uncertainties for each star in our study are given in Table~\ref{table: results}. 
	
	\begin{figure}
	\centering
	\includegraphics[width=0.49\textwidth]{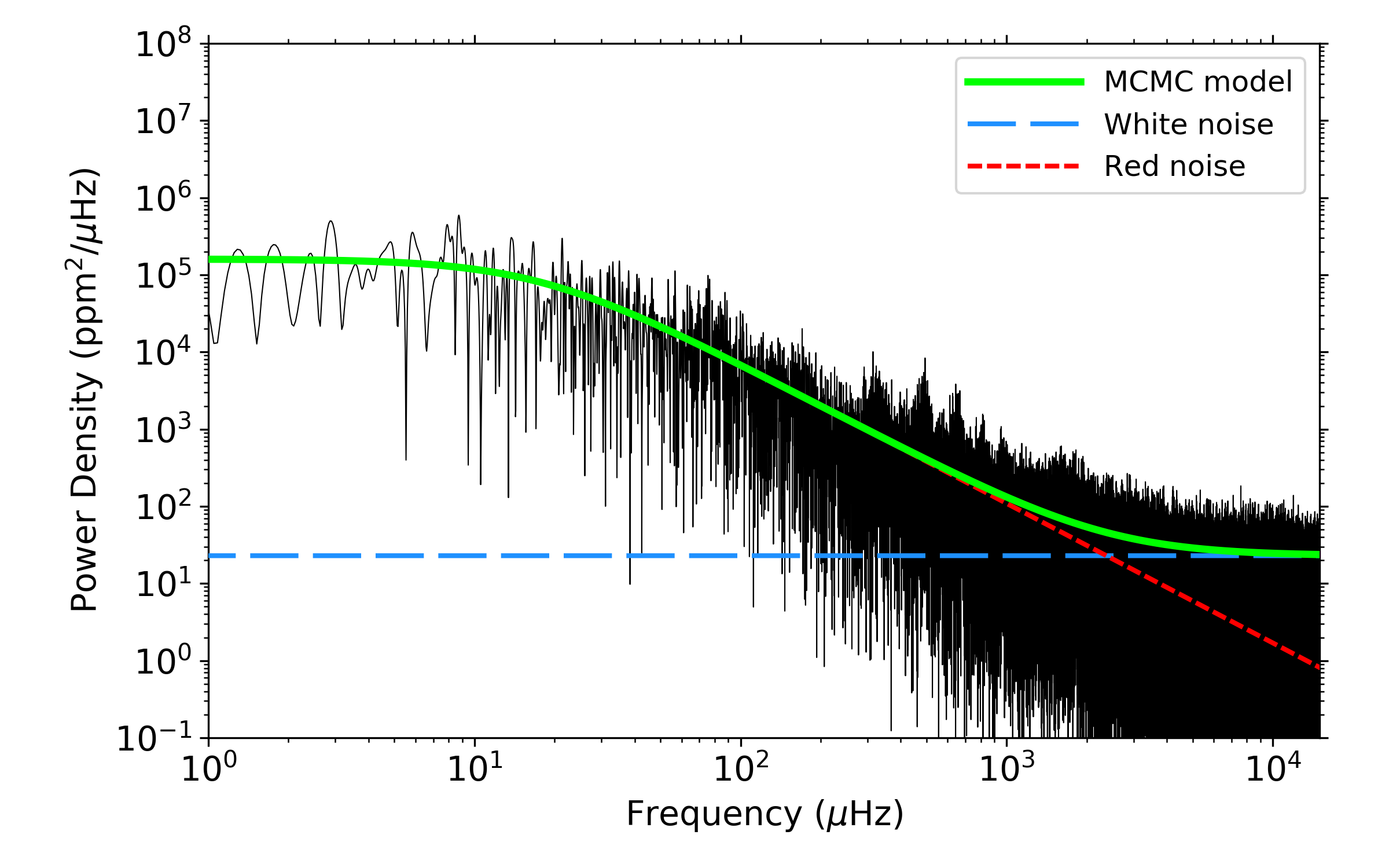}
	\includegraphics[width=0.49\textwidth]{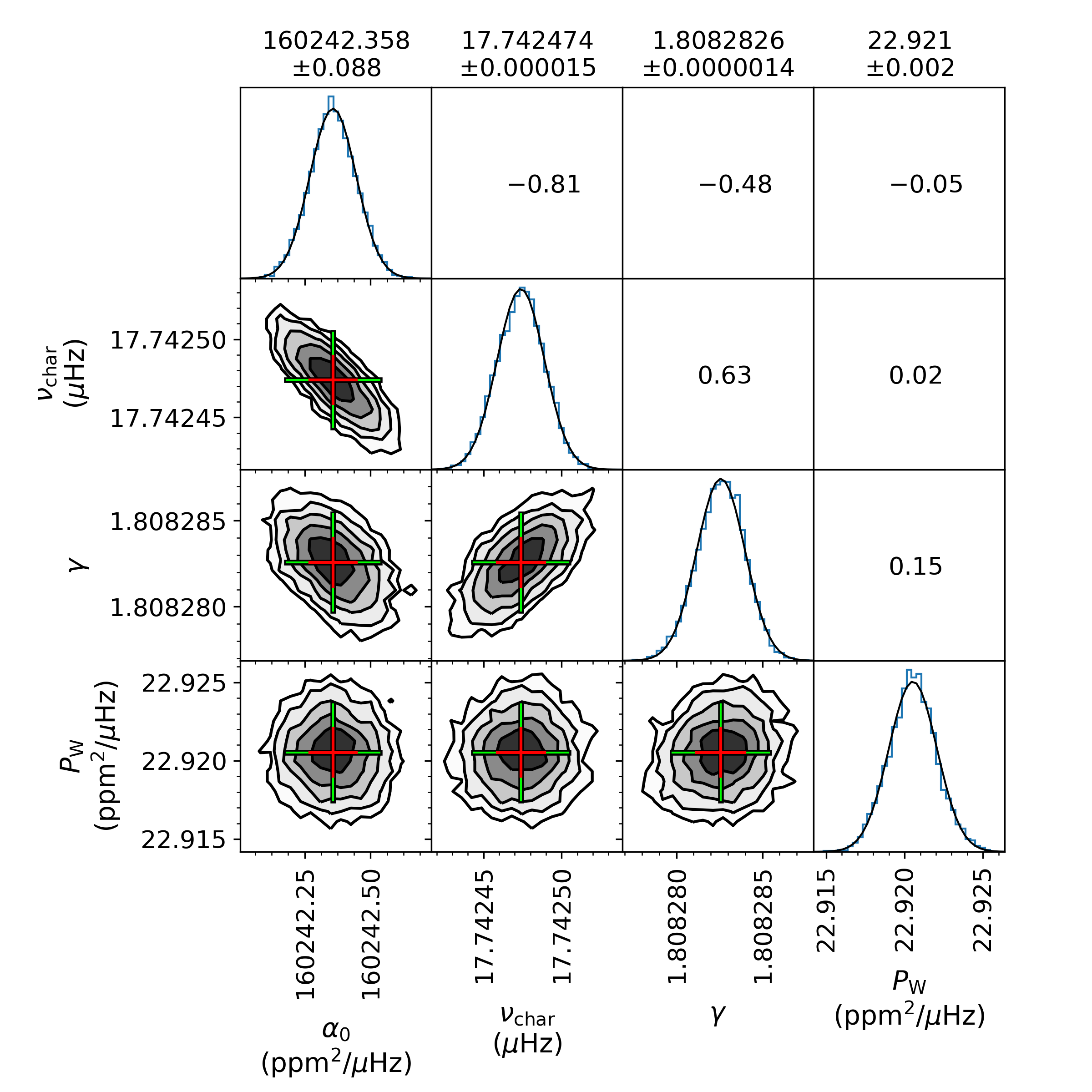}
	\includegraphics[width=0.49\textwidth]{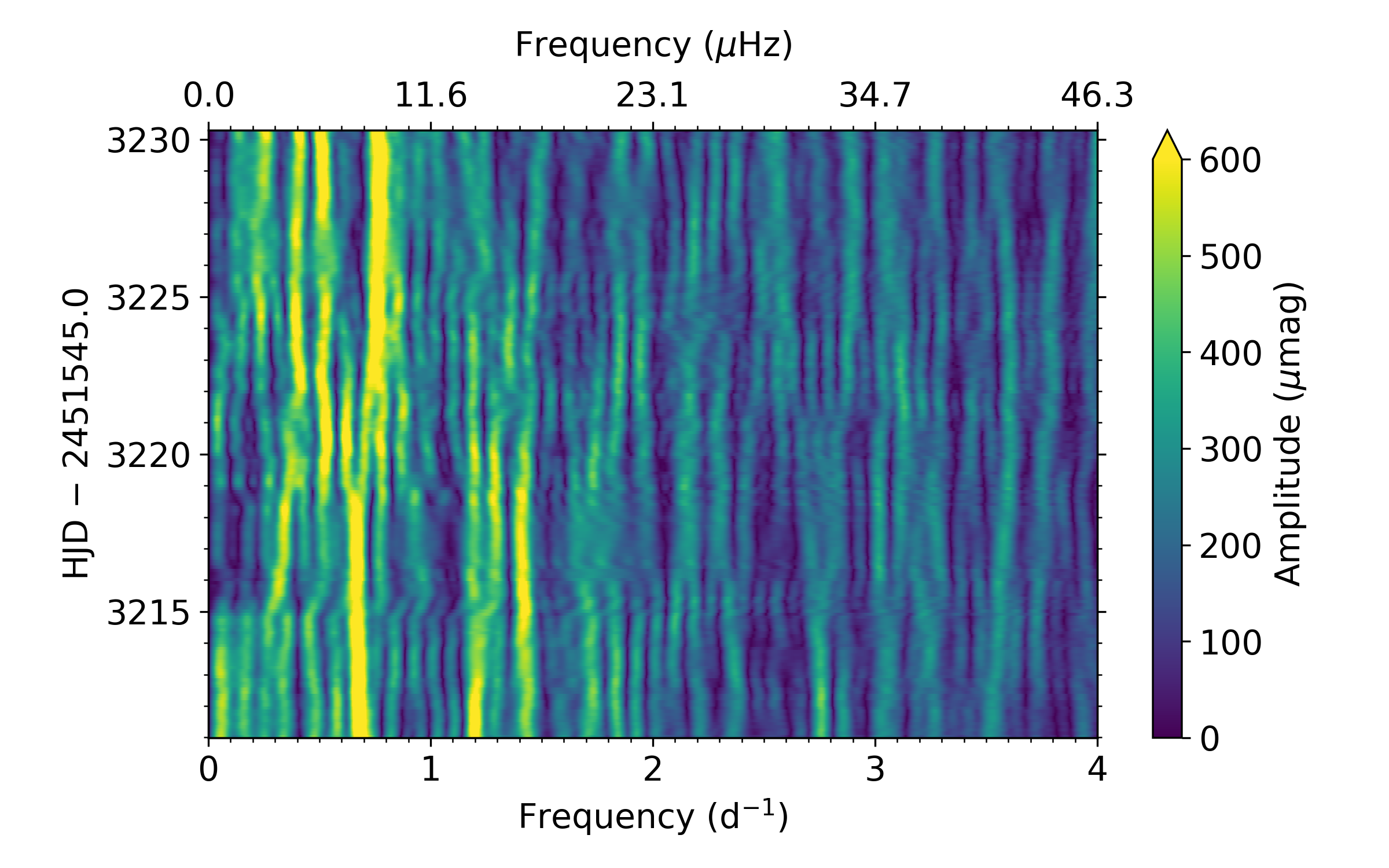}
	\caption{Summary figure for the O star HD~46223, which has a similar layout as shown in Fig.~\ref{figure: HD46150}.}
	\label{figure: HD46223}
	\end{figure}
	
	\begin{figure}
	\centering
	\includegraphics[width=0.49\textwidth]{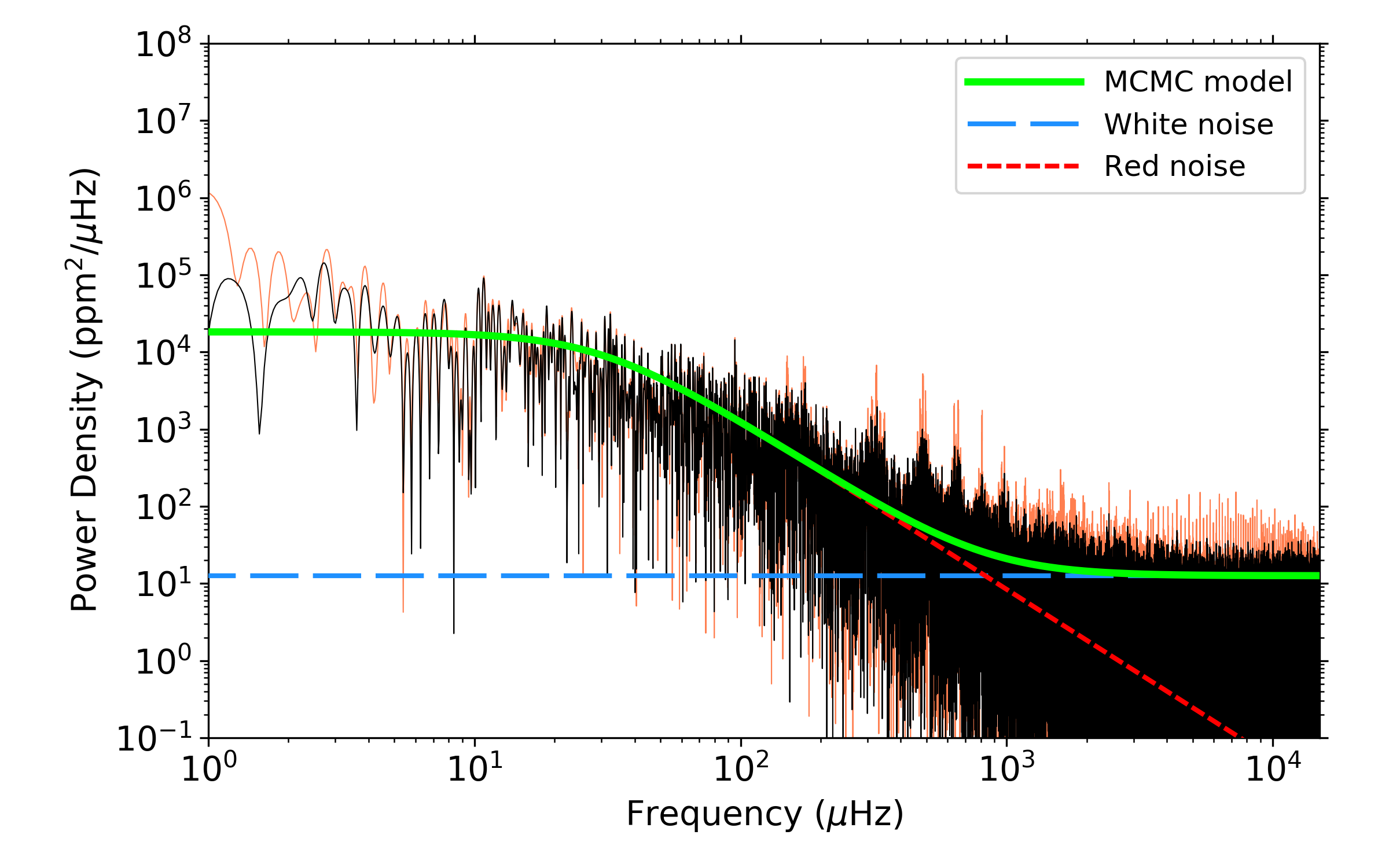}
	\includegraphics[width=0.49\textwidth]{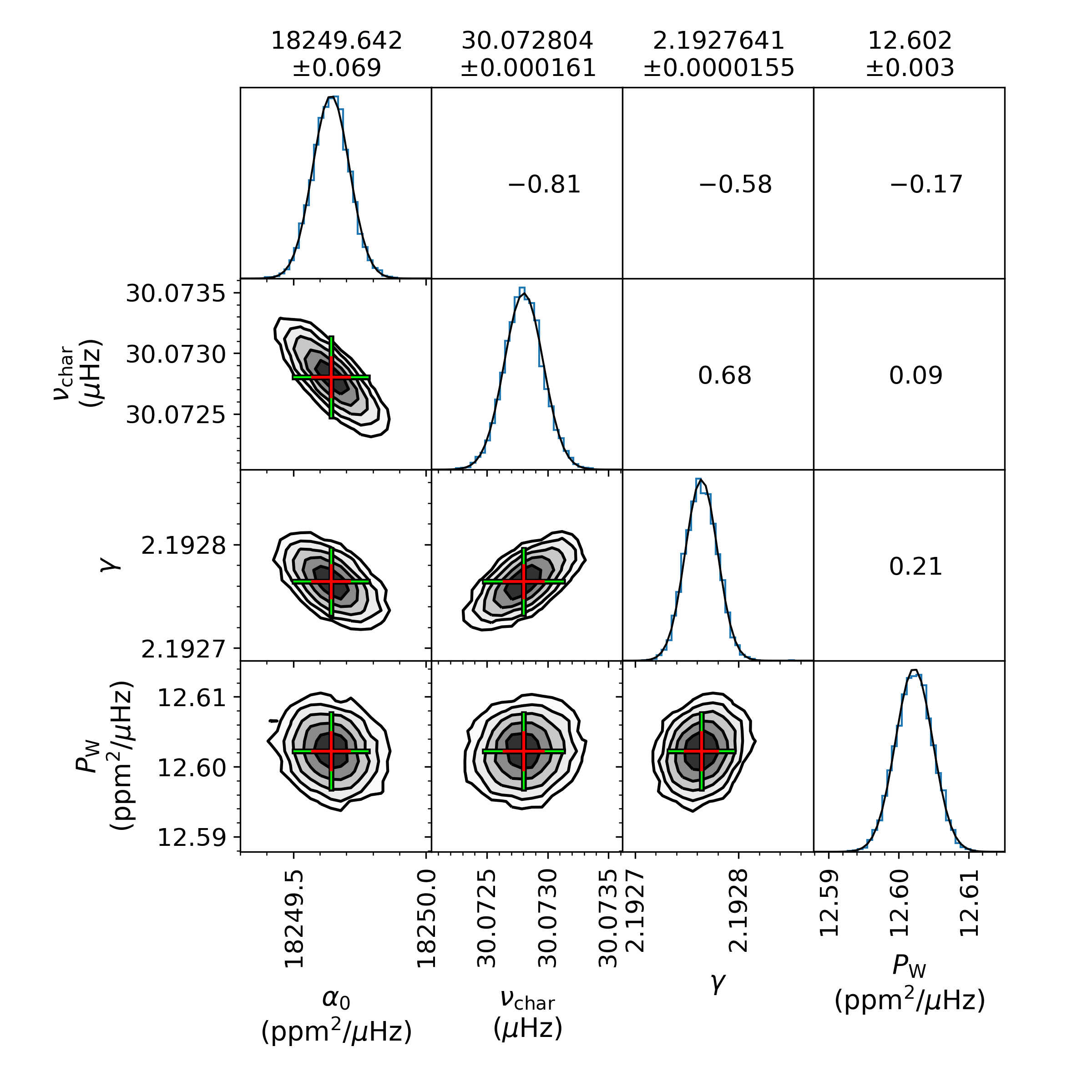}
	\includegraphics[width=0.49\textwidth]{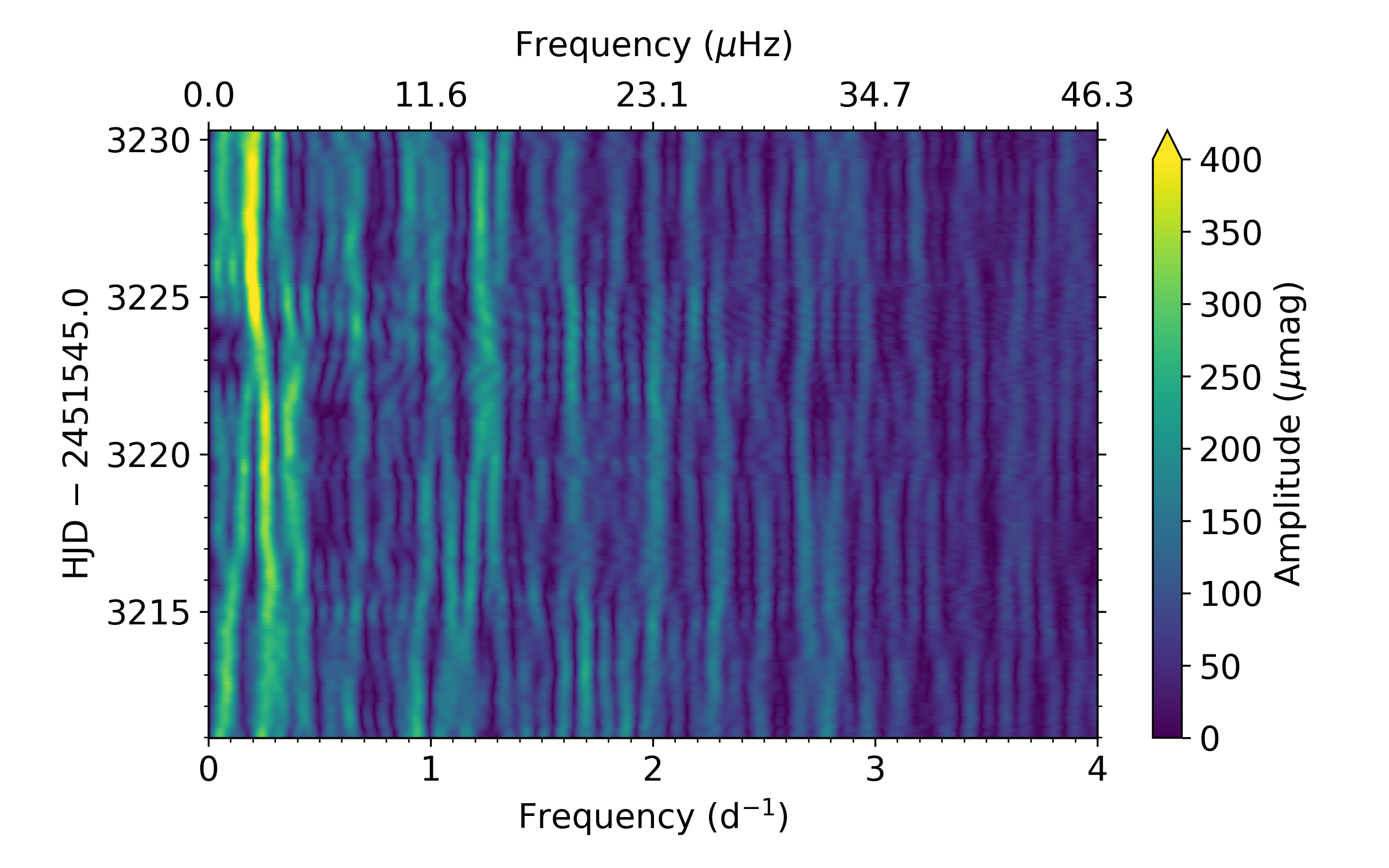}
	\caption{Summary figure for the O star HD~46966, which has a similar layout as shown in Fig.~\ref{figure: HD46150}, {except that the original and residual power density spectra are shown in orange and black, respectively, since HD~49666 underwent pre-whitening of peaks with amplitude $S/N \geq 4$.}}
	\label{figure: HD46966}
	\end{figure}
	
	Similar summary figures for the O stars HD~46223 and HD~46966 are shown in Figs~\ref{figure: HD46223} and \ref{figure: HD46966}, respectively, and all other summary figures are given in Appendix~\ref{section: appendix: MCMC results}. Most of our stars have undergone pre-whitening such that the original and residual spectra are shown as orange and black lines, respectively, expect for HD~46150 and HD~46233 in which the original spectra (shown in black) were used as the data as no significant ($S/N \geq 4$) peaks were present in their spectra.


	\subsection{Sliding Fourier Transforms}
	\label{subsection: sliding FTs}
	
	Another commonly used method to demonstrate stochastic signal is the use of sliding Fourier transforms (SFTs), in which frequency spectra for moving subsets of a time series are calculated and stacked into a 2D image. The length of CoRoT data for some of the stars in our sample is {$\sim$25}~d, so the choice of the bin length and step size in an SFT is a compromise between temporal and frequency resolution. If the bin length is too short then a poor frequency resolution is obtained, but if it is too long, then the number of steps in an SFT will be too few to investigate any time-dependent behaviour. The SFTs for the O-stars HD~46150, HD~46223 and HD~46966 are shown in the bottom panels of Figs~\ref{figure: HD46150}, \ref{figure: HD46223} and \ref{figure: HD46966}, respectively, in which 15-d bins and a step size of 0.1~d are used for the frequency range between $0 \leq \nu \leq 4$~d$^{-1}$ ($0 \leq \nu \leq 46.3$~$\mu$Hz). Clearly, as previously noted by \citet{Blomme2011b}, the dominant low-frequency peaks in these three O stars are not coherent and have short lifetimes of order hours and days. Similar SFTs for all other stars are given in Appendix~\ref{section: appendix: MCMC results}.


\section{Results: application to CoRoT stars}
\label{section: CoRoT results}

For each star given in Table~\ref{table: stars}, we applied our Bayesian MCMC methodology to determine the optimum values of $\alpha_0$, $\nu_{\rm char}$, $\gamma$ and $P_{\rm W}$ given in Eqn.~(\ref{equation: Blomme}) and their respective uncertainties using the available CoRoT photometry. However, some of the stars in our sample are (known) pulsators with multiple high-$S/N$ peaks in their frequency spectra representing coherent pulsation mode frequencies. Notable examples include the aforementioned \dsct stars HD~50844 and HD~174936, with each of these stars claimed to have hundreds of independent coherent pulsation modes \citep{Poretti2009, GH2009}. In this paper, we characterise the morphology of the background low-frequency power excess, so as previously discussed, coherent pulsation modes with $S/N \geq 4$ were removed by iterative pre-whitening prior to employing the Bayesian MCMC methodology, as they affect the parameter estimation of the best-fitting red noise model of the residual power density spectra. 

Our results are summarised in Table~\ref{table: results}, in which we also indicate the causes of the variability observed for each star, with specific details given in the text. However, for some of the stars in our sample, the observational time base and resultant frequency resolution was insufficient to resolve and reliably extract all the pulsation modes via iterative pre-whitening. This resulted in some stars having localised "bumps" of unresolved pulsation modes or a flat noise plateau in their residual power density spectra, which impact the reliability of the background fit. We provide the MCMC fits for all stars for completeness, but have separated those stars significantly affected by the limitations of CoRoT data in Table~\ref{table: results}.


	\begin{table*}
	\caption{The optimised parameters for the extracted red noise (c.f. Eqn.~\ref{equation: Blomme}) for the stars studied in this paper using our Bayesian MCMC method, and the dominant cause(s) that are likely contributing to the observed low-frequency power excess, which are not only sourced from this work but also from the literature. Note that pulsation can refer to coherent modes and/or damped modes (i.e. IGWs).}
	\centering
	\begin{tabular}{l r r r r r}
	\hline \hline
	\multicolumn{1}{c}{Name} & \multicolumn{1}{c}{$\alpha_0$} & \multicolumn{1}{c}{$\nu_{\rm char}$} & \multicolumn{1}{c}{$\gamma$} & \multicolumn{1}{c}{$P_{\rm W}$} & \multicolumn{1}{c}{Var. Type} \\
	\multicolumn{1}{c}{} & \multicolumn{1}{c}{(ppm$^2$/$\mu$Hz)} & \multicolumn{1}{c}{($\mu$Hz)} & \multicolumn{1}{c}{} & \multicolumn{1}{c}{(ppm$^2$/$\mu$Hz)} & \multicolumn{1}{c}{} \\
	\hline
	\multicolumn{6}{l}{O stars from \citet{Blomme2011b} inferred to have IGWs by \citet{Aerts2015c}:} \vspace{0.1cm} \\
	HD~46150	&	$49205.559 \pm 0.087$	&	$24.120096 \pm 0.000072$	&	$1.8777210 \pm 0.0000052$	&	$7.073 \pm 0.002$	&	puls + wind	\\
	HD~46223	&	$160242.358 \pm 0.088$	&	$17.742474 \pm 0.000015$	&	$1.8082826 \pm 0.0000014$	&	$22.921 \pm 0.002$	&	puls + wind	\\
	HD~46966	&	$18249.642 \pm 0.069$	&	$30.072804 \pm 0.000161$	&	$2.1927641 \pm 0.0000155$	&	$12.602 \pm 0.003$	&	puls + wind	\\
	\hline
	\multicolumn{6}{l}{CoRoT OBAF stars with detected stochastic low-frequency variability:} \vspace{0.1cm} \\
	HD~45418	&	$3203.510 \pm 0.034$	&	$30.607811 \pm 0.000361$	&	$4.8896087 \pm 0.0001728$	&	$5.590 \pm 0.001$	&	puls	\\
	HD~45517	&	$327.068 \pm 0.067$	&	$10.209637 \pm 0.002210$	&	$4.0164748 \pm 0.0025992$	&	$8.393 \pm 0.001$	&	puls	\\	
	HD~45546	&	$34626.797 \pm 0.054$	&	$24.157404 \pm 0.000046$	&	$3.2662493 \pm 0.0000125$	&	$9.865 \pm 0.001$	&	puls	\\
	HD~46149 	&	$4994.164 \pm 0.065$	&	$13.545526 \pm 0.000147$	&	$6.5129408 \pm 0.0004626$	&	$8.270 \pm 0.002$	&	puls	+ wind	\\
	HD~46179	&	$21.717 \pm 0.089$		&	$15.161279 \pm 0.084005$	&	$2.2933918 \pm 0.0167594$	&	$1.987 \pm 0.001$	&	puls	\\
	HD~46202 	&	$537.336 \pm 0.059$	&	$31.854515 \pm 0.004408$	&	$3.0236231 \pm 0.0008187$	&	$7.858 \pm 0.002$	&	puls	+ wind	\\
	HD~46769	&	$35.278 \pm 0.074$		&	$23.411051 \pm 0.068442$	&	$2.1318523 \pm 0.0090258$	&	$0.908 \pm 0.002$	&	gran	\\
	HD~48784	&	$2299.504 \pm 0.056$	&	$30.020785 \pm 0.000762$	&	$4.0222079 \pm 0.0003455$	&	$3.192 \pm 0.002$	&	gran + puls	\\
	HD~49677	&	$246.504 \pm 0.117$	&	$11.178924 \pm 0.006973$	&	$2.2987365 \pm 0.0016031$	&	$6.888 \pm 0.002$	&	puls	\\
	HD~50747	&	$2840.821 \pm 0.052$	&	$44.340794 \pm 0.001147$	&	$3.3346001 \pm 0.0002550$	&	$1.077 \pm 0.001$	&	puls	\\
	HD~51332 	&	$757.890 \pm 0.087$	&	$18.108611 \pm 0.003225$	&	$1.9357792 \pm 0.0003633$	&	$3.745 \pm 0.001$	&	puls + gran	\\
	HD~51756 	&	$6458.932 \pm 0.060$	&	$24.569608 \pm 0.000341$	&	$2.2164263 \pm 0.0000373$	&	$4.888 \pm 0.001$	&	puls + wind	\\
	HD~52130 	&	$2320.871 \pm 0.029$	&	$43.503073 \pm 0.000515$	&	$5.3152045 \pm 0.0002527$	&	$21.266 \pm 0.001$	&	puls + gran	\\
	HD~174589	&	$1856.345 \pm 0.179$	&	$8.984878 \pm 0.001486$		&	$1.4239502 \pm 0.0001289$	&	$2.688 \pm 0.002$	&	puls + gran	\\
	HD~174966	&	$401.056 \pm 0.101$	&	$33.149762 \pm 0.016483$	&	$1.2198259 \pm 0.0002660$	&	$4.006 \pm 0.003$	&	puls + gran	\\
	HD~174967	&	$404.584 \pm 0.074$	&	$15.102132 \pm 0.002910$	&	$4.2063454 \pm 0.0027710$	&	$32.480 \pm 0.002$	&	puls	\\
	HD~174990	&	$341.897 \pm 0.059$	&	$24.196653 \pm 0.004067$	&	$5.9565225 \pm 0.0042772$	&	$17.059 \pm 0.002$	&	puls	\\
	HD~175272	&	$884.411 \pm 0.135$	&	$10.357587 \pm 0.002366$	&	$2.0133028 \pm 0.0004521$	&	$3.792 \pm 0.002$	&	puls	\\
	HD~175542	&	$147.397 \pm 0.073$	&	$31.671226 \pm 0.019787$	&	$2.6875427 \pm 0.0030297$	&	$17.831 \pm 0.006$	&	puls	\\
	HD~175640	&	$23.213 \pm 0.071$		&	$35.170619 \pm 0.134604$	&	$2.8834658 \pm 0.0219825$	&	$1.496 \pm 0.006$	&	puls	\\
	HD~263425	&	$3155.639 \pm 0.535$	&	$3.267273 \pm 0.000919$		&	$1.2824296 \pm 0.0001254$	&	$25.100 \pm 0.002$	&	puls	\\
	
	\hline
	\multicolumn{6}{l}{Additional OBAF stars with insufficient CoRoT data for an accurate background fit:} \vspace{0.1cm} \\
	HD~47485	&	$31438.683 \pm 0.021$	&	$125.735343 \pm 0.000056$	&	$12.2180879 \pm 0.0000552$	&	$56.380 \pm 0.002$	&	gran	+ puls	\\
	HD~48977	&	$30314.154 \pm 0.064$	&	$34.301079 \pm 0.000087$	&	$3.2625348 \pm 0.0000179$	&	$7.380 \pm 0.002$	&	puls	+ wind	\\
	HD~50844	&	$33248.802 \pm 0.011$	&	$257.800434 \pm 0.000072$	&	$6.6387810 \pm 0.0000101$	&	$57.262 \pm 0.001$	&	gran + puls	\\
	HD~50870 	&	$4513.878 \pm 0.009$	&	$354.121974 \pm 0.000514$	&	$10.1627352 \pm 0.0001001$	&	$28.953 \pm 0.001$	&	gran + puls	\\
	HD~51193	&	$554739.578 \pm 0.053$	&	$25.007251 \pm 0.000007$	&	$3.0604732 \pm 0.0000007$	&	$88.077 \pm 0.001$	&	puls	\\
	HD~51452 	&	$880984.219 \pm 0.054$	&	$23.579036 \pm 0.000002$	&	$3.1299536 \pm 0.0000005$	&	$144.210 \pm 0.001$&	puls + wind	\\
	HD~51359	&	$2200.012 \pm 0.009$	&	$303.217065 \pm 0.000859$	&	$14.3801743 \pm 0.0003426$	&	$10.686 \pm 0.001$	&	gran	+ puls	\\
	HD~51722 	&	$8629.083 \pm 0.013$	&	$225.375146 \pm 0.000292$	&	$6.2225653 \pm 0.0000433$	&	$11.397 \pm 0.001$	&	gran	+ puls	\\
	HD~174532	&	$1941.594 \pm 0.012$	&	$402.953130 \pm 0.001633$	&	$13.9727879 \pm 0.0005824$	&	$9.410 \pm 0.002$	&	gran	+ puls	\\
	HD~174936	&	$625.629 \pm 0.011$	&	$628.365922 \pm 0.005419$	&	$17.0182510 \pm 0.0018970$	&	$25.638 \pm 0.006$	&	gran	+ puls 	\\
	HD~175445	&	$280.216 \pm 0.026$	&	$279.849349 \pm 0.025642$	&	$3.6303503 \pm 0.0010499$	&	$4.030 \pm 0.008$	&	puls + gran	\\	
	\hline \hline
	\end{tabular}
	\label{table: results}
	\end{table*}


	\subsection{HD~46150}
	\label{subsection: HD46150}

	HD~46150 is a young main-sequence O dwarf, which was first observed by \citet{Plaskett_J_1931}, and is the second hottest star in the cluster NGC~2244. It has an age of a few Myr \citep{Bonatto2009b, Martins2012b} and a spectral type of O5\,V((f))z \citep{Sota2011a, Sota2011b}. HD~46150 is suspected of being a long-period ($P_{\rm orb} \sim 100\,000$~yr) binary system, because multiple nearby faint ($\Delta\,V \geq 5$~mag) stars having been detected in interferometry \citep{Mason1998a, Mason2001c, Maiz2010b, Sana2014, Aldoretta2015}, and large radial velocity shifts in spectroscopy (e.g. \citealt{Abt1972a, Liu_T_1989b, Underhill1990d, Fullerton_PhD, Fullerton2006a, Mahy2009, Martins2012b, Chini2012}).	
	
	The known spectroscopic parameters of HD~46150 include an effective temperature of $T_{\rm eff} = 42\,000$~K and a surface gravity of $\log\,g = 4.0$ \citep{Martins2012b, Martins2015d}, but its projected surface rotational velocity varies in the literature between 66 and 100~km\,s$^{-1}$ \citep{Mahy2009, Martins2012b, Simon-Diaz2014a, Martins2015d, Grunhut2017}, with an average value of $v\,\sin\,i \simeq 80$~km\,s$^{-1}$. Similarly, the macroturbulent velocities of HD~46150 also range between 37 and 176~km\,s$^{-1}$ \citep{Martins2012b, Martins2015d, Simon-Diaz2014a, Grunhut2017}. HD~46150 was included as a target in the BOB campaign \citep{Morel2014c, Morel2015b} and MiMeS survey \citep{Wade2016a}, but null detections of a large-scale magnetic field were obtained by \citet{Fossati2015b} and \citet{Grunhut2017}.
		
	As previously discussed by \citet{Blomme2011b} and demonstrated in Fig.~\ref{figure: HD46150}, HD~46150 clearly has a broad low-frequency power excess in its power density spectrum. Theoretical models were unable to explain the astrophysical red noise as coherent pulsation modes because of the inherent stochastic variability and \citet{Blomme2011b} concluded that sub-surface convection, granulation, or inhomogeneities in the stellar wind were plausible explanations. Later, \citet{Aerts2015c} interpreted the stochastic variability and red noise as IGWs. The summary figure showing the morphology of the red noise in the CoRoT photometry and the SFT for HD~46150 is given as Fig.~\ref{figure: HD46150}.


	\subsection{HD~46223}
	\label{subsection: HD46223}
	
	HD~46223 is the hottest star in the cluster NGC~2244 with an age of a few Myr \citep{Hensberge2000b, Bonatto2009b, Martins2012b}, and a spectral type of O4\,V((f)) \citep{Underhill1990d, Massey1995c, Sota2011a, Sota2011b} or O4\,V((f+)) \citep{Maiz2004a, Martins2005b, Mokiem2007b, Mahy2009}. Although a faint companion star has been detected for HD~46223 \citep{Mason2001c, Turner2008}, long-term spectroscopic campaigns have not revealed any significant radial velocity variations (e.g. \citealt{Maiz2004a, Fullerton2006a, Mahy2009, Maiz2010b, Chini2012, Scholler2017}).
	
	The known spectroscopic parameters of HD~46223 include an effective temperature of $T_{\rm eff} = 43\,000$~K and a surface gravity of $\log\,g = 4.0$ \citep{Martins2012b, Martins2015d}. The projected surface rotational velocity of HD~46223 has a large variance in the literature with values that range from 58 to 100~km\,s$^{-1}$ \citep{Mahy2009, Martins2012b, Simon-Diaz2014a, Martins2015d, Grunhut2017}, with an average value of approximately $v\,\sin\,i \simeq 80$~km\,s$^{-1}$. Similarly, the macroturbulent velocities also range from 32 to 156~km\,s$^{-1}$ \citep{Martins2012b, Martins2015d, Simon-Diaz2014a, Grunhut2017}. Estimates of the mass-loss rate for HD~46223 using H$\alpha$ include $\log\,\dot{M} \simeq -5.8$ \citep{Puls1996, Fullerton2006a} and $\log\,\dot{M} \simeq -6.2$ \citep{Martins2012b}. HD~46223 was included as a target in the BOB campaign \citep{Morel2014c, Morel2015b} and MiMeS survey \citep{Wade2016a}, but null detections of a large-scale magnetic field were obtained by \citet{Fossati2015b} and \citet{Grunhut2017}.
	
	Similarly to HD~46150, HD~46223 also has a significant low-frequency power excess in its power density spectrum and stochastic variability in its SFT, which was interpreted as IGWs by \citet{Aerts2015c}. The summary figure showing the morphology of the red noise in the CoRoT photometry and the SFT for HD~46223 is given as Fig.~\ref{figure: HD46223}.


	\subsection{HD~46966}
	\label{subsection: HD46966}
	
	HD~46966 is a slightly evolved O star in the Mon OB2 association \citep{Mahy2009}, and has a spectral type of O8.5\,IV \citep{Mahy2009, Sota2011a, Sota2011b}. Multiple faint objects ($\Delta\,V \geq 5$~mag) have been detected near to HD~46966 \citep{Mason2001c, Turner2008, Sana2014}, although multi-epoch spectroscopy has yet to reveal strong evidence of binarity (see, e.g. \citealt{Munari1999e, Maiz2004a, Mahy2009, DeBruijne2012b, Chini2012}).
	
	The known spectroscopic parameters of HD~46966 include an effective temperature of $T_{\rm eff} = 35\,000$~K and a surface gravity of $\log\,g = 3.9$ \citep{Martins2012b, Martins2015d}. The projected surface rotational velocity of HD~46966 ranges from 33 to 50~km\,s$^{-1}$ in the literature \citep{Mahy2009, Martins2012b, Simon-Diaz2014a, Martins2015d, Grunhut2017}, with an average value of $v\,\sin\,i \simeq 42$~km\,s$^{-1}$. Similarly, the macroturbulent velocities of HD~46966 range from 27 to 90 km\,s$^{-1}$ \citep{Martins2012b, Martins2015d, Simon-Diaz2014a, Grunhut2017}. HD~46966 was included as a target in the BOB campaign \citep{Morel2014c, Morel2015b} and MiMeS survey \citep{Wade2016a}, but null detections of a large-scale magnetic field were obtained by \citet{Fossati2015b} and \citet{Grunhut2017}.
	
	Similarly to HD~46150 and HD~46223, HD~46966 has a significant low-frequency power excess interpreted as evidence for IGWs \citep{Aerts2015c}, but there is also a dominant low-frequency peak in its spectrum which could be caused by rotational modulation. The summary figure showing the morphology of the red noise in the residual power density spectrum and the SFT after extracting this dominant peak for HD~46966 is given as Fig.~\ref{figure: HD46966}.


	\subsection{HD~45418}
	\label{subsection: HD45418}
	
	HD~45418 (9~Mon) is a B5\,V star, which is a member of the NGC~2232 cluster \citep{Levato1974a, Baumgardt2000}, and has an effective temperature of $T_{\rm eff} = 16\,750 \pm 130$~K, a surface gravity of $\log\,g = 4.27 \pm 0.04$, and a projected surface rotational velocity of $v\,\sin\,i = 247 \pm 8$~km\,s$^{-1}$ \citep{Lefever2010, Tetzlaff2011, McDonald2012a, Braganca2012, Mugnes2015}. Previous investigations of the spectroscopic variability of HD~45418 have indicated that it may be a non-pulsating star \citep{Adelman2001d, Lefever2010}. 

Our study is the first to analyse the CoRoT light curve of HD~45418, with its power density spectrum and SFT shown in Fig.~\ref{figure: HD45418}. Clearly, HD~45418 has multiple high-$S/N$ low-frequency peaks in its power density spectrum, stochastic low-frequency variability, and a red noise background. Although it is not classified as a Be star, HD~45418 is a fast rotator and its pulsation modes between $0.3 \leq \nu \leq 3$~d$^{-1}$ ($3 \leq \nu \leq 30$~$\mu$Hz) may be stochastically-excited gravito-inertial modes, similar to those found in the Be star HD~51452 by \citet{Neiner2012d}. The summary figure showing the morphology of the red noise in the CoRoT photometry and the SFT for HD~45418 is given as Fig.~\ref{figure: HD45418}.


	\subsection{HD~45517}
	\label{subsection: HD45517}
	
	HD~45517 is a member of the cluster NGC~2232 \citep{Claria1972a}, and the only literature reference concerning its fundamental parameters is the determination is a spectral type as A0/2\,V by \citet{Houk1999}. From analysis of its CoRoT photometry, we determine that HD~45517 has a low-frequency power excess and stochastic variability of order a few tens of $\mu$mag between $0 \leq \nu \leq 20$~d$^{-1}$ ($0 \leq \nu \leq 231$~$\mu$Hz). The summary figure showing the morphology of the red noise in the CoRoT photometry and the SFT for HD~45517 is given as Fig.~\ref{figure: HD45517}.


	\subsection{HD~45546}
	\label{subsection: HD45546}
	
	HD~45546 (10~Mon) is a member of the cluster NGC~2232 \citep{Claria1972a}, and has a spectral type of B2\,V, an effective temperature of $T_{\rm eff} \simeq 19\,000$~K, a surface gravity of $\log\,g \simeq 4.0$, and a projected surface rotational velocity of $v\,\sin\,i = 61 \pm 3$~km\,s$^{-1}$ \citep{Lyubimkov2000, Lyubimkov2002, Lyubimkov2013a, Abt2002, Tetzlaff2011, Braganca2012, Silaj2014a}. HD~45546 has no strong evidence of being a binary or multiple system \citep{Abt1984b, Eggleton2008e}, but its spectroscopic variability revealed line profile variability that was indicative of high-$\ell$ pulsation modes \citep{Telting2006c}. As the first study to investigate the photometric variability of HD~45546 using CoRoT observations, we confirm the presence of significant low-frequencies in its power density spectrum, as shown in Fig.~\ref{figure: HD45546}. Furthermore, we find that the low-frequency power excess in HD~45546 is stochastic, which could be caused by either unresolved coherent pulsation modes or damped pulsation modes. The summary figure showing the morphology of the red noise in the CoRoT photometry and the SFT for HD~45546 is given as Fig.~\ref{figure: HD45546}.


	\subsection{HD~46149}
	\label{subsection: HD46149}
	
	HD~46149 is a member of the young cluster NGC~2244 and was first observed by \citet{Plaskett_J_1931}. Later it was suspected of being a binary system because of variance in its radial velocity \citep{Abt1972a, Liu_T_1989b, Maiz2004a, Turner2008}. More recent studies have confirmed HD~46149 as an O8\,V $+$ B0/1\,V long-period binary system with an orbital period of $P_{\rm orb} = 829 \pm 4$~d and an eccentricity of $e = 0.59 \pm 0.02$ \citep{Mahy2009, Degroote2010b, Sota2011a, Sota2011b, Chini2012, Martins2012b}.
		
	The known spectroscopic parameters of the O8\,V primary include an effective temperature of $T_{\rm eff} = 36\,000$~K and a surface gravity of $\log\,g = 3.7$, and an effective temperature of $T_{\rm eff} = 33\,000$~K and a surface gravity of $\log\,g = 4.0$ for the B0/1\,V secondary \citep{Degroote2010b}. The projected surface rotational velocity of the primary ranges between $\sim$0 and 78~km\,s$^{-1}$ in the literature \citep{Degroote2010b, Martins2012b, Martins2015d, Simon-Diaz2014a}, with an average value of $v\,\sin\,i \simeq 30$~km\,s$^{-1}$, and approximate value of $v\,\sin\,i \simeq 100$~km\,s$^{-1}$ for the secondary \citep{Martins2012b}. Similarly, macroturbulent velocities range from 24 to 61~km\,s$^{-1}$ for the primary \citep{Martins2012b, Martins2015d, Simon-Diaz2014a} and approximately 27~km\,s$^{-1}$ for the secondary \citep{Martins2012b}. HD~46149 was included as a target in the BOB campaign \citep{Morel2014c, Morel2015b} and MiMeS survey \citep{Wade2016a}, but null detections of a large-scale magnetic field were obtained by \citet{Fossati2015b} and \citet{Grunhut2017}.
	
	Previously, the 34-d CoRoT light curve of HD~46149 was analysed by \citet{Degroote2010b}, who found evidence for pulsation modes between $3.0 < \nu < 7.5$~d$^{-1}$ ($35 < \nu < 87$~$\mu$Hz) that have have variable amplitudes, short lifetimes, and a characteristic frequency spacing of $\Delta\,\nu = 0.48 \pm 0.02$~d$^{-1}$ ($5.6 \pm 0.2$~$\mu$Hz). These pulsation modes were postulated to be p~modes excited by turbulent pressure in a sub-surface convection zone since coherent p~modes excited by the opacity mechanism are not expected for such a star \citep{Degroote2010b, Belkacem2010a}. However, as demonstrated in Fig.~\ref{figure: HD46149}, HD~46149 has an underlying stochastic low-frequency background in its residual power density spectrum, which is indicative of IGWs.


	\subsection{HD~46179}
	\label{subsection: HD46179}
		
	First classified by \citet{Merrill1925}, HD~46179 is a known binary system with a B9\,V primary \citep{Dommanget2000b, Mason2001c, Fabricius2002a, Gontcharov2006b}, an effective temperature of $T_{\rm eff} = 10500 \pm 500$~K, a surface gravity of $\log\,g = 4.0 \pm 0.2$, and a projected surface rotational velocity of $v\,\sin\,i = 152 \pm 13$~km\,s$^{-1}$ \citep{Niemczura2009b, Degroote2011}. Using $\sim$30~d of CoRoT photometry, \citet{Degroote2010c, Degroote2011} found no significant periodicity indicative of coherent pulsation modes in HD~46179, which is not surprising since HD~46179 is located between the SPB and \dsct instability regions and is not predicted to pulsate. In this work, we confirm the lack of coherent pulsation modes in HD~46179, but demonstrate that it has stochastic low-frequency variability between $0 \leq \nu \leq 2$~d$^{-1}$ ($0 \leq \nu \leq 23$~$\mu$Hz) in its power density spectrum, as shown in Fig.~\ref{figure: HD46179}. We also note, like \citet{Degroote2010c, Degroote2011}, that there is a dominant peak at an approximate frequency of $\nu \simeq 0.07$~d$^{-1}$ (0.8~$\mu$Hz) which could be caused by rotational modulation. HD~46179 is an interesting case study of an almost constant intermediate-mass star in our sample, thus its low-frequency power excess provides a useful upper constraint of the photometric amplitudes of order a few $\mu$mag for IGWs in a main sequence {$\sim$3}-M$_{\rm \odot}$ late-B star.


	\subsection{HD~46202}
	\label{subsection: HD46202}

	HD~46202 is a young late-O dwarf and was first observed by \citet{Plaskett_J_1931}. It has a spectral type of O9.5\,V \citep{Walborn1990b, Martins2005b, Walborn2011b, Sota2011a, Sota2011b} or O9.2\,V \citep{Sota2014}. Some studies inferred HD~46202 to have variable radial velocities (e.g. \citealt{Abt1972a}), yet more recent dedicated spectroscopic campaigns show that HD~46202 is likely a single star \citep{Underhill1990d, Mahy2009, Chini2012, Scholler2017}. The spectroscopic parameters of HD~46202 include an effective temperature of $T_{\rm eff} = 33\,000$~K and a surface gravity of $\log\,g = 4.0$ \citep{Martins2005b, Martins2012b, Martins2015d}. The projected surface rotational velocity of HD~46202 ranges between 15 and 21~km\,s$^{-1}$ \citep{Simon-Diaz2014a, Martins2012b, Martins2015d} with an average value of $v\,\sin\,i \simeq 19$~km\,s$^{-1}$. Similarly, the macroturbulent velocities of HD~46202 range between 13 and 58~km\,s$^{-1}$ \citep{Simon-Diaz2014a, Martins2012b, Martins2015d}. HD~46202 was included as a target in the BOB campaign \citep{Morel2014c, Morel2015b} and MiMeS survey \citep{Wade2016a}, but null detections of a large-scale magnetic field were obtained by \citet{Fossati2015b} and \citet{Grunhut2017}.
		
	Previously, \citet{Briquet2011} analysed the CoRoT photometry of HD~46202 and performed an asteroseismic analysis to identify p~modes and perform forward seismic modelling. The authors determined a step convective core overshooting parameter of $\alpha_{\rm ov} = 0.10 \pm 0.05$, but emphasise that none of the observed pulsation modes were predicted to be excited by theoretical models. More recently, \citet{Moravveji2016a} investigated the changes in the location of \bcep instability region after artificially enhancing the iron opacity by $\sim$75~per~cent. This alternative instability region includes HD~46202 and may provide an explanation for the observed pulsation mode frequencies.
	
	HD~46202 is one of the more interesting stars in our study, with its summary figure shown in Fig.~\ref{figure: HD46202}. Its power density spectrum contains a few significant peaks, including $\nu \simeq 2.29$, 2.64, and 3.00~d$^{-1}$ (26.5, 30.6 and 34.7~$\mu$Hz, respectively), which \citet{Briquet2011} interpret as \bcep pulsations. These frequencies are similar to the dominant IGW frequencies shown in Fig.~\ref{figure: IGW simulations}. After pre-whitening these significant peaks, HD~46202 also has an underlying background red noise profile and low-frequency stochastic variability in its residual power density spectrum, as shown in Fig.~\ref{figure: HD46202}, which is indicative of IGWs.


	\subsection{HD~46769}
	\label{subsection: HD46769}
	
	HD~46769 is a blue supergiant and a confirmed non-magnetic single star with a spectral type of B5\,II, an effective temperature of $T_{\rm eff} = 13\,000 \pm 1000$~K, a surface gravity of $\log\,g = 2.7 \pm 0.1$, and a projected surface rotational velocity of $v\,\sin\,i = 72 \pm 2$~km\,s$^{-1}$ \citep{Houk1999, Abt2002, Lefever2007a, Zorec2009, Aerts2013, Wade2016a}. The CoRoT light curve and high-$S/N$, high-resolution spectroscopy of HD~46769 were previously analysed by \citet{Aerts2013}, which revealed rotational modulation with multiple harmonics and a rotation period of 9.69~d. Their analysis also revealed a lack of coherent variability. The macroturbulence of HD~46769 was determined to be less than 10~km\,s$^{-1}$, which is much smaller than the measured rotational broadening and is atypically small for a blue supergiant \citep{Aerts2013}.

	Similarly to other massive stars in our study, the lack of coherent periodicity in HD~46769 provides the opportunity to place constraints on the photometric low-frequency power excess in this blue supergiant (e.g. \citealt{Aerts2018a}). The morphology of the red noise in the CoRoT photometry after extracting the rotational modulation and the SFT for HD~46769 are shown in Fig.~\ref{figure: HD46769}. Our analysis of the blue supergiant HD~46769 revealed similar stochastic variability between $0 \leq \nu \leq 3$~d$^{-1}$ ($0 \leq \nu \leq 34.7$~$\mu$Hz) to that observed in the other blue supergiants HD~188209 \citep{Aerts2017a} and HD~91316 \citep{Aerts2018a}, which suggests granulation from a surface convection zone or IGWs as plausible explanations.


	\subsection{HD~47485}
	\label{subsection: HD47485}
	
	HD~47485 has a spectral type of A5\,IV/V \citep{Houk1999}, an effective temperature of $T_{\rm eff} \simeq 7300$~K, a surface gravity of $\log\,g \simeq 2.9$, and a projected surface rotational velocity of $v\,\sin\,i \simeq 40$~km\,s$^{-1}$ \citep{Gebran2016}. Our analysis of the CoRoT photometry of HD~47485 is the first analysis in the literature, and revealed intrinsic variability indicative of \dsct pulsation modes between $6 \leq \nu \leq 10$~d$^{-1}$ ($69 \leq \nu \leq 116$~$\mu$Hz) as shown in its power density spectrum in Fig.~\ref{figure: HD47485}. However, the density of the peaks in this frequency range is extremely high and only {$\sim$26}~d of CoRoT photometry is not sufficient to resolve the beating and identify close-frequency pulsation modes \citep{Bowman2016a, Bowman_BOOK, Buysschaert2018c}. The residual pulsation signal in HD~47485 significantly affects the accuracy of our background fit. Therefore, we have labelled HD~47485 as having insufficient CoRoT data for an accurate background fit in Table~\ref{table: results}. Nonetheless, we note the stochastic variability between $0 \leq \nu \leq 3$~d$^{-1}$ ($0 \leq \nu \leq 34.7$~$\mu$Hz) in HD~47485 which may be caused by granulation or IGWs.

	
	\subsection{HD~48784}
	\label{subsection: HD48784}
	
	HD~48784 has a literature spectral type of F0~V \citep{Iijima1978} and an effective temperature of approximately $T_{\rm eff} \simeq 6900$~K \citep{McDonald2012a}. The CoRoT light curve of HD~48784 was previously studied by \citet{Barcelo2017a}, who extracted 163 significant frequencies and detected a regular frequency spacing of $\Delta\,\nu = 40.3 \pm 0.6~\mu$Hz ($\simeq 3.5$~d$^{-1}$). This was interpreted as the large frequency separation between consecutive radial order p modes, and from this an average density of $\bar{\rho} = 0.12$~g\,cm$^{-3}$ and a mass of $M = 2.0 \pm 0.1$~M$_{\rm \odot}$ were derived for HD~48784 using scaling relations \citep{Barcelo2017a}. The low-frequency power excess in HD~48784 was interpreted to be related to the oblateness and rotation of the star, or the efficiency of the convective envelope \citep{Barcelo2017a}. 
	
	From the approximate mass of $2$~M$_{\rm \odot}$ and radius $R = 2.8$~R$_{\rm \odot}$ from \citet{Barcelo2017a}, one can easily calculate an approximate surface gravity of $\log\,g = 3.5$, which we have included in Table~\ref{table: stars}. In our analysis of the CoRoT photometry of HD~48784, we extract 11 peaks that have $S/N \geq 4$ between $0.5 \leq \nu \leq 16.5$~d$^{-1}$ ($5.8 \leq \nu \leq 191.0$~$\mu$Hz), which is far fewer than previous studies because our approach to iterative pre-whitening assumes non-white noise being present in the spectrum. The summary figure showing the morphology of the red noise in the CoRoT photometry and the SFT for HD~48784 is given as Fig.~\ref{figure: HD48784}.

	
	\subsection{HD~48977}
	\label{subsection: HD48977}
	
	HD~48977 (16~Mon) has a spectral type of B2.5\,V, an effective temperature of $T_{\rm eff} \simeq 20\,000$~K, a surface gravity of $\log\,g \simeq 4.2$, and a projected surface rotational velocity of $v\,\sin\,i \simeq 25$~km\,s$^{-1}$ \citep{Abt2002, Lyubimkov2005a, Telting2006c, Zorec2009, Hohle2010, Lefever2010, Thoul2013, Ahmed2017}. In their previous analysis of the CoRoT photometry, \citet{Thoul2013} detect rotational modulation and infer a rotation frequency of $\nu = 0.6372$~d$^{-1}$ (7.375~$\mu$Hz), but also detect multiple g-mode pulsation frequencies. However, the {$\sim$30}-d length of the available CoRoT photometry limits the capability of identifying pulsation modes and subsequent modelling because of the poor frequency resolution, and since independent pulsation mode frequencies can be as close as 0.001~d$^{-1}$; 0.011~$\mu$Hz \citep{Bowman2016a, Bowman_BOOK, Buysschaert2018c}. 
	
	In our analysis of the CoRoT photometry of HD~48977, we extract 12 peaks that have $S/N \geq 4$ between $0.2 \leq \nu \leq 1.3$~d$^{-1}$ ($2.3 \leq \nu \leq 15.0$~$\mu$Hz), which are indicative of (unresolved) g-mode pulsation frequencies given that HD~48977 has a spectral type of B2.5\,V. The low-frequency power excess in both the original and residual power density spectra of HD~48977 is clear, as demonstrated in the summary figure for HD~48977 in Fig.~\ref{figure: HD48977}, which supports the interpretation of damped pulsation modes in HD~48977 similar to those discussed by \citet{Aerts2017a, Aerts2018a}. However, we note that the complex CoRoT window pattern in HD~48977 results in a background fit which overestimates the white noise contribution. Thus we label HD~48977 as having insufficient data for an accurate background fit in Table~\ref{table: results}, yet it is clear that this star still has a clear low-frequency power excess as shown in Fig.~\ref{figure: HD48977}.

	
	\subsection{HD~49677}
	\label{subsection: HD49677}
	
	HD~49677 is a slowly-rotating B9\,V star, with an effective temperature of $T_{\rm eff} = 9200 \pm 250$~K and a surface gravity of $\log\,g = 4.0 \pm 0.2$, which were estimated using its spectral energy distribution (SED) by \citet{Degroote2011}. If one assumes that the peak at $\nu \simeq 0.1$~d$^{-1}$ (1.2~$\mu$Hz) is the rotation frequency of HD~49677, a surface rotational velocity of order 10~km\,s$^{-1}$ can be derived, which is in agreement with the determination of HD~49677 as a slow rotator by \citet{Degroote2011}. HD~49677 was also determined to be a candidate pulsator since it has frequencies between $0 \leq \nu \leq 2$~d$^{-1}$ ($0 \leq \nu \leq 23.1$~$\mu$Hz) in its spectrum \citep{Degroote2011}. However, HD~49677 lies outside of the \dsct and SPB instability regions and is not expected to pulsate with coherent g modes. With no further analysis of HD~49677 having been performed since the study by \citet{Degroote2011}, HD~49677 is another good example of stochastic low-frequency variability in a late-B star that cannot be explained as coherent pulsation modes, but can be explained by IGWs. The summary figure showing the morphology of the red noise in the CoRoT photometry and the SFT for HD~49677 is given as Fig.~\ref{figure: HD49677}.


	\subsection{HD~50747}
	\label{subsection: HD50747}
	
	HD~50747 has a spectral type of A4\,IV, an effective temperature of $T_{\rm eff} \simeq 7500$~K, a surface gravity of $\log\,g \simeq 3.5$ and a projected surface rotational velocity of $v\,\sin\,i \simeq 70$~km\,s$^{-1}$ \citep{Royer2007a, Zorec2012, McDonald2012a}. However, \citet{Renson2009} list a spectral type in the range of A4--A9 and flag HD~50747 as having possible chemical peculiarities. \citet{Dolez2009} previously analysed the CoRoT photometry of HD~50747 and found pulsation mode frequencies that they interpreted to be g~modes. This star is also known to be a triple system from analysis of high-resolution spectroscopy as its spectra contain two narrow-line component stars and an additional broad-line component star \citep{Eggleton2008e, Dolez2009}. Thus, \citet{Dolez2009} concluded that HD~50747 is a triple system containing a \gdor star because of the spectroscopic parameters and observed range of pulsation mode frequencies. 
	
	From our analysis of the CoRoT light curve of HD~50747, we confirm the presence of significant low-frequency peaks in its power density spectrum, many of which are variable in amplitude and frequency as demonstrated in the SFT in the bottom panel of Fig.~\ref{figure: HD50747}. If the low-frequency power excess observed in HD~50747 is caused by coherent pulsation modes, then they are clearly not resolved with only {$\sim60$~d} of CoRoT observations, which would explain their apparent stochastic nature. On the other hand, IGWs can also explain the low-frequency power excess in HD~50747, although we cannot rule out granulation for HD~50747 since it is cool and evolved enough to have a surface convection zone. The summary figure showing the morphology of the red noise in the CoRoT photometry and the SFT for HD~50747 is given as Fig.~\ref{figure: HD50747}.


	\subsection{HD~50844}
	\label{subsection: HD50844}
	
	HD~50844 has a spectral type of A2~II \citep{Houk1999} and is one of the well-known \dsct stars previously studied using CoRoT photometry \citep{Poretti2005a, Poretti2009}. From high-resolution spectroscopy, \citet{Poretti2009} derive an effective temperature of $T_{\rm eff} = 7400 \pm 200$~K, a surface gravity of $\log\,g = 3.6 \pm 0.2$ and a projected surface rotational velocity of $v\,\sin\,i = 58 \pm 2$~km\,s$^{-1}$. The frequency spectrum of HD~50844 is claimed to be dense enough to contain thousands of peaks, some of which were interpreted as prograde high-$\ell$ p~modes from spectroscopic line profile observations \citep{Poretti2009}. Later the extremely high number of peaks extracted by \citet{Poretti2009} were interpreted by \citet{Kallinger2010c} as part of a red noise background caused by granulation. 
	
	If the mode density in an frequency spectrum is sufficiently high such that peaks are not well-resolved from each other and/or the intrinsic pulsation modes exhibit amplitude or frequency modulation, the extraction of non-periodic variability using a pure (co)sinusoid leads to a vast over-estimate in the number of significant frequencies. This is because many fictitious and spurious frequencies will be injected into the light curve as part of the iterative pre-whitening process --- see \citet{Balona2014b} for a detailed discussion of this problem using HD~50844 as a case study. Therefore, in cases where mode density is high it is more prudent to be conservative in the extraction of significant peaks and always check that individual frequencies are resolved from each other and that they are statistically significant beyond a simplistic $S/N \geq 4$ criterion which assumes only white noise in a spectrum. 
		
	In our analysis of the CoRoT light curve of HD~50844, we extract 54 significant frequencies using our conservative approach to iterative pre-whitening. The summary figure showing the morphology of the red noise in the CoRoT photometry extracted using our Bayesian MCMC fit and the SFT for HD~50844 is given as Fig.~\ref{figure: HD50844}. We emphasise that the remaining variance in the residual power density spectrum, shown in black in Fig.~\ref{figure: HD50844} has an amplitude $S/N < 4$. As discussed by \citet{Poretti2009}, the high mode density in HD~50844 produces an almost flat noise plateau below $\nu \lesssim 30$~d$^{-1}$ ($\nu \lesssim 347$~$\mu$Hz) after the dominant peaks have been extracted, as shown in the top panel of Fig.~\ref{figure: HD50844}. The background signal in the residual power density spectrum, interpreted as granulation signal by \cite{Kallinger2010c}, may explain the steep red noise morphology obtained from our Bayesian MCMC red noise fit. Since the flat noise plateau in HD~50870 extends to approximately $100$~$\mu$Hz, it impacts the determination of the parameter $\gamma$ when fitting any red noise background caused by IGWs that may be in this star. Thus we label HD~50844 in Table~\ref{table: results} as having insufficient CoRoT data to remove all pulsation modes, which subsequently affects the accurate determination of the fitting parameter $\gamma$.


	\subsection{HD~50870}
	\label{subsection: HD50870}
	
	HD~50870 has spectral types in the literature that range between F0\,IV and A8\,V \citep{McCuskey1956b, Houk1999, Mantegazza2012}. High-resolution and high-$S/N$ spectroscopy of HD~50870 revealed it to be long-period SB2 system with primary star having a spectral type of A8\,III, an effective temperature of $7660 \pm 250$~K, a surface gravity of $\log\,g = 3.68 \pm 0.25$ and a projected surface rotational velocity of $v\,\sin\,i = 37.5 \pm 2.5$~km\,s$^{-1}$ \citep{Mantegazza2012}. The secondary star has a spectral type of F2\,III, an effective temperature of $7077 \pm 250$~K, surface gravity of $\log\,g = 3.74 \pm 0.25$ and a projected surface rotational velocity of $v\,\sin\,i = 8.0 \pm 2.5$~km\,s$^{-1}$ \citep{Mantegazza2012}. With both stars having similar fundamental parameters, they are also in a very similar location within the classical instability strip so it cannot be uniquely claimed that the primary or secondary, or both stars are pulsating.
		
	Over 1000 frequencies were extracted from the CoRoT light curve of HD~50870 by \citet{Mantegazza2012} and independently by \citet{Barcelo2017a}. These studies found a regular frequency spacing of approximately 55~$\mu$Hz (4.8~d$^{-1}$), which was interpreted as the large frequency separation and used to derive an average density of $\bar{\rho} = 0.17$~g\,cm$^{-3}$ and a mass of $M = 1.5$~M$_{\rm \odot}$ \citep{Suarez2014b, Barcelo2017a}. The numerous frequencies and the presence of a low frequency power excess in HD~50870 was interpreted to be either caused by granulation \citep{Mantegazza2012}, or related to the oblateness and rotation of the star \citep{Barcelo2017a}. It should be noted that \citet{Mantegazza2012} binned the high-cadence CoRoT light curve into 20535 data points with an effective Nyquist frequency of 100~d$^{-1}$ (1157~$\mu$Hz) as part of their analysis. Consequently, such a significantly lower Nyquist frequency dramatically increases the amplitude suppression within a "long" cadence data set and suppresses astrophysical signal (see section~2.3.3 of \citealt{Bowman_BOOK}). 
	
	Similarly to HD~50844, the high mode density of HD~50870 produces an almost flat noise plateau in the residual power density spectrum, which is shown in Fig.~\ref{figure: HD50870}. Since the flat noise plateau in HD~50870 extends to high frequencies, this impacts the determination of the fitting parameter $\gamma$, and limits the interpretation of any low-frequency power excess being caused by IGWs. Thus we label HD~50870 as having insufficient CoRoT data for an accurate determination of the fitting parameter $\gamma$ in Table~\ref{table: results}.


	\subsection{HD~51193}
	\label{subsection: HD51193}
	
	HD~51193 (V746~Mon) has a spectral type of B1.5\,IVe, an effective temperature of $T_{\rm eff} \simeq 23\,000$~K, a surface gravity of $\log\,g \simeq 3.6$ and a projected surface rotational velocity of $220 \pm 25$~km\,s$^{-1}$ \citep{Fremat2006c, Gutierrez-Soto2011a}. From photometric and spectroscopic data sets, HD~51193 is known to be a multi-periodic pulsating star with its low-frequency variability previously interpreted as non-radial g~modes \citep{Gutierrez-Soto2007c, Gutierrez-Soto2011a, Lefevre2009c}. The low-frequency power excess in the power density spectrum of HD~51193 is clear in addition to strong beating of unresolved frequencies, as demonstrated in Fig.~\ref{figure: HD51193}. Similarly to HD~45418, HD~51193 is a rapid rotator and not known to be a Be star, but the high-$S/N$ peaks in the frequency range $1 \leq \nu \leq 2$~d$^{-1}$ ($11.6 \leq \nu \leq 23.1$~$\mu$Hz) may be stochastically-excited gravito-inertial modes like those found in the Be star HD~51452 by \citet{Neiner2012d}. The summary figure showing the morphology of the red noise in the CoRoT photometry and the SFT for HD~51193 is given as Fig.~\ref{figure: HD51193}. We note that the complex CoRoT window pattern in HD~51193 results in a background fit which overestimates the white noise contribution. Thus we label HD~51193 as having insufficient CoRoT data for an accurate background fit in Table~\ref{table: results}, yet it is clear that this star still has a clear low-frequency power excess.


	\subsection{HD~51332}
	\label{subsection: HD51332}
	
	HD~51332 has a spectral type of F0\,V \citep{Houk1999}, an effective temperature of $T_{\rm eff} \simeq 6700$~K and a surface gravity of $\log\,g \simeq 4.0$ \citep{Masana2006, Holmberg2009, Casagrande2011, Bailer-Jones2011a, McDonald2012a, David2015a}. The summary figure showing the morphology of the red noise in the CoRoT photometry after the significant peaks in the power density spectrum have been extracted and the SFT for HD~51332 is given as Fig.~\ref{figure: HD51332}. We also note that HD~51332 has a series of low-frequency harmonics in its power density spectrum, which suggests that it may be a binary system. The binarity of HD~51332 has also been inferred from radial velocity measurements \citep{Frankowski2007b}.


	\subsection{HD~51359}
	\label{subsection: HD51359}
	
	HD~51359 has a spectral type of A9/F0\,IV \citep{Houk1999} and an approximate effective temperature of $T_{\rm eff} \simeq 6800$~K \citep{McDonald2012a}. It is also a known visual double star determined using speckle interferometry at the US Naval Observatory, with a companion star that is $\Delta\,V \simeq 2$~mag fainter \citep{Germain1999b, Douglass1999}. HD~51359 is a \dsct star with numerous p-mode frequencies, which similarly to HD~50844 and HD~50870, produce an almost flat noise plateau in the residual power density spectrum, as shown in Fig.~\ref{figure: HD51359}. This low-frequency power excess could be evidence of granulation since HD~51359 is a late-A/early-F star with a thin surface convection zone, or low-$S/N$ p modes. The summary figure showing the morphology of the red noise in the CoRoT photometry and the SFT for HD~51359 is given as Fig.~\ref{figure: HD51359}. Since the flat noise plateau in HD~51359 impacts the determination of the fitting parameter $\gamma$, we label HD~51359 as having insufficient CoRoT data in Table~\ref{table: results}.


	\subsection{HD~51452}
	\label{subsection: HD51452}
	
	HD~51452 is a rapidly-rotating Be star with a spectral type of B0\,IVe, an effective temperature of $T_{\rm eff} = 29\,000 \pm 500$~K, a surface gravity of $\log\,g = 3.95 \pm 0.04$, and a projected surface rotational velocity of $v\,\sin\,i = 315.5 \pm 9.5$~km\,s$^{-1}$ \citep{Fremat2006c, Neiner2012d}. From a combined analysis of CoRoT photometry and high-resolution, high-$S/N$ spectroscopy, \citet{Neiner2012d} discovered that HD~51452 exhibits stochastically-excited gravito-inertial pulsation modes that can be explained by the rapid rotation of this star. The stochastic excitation of gravito-inertial waves was predicted by \citet{Rogers2013b} and \citet{Mathis2014a}. Inspired by these findings, it was suggested that the outburst phenomena seen in Be stars may be explained by angular momentum transport by IGWs in rapidly-rotating B stars \citep{Neiner2013b, Lee_U_2014b}. The summary figure showing the morphology of the red noise in the CoRoT photometry and the SFT for HD~51452 is given as Fig.~\ref{figure: HD51452}. However, we note that the complex CoRoT window pattern in HD~51452 results in a background fit which overestimates the white noise contribution. Thus we label HD~51452 as having insufficient CoRoT data for an accurate background fit in Table~\ref{table: results}, yet it is clear that this star still has a clear low-frequency power excess caused by pulsations.


	\subsection{HD~51722}
	\label{subsection: HD51722}
	
	HD~51722 has a spectral type of A9\,V \citep{Houk1999} and an approximate effective temperature of $T_{\rm eff} \simeq 7000$~K \citep{McDonald2012a}. It is also known to a be a double star in the Tycho double star catalogue \citep{Fabricius2002a, Roberts2015d}, and has possible transit signatures in its CoRoT light curve \citep{Liu_C_2015}. A possible detection of variability in Hipparcos photometry was made by \citet{Koen2002c}, but no studies since have used the available CoRoT light curve to investigate if HD~51722 exhibits coherent and/or damped pulsation modes. In our analysis of the CoRoT light curve of HD~51722, we extract almost 100 significant frequencies between $0 \leq \nu \leq 20$~d$^{-1}$ ($0 \leq \nu \leq 231$~$\mu$Hz) using our conservative approach to iterative pre-whitening, which makes this star similar to the other high-mode density \dsct stars including HD~50844, HD~50870 and HD~51359, since it has an almost flat noise plateau in its residual power density spectrum. This background signal explains the steep red noise morphology obtained from our Bayesian MCMC fit and could be caused by granulation and/or low-$S/N$ pulsation modes. The summary figure showing the morphology of the red noise in the CoRoT photometry and the SFT for HD~51722 is given as Fig.~\ref{figure: HD51722}. The flat noise plateau in HD~51722 extends to high frequencies, which affects the determination of the fitting parameter $\gamma$. Thus we label HD~51722 as having insufficient CoRoT data for an accurate determination of the fitting parameter $\gamma$ in Table~\ref{table: results}.


	\subsection{HD~51756}
	\label{subsection: HD51756}
	
	HD~51756 has a spectral type of B0.5\,IV \citep{Papics2011} and is a visual double star in the Tycho double star catalogue \citep{Fabricius2002a}. \citet{Papics2011} previously studied the CoRoT photometry, combined it with high-resolution, high-$S/N$ spectroscopy and discovered HD~51756 to be a long-period SB2 system. The primary is slow rotator ($v\,\sin\,i = 28 \pm 4$~km\,s$^{-1}$) and the secondary is a fast rotator ($v\,\sin\,i \simeq 170 \pm 15$~km\,s$^{-1}$), but both components have the same spectral type, similar effective temperatures of $T_{\rm eff} = 30\,000 \pm 1000$~K and surface gravities of $\log\,g = 3.75 \pm 0.25$. Coherent heat-driven pulsation modes are expected for these stars but none were detected by \citet{Papics2011}. The CoRoT photometry of HD~51756 does contain signatures of rotational modulation which \citet{Papics2011} speculated to be caused by surface abundance inhomogeneities. In our study using the latest CoRoT light curve of HD~51756, we confirm the absence of coherent pulsation modes. However, we note the presence of a significant low-frequency power excess which is indicative of IGWs, because both components of HD~51756 have a spectral type of B0.5\,IV and are thus too hot to have granulation from a surface convection envelope. The summary figure showing the morphology of the red noise in the CoRoT photometry and the SFT for HD~51756 is given as Fig.~\ref{figure: HD51756}.


	\subsection{HD~52130}
	\label{subsection: HD52130}
	
	The only literature reference of HD~52130 is the determination of its spectral type as A2\,III/IV by \citet{Houk1999}. Unfortunately, fundamental parameters of HD~52130 are not available, but from its spectral type we can estimate an effective temperature of $T_{\rm eff} = 8200$~K. The summary figure showing the morphology of the red noise in the CoRoT photometry and the SFT for HD~52130 is given as Fig.~\ref{figure: HD52130}, which clearly shows stochastic low-frequency variability consistent with either granulation or IGWs.


	\subsection{HD~174532}
	\label{subsection: HD174532}
	
	HD~174532 (7~Aql) has a spectral type of A9\,IV listed by \citet{Houk1999}, although later studies revealed it to be an evolved star with a spectral type of A2\,III/IV, an effective temperature of approximately $T_{\rm eff} \simeq 7200$~K, a surface gravity of $\log\,g \simeq 3.6$ and a projected surface rotational velocity of $v\,\sin\,i \simeq 32$~km\,s$^{-1}$ \citep{Poretti2003a, Fox-Machado2008a, Fox-Machado2010a, McDonald2012a}. HD~174532 was previously known to be a multi-periodic \dsct star \citep{Fox-Machado2008a, Fox-Machado2010a}, and in our analysis of the CoRoT photometry of HD~174532, we extract 33 frequency peaks with a $S/N \geq 4$ using our conservative approach to iterative pre-whitening. As shown in Fig.~\ref{figure: HD174532}, HD~174532 has a bump excess of unresolved pulsation modes at $\sim$~$200~\mu$Hz in its residual power density spectrum. This is because only 26.24~d of CoRoT data are available making the removal of close-frequency (hence unresolved) pulsation modes practically impossible. Since this unresolved remaining signal impacts the determination of the fitting parameter $\gamma$, we label HD~174532 as having insufficient CoRoT data for an accurate fit in Table~\ref{table: results}.


	\subsection{HD~174589}
	\label{subsection: HD174589}

	HD~174589 (8~Aql) has a spectral type of F0\,IV listed by \citet{Houk1999}, although later studies revealed it to be an evolved star with a spectral type of F2\,III, an effective temperature of approximately $T_{\rm eff} \simeq 7000$~K, a surface gravity of approximately $\log\,g \simeq 3.5$ and a projected surface rotational velocity of $v\,\sin\,i \simeq 97$~km\,s$^{-1}$ \citep{Poretti2003a, Fox-Machado2008a, Fox-Machado2010a, McDonald2012a}. Similarly to HD~174532, HD~174589 is also a known multi-periodic \dsct star \citep{Fox-Machado2008a, Fox-Machado2010a}. In our analysis of the CoRoT photometry, we extract 17 frequency peaks with a $S/N \geq 4$ using our conservative approach to iterative pre-whitening, which reveals a low-frequency power excess in the residual power density spectrum that is indicative of granulation or IGWs. The summary figure showing the morphology of the red noise in the CoRoT photometry and the SFT for HD~174589 is given as  Fig.~\ref{figure: HD174589}.

	
	\subsection{HD~174936}
	\label{subsection: HD174936}

	HD~174936 has a spectral type of A3\,IV \citep{Houk1999}, and Str{\" o}mgren photometry indicates an effective temperature of $T_{\rm eff} = 8000 \pm 200$~K and surface gravity of $\log\,g = 4.1 \pm 0.2$ \citep{Hauck1998b}. Later, \citet{GH2009} obtained high-resolution spectroscopy, confirmed its fundamental parameters and measured a projected surface rotational velocity of $v\,\sin\,i \simeq 170$~km\,s$^{-1}$. The CoRoT light curve of HD~174936 was previously analysed by \citet{GH2009}, who extracted 422 frequencies and found a regular frequency spacing of approximately 52~$\mu$Hz ($\simeq 4.5$~d$^{-1}$). This regular spacing was interpreted as the large frequency separation and used it to derive an average density of $\bar{\rho} = 0.33$~g\,cm$^{-3}$ and a mass of $M = 1.6$~M$_{\rm \odot}$ \citep{GH2009, Suarez2014b}. 
	
	In our analysis of the CoRoT photometry of HD~174936, we extract 64 frequency peaks with a $S/N \geq 4$ using our conservative approach to iterative pre-whitening, which reveals an almost flat noise plateau in the residual power density spectrum up to $50$~d$^{-1}$ (579~$\mu$Hz) that could be caused by granulation and/or low-$S/N$ pulsations. The summary figure showing the morphology of the red noise in the CoRoT photometry and the SFT for HD~174936 is given as Fig.~\ref{figure: HD174936}. The flat noise plateau likely caused by a high density of unresolved pulsation modes in HD~174936 and the insufficient frequency resolution to remove them via pre-whitening impacts the determination of the fitting parameter $\gamma$. Thus we label HD~174936 as having insufficient CoRoT data for an accurate determination of the fitting parameter $\gamma$ in Table~\ref{table: results}.

	
	\subsection{HD~174966}
	\label{subsection: HD174966}
	
	HD~174966 has a spectral type of A7\,III/IV \citep{Houk1999}, an effective temperature of $T_{\rm eff} = 7555 \pm 50$~K, a surface gravity of $\log\,g = 4.2 \pm 0.1$, and a projected surface rotational velocity of $v\,\sin\,i \simeq 126$~km\,s$^{-1}$ determined from high-resolution, high-$S/N$ spectroscopy \citep{GH2013}. The CoRoT light curve of HD~174966 was previously analysed by \citet{GH2013}, who extracted 185 frequencies and detected a regular frequency spacing of approximately 65~$\mu$Hz ($\simeq 5.6$~d$^{-1}$), which was interpreted as the large frequency separation and used to derive an average density of $\bar{\rho} = 0.51$~g\,cm$^{-3}$ and a mass of $M = 1.7 \pm 0.2$~M$_{\rm \odot}$. The combination of spectroscopy and Str{\" o}mgren photometry also allowed \citet{GH2013} to identify 18 frequencies between $18 \leq \nu \leq 33$~d$^{-1}$ ($208.3 \leq \nu \leq 381.9$~$\mu$Hz) as low-degree non-radial p modes. 
	
	In our analysis of the CoRoT photometry of HD~174966, we extract 42 significant frequencies with $S/N \geq 4$ between $0 \leq \nu \leq 51$~d$^{-1}$ (590~$\mu$Hz) using our conservative approach to iterative pre-whitening, which reveals a low-frequency power excess in the residual power density spectrum that could be caused by granulation and/or low-$S/N$ pulsations. The summary figure showing the morphology of the red noise in the CoRoT photometry and the SFT for HD~174966 is given as Fig.~\ref{figure: HD174966}.


	\subsection{HD~174967}
	\label{subsection: HD174967}
	
	The only literature reference to HD~174967 is a spectral type of B9.5\,V by \citet{Houk1999}, which we use to estimate an approximate effective temperature of $T_{\rm eff} \simeq 11\,000$~K. In our analysis of the CoRoT photometry for HD~174967, we note the absence of coherent pulsation modes in the power density spectrum, which is in agreement with the expectation from theory for a lack of pulsation in stars between the SPB and \dsct instability regions. However, as shown in the summary figure of HD~174967 in Fig.~\ref{figure: HD174967}, there is a low-frequency power excess below $\nu \lesssim 2$~d$^{-1}$ (23~$\mu$Hz) in the power density spectrum. The morphology of the red noise in the CoRoT photometry and the SFT for HD~174967 are shown in Fig.~\ref{figure: HD174967}.


	\subsection{HD~174990}
	\label{subsection: HD174990}
	
	The only literature reference to HD~174990 is a spectral type of A3\,V by \citet{Houk1999}, which we use to estimate an approximate effective temperature of $T_{\rm eff} \simeq 8500$~K. Similarly to HD~174967, the early-A star HD~174990 also lacks coherent pulsation modes but does have a clear low-frequency power excess below $\nu \lesssim 2$~d$^{-1}$ (23~$\mu$Hz) in its power density spectrum. Since neither coherent gravity mode pulsations or granulation from a surface convection zone are not expected in early-A main-sequence stars, we interpret this stochastic variability as IGWs. The summary figure showing the morphology of the red noise in the CoRoT photometry and the SFT for HD~174990 is given as Fig.~\ref{figure: HD174990}.

	
	\subsection{HD~175272}
	\label{subsection: HD175272}
	
	The spectral type of HD~175272 is listed as A0\,V and flagged as being a member of a multiple system by \citet{Houk1999}. Using medium-resolution spectroscopy, \citet{Prugniel2001} determined a spectral type of F5\,V and estimated an effective temperature of 6500~K and a surface gravity of $\log\,g = 4.1$ for HD~175272. Later, \citet{Poretti2003a} measured a projected surface rotational velocity of $v\,\sin\,i \simeq 23$~km\,s$^{-1}$ and noted a lack of periodic photometric variability. In their catalogue of stellar parameters for stars observed by Hipparcos, \citet{McDonald2012a} list an effective temperature of 6900~K for HD~175272. More recently, HD~175272 has been claimed to show solar-like pulsations by \citet{Ozel2013a} and \citet{Hekker2014b}, who derived an effective temperature of $T_{\rm eff} \simeq 6700$~K, a mass of $M \simeq 1.5$~M$_{\rm \odot}$, and a frequency of maximum amplitude $\nu_{\rm max} \simeq 1500$~$\mu$Hz ($\simeq 130$~d$^{-1}$). However, these parameters were determined by comparing the low-$S/N$ detection of solar-like oscillations in HD~175272 to the "seismically-similar" star HD~181420.
	
	HD~175272 is amongst the coolest stars in our sample with an effective temperature of $T_{\rm eff} \simeq 6700$~K, which is certainly cool enough for it to have a surface convection zone. However, it is still massive enough to host a convective core that could excite IGWs. It is possible for stars of this temperature to exhibit coherent g-mode pulsation frequencies, i.e. \gdor stars, but we do not detect any significant periodic variability in the CoRoT observations of HD~175272. We do find a significant low-frequency power excess, which is presumably dominated by granulation. The summary figure showing the morphology of the red noise in the CoRoT photometry and the SFT for HD~175272 is given as Fig.~\ref{figure: HD175272}.


	\subsection{HD~175445}
	\label{subsection: HD175445}
	
	HD~175445 is listed as an A1\,V star by \citet{Houk1999}. Later, \citet{Paunzen2001c} and \citet{Paunzen2002c} classified HD~175445 as a $\lambda$~Boo star with a spectral type of hA5mA2V, an effective temperature of $T_{\rm eff} = 8520 \pm 200$~K and a surface gravity of $\log\,g = 3.96 \pm 0.10$. Although, the effective temperature may be as high as $T_{\rm eff} = 9060 \pm 330$~K \citep{Murphy2017a}. In our analysis of the CoRoT photometry, we extract 32 frequency peaks with a $S/N \geq 4$ using our conservative approach to iterative pre-whitening, which reveals an almost flat noise plateau below $\nu \lesssim 30$~d$^{-1}$ (347~$\mu$Hz) in the residual power density spectrum, which is indicative of granulation, low-$S/N$ pulsation modes or IGWs. The summary figure showing the morphology of the red noise in the CoRoT photometry and the SFT for HD~175445 is given as Fig.~\ref{figure: HD175445}. Motivated by having only 26.24~d of CoRoT photometry, we label HD~175445 as having insufficient data for an accurate determination of the fitting parameter $\gamma$ in Table~\ref{table: results}, because its residual power density spectrum likely contains unresolved close-frequency pulsation modes.

	
	\subsection{HD~175542}
	\label{subsection: HD175542}
	
	The only literature reference of HD~175542 is the determination of its spectral type as A1\,V by \citet{Houk1999}, but from its spectral type we can estimate an effective temperature of $T_{\rm eff} \simeq 8500$~K. In our analysis of the CoRoT photometry of HD~175542, we find a lack of coherent pulsation modes but do find a clear low-frequency power excess below $\nu \lesssim 2$~d$^{-1}$ (23~$\mu$Hz) in the power density spectrum. The morphology and stochastic nature of this low-frequency signal is similar to other early-A stars in our sample, which are not expected to have coherent g modes or granulation from a surface convection zone. The summary figure showing the morphology of the red noise in the CoRoT photometry and the SFT for HD~175542 is given as Fig.~\ref{figure: HD175542}, which clearly show evidence of low-frequency stochastic variability that we interpret as being caused by IGWs.

	
	\subsection{HD~175640}
	\label{subsection: HD175640}
	
	HD~175640 is a known non-magnetic chemically peculiar star with a spectral type of B9\,V\,HgMn, an effective temperature of $T_{\rm eff} = 12\,000 \pm 100$~K, a surface gravity of $\log\,g = 3.95 \pm 0.15$ and a projected surface rotational velocity of order a few km\,s$^{-1}$ \citep{Hubrig2001a, Dolk2002a, Castelli2004c, Renson2009, Landstreet2009b, Lefever2010, Makaganiuk2011a}. In their previous analysis of the CoRoT light curve of HD~175640, \citet{Ghazaryan2013b} concluded that there were no significant coherent pulsation modes. In our analysis, having excluded known CoRoT alias frequencies (e.g. see Fig.~\ref{figure: CoRoT window}), we confirm that lack of significant coherent pulsation modes, but do detect a low-frequency power excess in the power density spectrum. The summary figure showing the morphology of the red noise in the CoRoT photometry and the SFT for HD~175640 is given Fig.~\ref{figure: HD175640}, which clearly show evidence of low-frequency stochastic variability.

	
	\subsection{HD~263425}
	\label{subsection: HD263425}
	
	The only literature reference of HD~263425 is the determination of its spectral type as A0\,V by \citet{Voroshilov1985b}, which we use to estimate an effective temperature of approximately $T_{\rm eff} \simeq 9500$~K. Similar to other early-A stars, which do not have coherent pulsation modes nor expected to have granulation from a surface convection zone, we detect a low-frequency power excess in the power density spectrum of HD~263425. The summary figure showing the morphology of the red noise in the CoRoT photometry and the SFT for HD~263425 is given as Fig.~\ref{figure: HD263425}, which clearly show evidence of low-frequency stochastic variability that we interpret as IGWs.


\section{Discussion}
\label{section: discussion}

Within our sample of O, B, A and F stars are pulsating stars that underwent pre-whitening to remove high-$S/N$ peaks since the physical driving mechanism of these dominant frequencies cannot be determined from observations alone. For stars without significant limitations imposed by the length and quality of CoRoT photometry, the morphology of the background low-frequency power excess in stars after pre-whitening standing waves allows inference of the physics that is responsible for the stochastic low-frequency variability in photometric observations of early-type stars.

In addition to the three O stars studied by \citet{Blomme2011b} and \citet{Aerts2015c}, there are two other O stars in our sample: HD~46149 and HD~46202. The CoRoT photometry of these stars have been previously analysed by \citet{Degroote2010b} and \citet{Briquet2011}, respectively, with both stars showing evidence for intrinsic variability attributed to pulsation modes. However, these stars also have low-frequency stochastic variability in their power density spectra and a significant low-frequency power excess. Individual amplitudes of peaks in the background low-frequency power excess are of order tens and hundreds of $\mu$mag for these five O stars. Further down the main sequence are early-B stars, such as HD~45546, which also have similar characteristic frequencies, $\nu_{\rm char}$, and amplitudes, $\alpha_{0}$, in the morphology of their red noise compared to the O stars. The similar morphology in the power density spectra of these O and B stars, is evidence for a common physical mechanism acting in these stars.

Amongst the A and F stars in our sample are several \dsct stars, which contain dozens of high-frequency coherent standing pulsation modes that we removed via pre-whitening. However, these intermediate-mass stars also have a background low-frequency power excess of order 10~$\mu$mag --- i.e. $\alpha_0 \simeq$ a few hundred ppm$^2/\mu$Hz in their residual power density spectra. For late-A and F stars it is plausible that the observed stochastic low-frequency variability is caused by granulation, but this cannot be true for O and B stars as these do not have surface convection zones whilst on the main sequence. We explore why the observed low-frequency power excess for many of the stars in our sample is not consistent with granulation in section~\ref{subsection: granulation}. 

On the other hand, we clearly have early-A stars in our sample that are not expected to pulsate in coherent g-modes, such as HD~45517 and HD~263425, which have a significant low-frequency power excess in their power density spectra. It has been investigated using observations of variable stars in clusters and later followed up with theoretical work, that the rapid rotation of some SPB stars is sufficient to distort a star such that its measured parameters place it between the SPB and \dsct instability regions in the HR~diagram \citep{Mowlavi2013, Salmon_S_2014}. However, the stars in our sample that fall within this region are slow or moderate rotators.


	\subsection{Asteroseismic HR diagrams}
	\label{subsection: HR}
	
	\begin{figure*}
	\centering
	\includegraphics[width=0.99\textwidth]{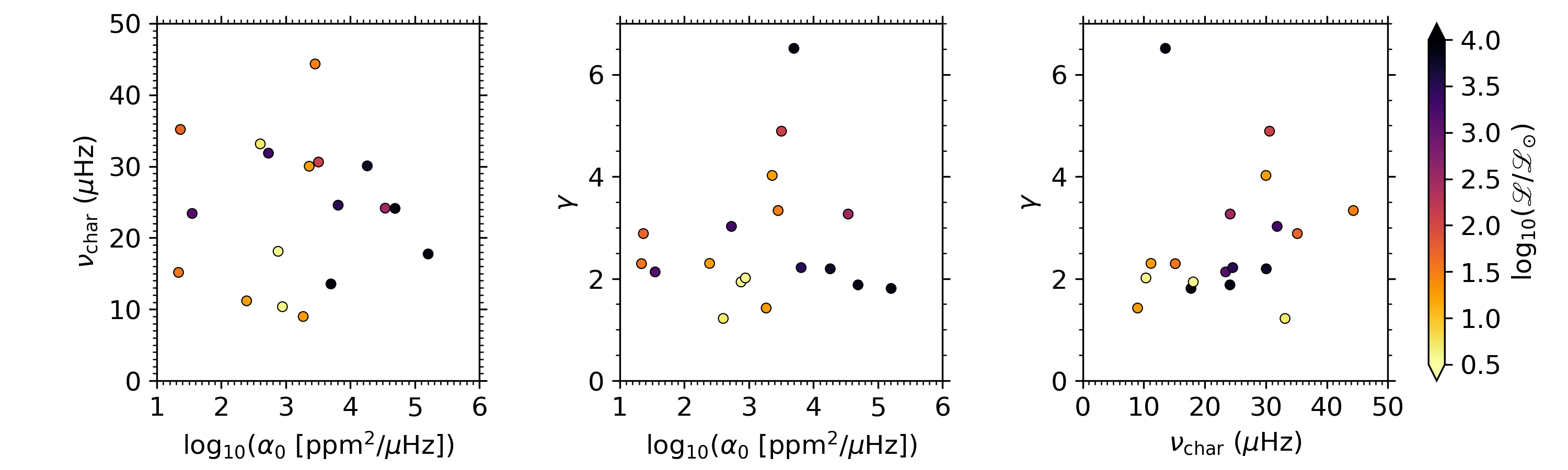}
	\includegraphics[width=0.49\textwidth]{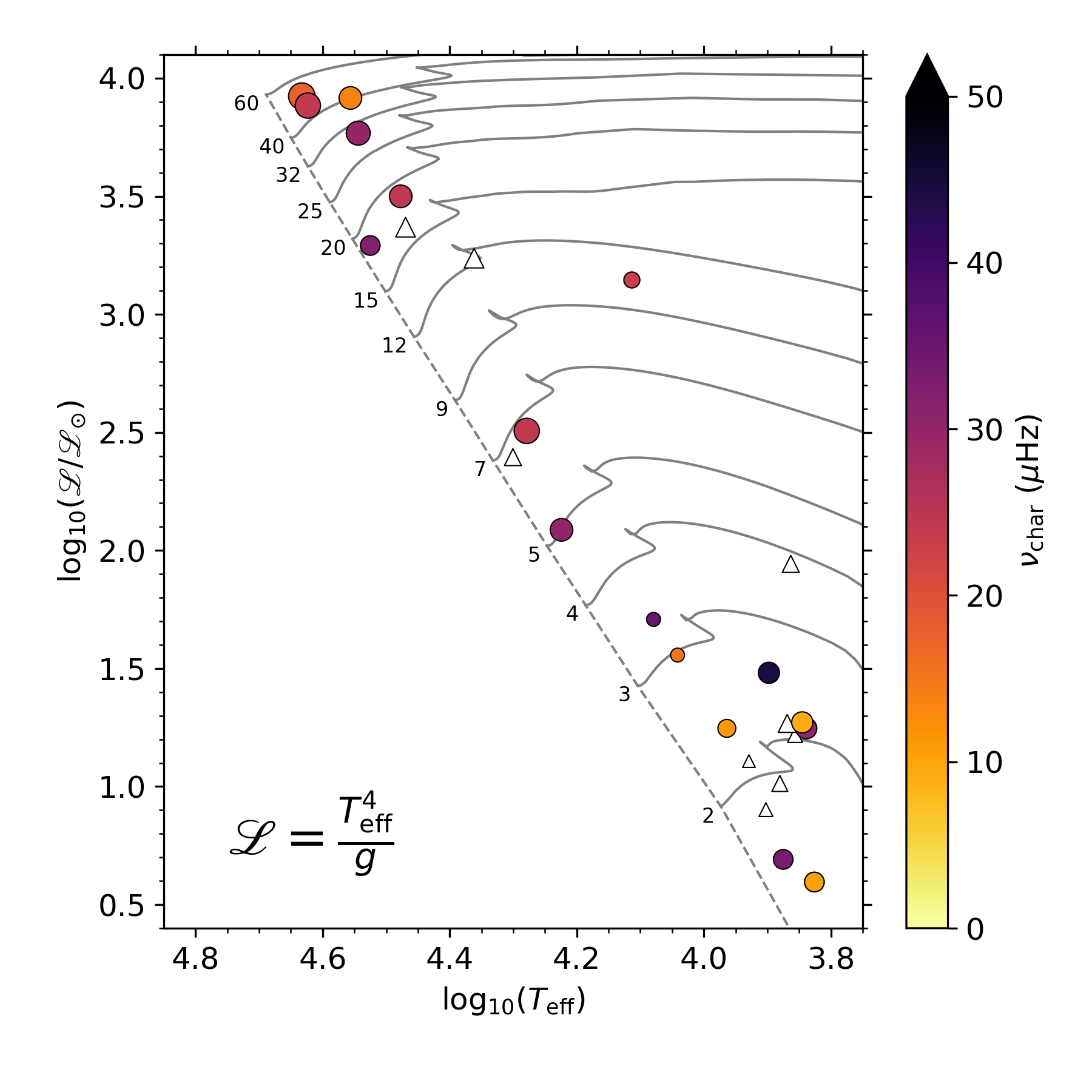}
	\includegraphics[width=0.49\textwidth]{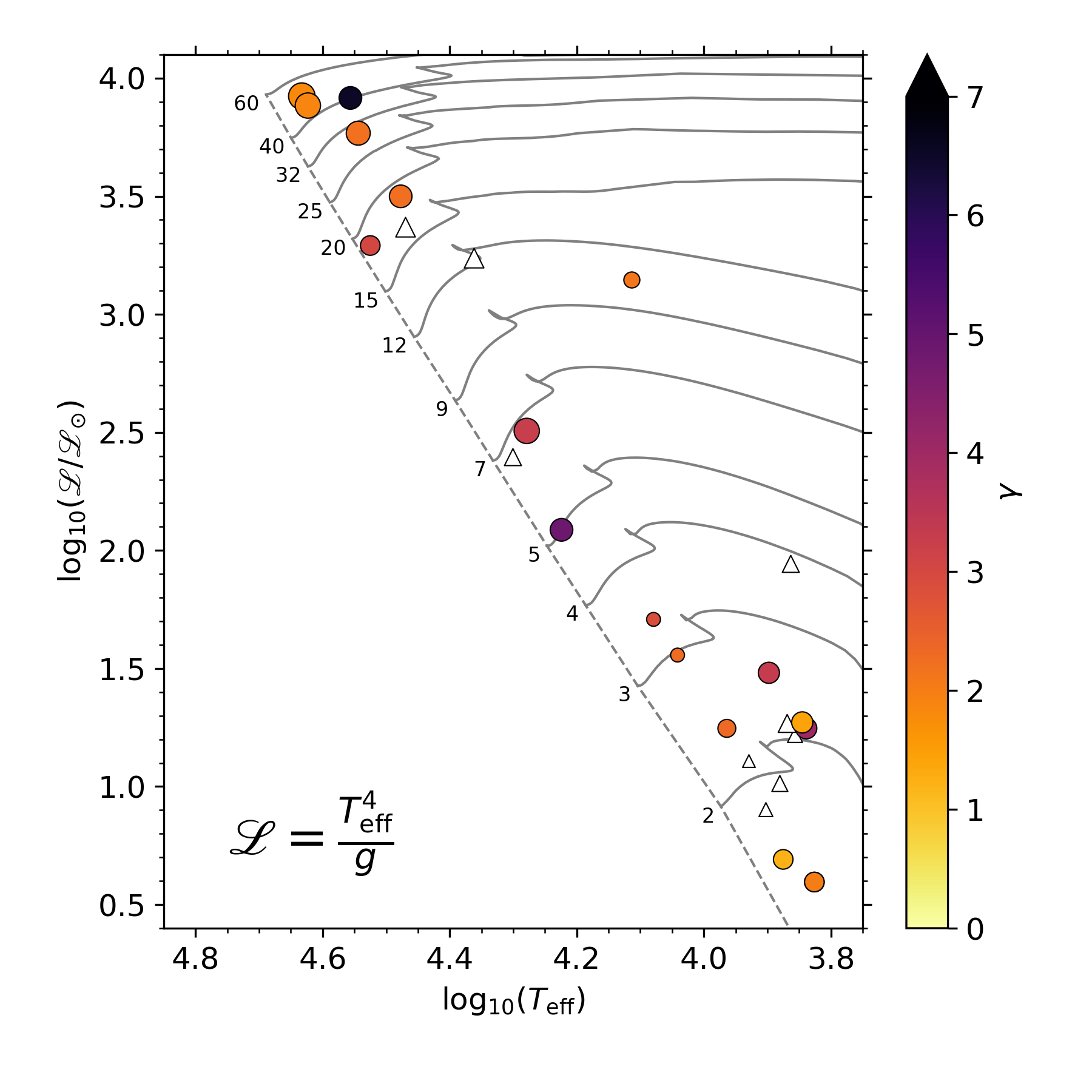}
	\caption{The top row shows the pair-wise relationship between the fitting parameters, $\alpha_0$, $\nu_{\rm char}$ and $\gamma$ given in Eqn.~(\ref{equation: Blomme}), for stars in our sample with spectroscopic parameters and sufficient CoRoT data for an accurate background fit. The bottom row shows the location of stars in the (spectroscopic) HR~diagram as filled circles that have been colour-coded by the red-noise fit (Eqn.~\ref{equation: Blomme}) parameters $\nu_{\rm char}$ (left) and $\gamma$ (right), with the size of each symbol corresponding to the fit parameter $\alpha_0$. The location of stars with insufficient CoRoT data for an accurate background fit to the residual power density spectrum are shown as white triangles. Evolutionary tracks for representative masses (in units of M$_{\rm \odot}$) from \citet{Elkstrom2012a} are shown as solid grey lines, with the dashed grey line representing the ZAMS.}
	\label{figure: IGW HR}
	\end{figure*}
		
	\begin{figure}
	\centering
	\includegraphics[width=0.49\textwidth]{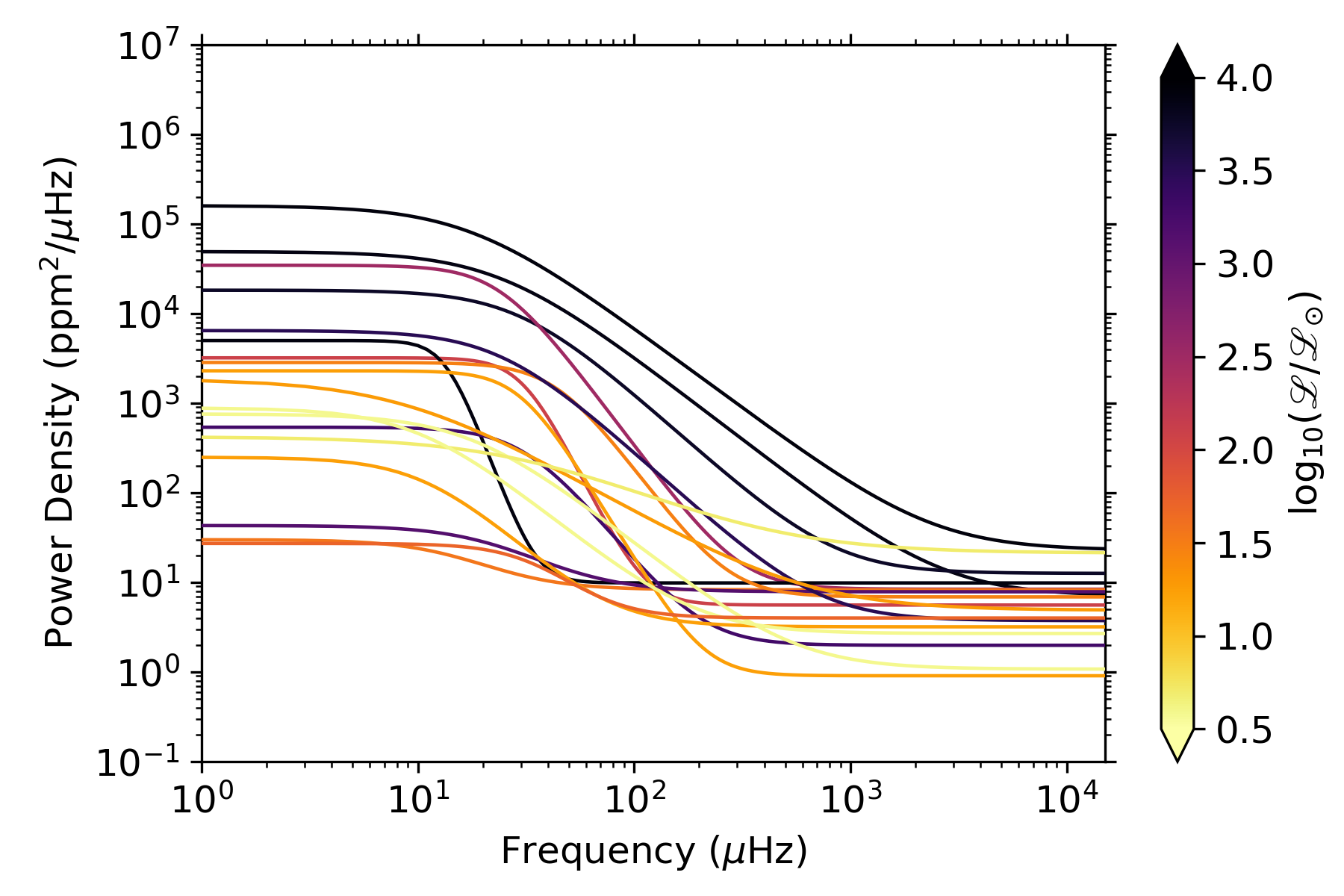}
	\caption{The measured profiles of red noise determined from the power density spectra of CoRoT photometry for stars with known stellar parameters from spectroscopy and sufficient CoRoT data for an accurate background fit to the residual power density spectrum, which have been colour-coded by their spectroscopic luminosity.}
	\label{figure: multi star FT}
	\end{figure}
		
	Following the methodology employed by \citet{Simon-Diaz2017a} and \citet{Godart2017}, who studied the distribution of macroturbulence in the spectroscopic HR~diagram of massive stars, we use a spectroscopic proxy for stellar luminosity defined as $\mathscr{L} = {T_{\rm eff}^4}/{g}$. Using the available spectroscopic parameters for the 18 stars in our sample with sufficient CoRoT data for an accurate background fit of the residual power density spectrum, we place them in (spectroscopic) HR~diagrams in the bottom row of Fig.~\ref{figure: IGW HR}, in which the {filled circles} in the left and right panels have been colour-coded by the red noise fit parameters $\nu_{\rm char}$ and $\gamma$, respectively, and the symbol size is proportional to the red-noise parameter $\alpha_0$. For completeness, we also include the location of stars with available spectroscopic parameters but insufficient CoRoT photometry for an accurate background fit of the residual power density spectrum as white triangles in Fig.~\ref{figure: IGW HR}. For illustrative purposes, the (non-rotating) evolutionary tracks for representative masses between 2 and 60~M$_{\rm \odot}$ from \citet{Elkstrom2012a} are shown as solid grey lines, with the ZAMS line indicated by the dashed grey line. A notable result from Fig.~\ref{figure: IGW HR} is the trend between the amplitude of the measured red noise, $\alpha_0$, and stellar mass on the main sequence.

	To compare the morphologies of the measured red noise profiles for the 18 stars with spectroscopic parameters and accurate background fits, we overplot them in Fig.~\ref{figure: multi star FT} and colour-code each profile by each star's spectroscopic luminosity. From inspection of Figs~\ref{figure: IGW HR} and \ref{figure: multi star FT} and of the individual summary figures of each star given in Appendix~\ref{section: appendix: MCMC results}, it is clear that the morphology of the observed red noise in upper-main sequence stars is quite diverse. This mirrors the diverse range of macroturbulence values found in the spectroscopy for stars on the upper-main sequence and is likely caused by the interplay of different dominant physical mechanisms causing the observed stochastic variability across the HR~diagram --- see the discussion from \citet{Simon-Diaz2017a}.

	
	\subsection{A universal scaling law for granulation?}
	\label{subsection: granulation}

	From the original work by \citet{Kjeldsen1995} the characteristic granulation frequency can be estimated from the scaling relation:
	
	\begin{equation}
	\nu_{\rm gran} \propto \frac{c_s}{H_p} \propto M \cdot R^{-2} \cdot T_{\rm eff}^{-1/2} ~ ,
	\label{equation: granulation}
	\end{equation}
	
	\noindent where $\nu_{\rm gran}$ is the granulation frequency, $c_s$ is the sound speed, $H_p$ is the pressure scale height, $M$ is mass, $R$ is radius, and $T_{\rm eff}$ is the effective temperature. The photometric amplitude of granulation is expected to inversely scale with the number of convective granules on a star's surface, with the size of an individual granule being proportional to the pressure scale height $H_p$, whereas the granulation timescale is expected to scale with the ratio of the size and (sound) speed of a granule under the assumption of an ideal adiabatic gas \citep{Stello2007a, Huber2009b}.
	
	As demonstrated by \citet{Kallinger2014} for thousands of pulsating solar-type and red giant stars, a tight correlation exists between $\nu_{\rm gran}$ and the stellar parameters, such that the physics of surface convection in the Sun appears to adequately scale to stars on the red giant branch. From a Lorentzian fit to the power density spectra of the \dsct stars HD~50844 and HD~174936, \citet{Kallinger2010c} demonstrated that the red noise backgrounds in these stars were also consistent with those expected for granulation in A stars, but two evolved A stars is hardly representative of all A stars. If the use of this scaling relation for granulation is physically applicable to the sub-surface convection zones in high-mass stars, one would expect a similar smooth and tight correlation between the characteristic frequency and the stellar parameters.
		
\begin{figure}
	\centering
	\includegraphics[width=0.49\textwidth]{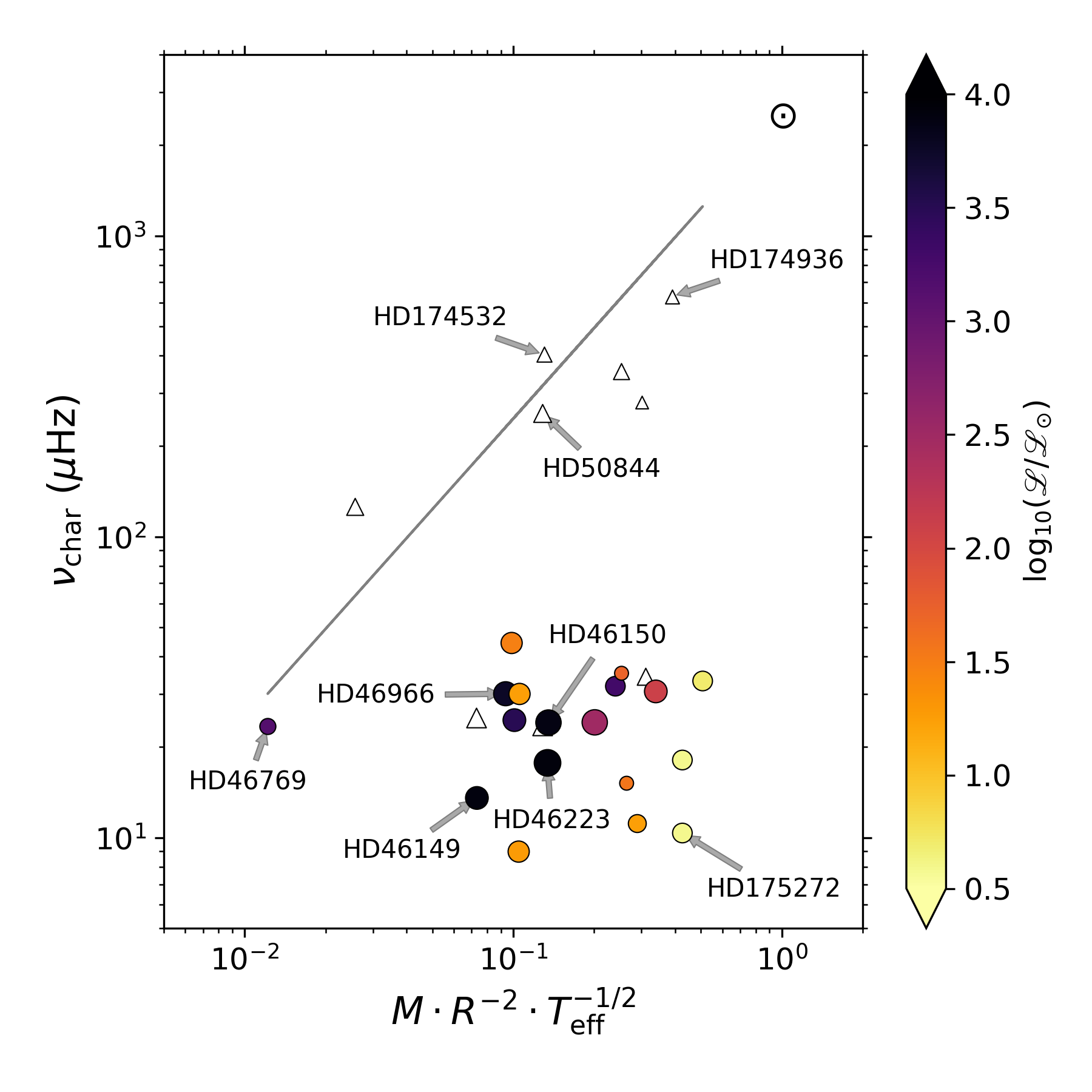}
	\caption{The distribution of measured characteristic frequencies, $\nu_{\rm char}$, as a function of stellar parameters given in solar units. The grey line represents the granulation frequency scaled using Eqn.~(\ref{equation: granulation}) with the Sun's location indicated by the $\odot$ symbol. The circles have been colour-coded by a star's spectroscopic luminosity and the size of the symbol is proportional to the fitting parameter, $\alpha_0$. White triangles represent stars which may have insufficient CoRoT data for an accurate determination for the fitting parameter $\gamma$ (c.f.~Eqn.~\ref{equation: granulation}).}
	\label{figure: granulation}
	\end{figure}

	To demonstrate the variance in the morphology of the observed red noise within our sample and specifically if the measured $\nu_{\rm char}$ values depend on the stellar parameters as predicted by the characteristic granulation frequency given by Eqn.~(\ref{equation: granulation}), we place all 27 stars with spectroscopic parameters in Fig.~\ref{figure: granulation}. Similarly to Fig.~\ref{figure: IGW HR}, each circle in Fig.~\ref{figure: granulation} represents a star with a reliable background fit whose size indicates the magnitude of $\alpha_0$ and the colour indicates $\log_{10}(\mathscr{L}/\mathscr{L_{\odot}})$. Stars with insufficient CoRoT data to accurately determine the fitting parameter $\gamma$ are shown in Fig.~\ref{figure: granulation} as white triangles. Our results for the previously studied \dsct stars, HD~50844 and HD~174936, are consistent with those of \citet{Kallinger2010c}, such that the characteristic frequencies measured in the residual power density spectra of these A stars are similar to those expected for granulation. Other stars in our study, e.g. HD~174532, also have a red noise profiles that are consistent with granulation. 
	
	However, the majority of our stars, which includes all of the main-sequence O and B stars and many of the main-sequence A and F stars, have characteristic frequencies that are approximately an order of magnitude smaller than the granulation frequency predicted by the scaling relation given in Eqn.~(\ref{equation: granulation}). If Eqn.~(\ref{equation: granulation}) is physically justified for main-sequence stars with masses $M \gtrsim 2$~M$_{\rm \odot}$ then the observed low-frequency power excess is unlikely to be caused by granulation. Therefore, we conclude that granulation is not the cause of the low-frequency variability for the stars in the bottom-right corner of Fig.~\ref{figure: granulation}. On the other hand, if the red noise is caused by granulation, our results show that the scaling relation is not appropriate for stars with masses above 2~M$_{\rm \odot}$. 
	
	Stellar winds are an important and possibly dominant cause of red noise in high-mass ($M \gtrsim 15$~M$_{\rm \odot}$) stars. The dynamic, aspherical and clumpy winds of high-mass stars modulate the flux of a star as it rotates causing stochastic and non-periodic variability, yet the exact physical cause of wind clumping is not known but has been shown to originate within the stellar photosphere \citep{Puls2006, Puls2008c}. Therefore as discussed by \citet{Aerts2018a}, there is clearly a strong interplay and possible causal relationship between photospheric variability caused by pulsations and wind variability in high-mass stars. Although high-mass stars do not have convective envelopes whilst on the main sequence, their dynamic winds may explain the scatter and the lack of a tight correlation in the red noise morphology for upper main-sequence stars.

	
	\subsection{Photometric evidence of IGWs in early-type stars}
	\label{subsection: evidence}	
	
	Our CoRoT targets range in mass between 1.5 and 50~M$_{\rm \odot}$, with all stars having characteristic frequencies that range approximately between $3 \leq \nu_{\rm char} \leq 630$~$\mu$Hz and $1.0 \leq \gamma \leq 17.0$. However, when the stars with unreliable fits caused by insufficient data for removing unresolved pulsation modes are excluded, then the remaining 18 stars with available spectroscopic parameters and a significant low-frequency power excess have $\nu_{\rm char} \leq 50$~$\mu$Hz, which are not consistent with granulation as shown in Fig.~\ref{figure: granulation}.
	
\begin{figure}
	\centering
	\includegraphics[width=0.49\textwidth]{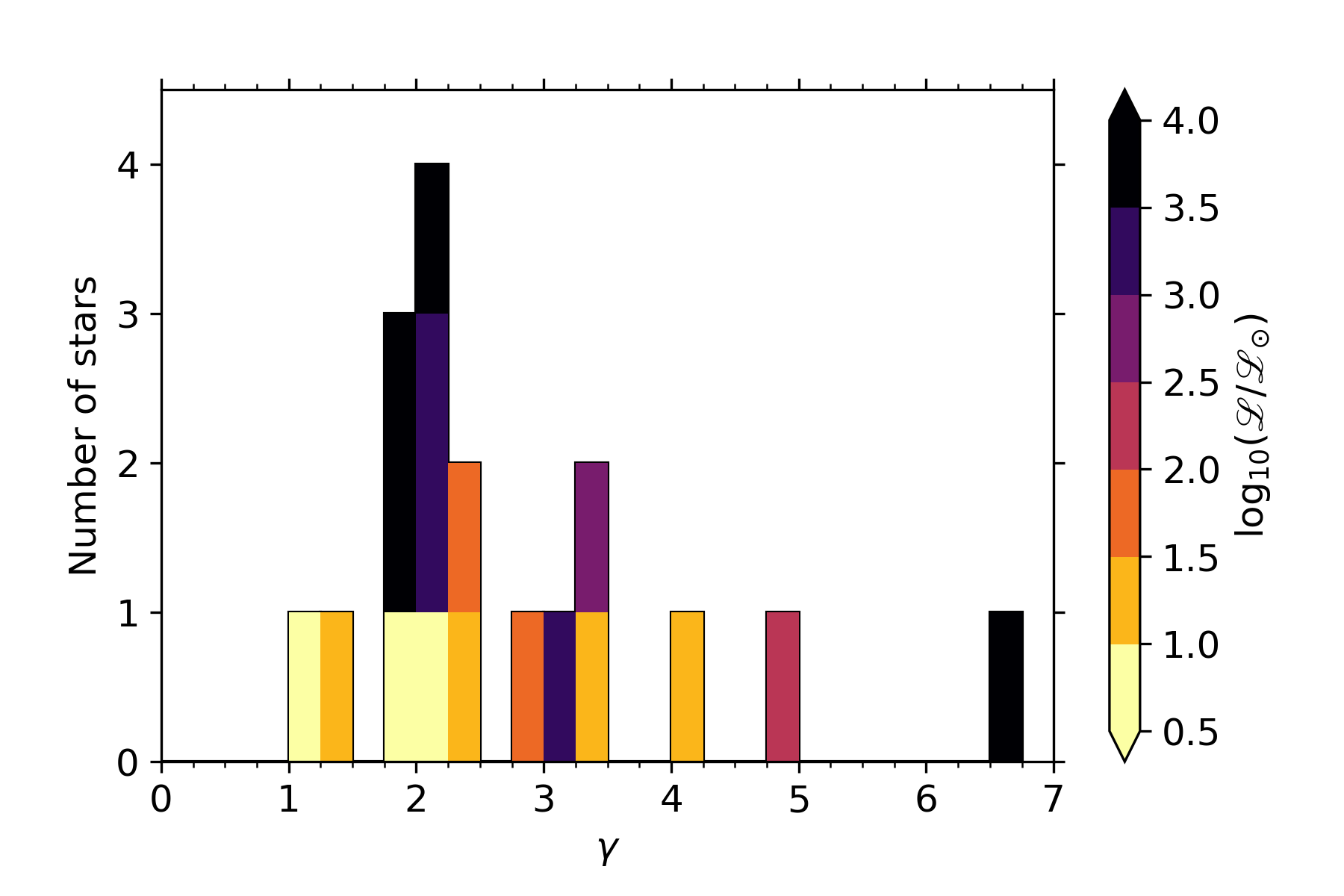}
	\caption{Histogram of fitting parameter $\gamma$ (c.f. Eqn.~\ref{equation: Blomme}) for all stars with spectroscopic parameters and sufficient CoRoT data for an accurate background fit to the residual power density spectrum}.
	\label{figure: gamma histogram}
	\end{figure}

	As discussed in section~\ref{section: IGW simulations}, the predicted morphology of IGWs is only weakly dependent on the stellar parameters for intermediate- and high-mass stars, such that we expect the dominant IGWs frequencies to be in the approximate frequency range of $1 \leq \nu \leq 3$~d$^{-1}$ ($11.6 \leq \nu \leq 34.7$~$\mu$Hz), as shown in Fig.~\ref{figure: IGW simulations}. Furthermore, numerical simulations of IGWs predict a low-frequency power excess with a frequency exponent between $-1$ and $-3$ dependent on the rotation rate and interior differential rotation profile of a star \citep{Rogers2013b, Rogers2015}. A histogram of the $\gamma$ values of the 18 stars with reliable background fits is shown in Fig.~\ref{figure: gamma histogram}, which is colour-coded by each star's spectroscopic effective luminosity, and demonstrates that almost all stars have a low-frequency power excess with $\gamma \leq 5$ as predicted by numerical simulations of IGWs \citep{Rogers2013b, Rogers2015}.
	
	Our results show that the majority of the stars in our sample have a low-frequency power excess, albeit with different amplitudes, which is consistent with the predictions of numerical simulations of a spectrum of IGWs with multiple values of $\ell$ and $m$ \citep{Rogers2013b, Rogers2015}. Therefore, from the grouping stars in Fig.~\ref{figure: granulation}, which implies that granulation is unlikely to be responsible, and because the distribution of the frequency exponent, $\gamma$, agrees with predictions by numerical simulations of IGWs, we conclude that the low-frequency variability in our sample of O, B, A and F stars is caused by IGWs. This is the only physical mechanism that can cause photometric red noise across the whole mass range, since all stars with a convective core are expected to excite IGWs \citep{Samadi2010c, Shiode2013, Rogers2013b, Rogers2015}. Furthermore, this conclusion is supported by the fact that not all stars in our sample are expected to have red noise caused by stellar winds, granulation or sub-surface convection \citep{Samadi2010c, Cantiello2009a}.
	
	There is an interesting outlier in the HR diagrams shown in Fig.~\ref{figure: IGW HR} and the measured characteristic frequency distribution shown in Fig.~\ref{figure: granulation}, which is the blue supergiant HD~46769. This star has a relatively low surface gravity of $\log\,g < 3.0$, such that this post-main sequence star likely has a surface convection zone. The measured characteristic frequency in HD~46769 lies close to the expected granulation frequency in Fig.~\ref{figure: granulation}. As stars evolve off the main sequence, they will generally move to the left in Fig.~\ref{figure: granulation} because of the large increase in radius. Moreover, they will develop an outer convection zone that becomes deeper as the star evolves further, which could plausibly cause granulation. Therefore, assuming that the granulation scaling relation holds for high-mass stars, it is plausible that the red noise detected in HD~46769 is caused by granulation.
	
	The blue supergiants HD~188209 \citep{Aerts2017a} and HD~91316 \citep{Aerts2018a} are known to have large values of macroturbulence, and low-frequency variability of order a few mmag. Conversely, the blue supergiant HD~46769 studied using spectroscopy by \citet{Aerts2013}, was found to have an unusually low macroturbulence of $v_{\rm macro} < 10$~km\,s$^{-1}$. The low-amplitude red noise of order a few $\mu$mag (i.e. $\alpha_0 \simeq 35$~ppm$^2/\mu$Hz) that we found in the CoRoT photometry of HD~46769 implies a common physical cause for these two observables and warrants further study (e.g. \citealt{Simon-Diaz2017a}). This work emphasises the complementary nature of studying stochastic low-frequency variability in blue supergiants with both photometry and spectroscopy. In a future paper, we will explore the complementarity of studying spectroscopic macroturbulence and photometric red noise using a large sample of blue supergiants observed by \Kepler and K2.


\section{Future prospects with TESS}
\label{section: future}

In the near future, we will be able to further populate our HR~diagram and characterise the morphology of IGWs in massive stars by including stars observed by the TESS mission \citep{Ricker2015}, which was launched on 18 April 2018. To achieve its primary goal of finding transiting exoplanets around the brightest stars in the sky, TESS has a wavelength passband that covers $6000-10\,000$~$\AA$, which is redder than the passbands of the CoRoT and \Kepler missions \citep{Auvergne2009, Borucki2010}. Eventually, TESS will observe a large fraction of the sky, with a cadence of 30~min and observing runs that last from 27~d up to 1~yr in the continuous viewing zones (CVZs) near the ecliptic polar regions \citep{Ricker2015}. Although TESS is optimised to study cool stars, asteroseismology of many different types of stars will be possible. 

\begin{figure}
\centering
\includegraphics[width=0.99\columnwidth]{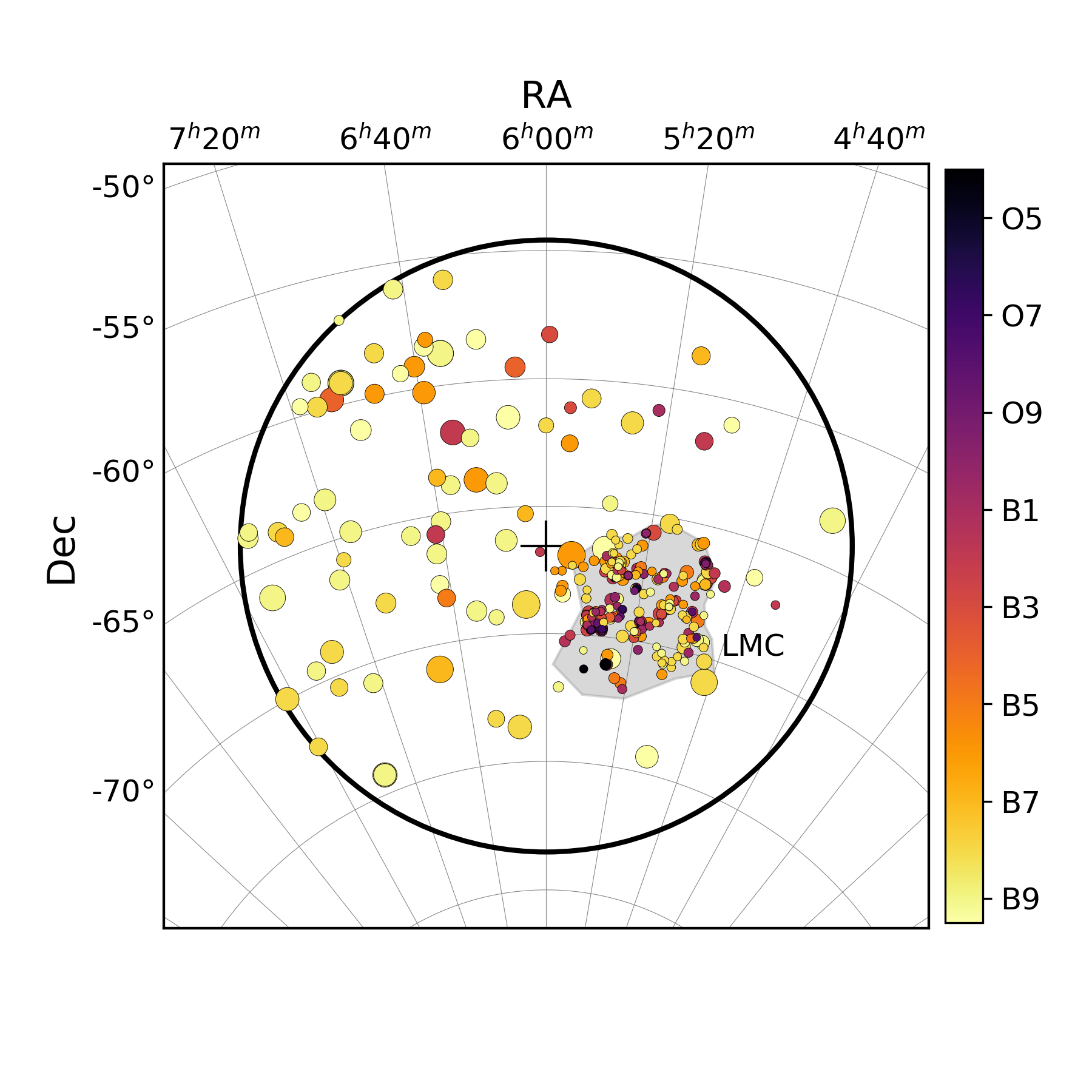}
\caption{The location of the O and B stars that TESS will observe for 1~yr in its southern continuous viewing zone, with the colour-scale indicating the spectral type and the size of the symbol indicating the $V$-band magnitude.}
\label{figure: TESS CVZ}
\end{figure}

The exciting prospect of TESS for upper-main sequence stars is that it will observe many more O and B stars compared to any other previous space-based telescope. Furthermore, TESS will provide continuous high-quality light curves for hundreds of early-O to late-B stars in its southern CVZ. We plot the distribution of these stars in the TESS southern-CVZ in Fig.~\ref{figure: TESS CVZ}, with each circle colour-coded by spectral type and the size denoting the $V$-band magnitude, which ranges between $5 \leq V \leq 13$~mag.

\begin{figure}
\centering
\includegraphics[width=0.99\columnwidth]{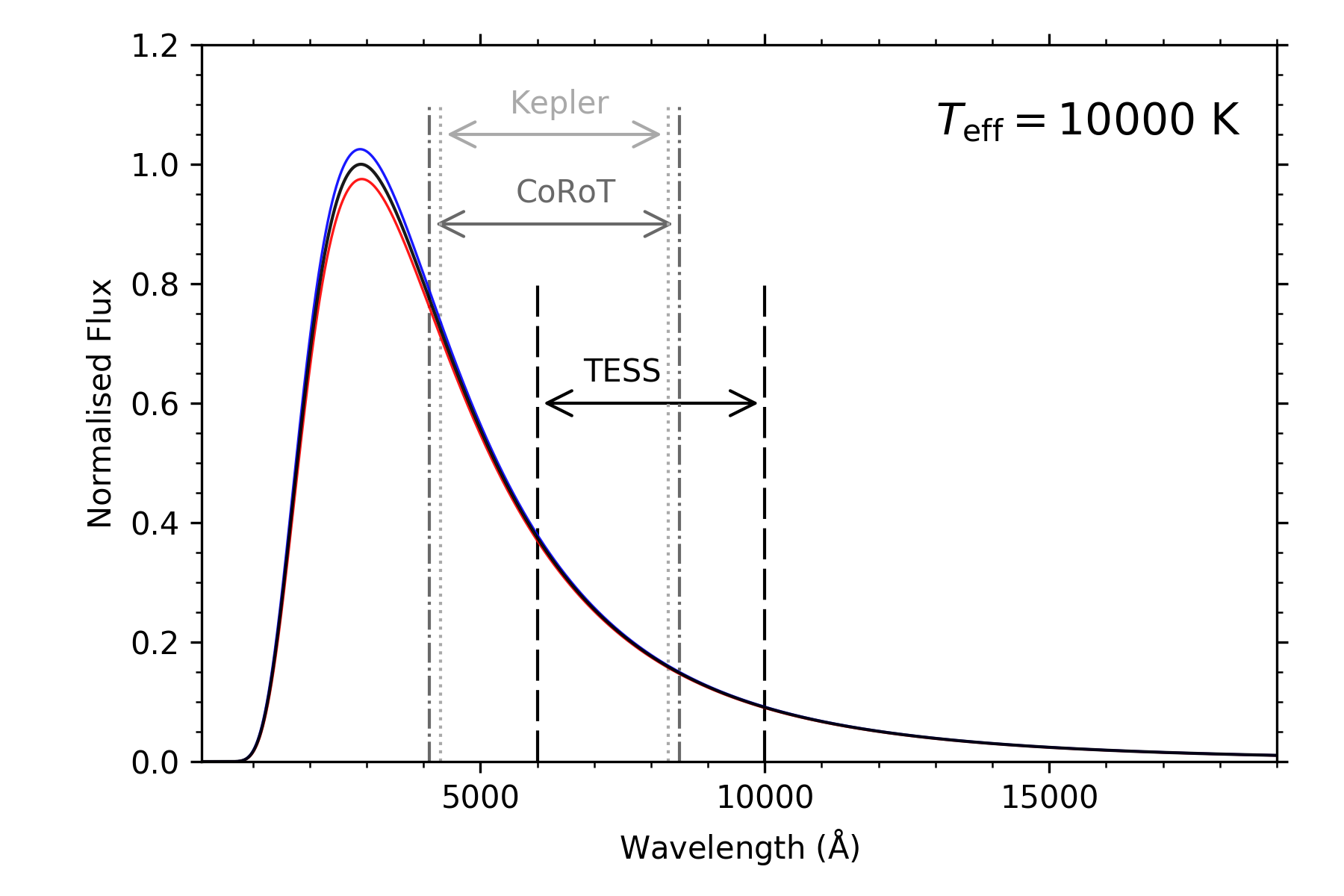}
\caption{The normalised Planck blackbody function for a $T_{\rm eff} = 10\,000$~K star is shown as the solid black line, with hotter and cooler perturbations caused by pulsations shown as the blue and red solid lines, respectively. The approximate wavelength ranges for TESS, CoRoT and \Kepler passbands are shown as dashed, dash-dot and dotted lines, respectively.}
\label{figure: TESS passband}
\end{figure}

However, an important consequence of the TESS wavelength passband for massive stars is the photometric suppression of a star's intrinsic variability. This is illustrated in Fig.~\ref{figure: TESS passband}, in which the Planck blackbody function for a representative $T_{\rm eff} = 10\,000$~K star is shown. The simulated flux perturbations caused by pulsations at the surface of a star perturb the $T_{\rm eff}$ to slightly hotter and cooler values during a pulsation cycle, which are represented by the blue and red solid lines in Fig.~\ref{figure: TESS passband}, respectively. The flux variations, ${\rm d}F/F$, caused by pulsations can be approximated as the difference between the area beneath the Blackbody curve at the temperature maximum and minimum within the wavelength range imposed by an instrument's passband. 

For a first-order estimate of the expected TESS pulsation amplitude suppression factor for massive stars, we use the wavelength range for which the quantum efficiency of an instrument's CCD is above 50~per~cent, which corresponds to an approximate wavelength of $4100-8500$~$\AA$ for CoRoT, with a similar range of $4300-8300$~$\AA$ for \Kepler since both instruments cover the visible wavelength range, and $6000-10\,000$~$\AA$ for TESS. Our calculations for the ratio of TESS and CoRoT, and TESS and \Kepler instruments yield an estimate of the photometric amplitudes that TESS will observe relative to the CoRoT and \Kepler telescopes, respectively. Individual values for representative effective temperatures are given in Table~\ref{table: TESS passband} but these values include large assumptions and are only given for illustrative purposes. Of course, the true test of these estimations will come when TESS observes stars that have previously been observed by CoRoT and {\it Kepler}, and the ratios can be measured directly --- see \citet{Bowman2015a} for an example.

\begin{table}
\caption{The expected TESS pulsation amplitudes for high-mass stars expressed as a relative percentage compared to the CoRoT and \Kepler passbands.} 
\centering
\begin{tabular}{c c c}
\hline \hline
$T_{\rm eff}$ & ${A_{\rm TESS}}/{A_{\rm CoRoT}}$ & ${A_{\rm TESS}}/{A_{\rm Kepler}}$ \\
(K) & $(\%)$ & ($\%$)   \\
\hline
7000 	&	47	&	53	\\
10\,000 	&	38	&	43	\\
15\,000 	&	32	&	38	\\
20\,000 	&	31	&	36	\\
30\,000 	&	29	&	35	\\
40\,000 	&	29	&	34	\\
\hline \hline
\end{tabular}
\label{table: TESS passband}
\end{table}

Taking the example of an early-F star with $T_{\rm eff} = 7000$~K star, photometric amplitudes will be approximately 50~per~cent lower than those observed by CoRoT and \Kepler --- i.e. a pulsation mode amplitude of 1~mmag in a star will be approximately 0.5~mmag as observed by TESS because of the redder passband. For a more massive, and hence hotter, star such as an O star with an effective temperature of $T_{\rm eff} = 30\,000$~K, the suppression factor is larger and hence the ratios $A_{\rm TESS}/A_{\rm CoRoT}$ and $A_{\rm TESS}/A_{\rm Kepler}$ become smaller because the peak of the Blackbody function is at bluer wavelengths. Therefore, the observational signatures of pulsations in massive stars, whether they are coherent or damped pulsation modes, will be suppressed in amplitude because of the TESS passband. Hence, this bias towards detecting photometric variability in cooler stars should be taken into account when studying the variability of massive stars, as redder passbands underestimate the true photometric variability.


\section{Conclusions}
\label{section: conclusions}

Previous studies that looked at individual or small samples of massive stars, i.e. $M \gtrsim 15$~M$_{\rm \odot}$, have noted the presence of astrophysical red noise in the photometry of these stars (e.g. \citealt{Blomme2011b}). Typically, the origin of a low-frequency power excess in massive stars has been interpreted as being caused by a dynamic stellar wind, granulation or sub-surface convection, or more recently IGWs. All of these physical explanations are plausible for massive stars, but are difficult to disentangle especially because they are not mutually exclusive for an individual star. However, recent studies have begun to unravel the different mechanisms causing stochastic variability that may be dominant in different mass regimes and at different evolutionary stages for massive stars (e.g. \citealt{Aerts2017a, Aerts2018a, Simon-Diaz2018a}).

In our study using CoRoT photometry, we demonstrate that the physical interpretation of stochastic variability in high-mass stars requires a long and continuous time series to provide the necessary frequency resolution to remove coherent modes via iterative pre-whitening. We have confronted the 2D numerical simulations of IGWs from \citet{Rogers2013b} with photometry of numerous O, B, A and F stars. We parameterise the low-frequency power excess in these stars and demonstrate the diverse range in the morphology of their stochastic low-frequency variability. For some of the A and F stars in our sample, we find that the morphology of the photometric red noise is consistent with that expected from the granulation scaling relation (e.g. \citealt{Kallinger2010c}). Yet this is certainly not the case for all A and F stars in our sample. Furthermore, we also demonstrate that the red noise in post-main sequence stars is also consistent with granulation including the blue supergiant HD~46769, which is expected to have a convective envelope.

The majority of the stars in our study required a significantly different parametrisation of the low-frequency power excess, which were not compatible with their expected characteristic granulation frequencies. As demonstrated in Fig.~\ref{figure: multi star FT}, the diversity in the observed red noise profiles is clearly large for the stars in our sample. Yet, the measured characteristic frequencies in the majority of our stars are between $1 \leq \nu \leq 4$~d$^{-1}$ ($11.6 \leq \nu \leq 46.3$~$\mu$Hz), which is approximately an order of magnitude smaller than expected for granulation and coincides within the frequency range of where the amplitudes of IGWs are predicted to be at their highest \citep{Rogers2013b, Rogers2015}. 

We demonstrate that the peaks in the residual background spectra are variable in amplitude and frequency, and have amplitudes that are of order 10~$\mu$mag in A and F stars and increase to a few hundred $\mu$mag for O and B stars. These values are comparable as those predicted by \citet{Shiode2013}, and are similarly correlated with stellar mass. Hence we find the first observational evidence of relationship between the photometric amplitudes of IGWs and stellar mass for main-sequence stars. This further supports the interpretation of IGWs as an explanation for the observed photometric red noise, since the larger convective cores in massive stars are predicted to excite IGWs with larger amplitudes \citep{Shiode2013, Aerts2015c}. 

Recent theoretical work and 3D numerical simulations of IGWs triggered by turbulent convection in a Cartesian model by \citet{Couston2018b} predict the same power-law frequency spectrum for IGWs as \citet{Lecoanet2013}, with a frequency exponent of $-13/2$ (corresponding to $\gamma=6.5$). This is steeper than the power-law predicted by 3D simulations of core convection in a 3-M$_{\rm \odot}$ star by \citet{Edelmann2018a**}. The latter simulations take into account the spherical configuration, as well as an appropriate density stratification from a realistic stellar model. These simulations rather predict an IGW frequency spectrum with a frequency exponent between $-1.5$ and $-3$ (i.e., $\gamma\in[1.5,3]$), which is similar to other simulations of core convection by \citet{Rogers2013b} and \citet{Augustson2016a}.  Our analysis of CoRoT stars with a convective core leads to $\gamma \leq 5$ for the majority of well-characterised stars.

A theoretically-predicted power-law excess caused by wind instabilities in the massive O stars HD~46150, HD~46223 and HD~46966 has also recently been investigated by \citet{Krticka2018d}. Whilst wind instabilities could potentially provide an plausible explanation for the most massive (i.e. $M > 15$~M$_{\rm \odot}$) stars in our sample, it is not a physical explanation for the similar low-frequency excess observed in intermediate-mass B, A and F stars.

In this work, we have demonstrated that the morphology of the low-frequency power excess observed in many O, B, A, and F stars, which we have parametrised using Eqn.~(\ref{equation: Blomme}), is not consistent with granulation (c.f. figure~\ref{figure: granulation}). Furthermore, as demonstrated in Fig.~\ref{figure: gamma histogram}, the majority of stars required $\gamma \leq 5$, which is consistent with the frequency exponent predictions of an entire frequency spectrum of IGWs of various length scales \citep{Rogers2013b, Rogers2015, Edelmann2018a**}. These observations are useful for constraining the driving and damping processes in numerical simulations of core convection and IGWs for various stellar masses and stages of stellar evolution.

Regardless of the interpretation of the low-frequency power excess found in many stars in our study, we provide valuable constraints on the maximum amplitude of the background IGW spectrum across the HR~diagram. Guided by the results of the present study that main-sequence massive stars represent the best targets for detecting IGWs, we can probe the differences between observations and the frequency spectra predicted by simulations of IGWs and sub-surface granulation. Our current study represents an important ensemble characterisation of variability in massive stars, which demonstrates the diverse range of physical phenomena at work in early-type stars. We are currently limited to a few dozen early-type stars by CoRoT, which were only observed for a relatively short time span, until TESS observations of O and B stars become available. In a subsequent paper, we will further populate our asteroseismic HR~diagram with many more O and B stars using photometry from the K2 and TESS missions. The future of asteroseismology of massive O and B stars is bright, with our study representing a necessary first-step in understanding the parameter space of stochastic variability in early-type stars.


\begin{acknowledgements}
The authors thank the \textit{CoRoT} science team for the excellent data and the referee for the useful comments that improved the manuscript. D.\,M\,.B. is grateful to Dr. Kelly Hambleton for useful discussions. The CoRoT space mission was developed and operated by the French space agency CNES, with participation of ESA's RSSD and Science Programmes, Austria, Belgium, Brazil, Germany, and Spain. This research has made use of the \texttt{SIMBAD} database, operated at CDS, Strasbourg, France; the SAO/NASA Astrophysics Data System; and the VizieR catalogue access tool, CDS, Strasbourg, France. The research leading to these results has received funding from the European Research Council (ERC) under the European Union's Horizon 2020 research and innovation programme (grant agreement N$^{o}$670519: MAMSIE). S.\,S.-D. acknowledges financial support from the Spanish Ministry of Economy and Competitiveness (MINECO) through grants AYA2015-68012-C2-1 and Severo Ochoa SEV-2015-0548, and grant ProID2017010115 from the Gobierno de Canarias. Support for this research was provided by STFC grant ST/L005549/1 and NASA grant NNX17AB92G. T.\,V.\,R. gratefully acknowledges support from the Australian Research Council, and from the Danish National Research Foundation (Grant DNRF106) through its funding for the Stellar Astrophysics Centre (SAC).
\end{acknowledgements}


\bibliographystyle{aa}
\bibliography{/Users/Dom/Documents/RESEARCH/Bibliography/master_bib}


\begin{appendix}

\section{Summary figures for all stars}
\label{section: appendix: MCMC results}

Summary figures containing the power density spectra, results of the MCMC analysis and SFTs are presented here. For stars that underwent pre-whitening of high-$S/N$ peaks, the original power density spectrum is shown in orange and the residual power density spectrum used in our analysis is shown in black for all stars except HD~46150 and HD~46223, in which no significant peaks were found in their frequency spectra.





\begin{figure}
\centering
\includegraphics[width=0.49\textwidth]{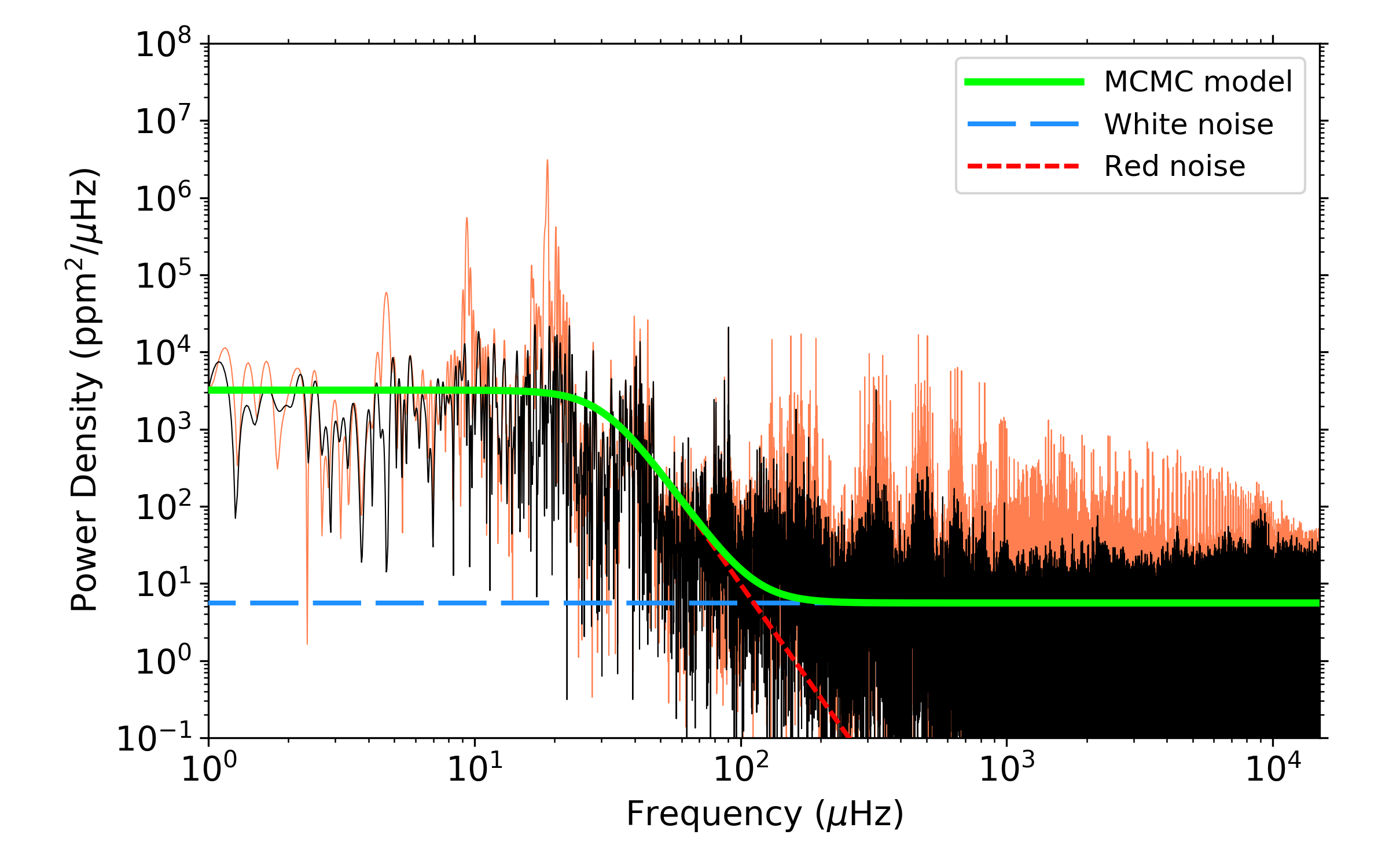}
\includegraphics[width=0.49\textwidth]{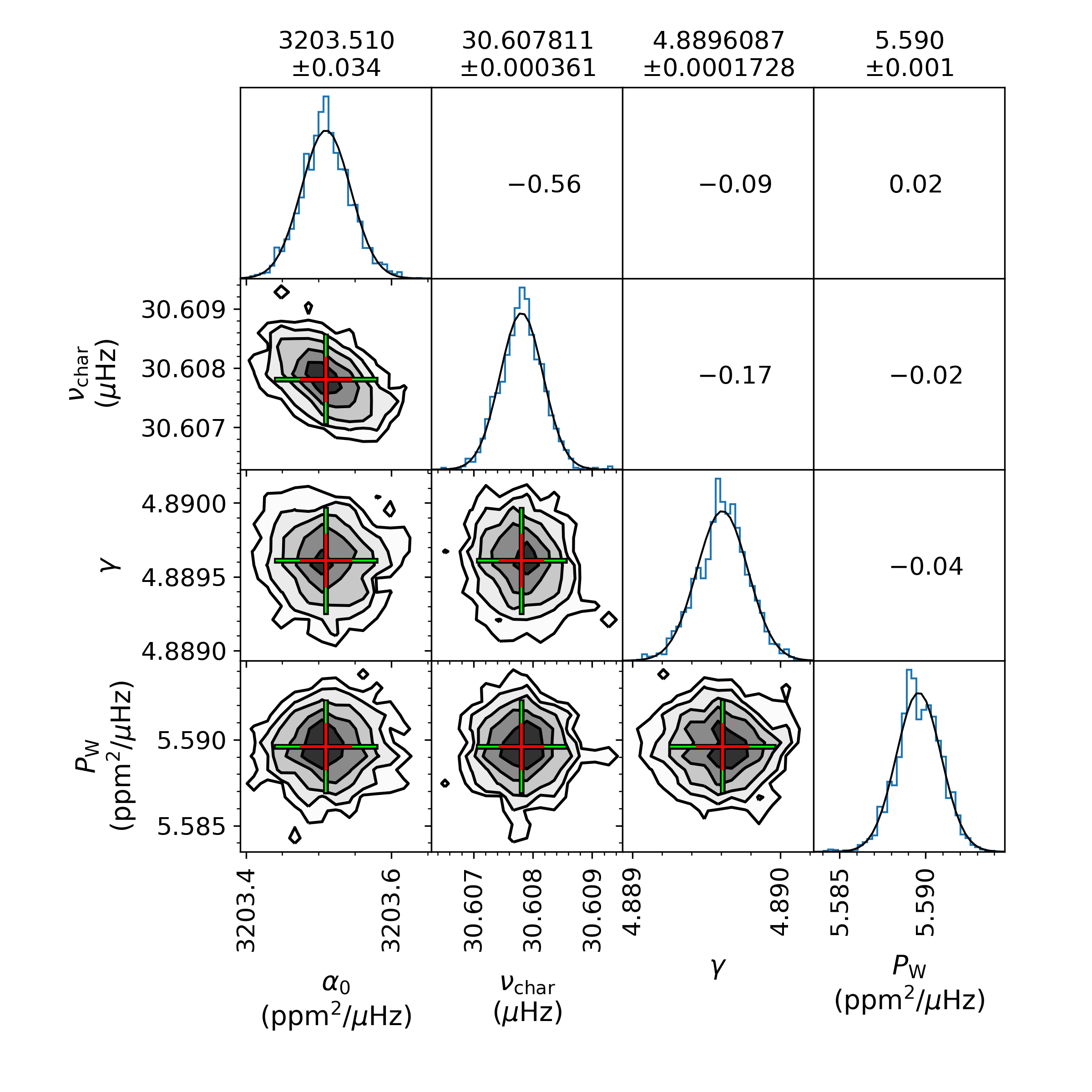}
\includegraphics[width=0.49\textwidth]{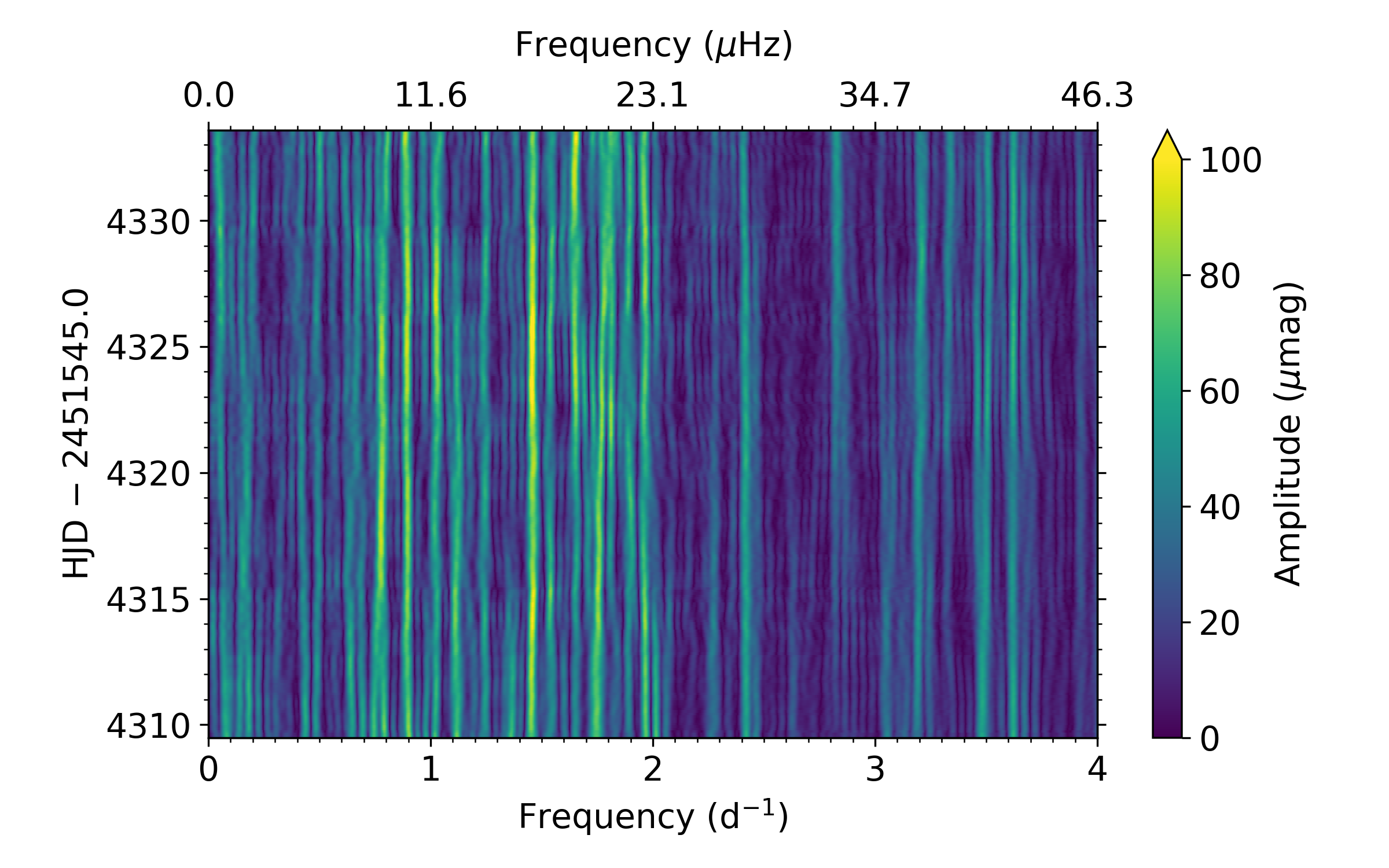}
\caption{Summary figure for the B star HD~45418, which has a similar layout as shown in Fig.~\ref{figure: HD46150}.}
\label{figure: HD45418}
\end{figure}


\begin{figure}
\centering
\includegraphics[width=0.49\textwidth]{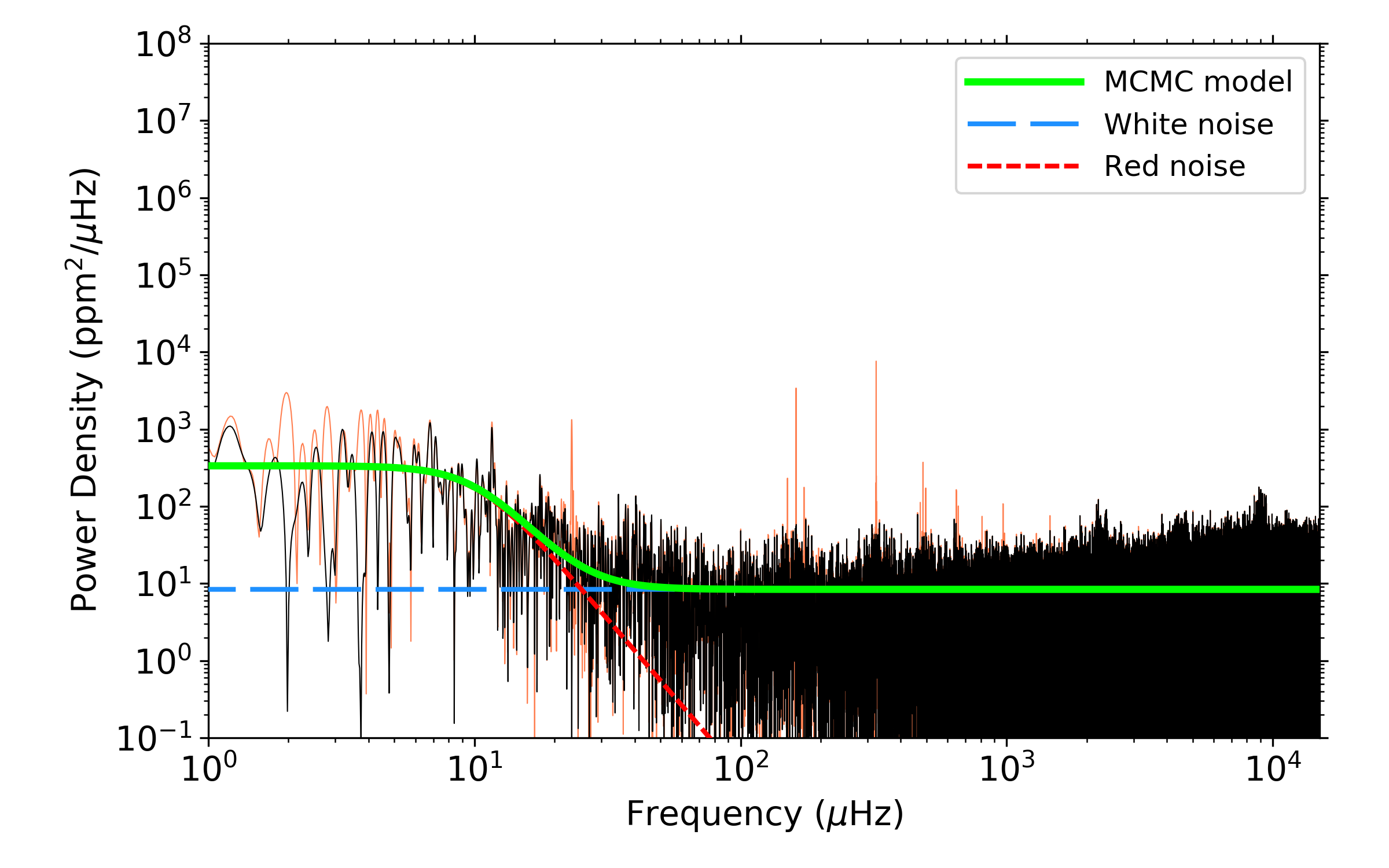}
\includegraphics[width=0.49\textwidth]{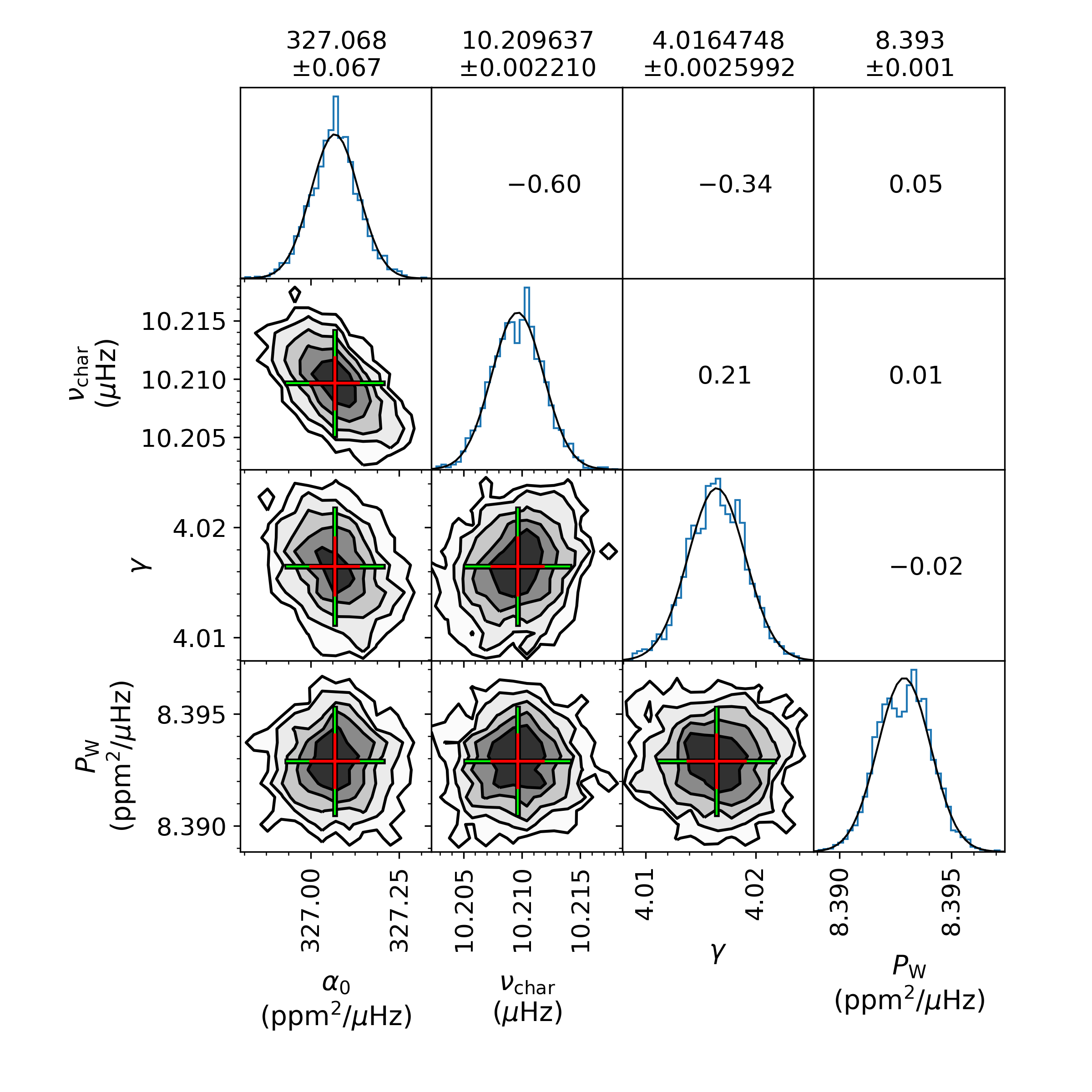}
\includegraphics[width=0.49\textwidth]{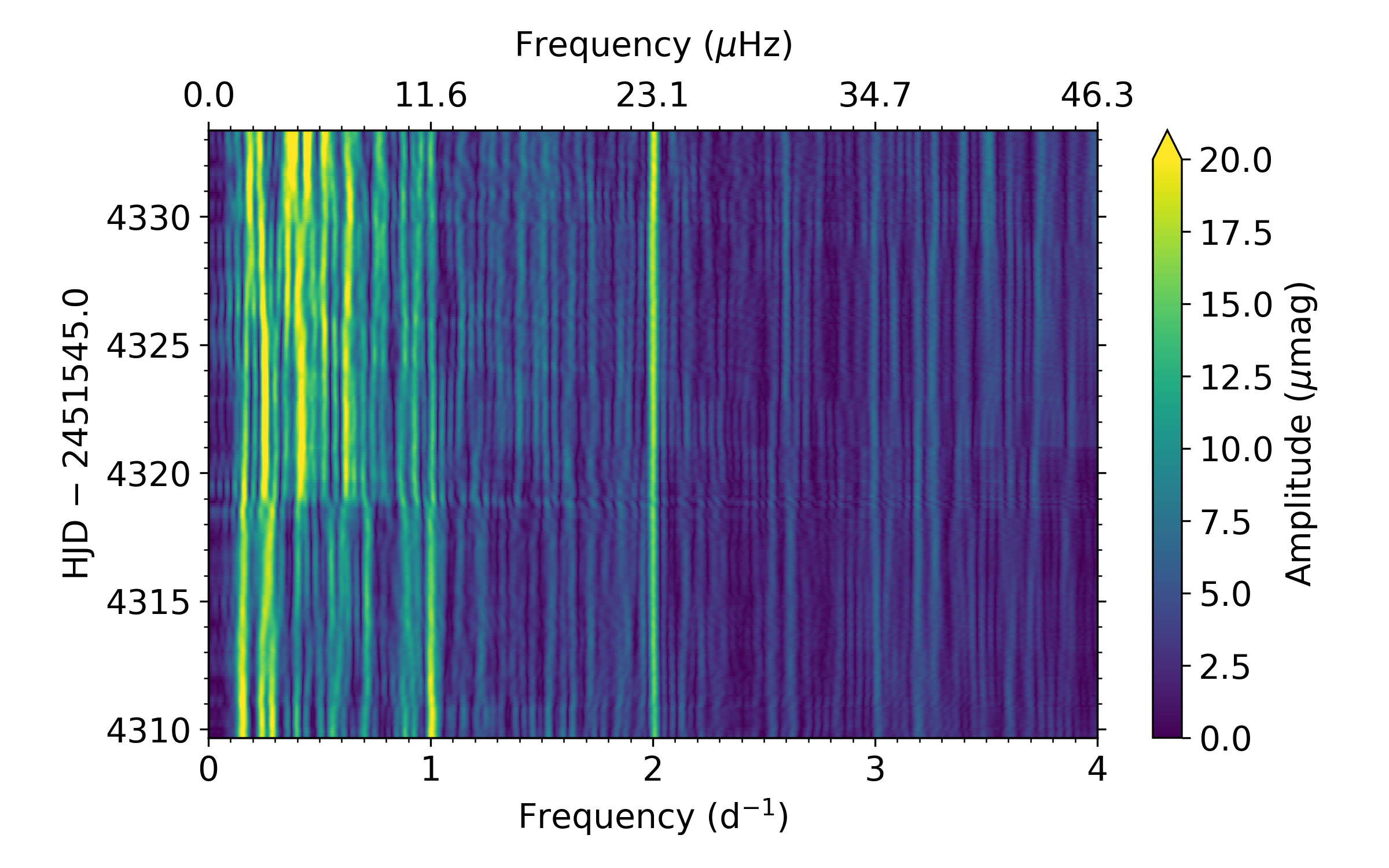}
\caption{Summary figure for the A star HD~45517, which has a similar layout as shown in Fig.~\ref{figure: HD46150}.}
\label{figure: HD45517}
\end{figure}

\clearpage 

\begin{figure}
\centering
\includegraphics[width=0.49\textwidth]{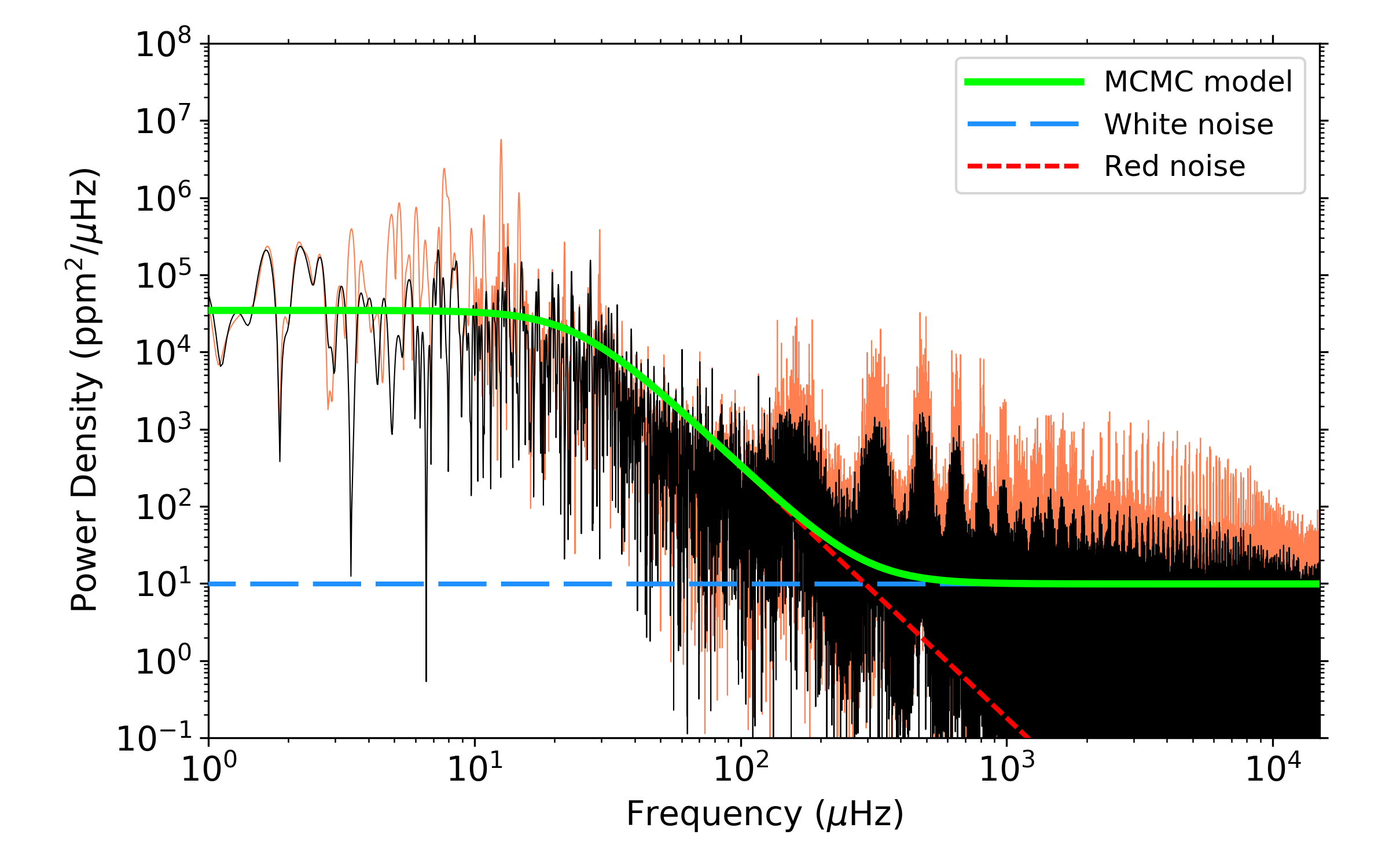}
\includegraphics[width=0.49\textwidth]{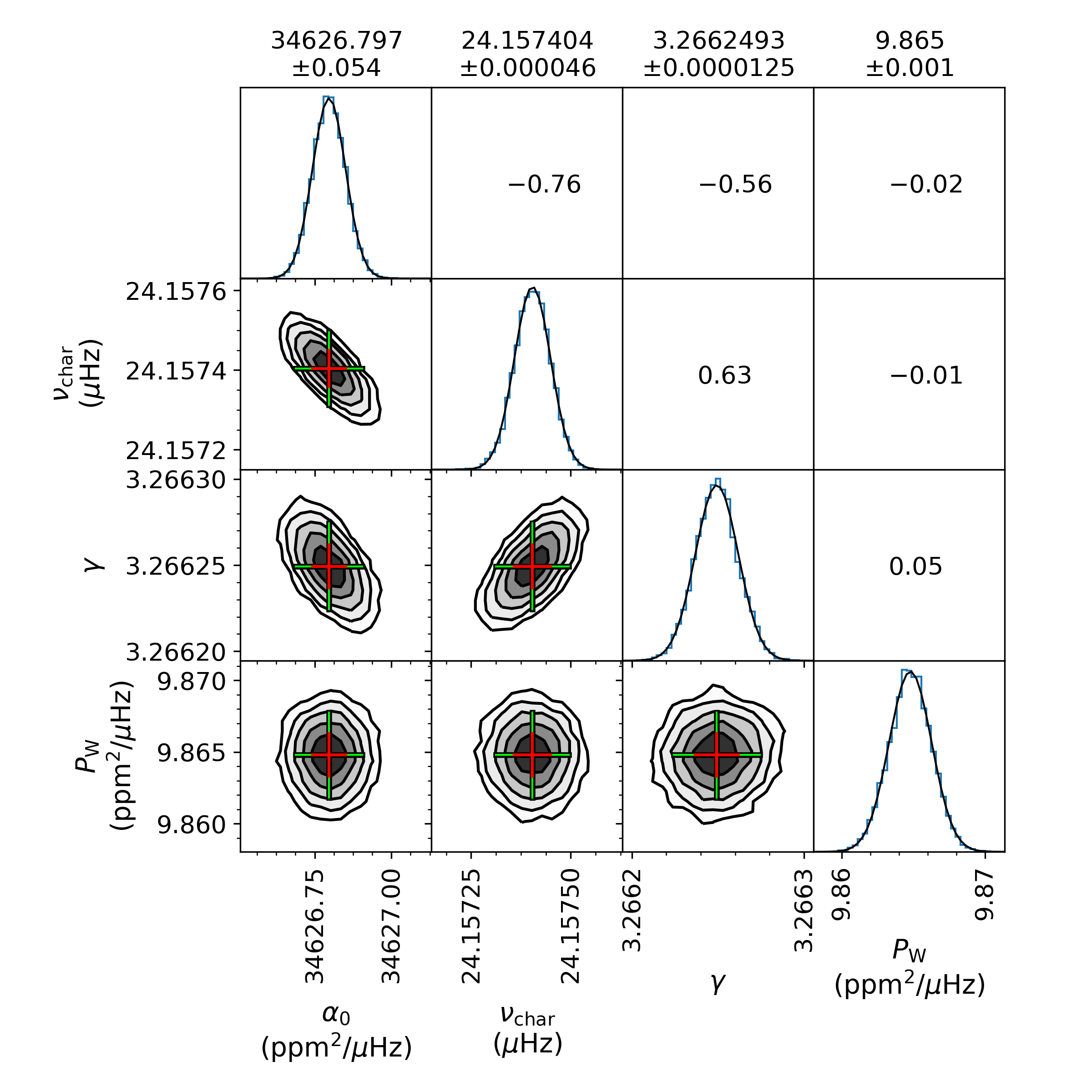}
\includegraphics[width=0.49\textwidth]{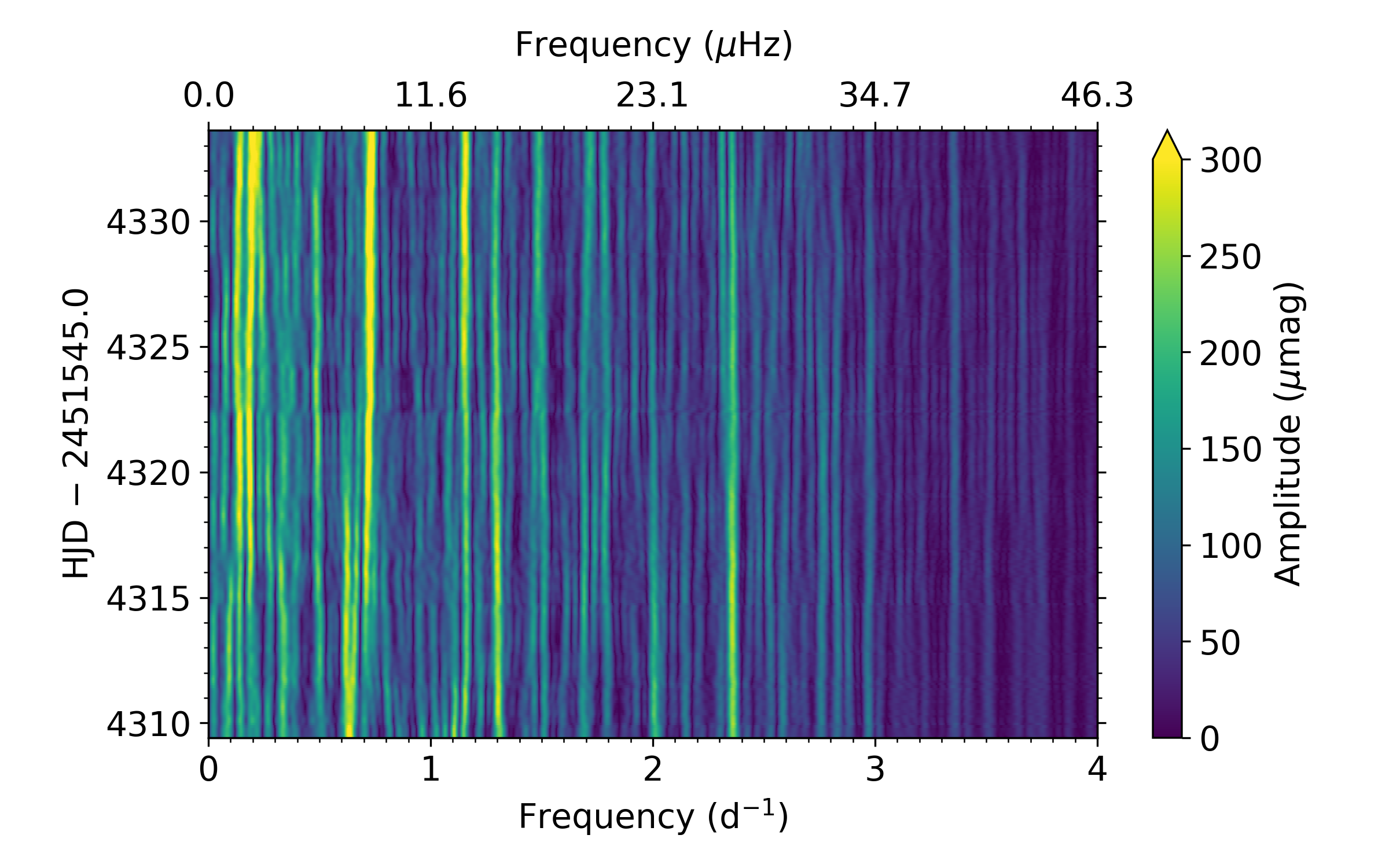}
\caption{Summary figure for the B star HD~45546, which has a similar layout as shown in Fig.~\ref{figure: HD46150}.}
\label{figure: HD45546}
\end{figure}


\begin{figure}
\centering
\includegraphics[width=0.49\textwidth]{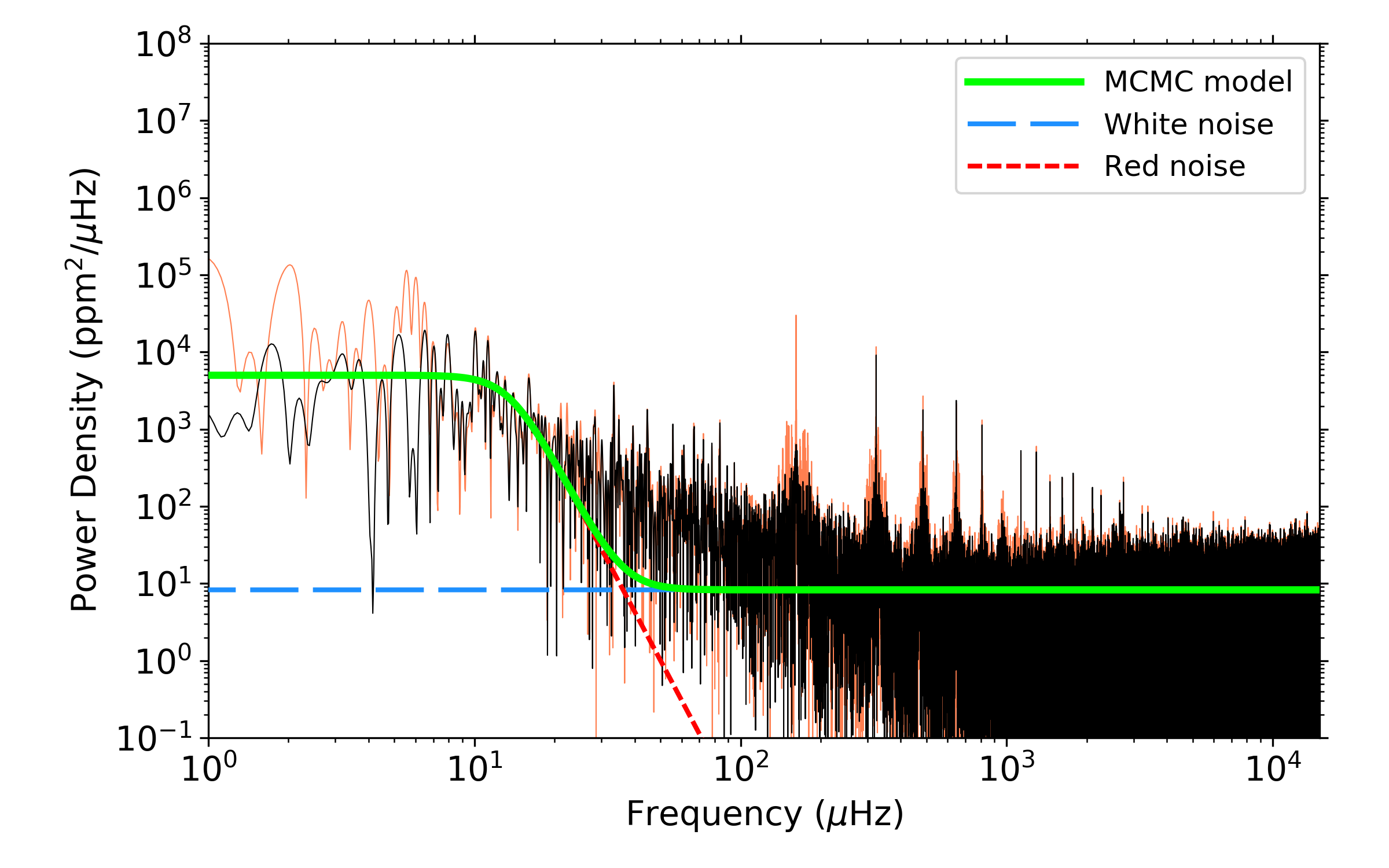}
\includegraphics[width=0.49\textwidth]{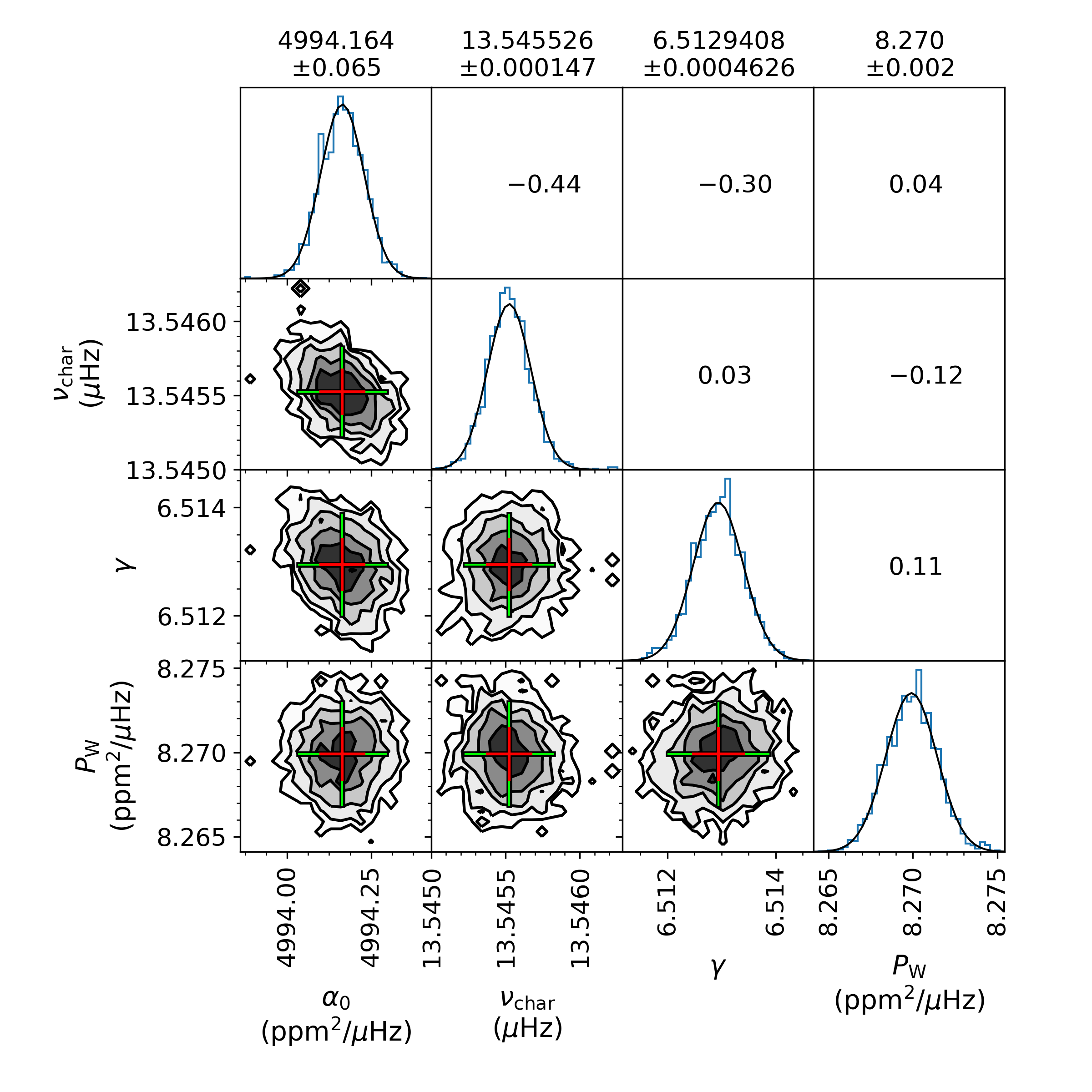}
\includegraphics[width=0.49\textwidth]{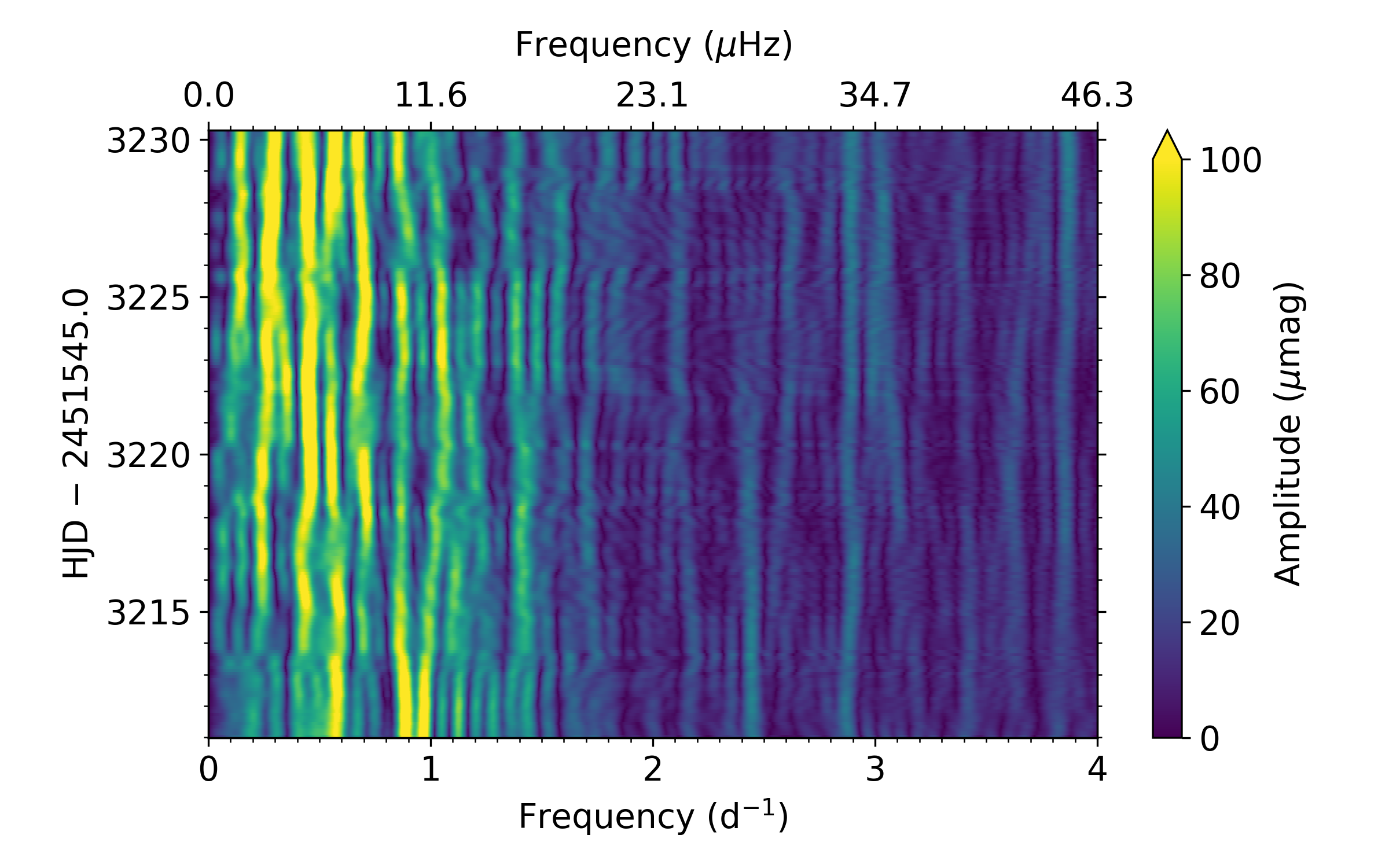}
\caption{Summary figure for the O star HD~46149, which has a similar layout as shown in Fig.~\ref{figure: HD46150}.}
\label{figure: HD46149}
\end{figure}

\clearpage 

\begin{figure}
\centering
\includegraphics[width=0.49\textwidth]{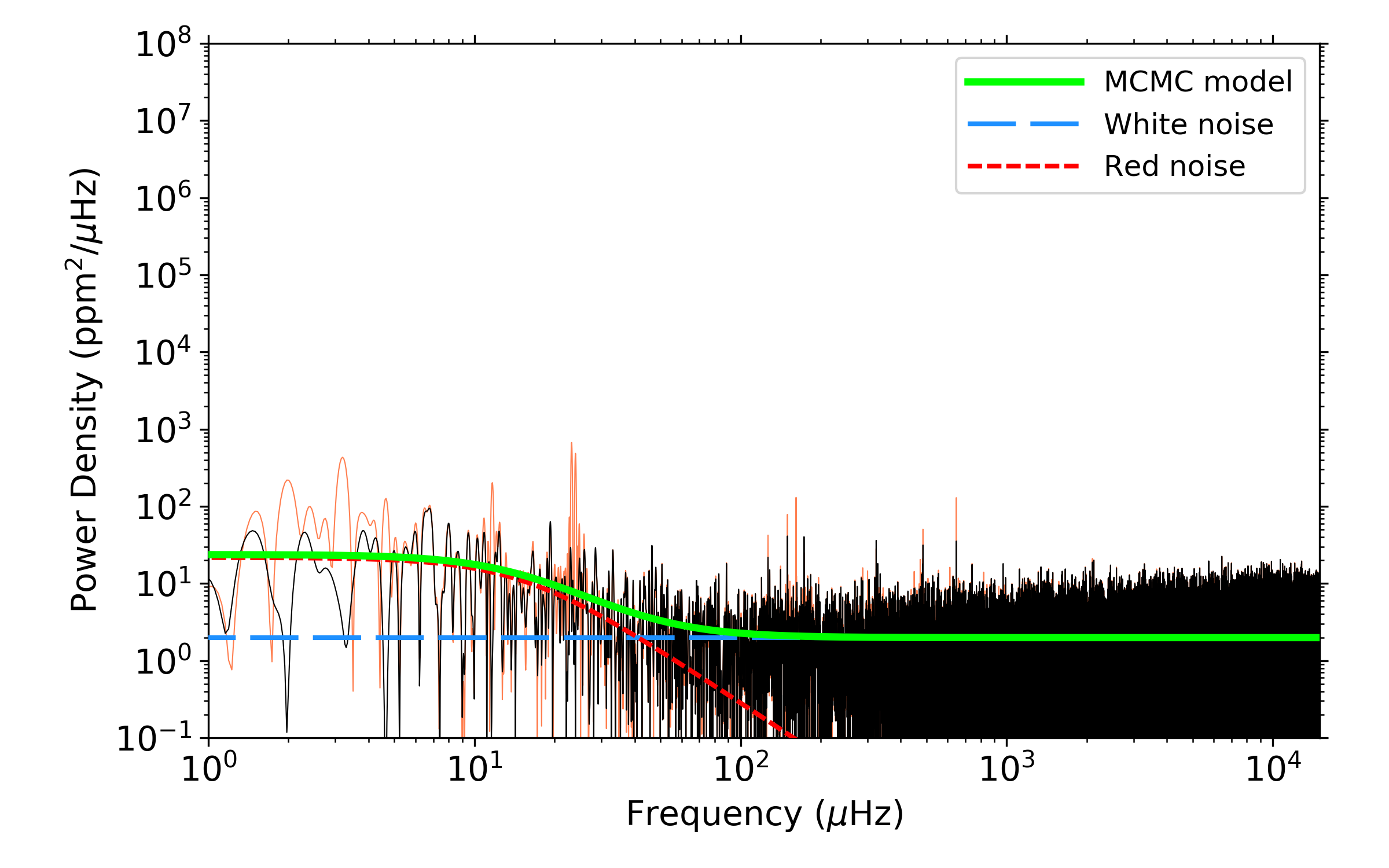}
\includegraphics[width=0.49\textwidth]{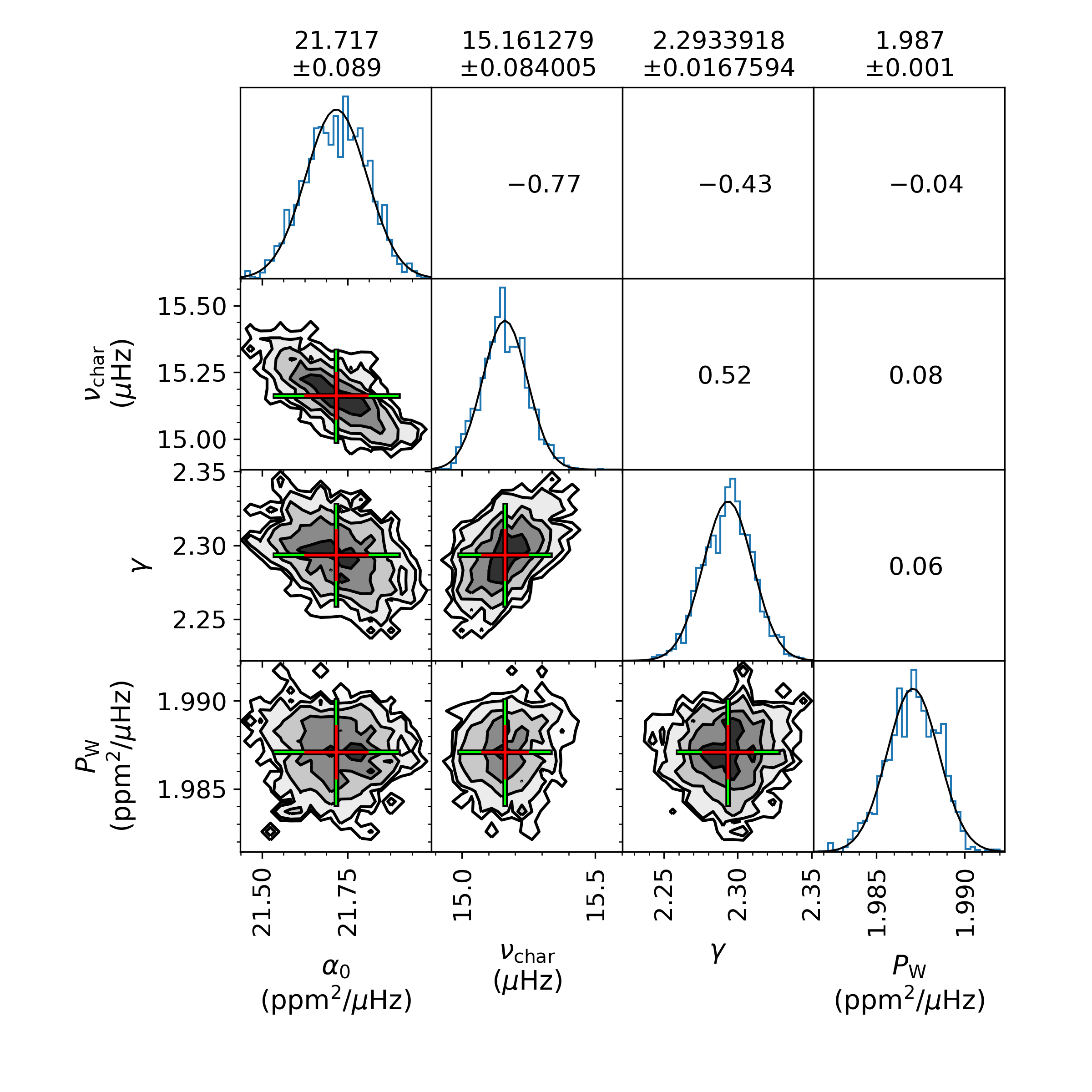}
\includegraphics[width=0.49\textwidth]{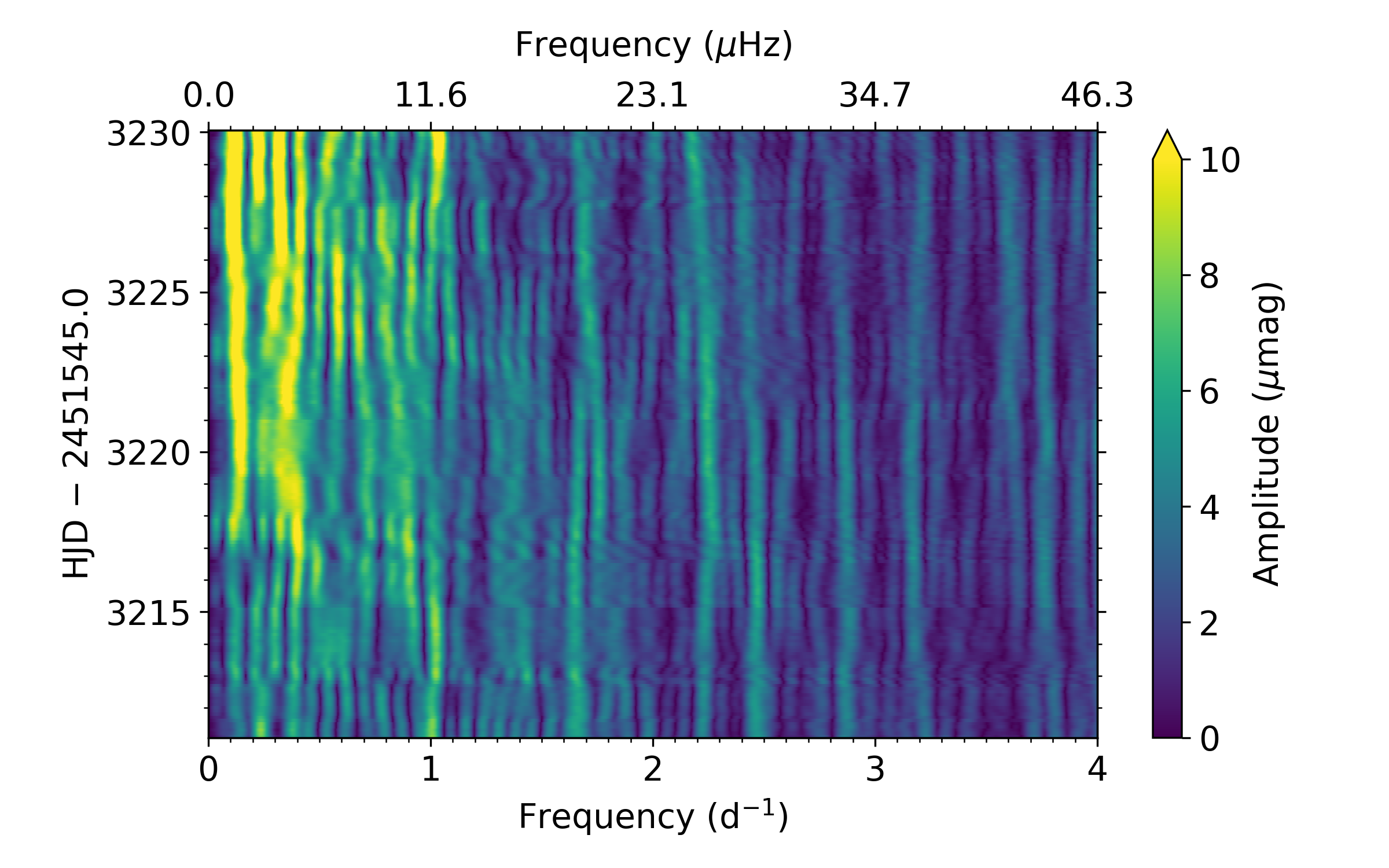}
\caption{Summary figure for the B star HD~46179, which has a similar layout as shown in Fig.~\ref{figure: HD46150}.}
\label{figure: HD46179}
\end{figure}


\begin{figure}
\centering
\includegraphics[width=0.49\textwidth]{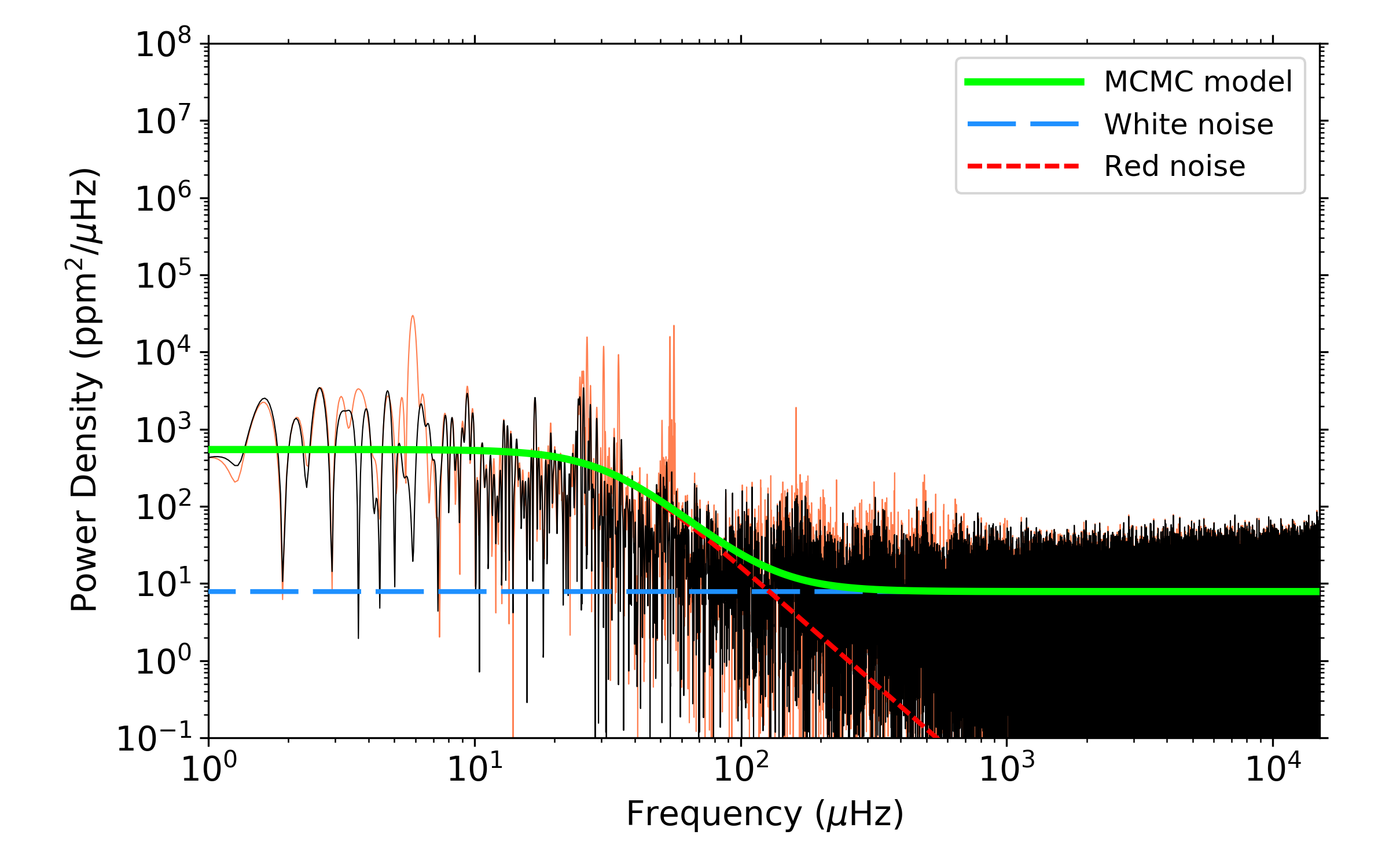}
\includegraphics[width=0.49\textwidth]{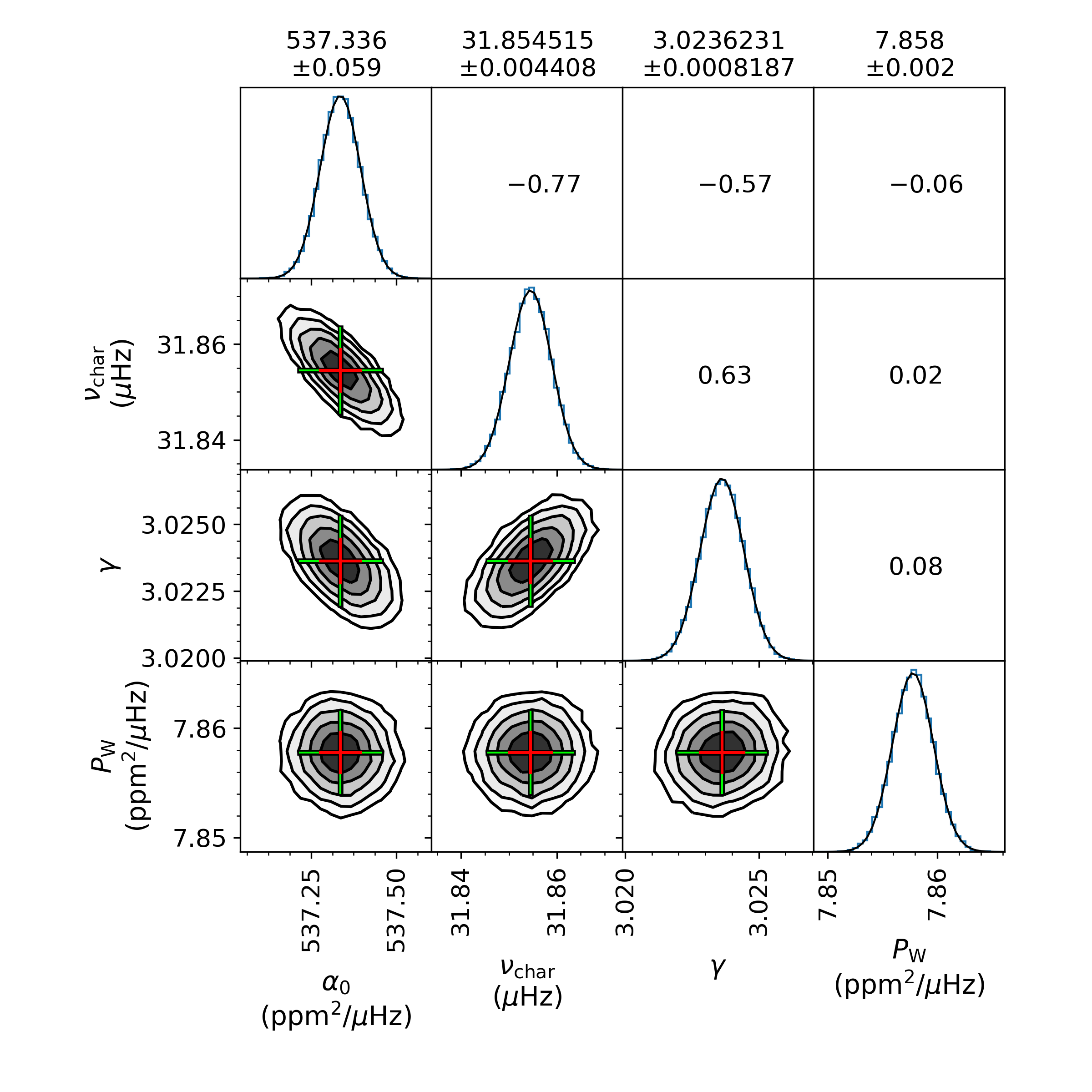}
\includegraphics[width=0.49\textwidth]{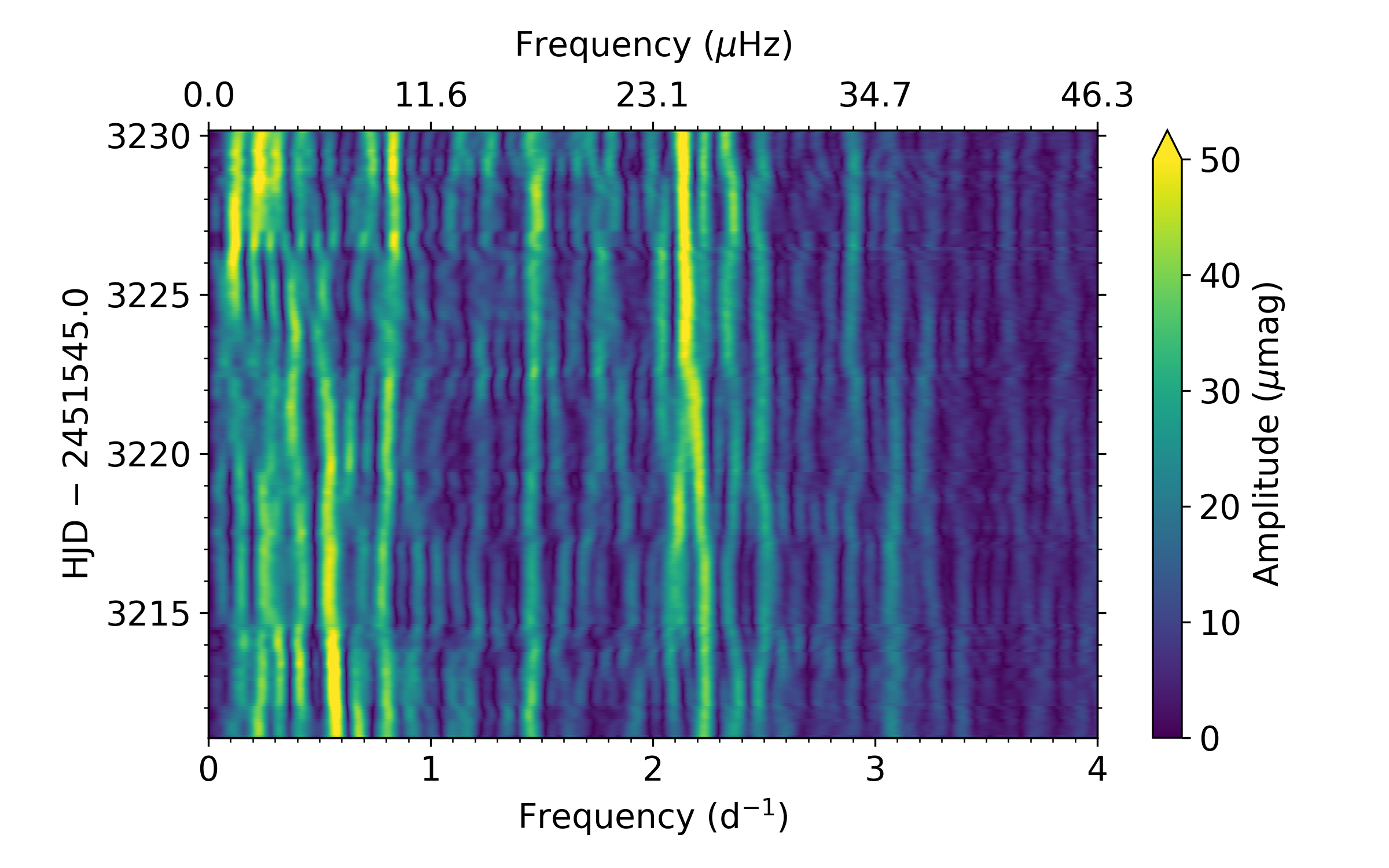}
\caption{Summary figure for the O star HD~46202, which has a similar layout as shown in Fig.~\ref{figure: HD46150}.}
\label{figure: HD46202}
\end{figure}

\clearpage 

\begin{figure}
\centering
\includegraphics[width=0.49\textwidth]{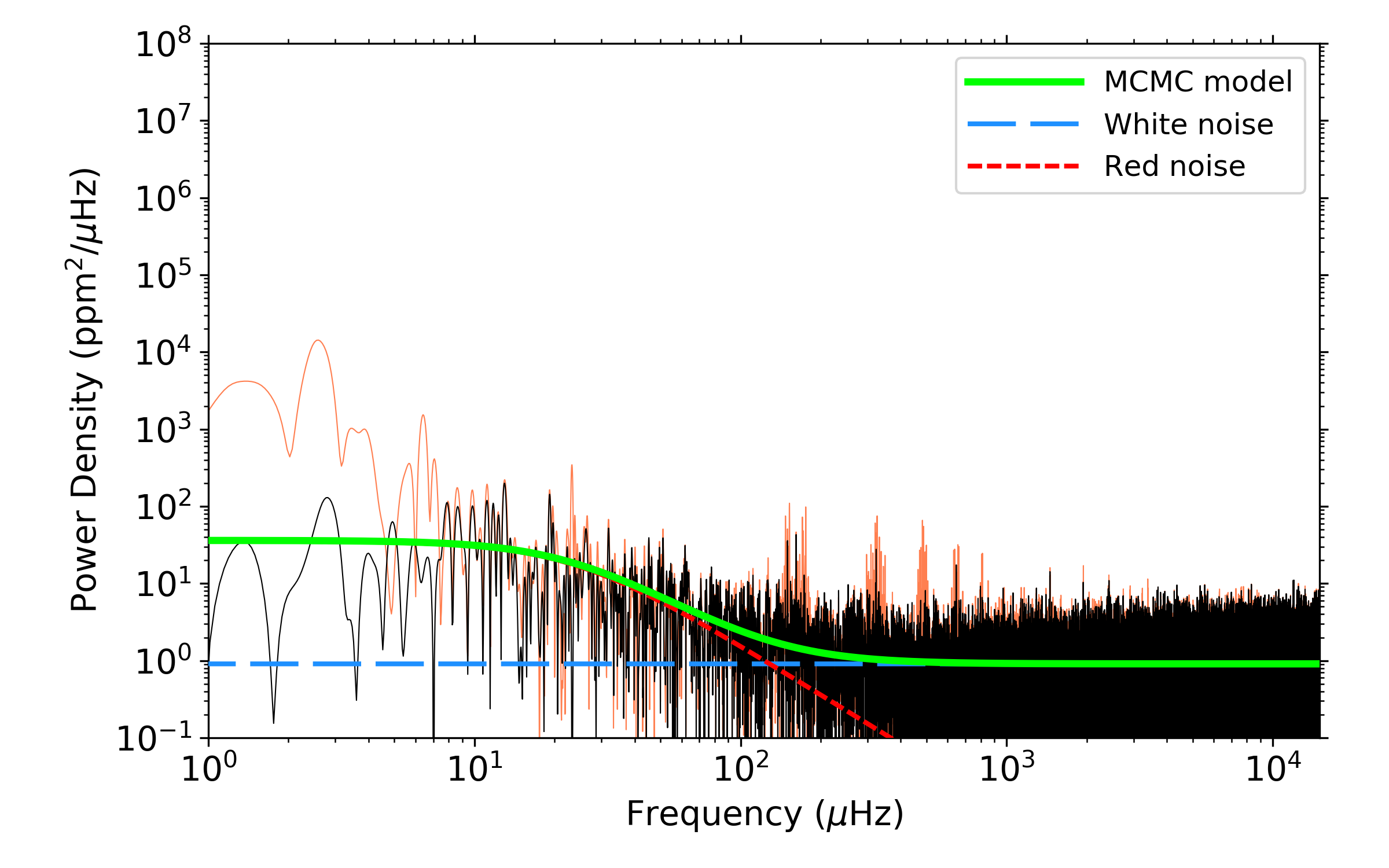}
\includegraphics[width=0.49\textwidth]{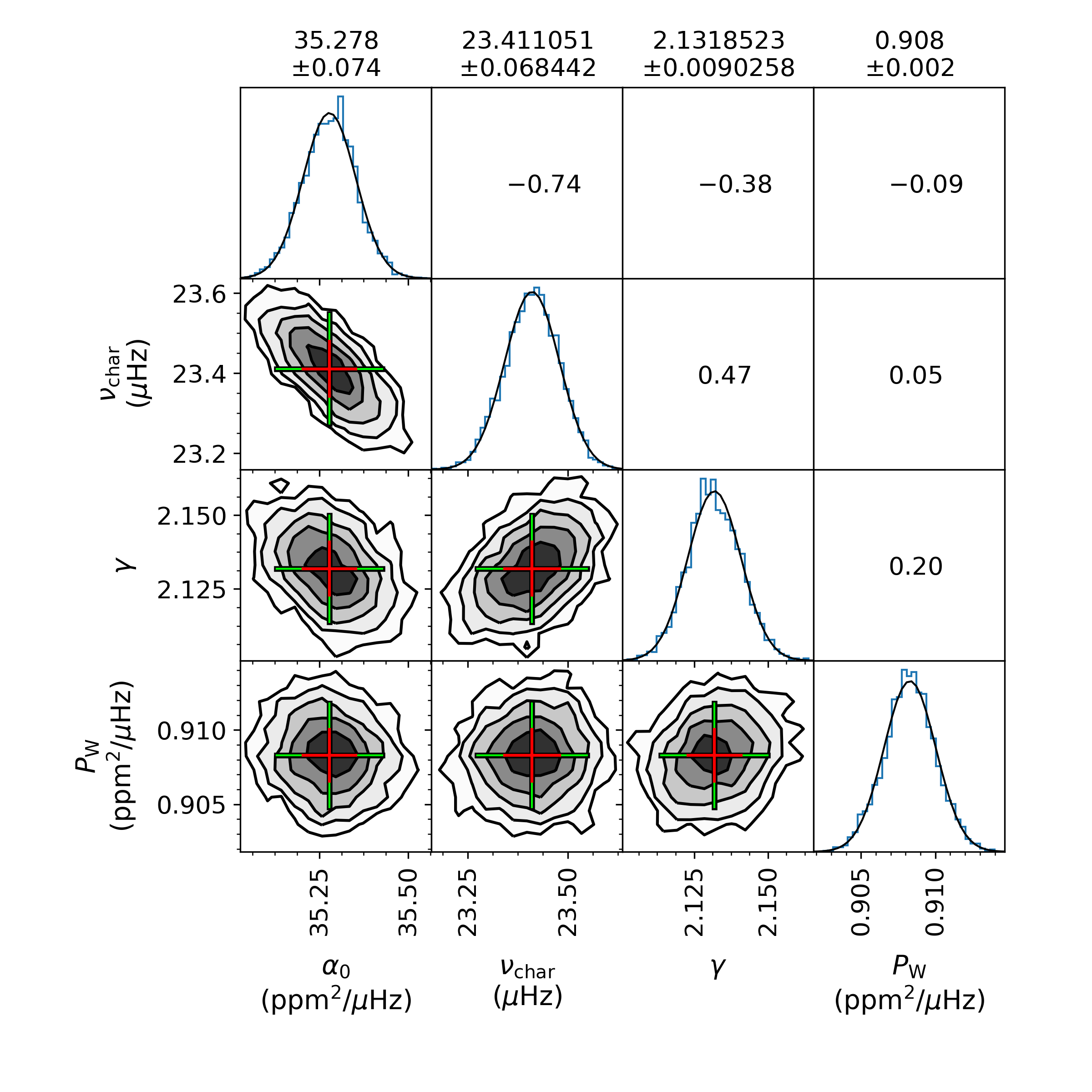}
\includegraphics[width=0.49\textwidth]{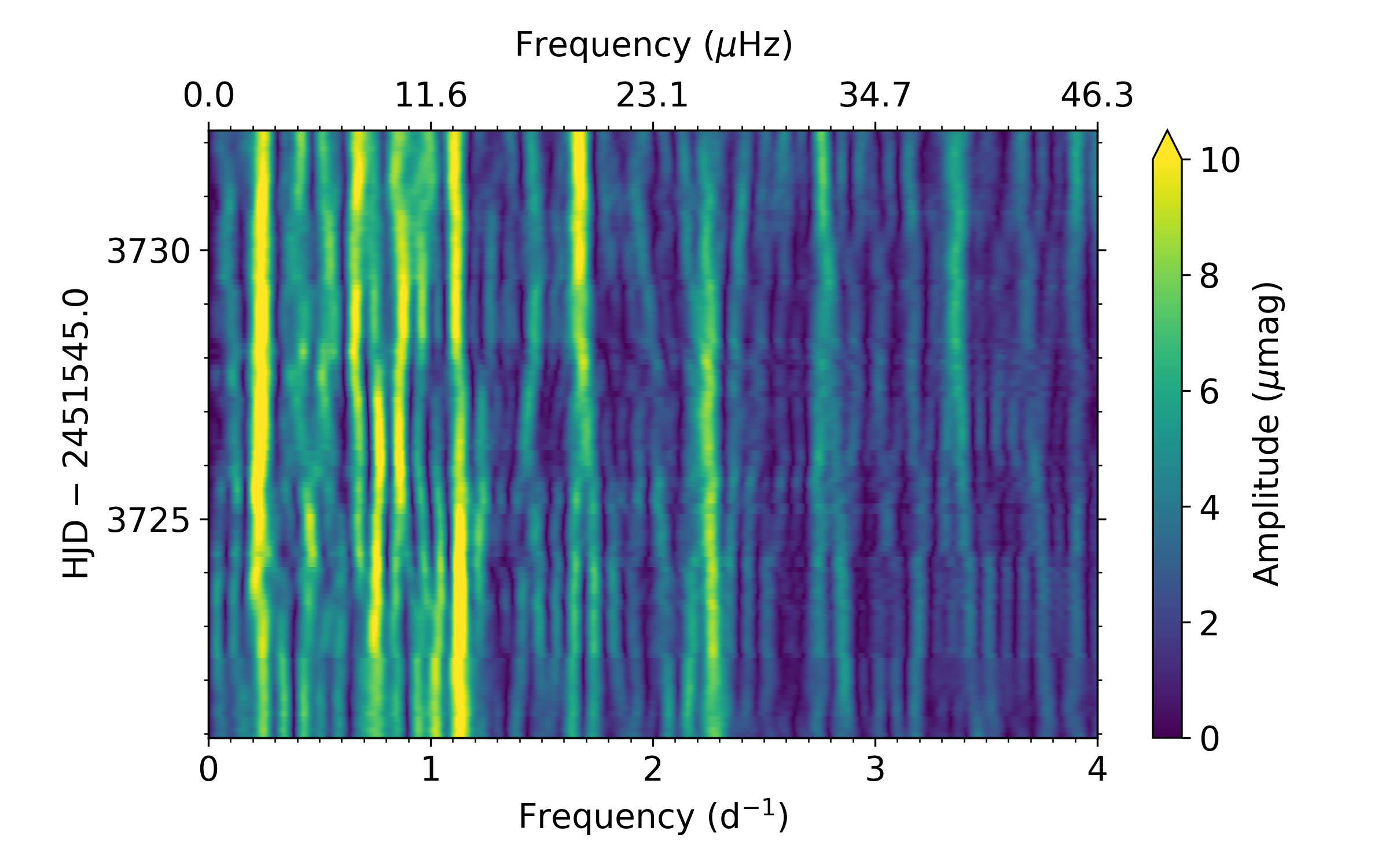}
\caption{Summary figure for the B star HD~46769, which has a similar layout as shown in Fig.~\ref{figure: HD46150}.}
\label{figure: HD46769}
\end{figure}


\begin{figure}
\centering
\includegraphics[width=0.49\textwidth]{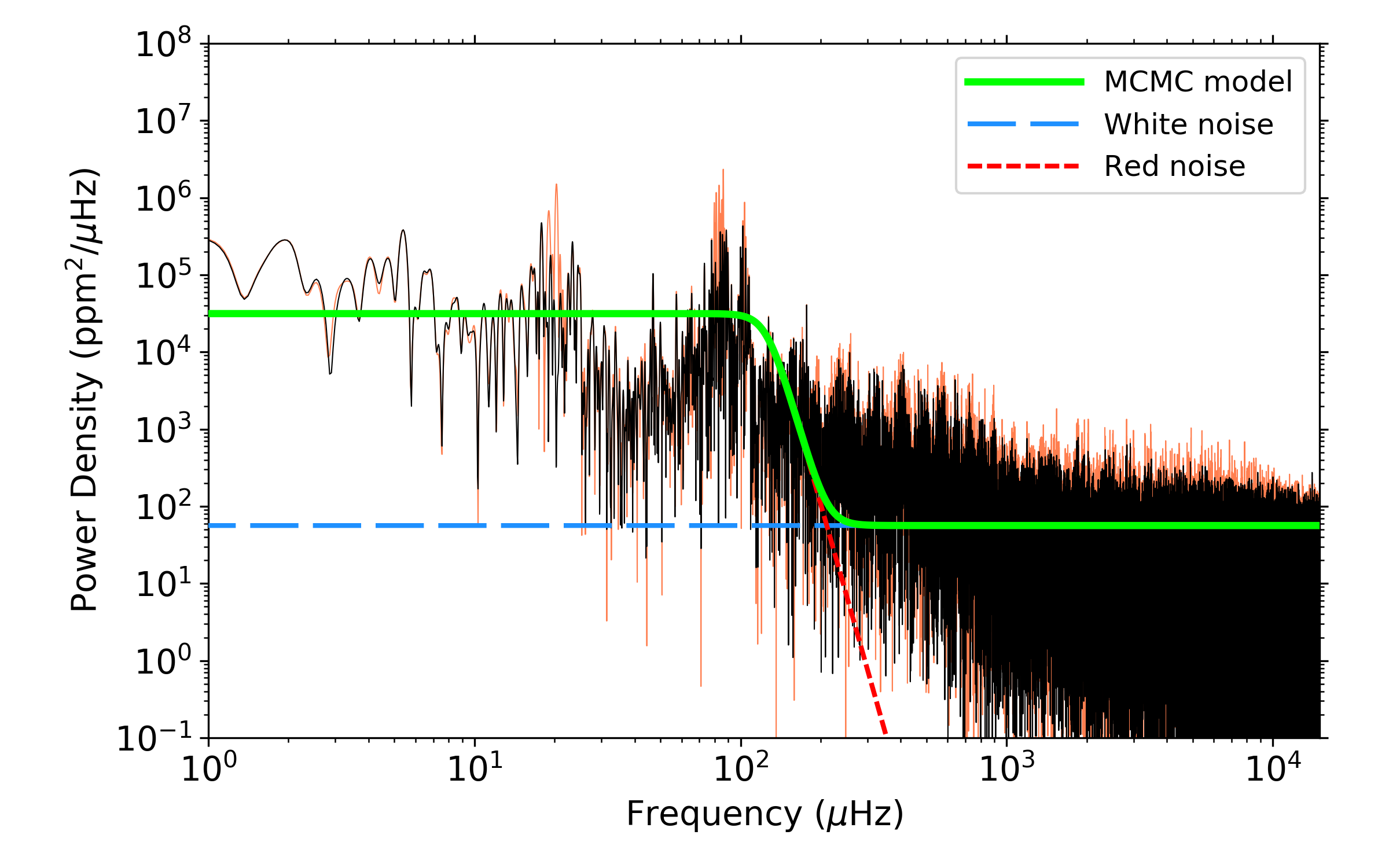}
\includegraphics[width=0.49\textwidth]{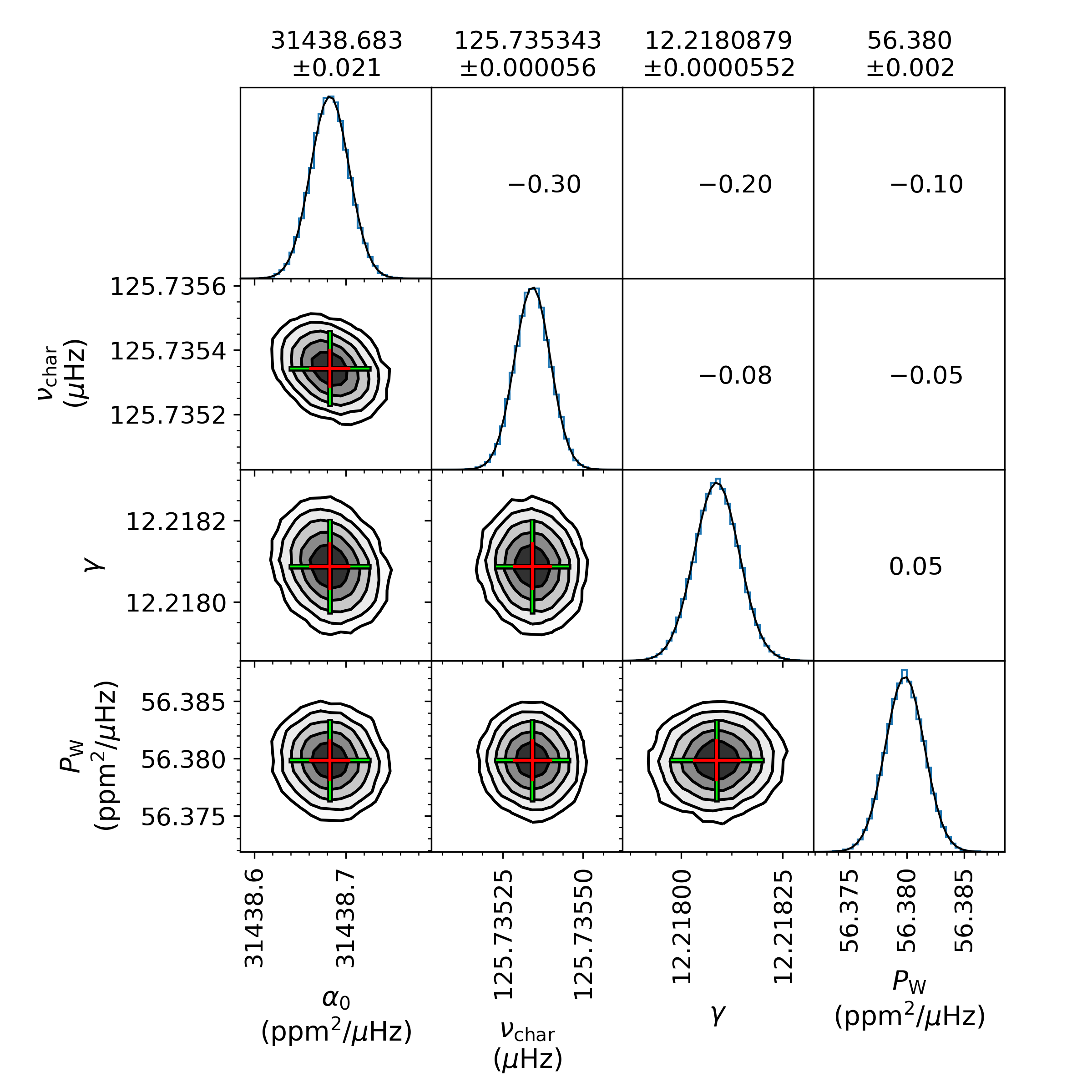}
\includegraphics[width=0.49\textwidth]{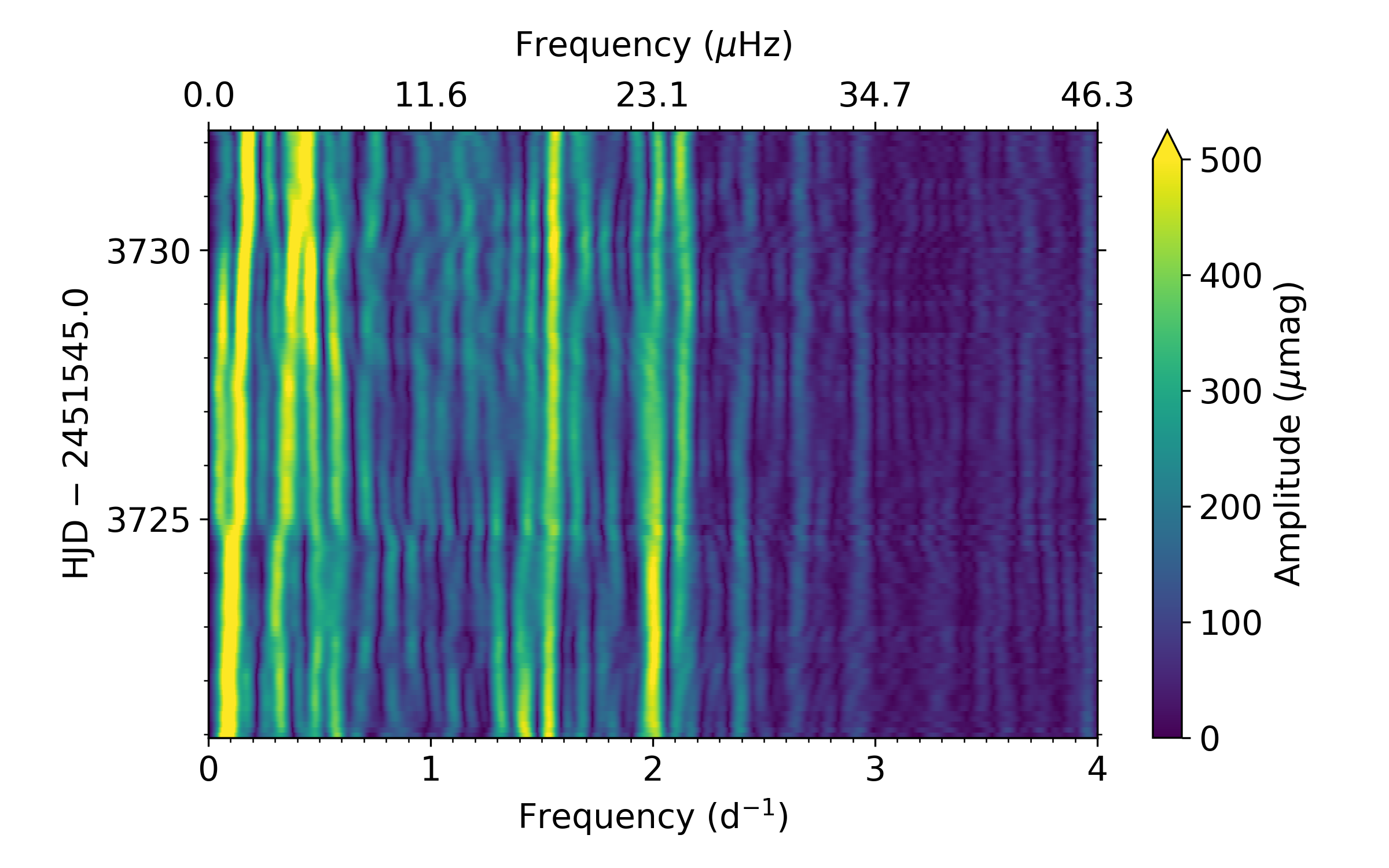}
\caption{Summary figure for the A star HD~47485, which has a similar layout as shown in Fig.~\ref{figure: HD46150}.}
\label{figure: HD47485}
\end{figure}
	
\clearpage 

\begin{figure}
\centering
\includegraphics[width=0.49\textwidth]{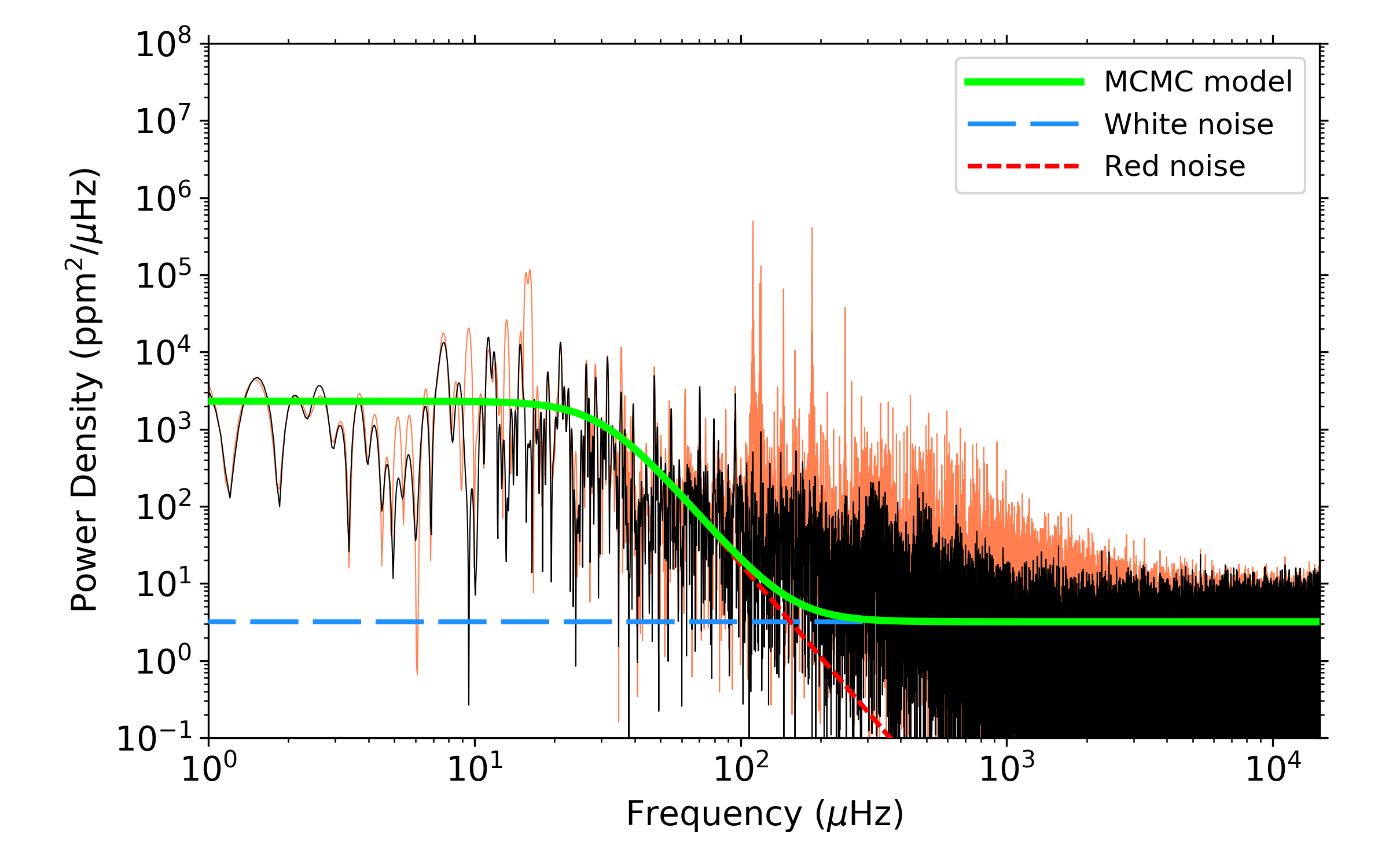}
\includegraphics[width=0.49\textwidth]{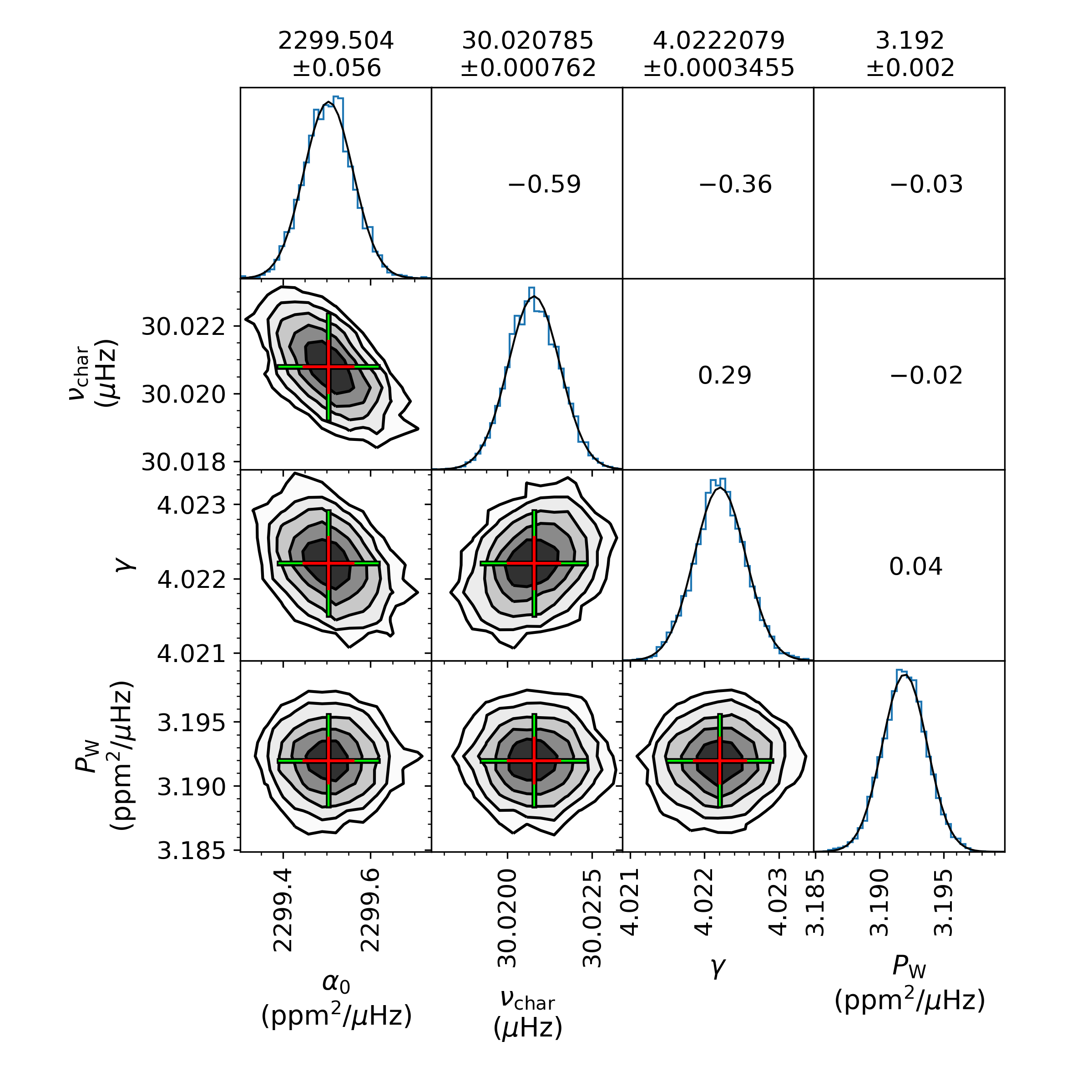}
\includegraphics[width=0.49\textwidth]{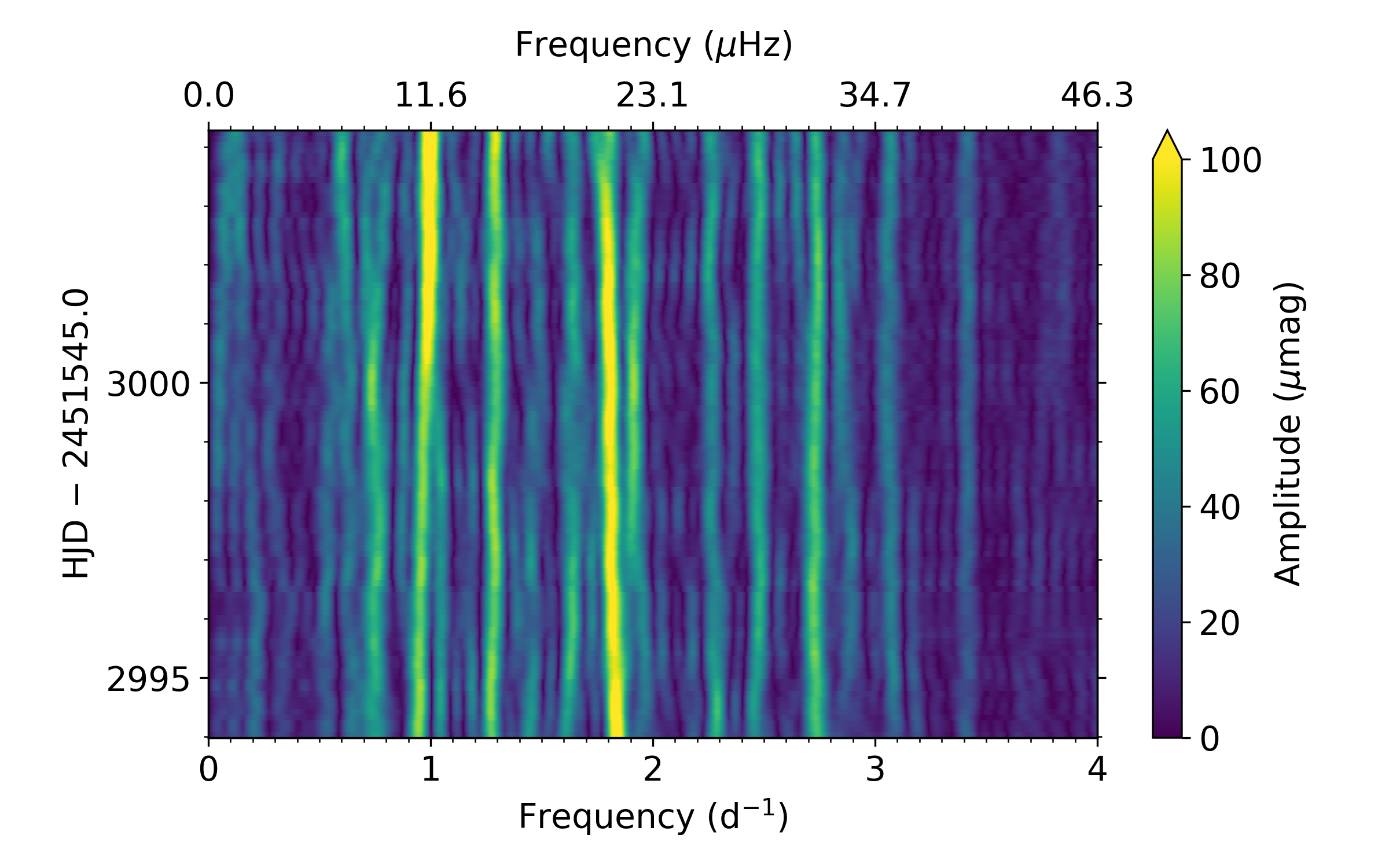}
\caption{Summary figure for the F star HD~48784, which has a similar layout as shown in Fig.~\ref{figure: HD46150}.}
\label{figure: HD48784}
\end{figure}


\begin{figure}
\centering
\includegraphics[width=0.49\textwidth]{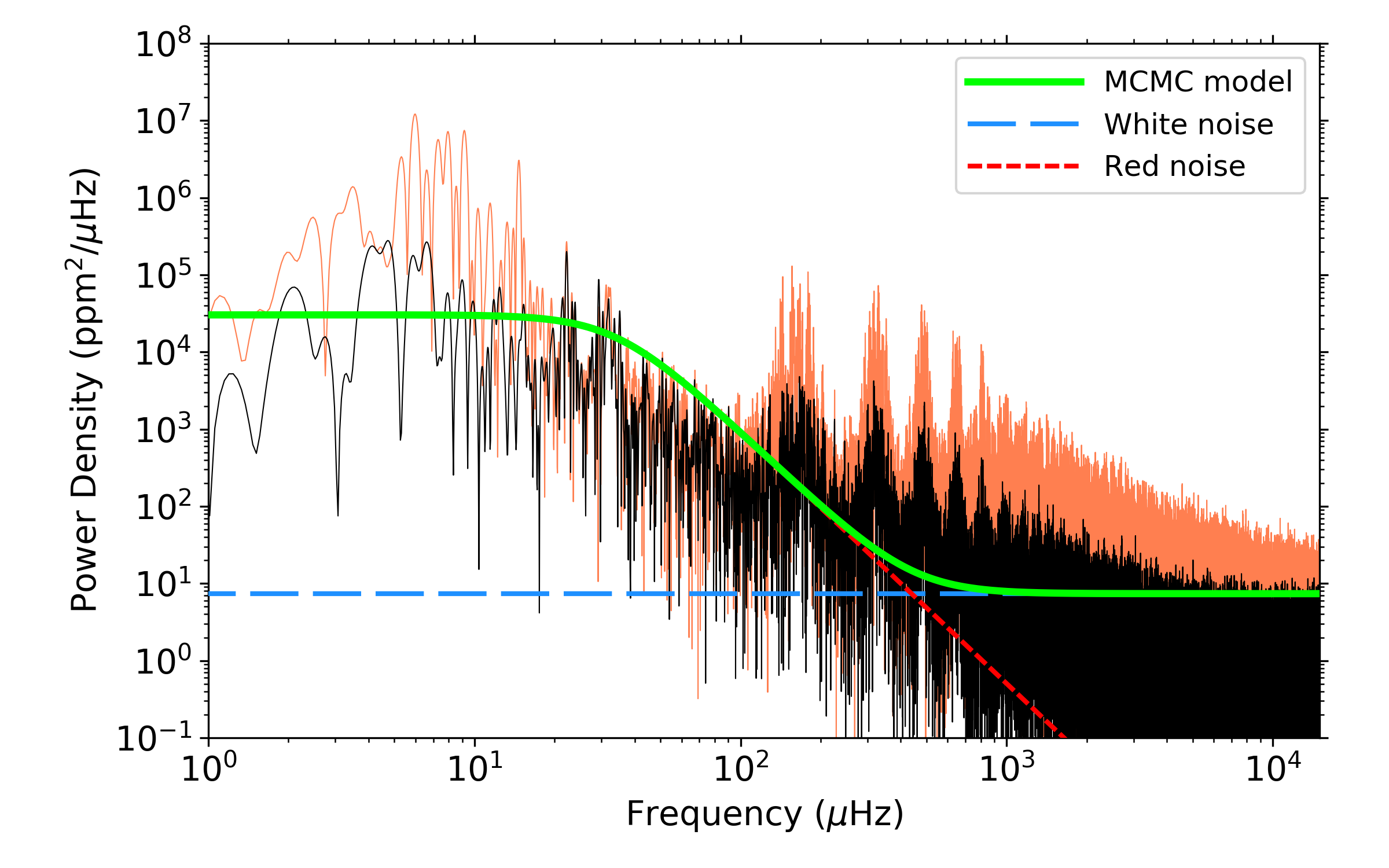}
\includegraphics[width=0.49\textwidth]{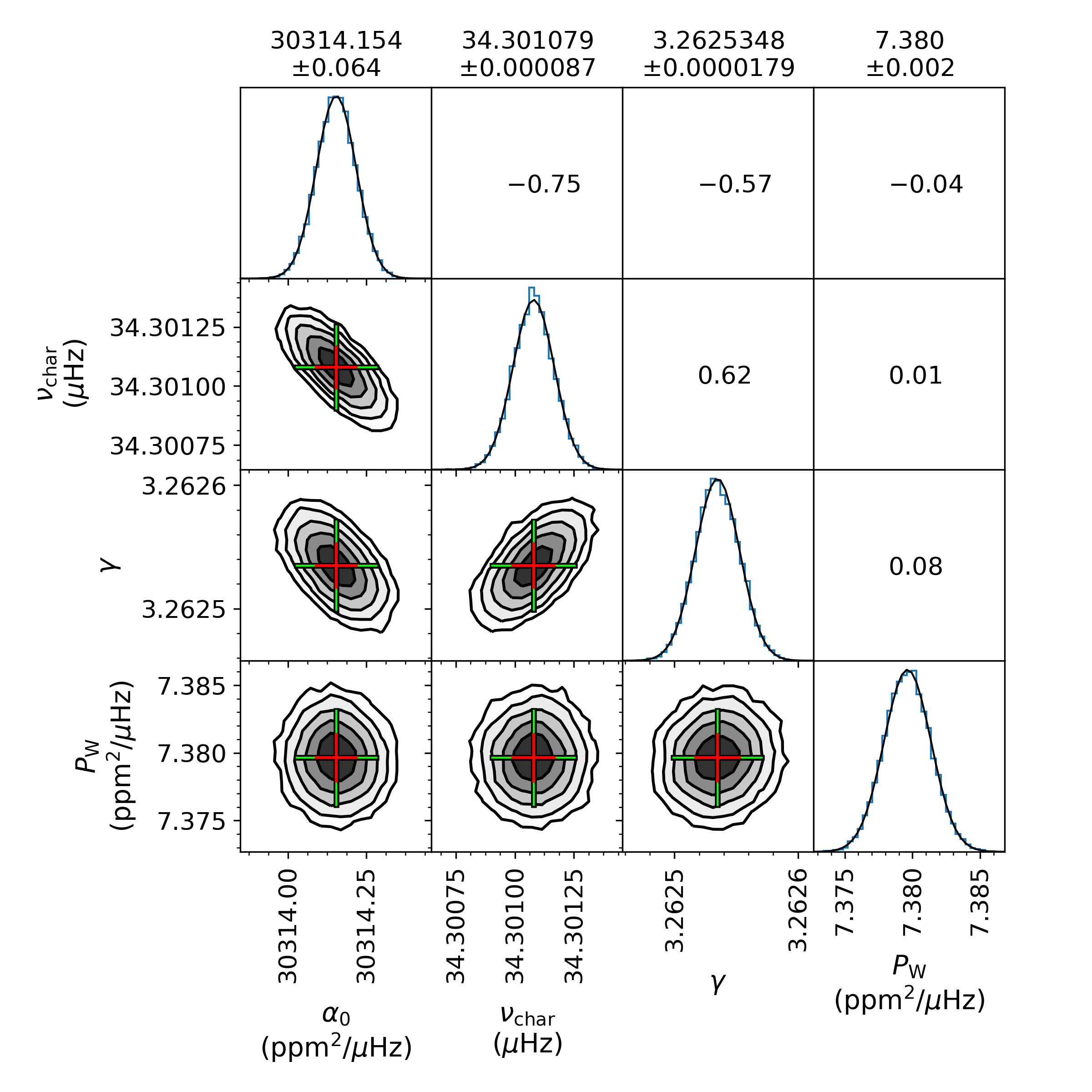}
\includegraphics[width=0.49\textwidth]{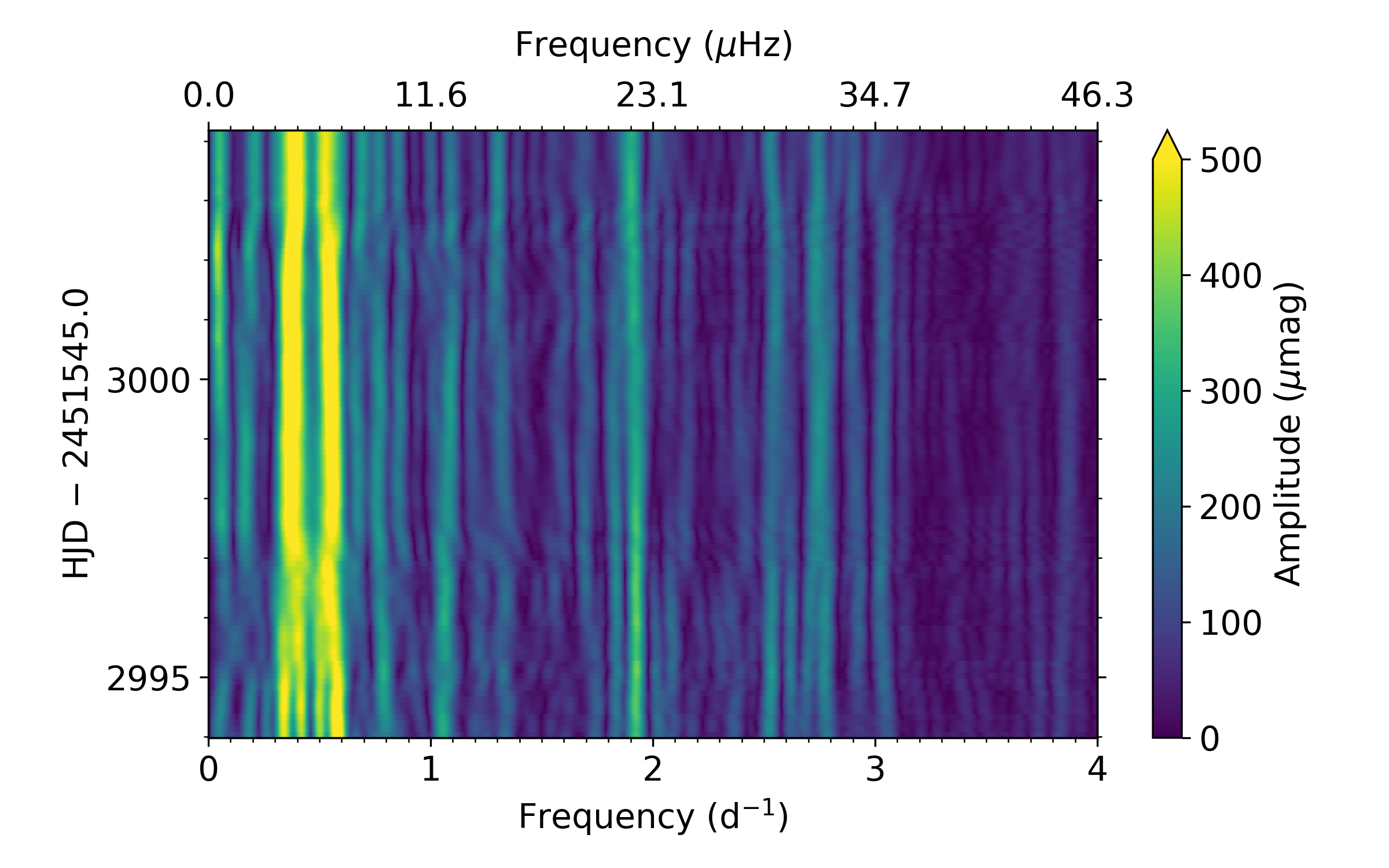}
\caption{Summary figure for the B star HD~48977, which has a similar layout as shown in Fig.~\ref{figure: HD46150}.}
\label{figure: HD48977}
\end{figure}

\clearpage 

\begin{figure}
\centering
\includegraphics[width=0.49\textwidth]{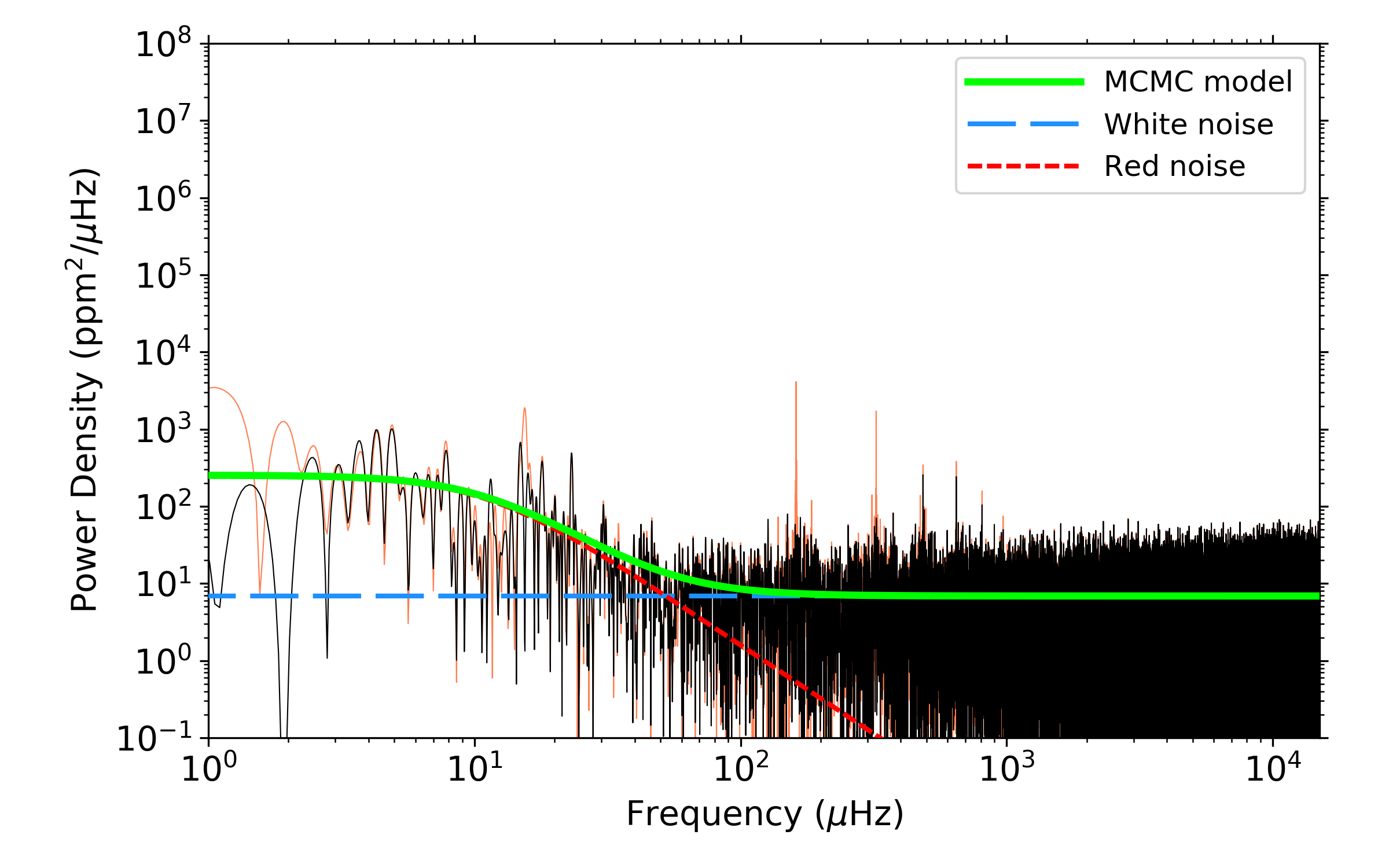}
\includegraphics[width=0.49\textwidth]{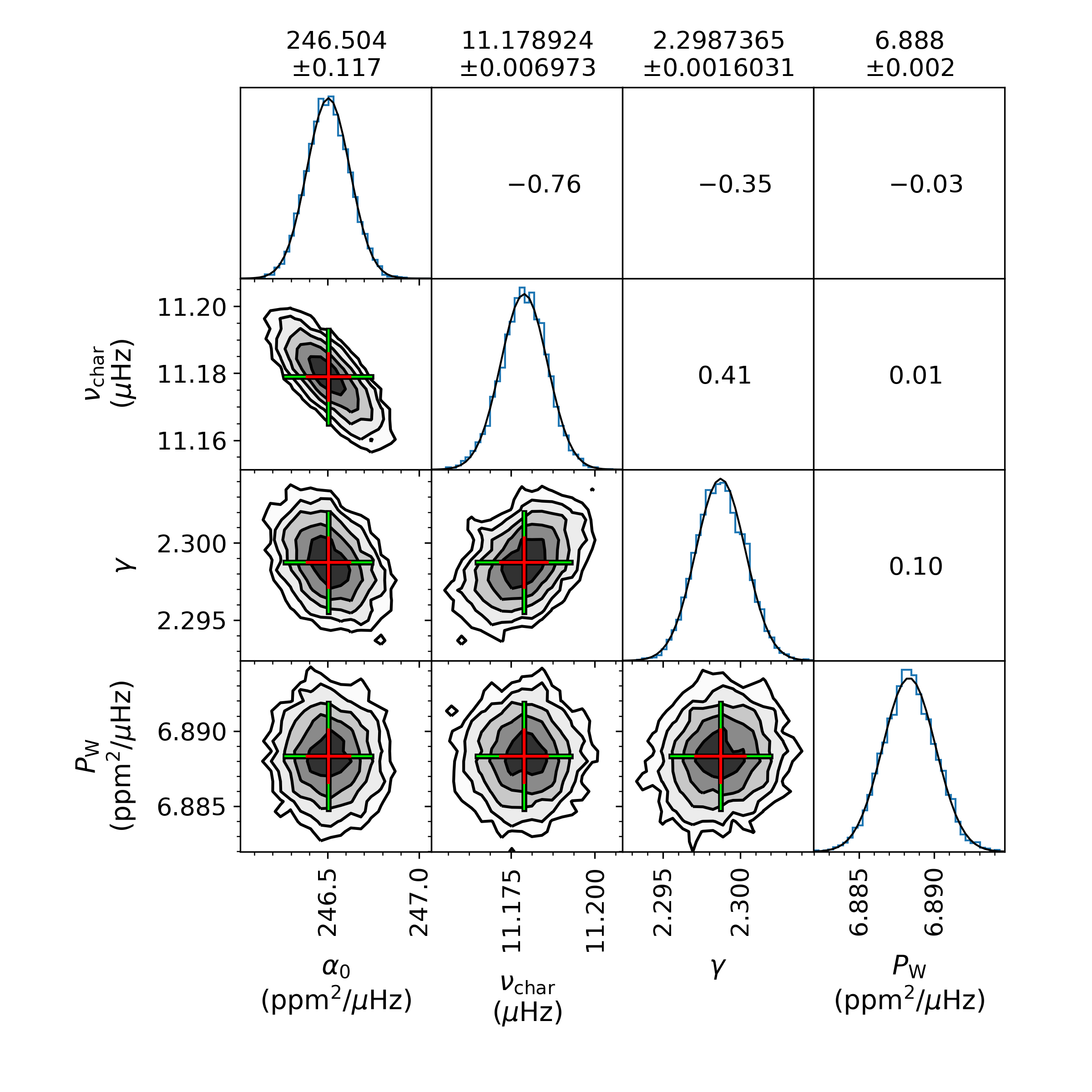}
\includegraphics[width=0.49\textwidth]{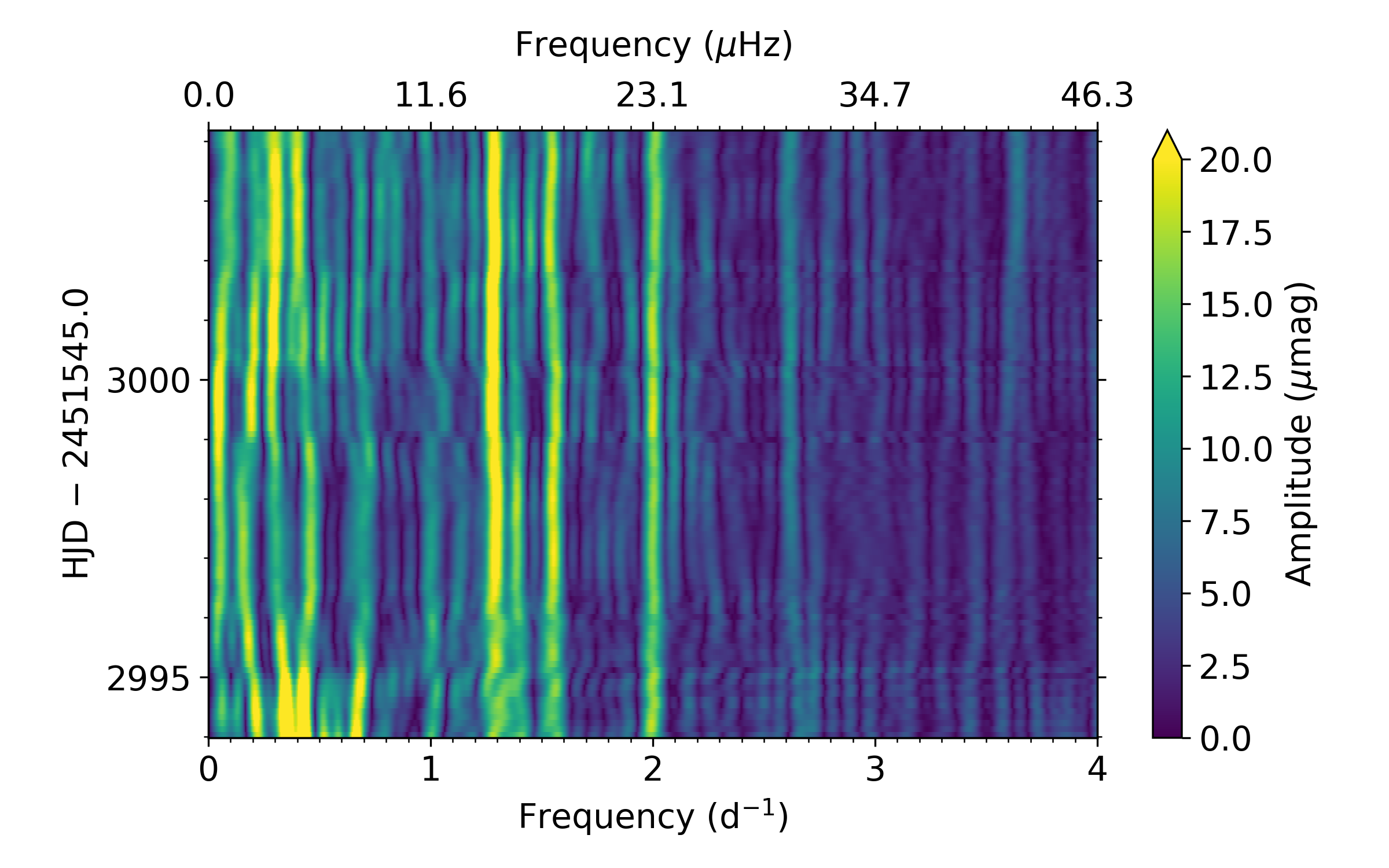}
\caption{Summary figure for the B star HD~49677, which has a similar layout as shown in Fig.~\ref{figure: HD46150}.}
\label{figure: HD49677}
\end{figure}


\begin{figure}
\centering
\includegraphics[width=0.49\textwidth]{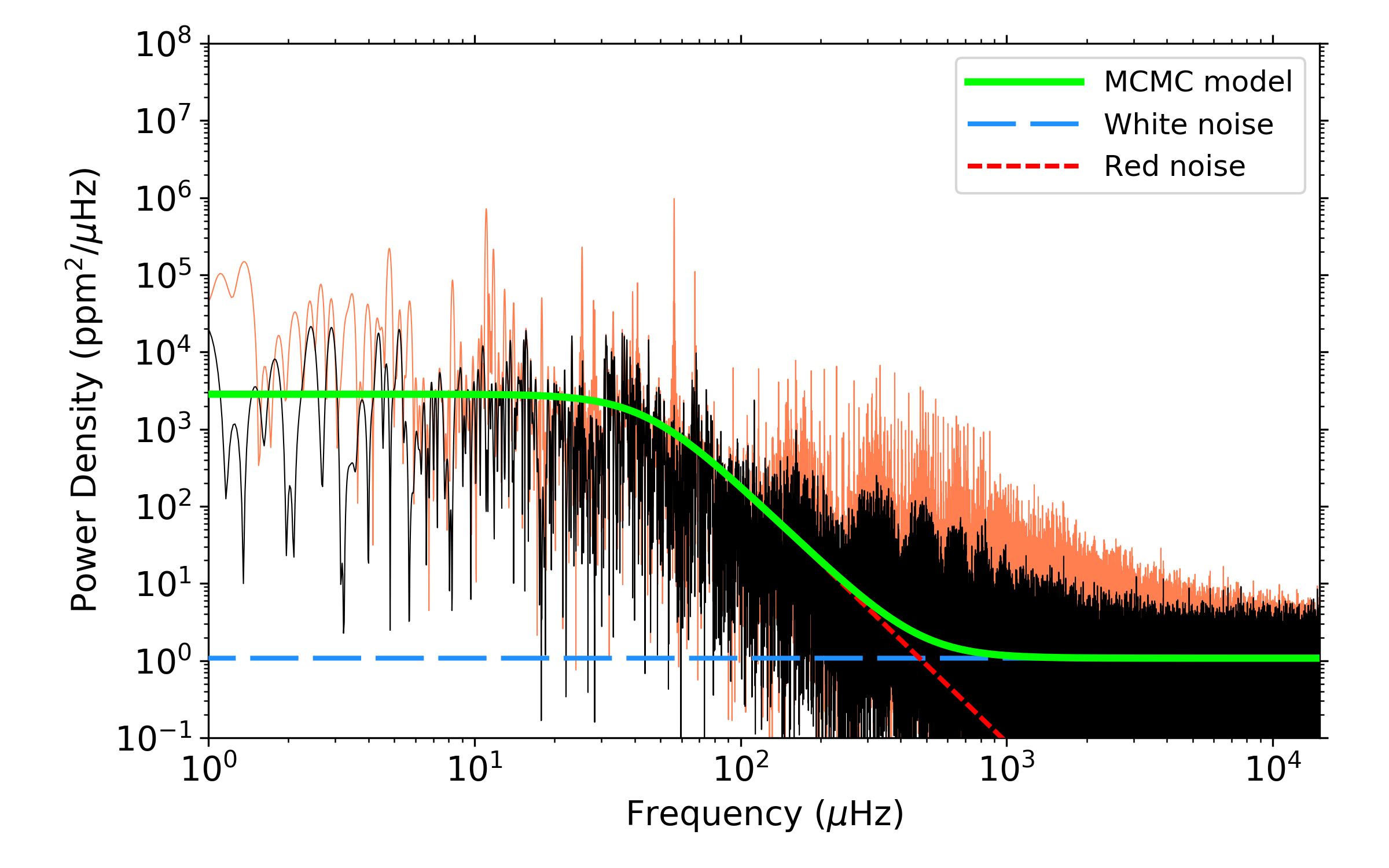}
\includegraphics[width=0.49\textwidth]{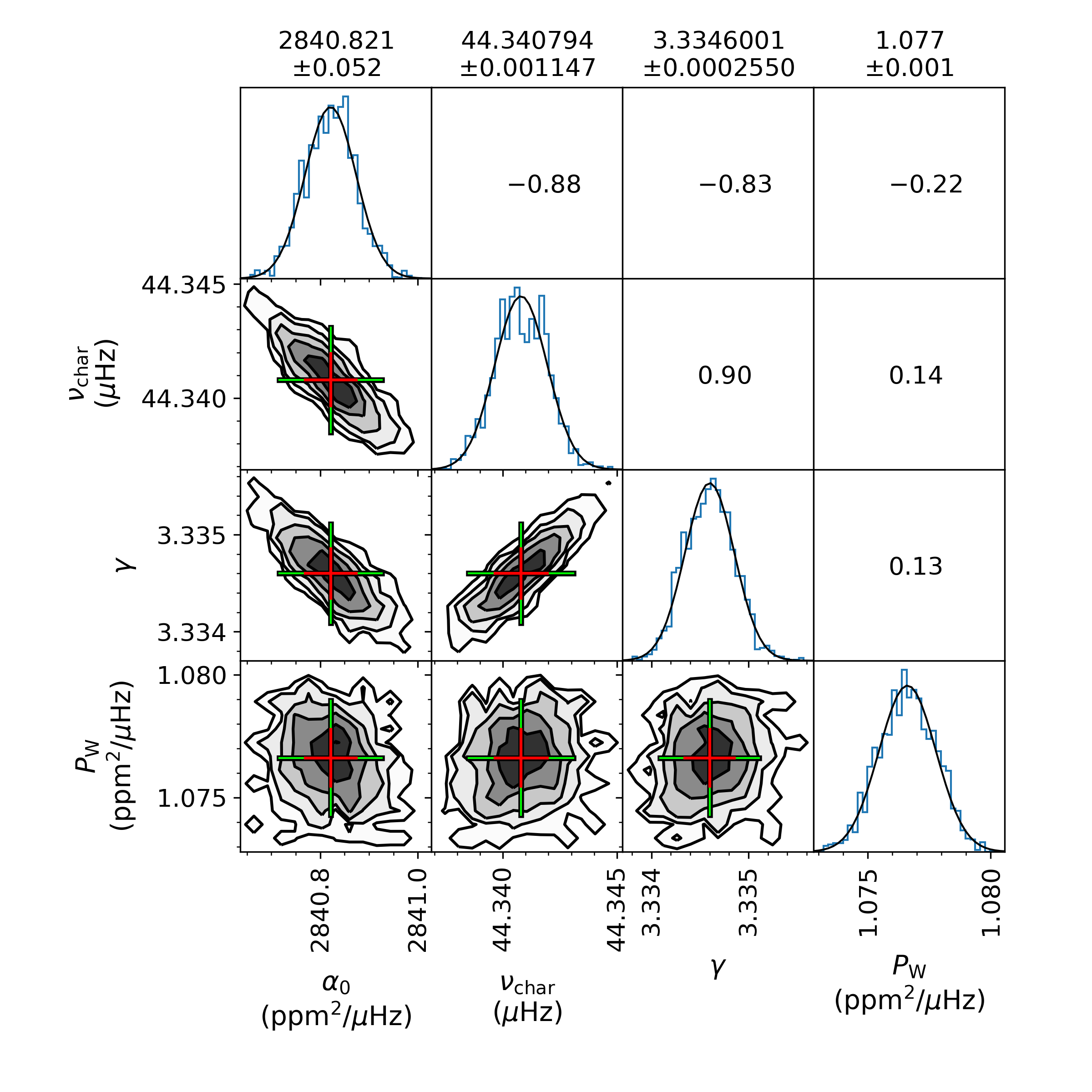}
\includegraphics[width=0.49\textwidth]{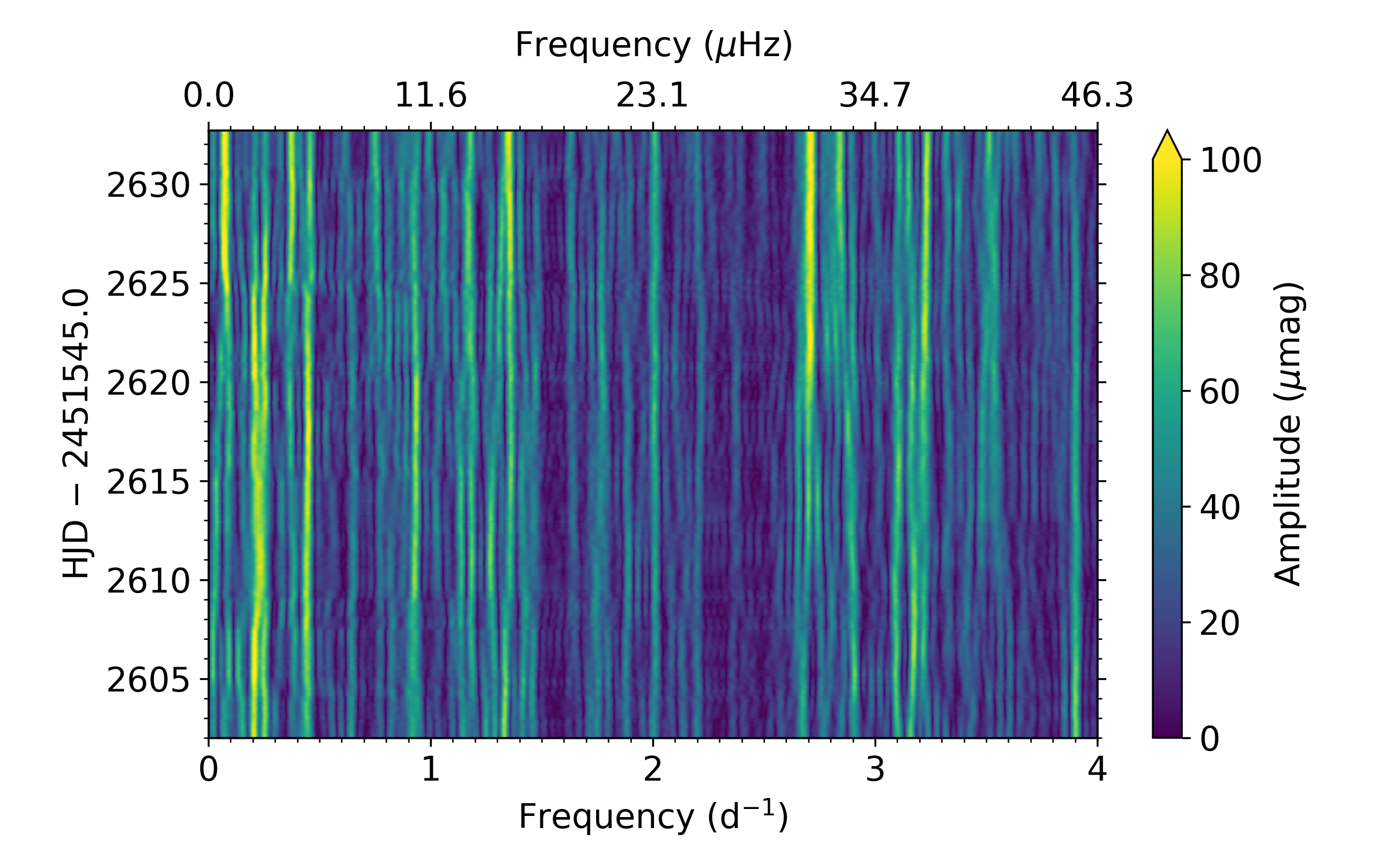}
\caption{Summary figure for the A star HD~50747, which has a similar layout as shown in Fig.~\ref{figure: HD46150}.}
\label{figure: HD50747}
\end{figure}

\clearpage 

\begin{figure}
\centering
\includegraphics[width=0.49\textwidth]{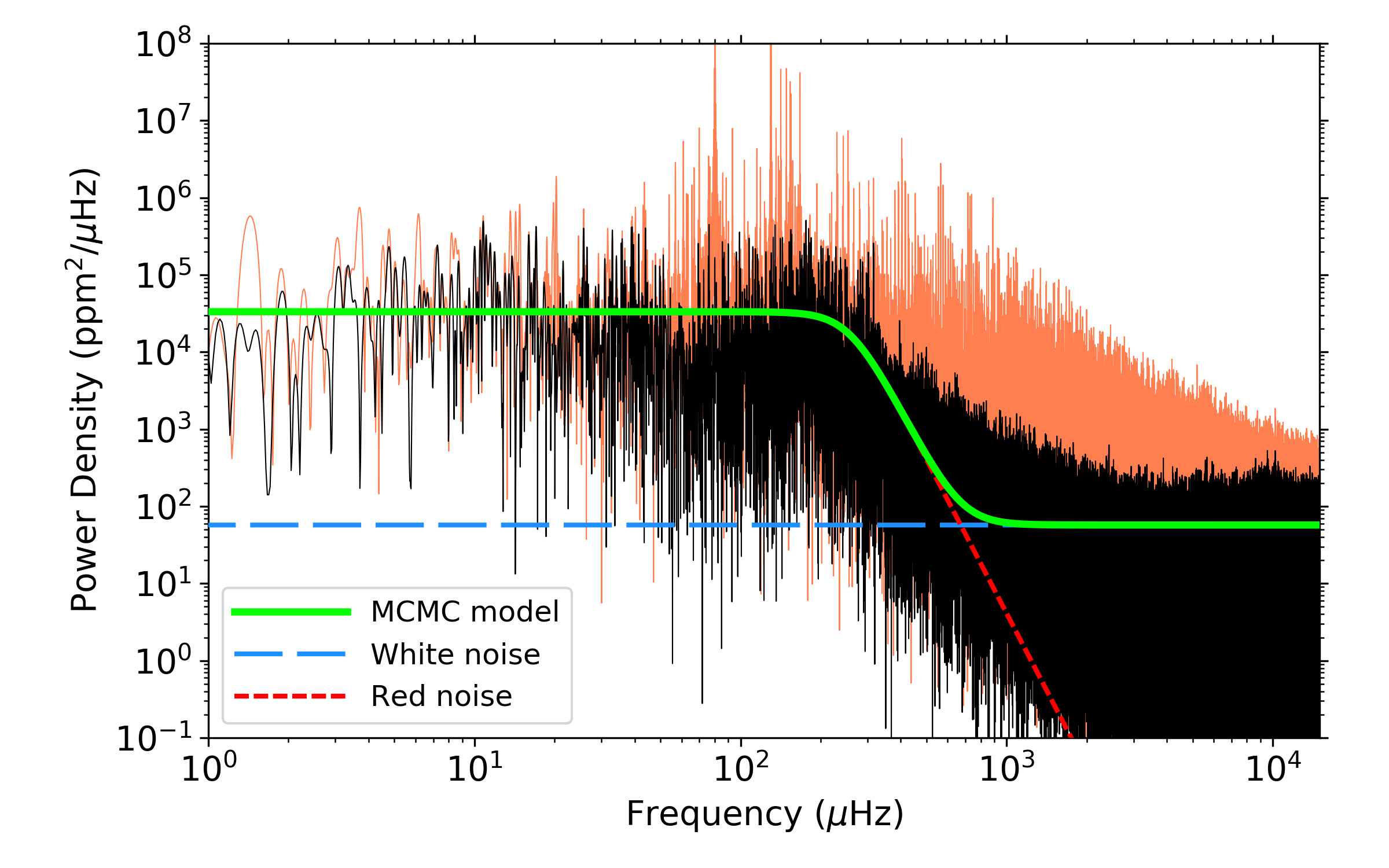}
\includegraphics[width=0.49\textwidth]{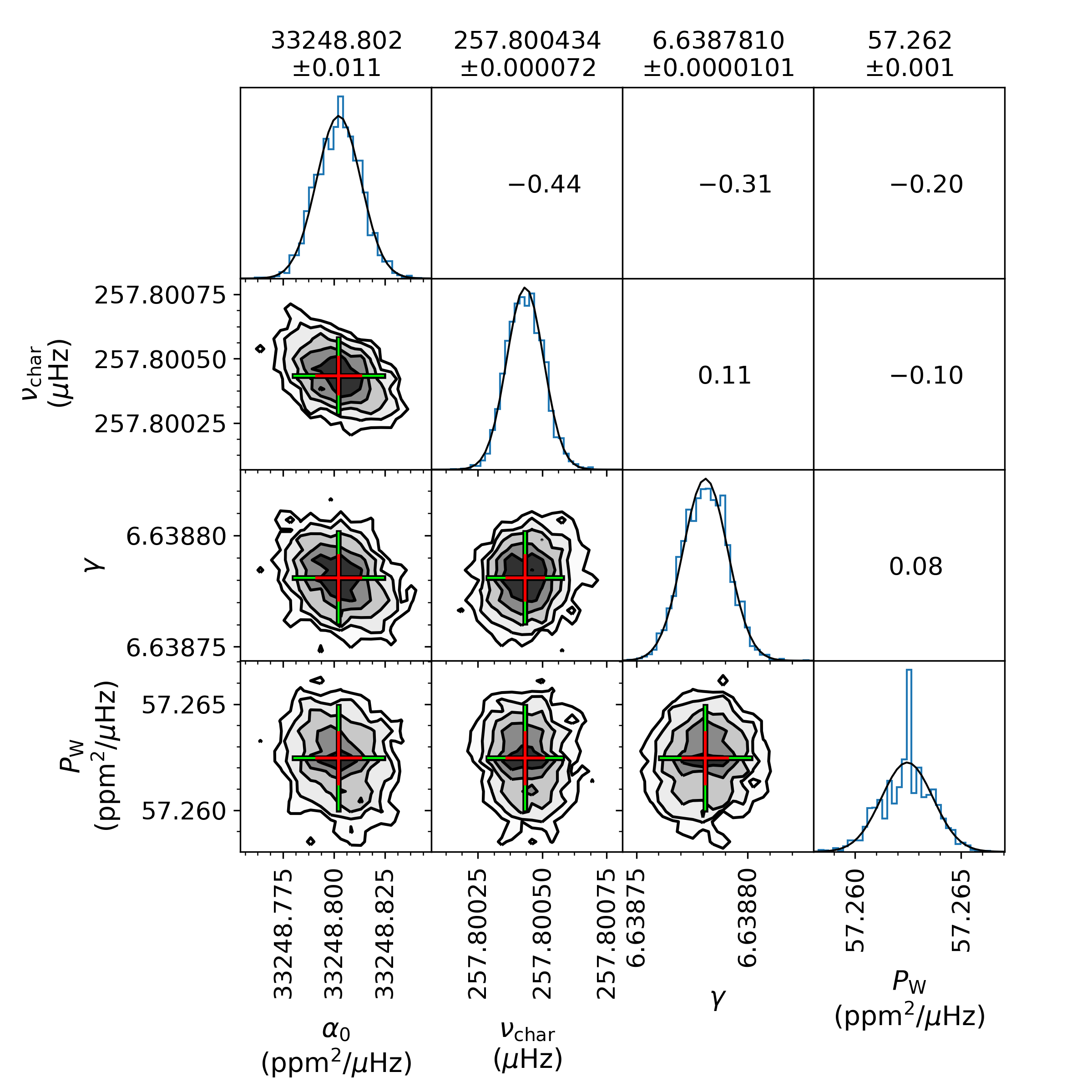}
\includegraphics[width=0.49\textwidth]{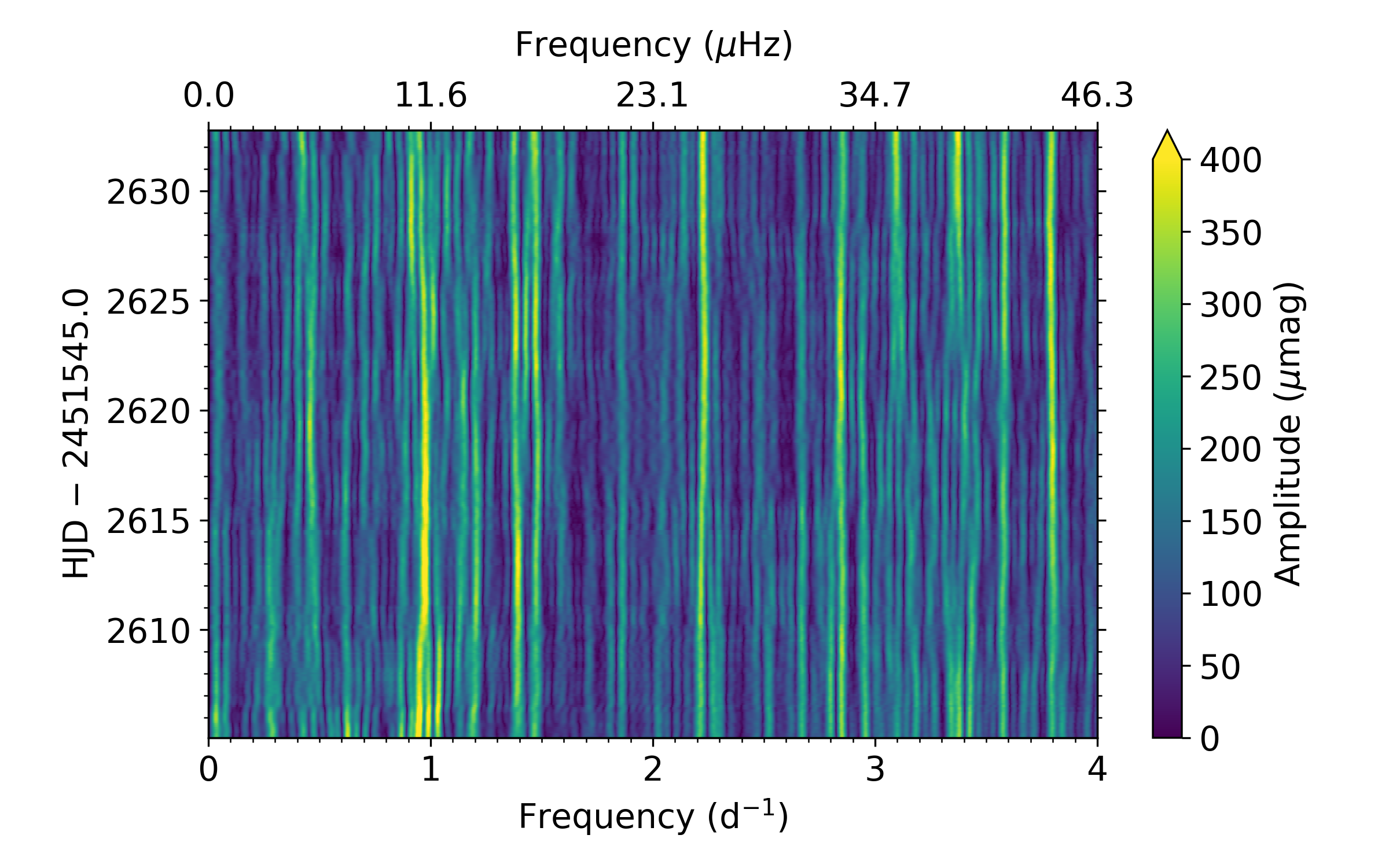}
\caption{Summary figure for the A star HD~50844, which has a similar layout as shown in Fig.~\ref{figure: HD46150}.}
\label{figure: HD50844}
\end{figure}


\begin{figure}
\centering
\includegraphics[width=0.49\textwidth]{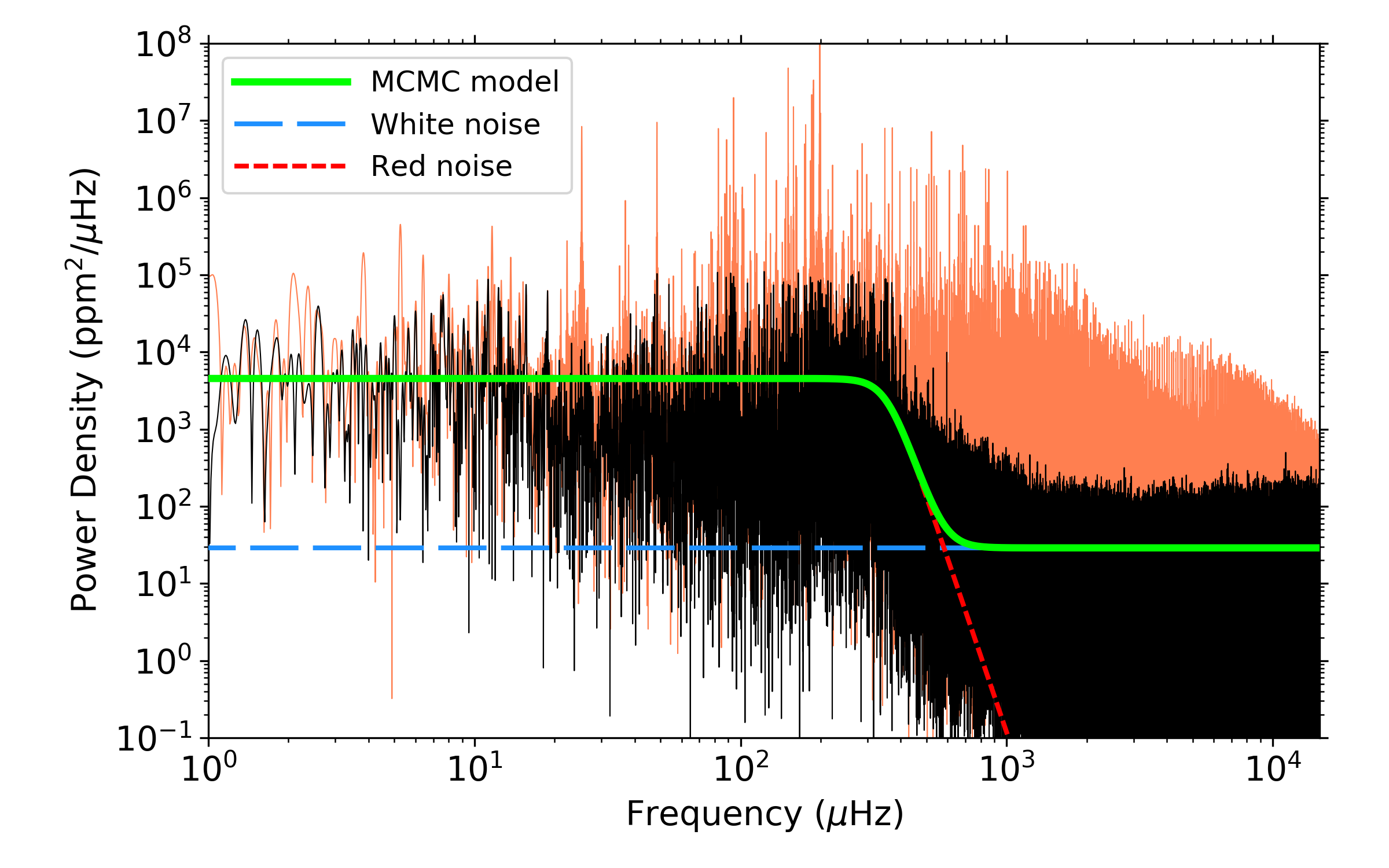}
\includegraphics[width=0.49\textwidth]{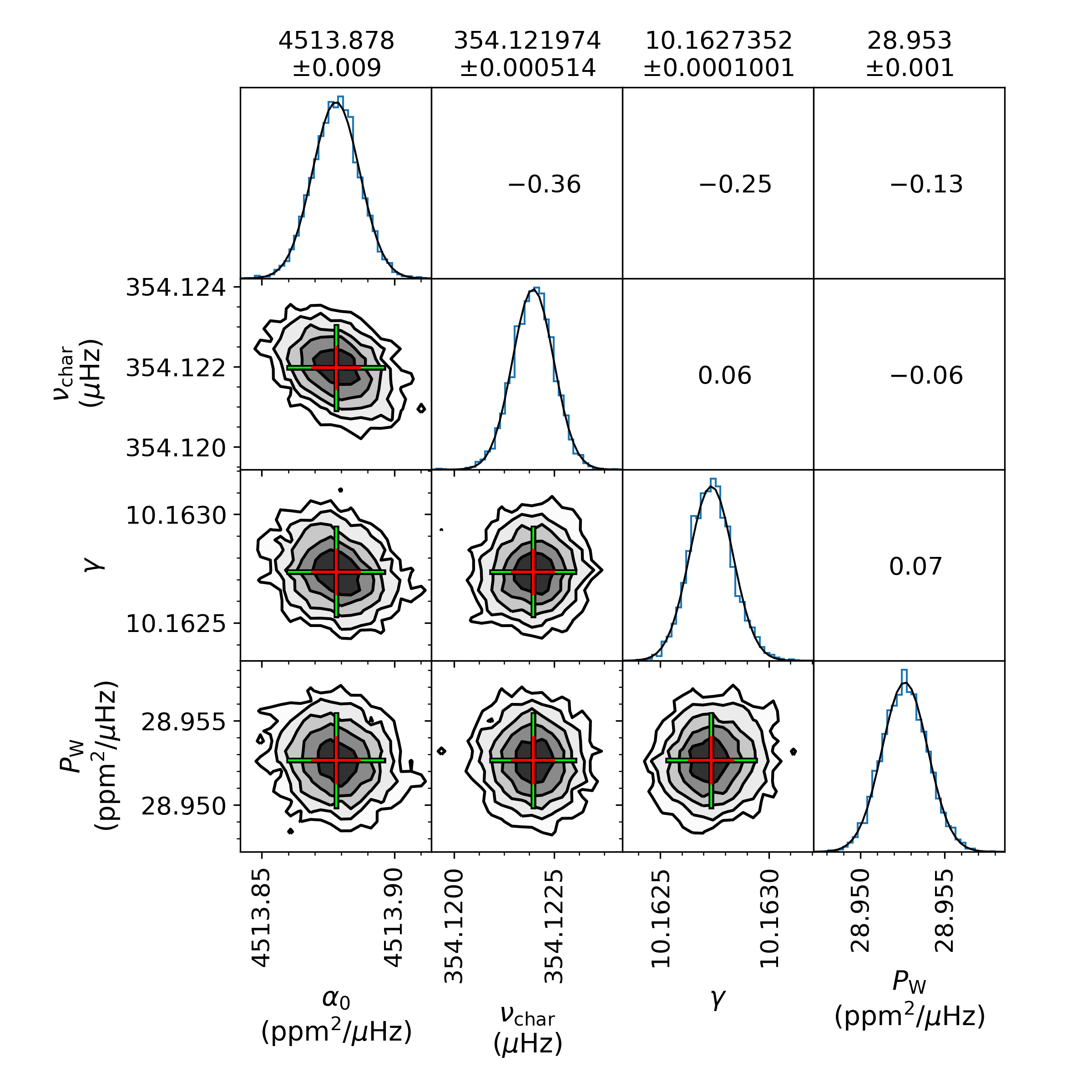}
\includegraphics[width=0.49\textwidth]{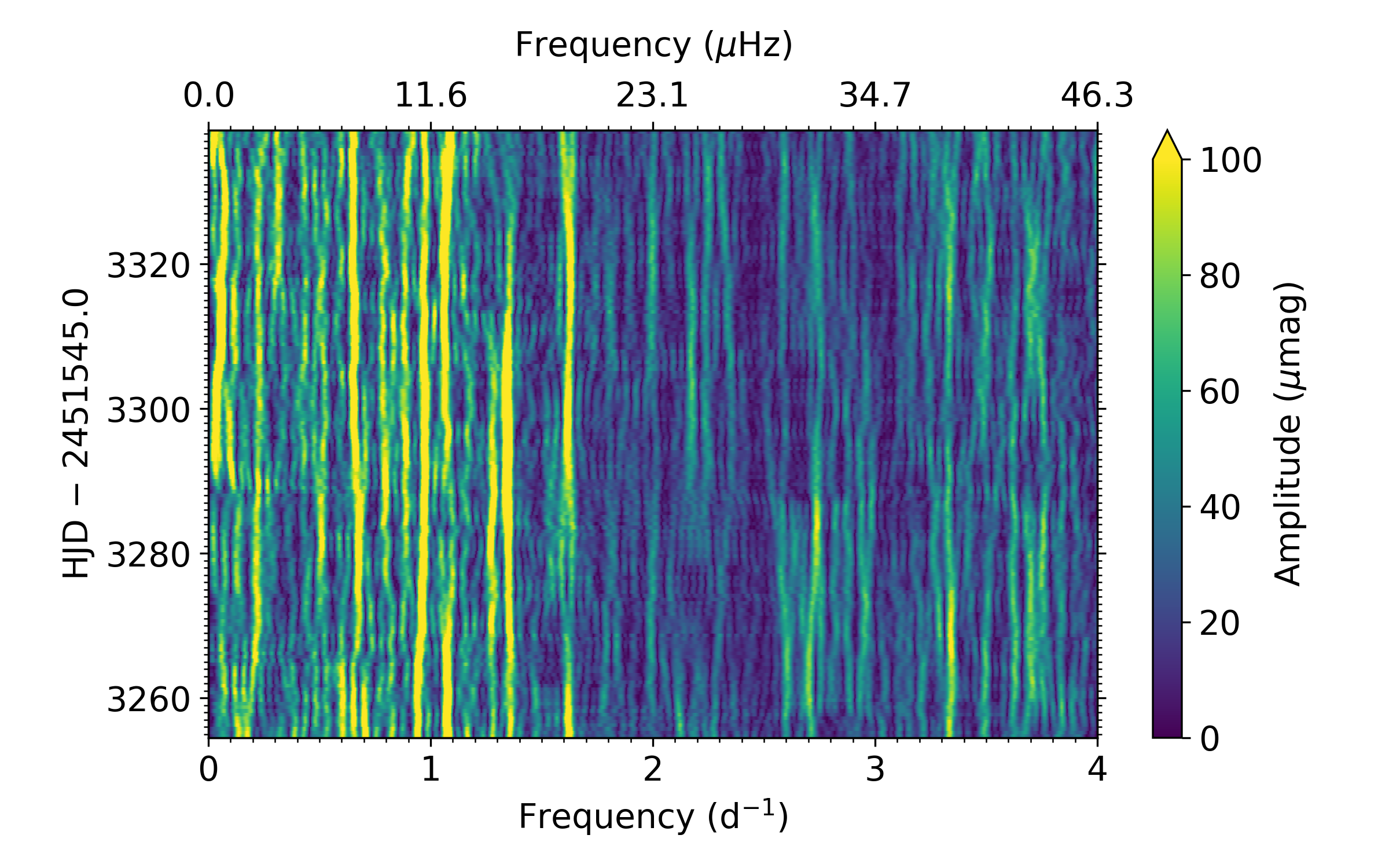}
\caption{Summary figure for the A star HD~50870, which has a similar layout as shown in Fig.~\ref{figure: HD46150}.}
\label{figure: HD50870}
\end{figure}
	
\clearpage 

\begin{figure}
\centering
\includegraphics[width=0.49\textwidth]{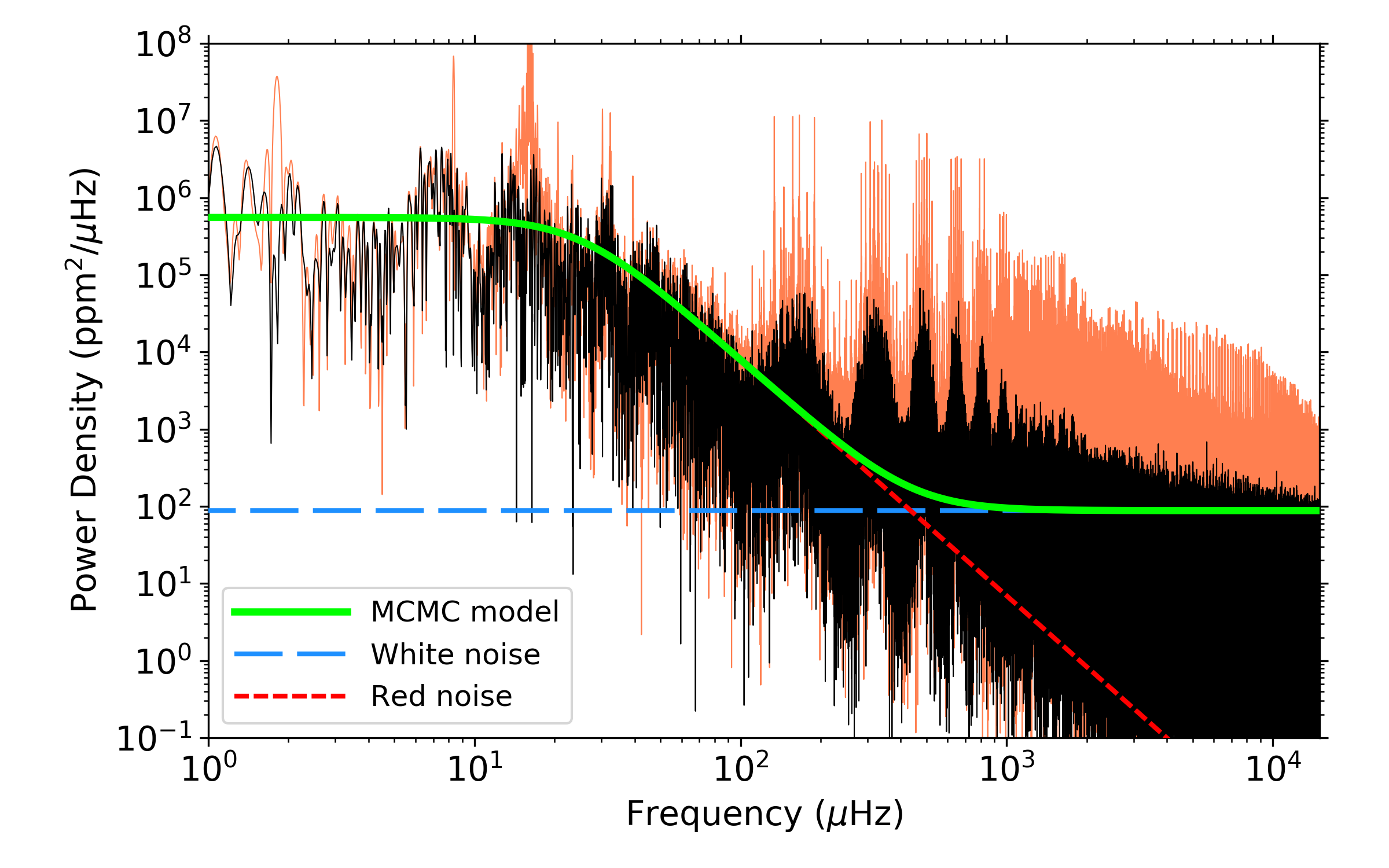}
\includegraphics[width=0.49\textwidth]{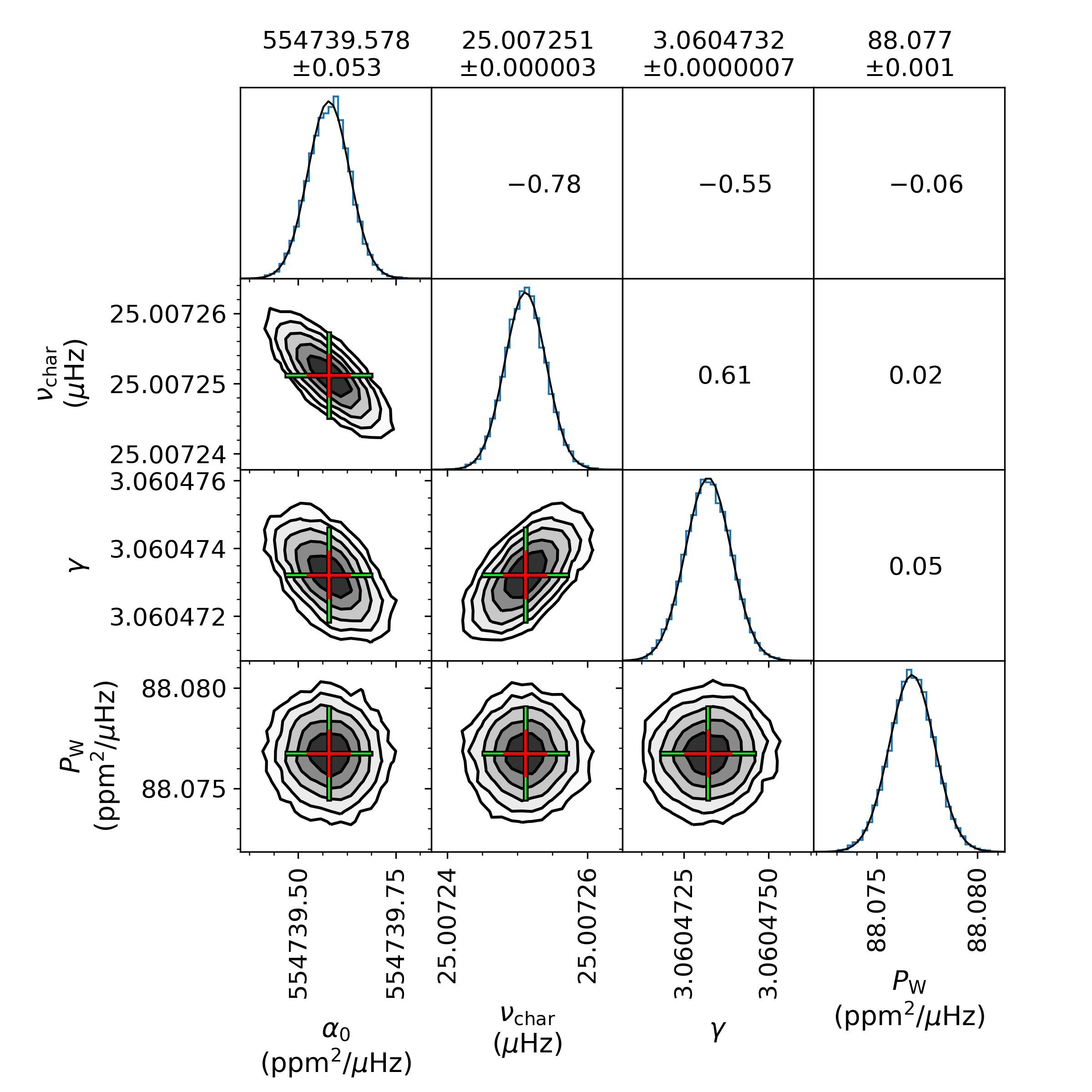}
\includegraphics[width=0.49\textwidth]{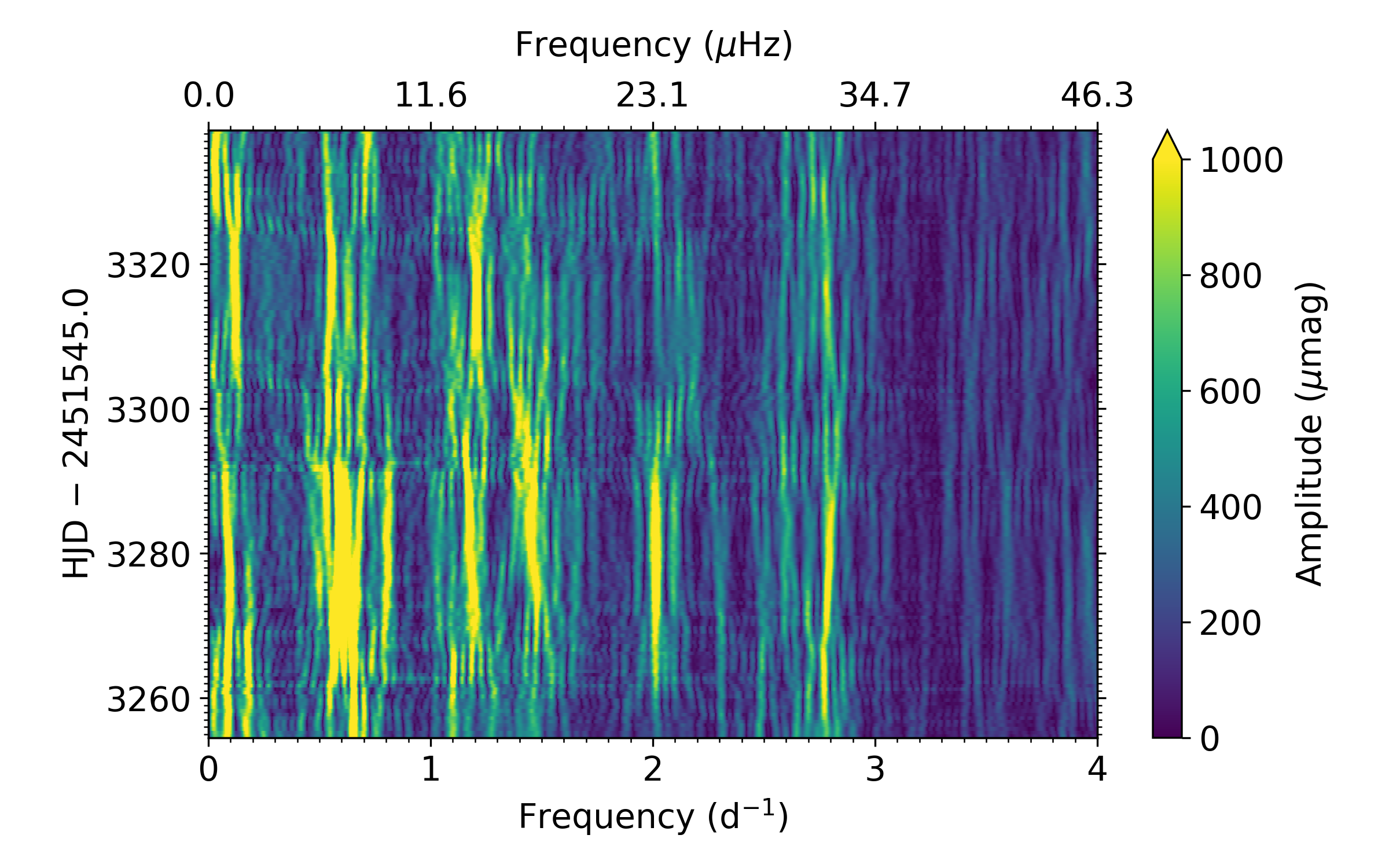}
\caption{Summary figure for the B star HD~51193, which has a similar layout as shown in Fig.~\ref{figure: HD46150}.}
\label{figure: HD51193}
\end{figure}


\begin{figure}
\centering
\includegraphics[width=0.49\textwidth]{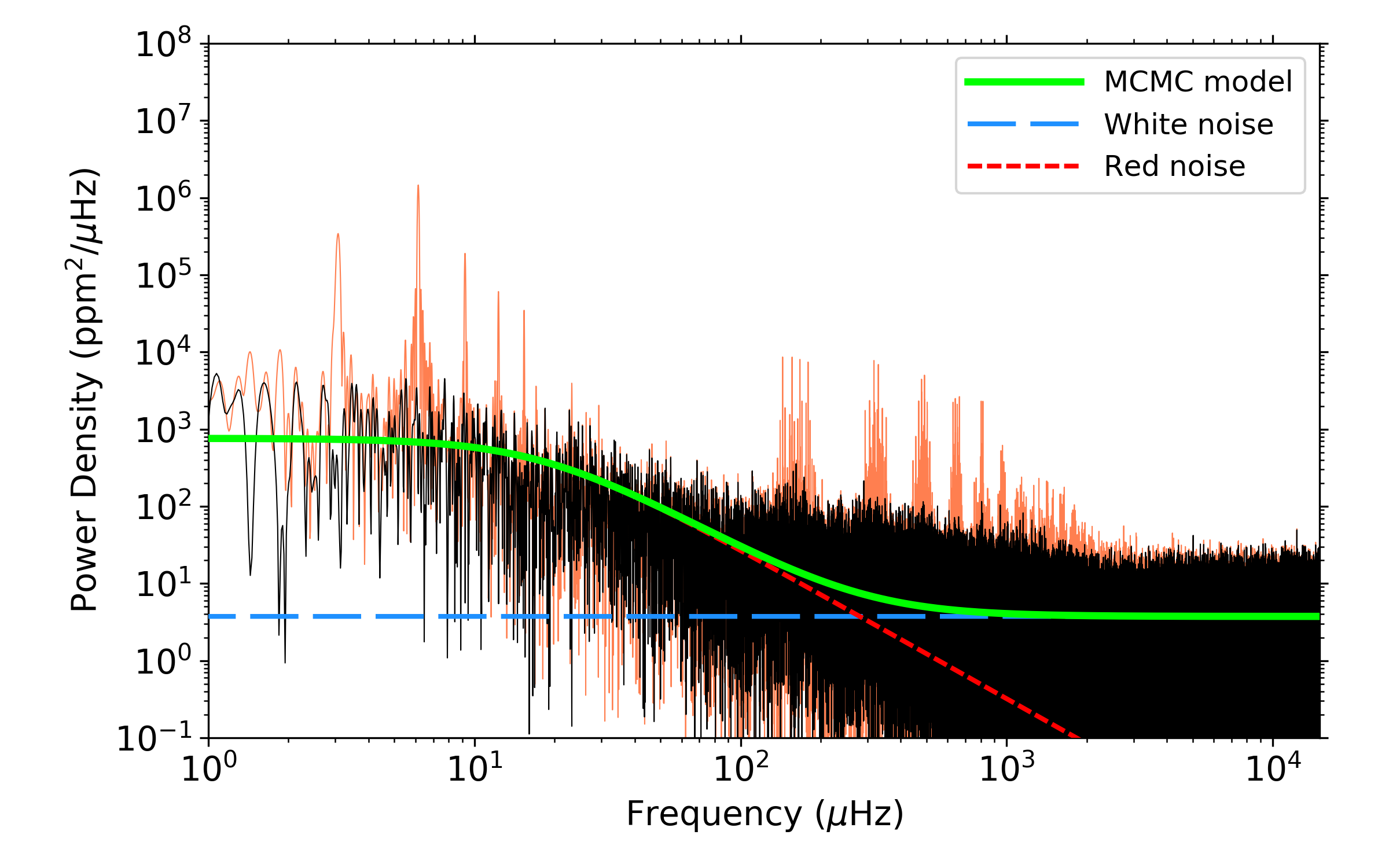}
\includegraphics[width=0.49\textwidth]{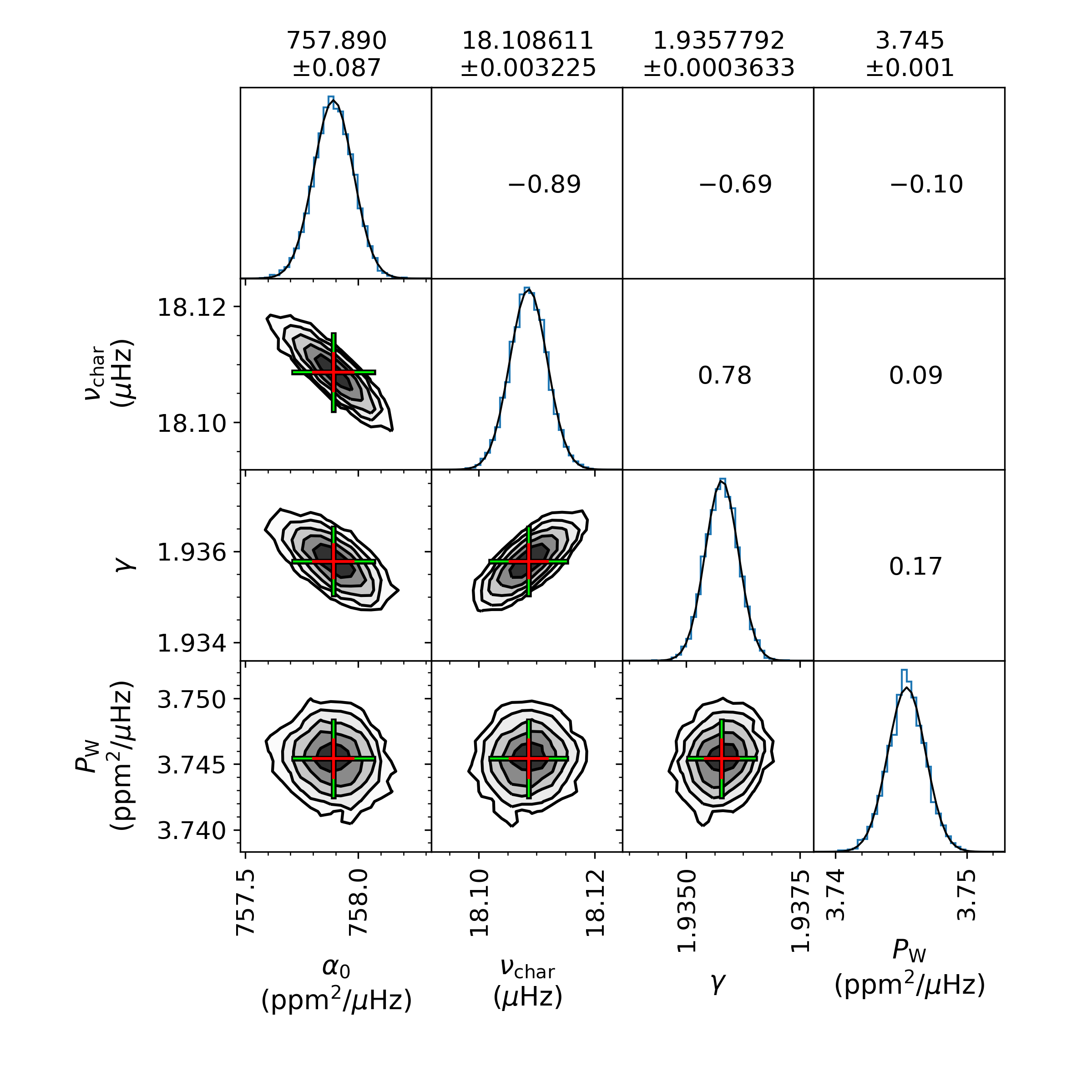}
\includegraphics[width=0.49\textwidth]{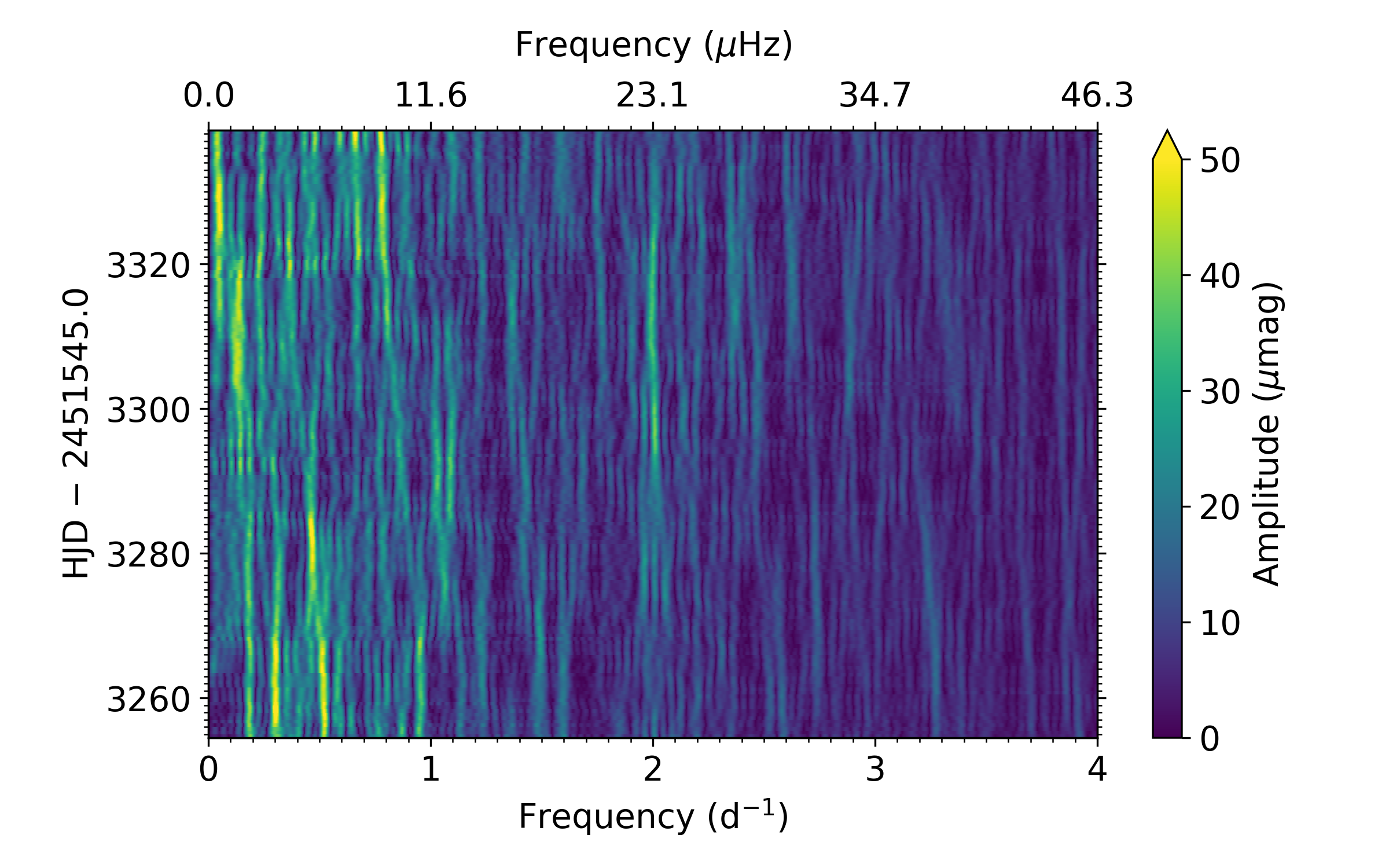}
\caption{Summary figure for the F star HD~51332, which has a similar layout as shown in Fig.~\ref{figure: HD46150}.}
\label{figure: HD51332}
\end{figure}	
	
\clearpage 

\begin{figure}
\centering
\includegraphics[width=0.49\textwidth]{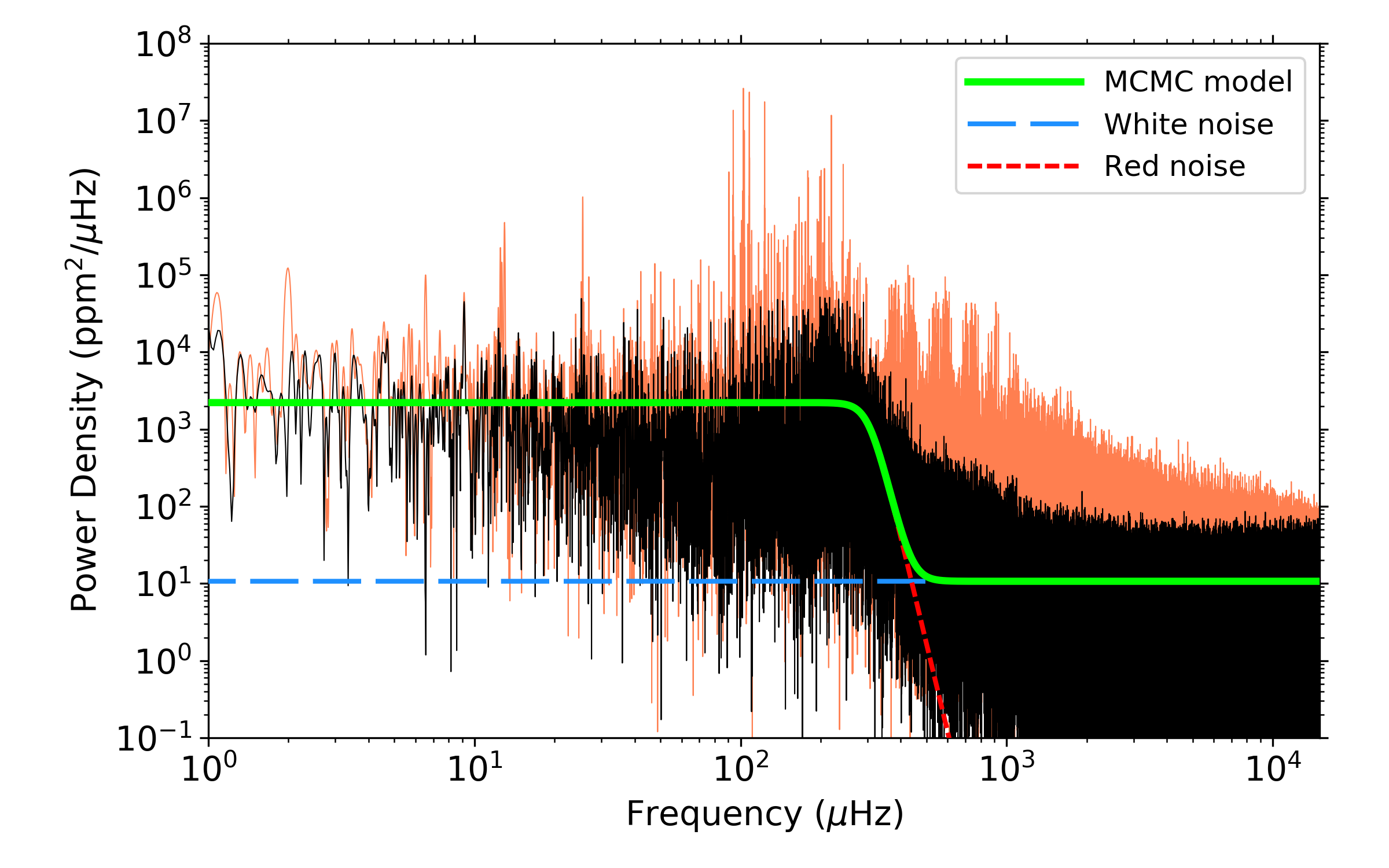}
\includegraphics[width=0.49\textwidth]{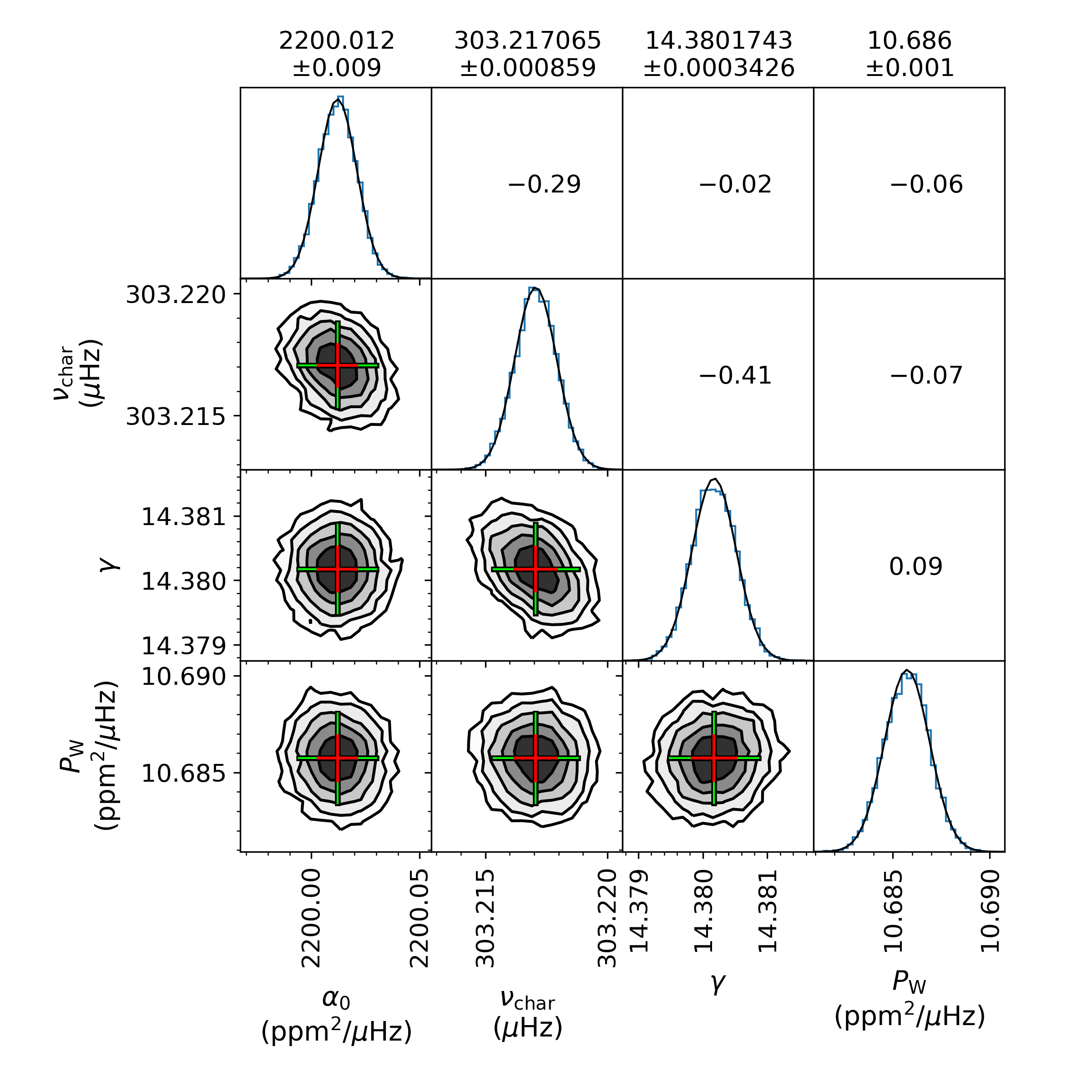}
\includegraphics[width=0.49\textwidth]{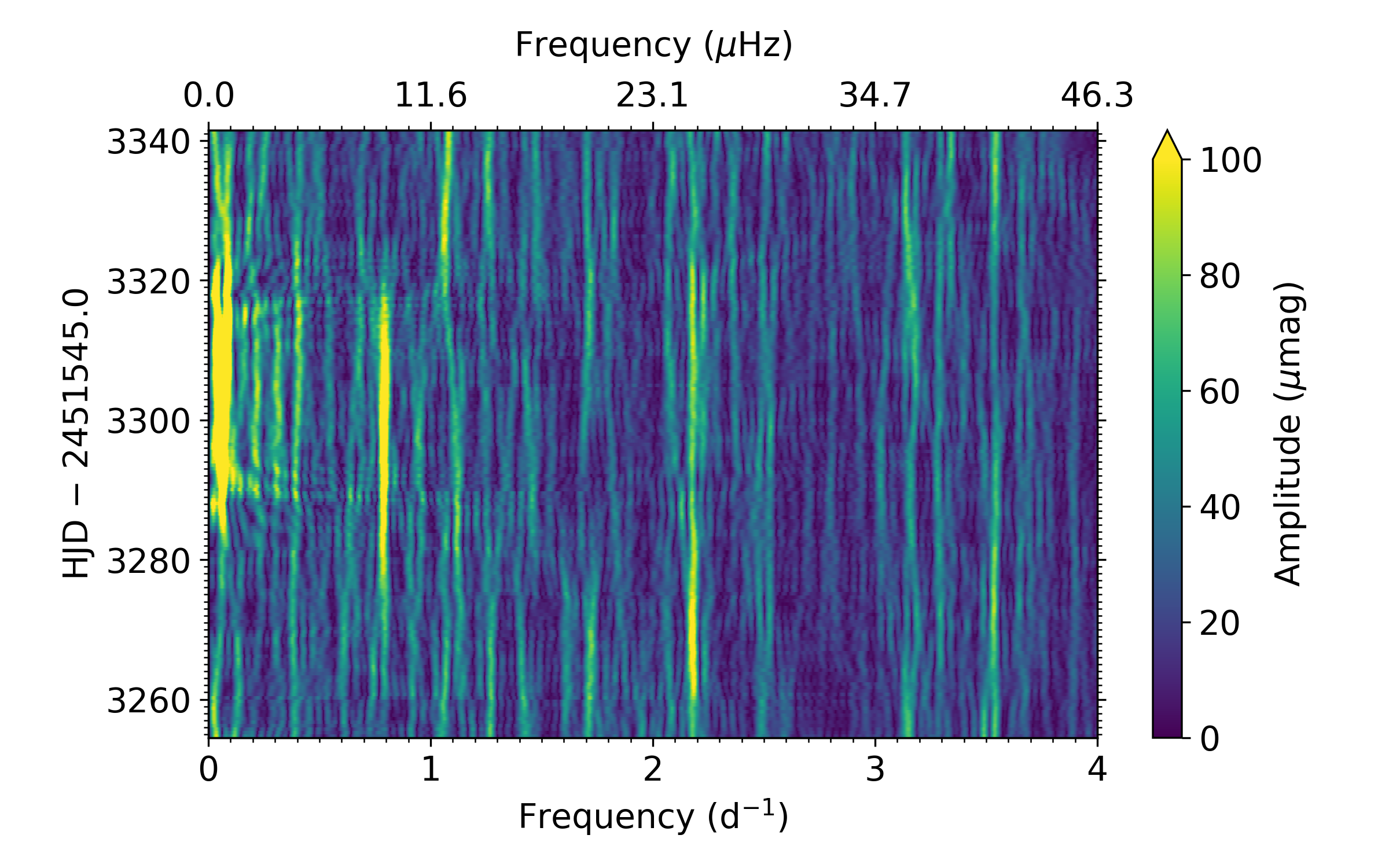}
\caption{Summary figure for the F star HD~51359, which has a similar layout as shown in Fig.~\ref{figure: HD46150}.}
\label{figure: HD51359}
\end{figure}


\begin{figure}
\centering
\includegraphics[width=0.49\textwidth]{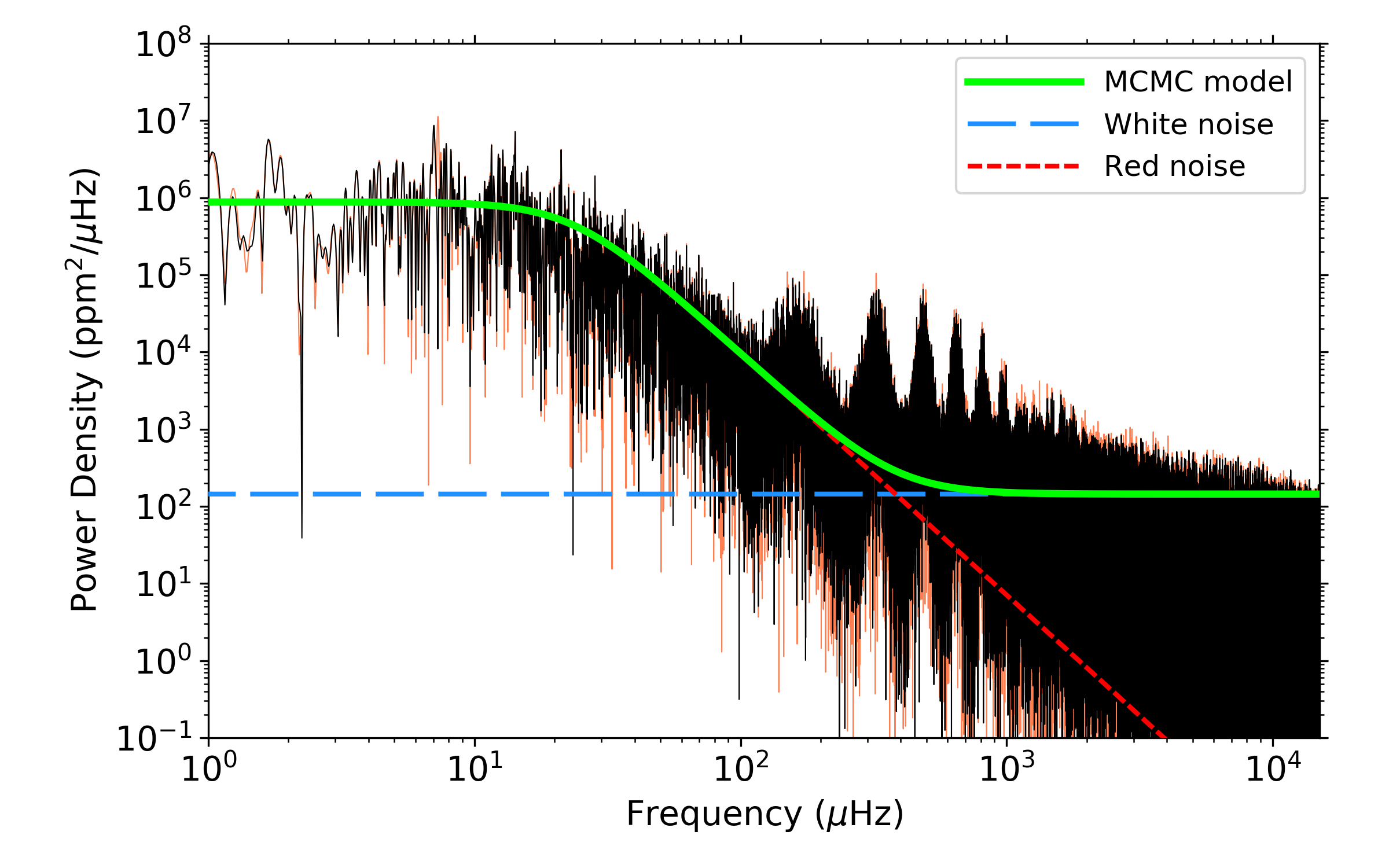}
\includegraphics[width=0.49\textwidth]{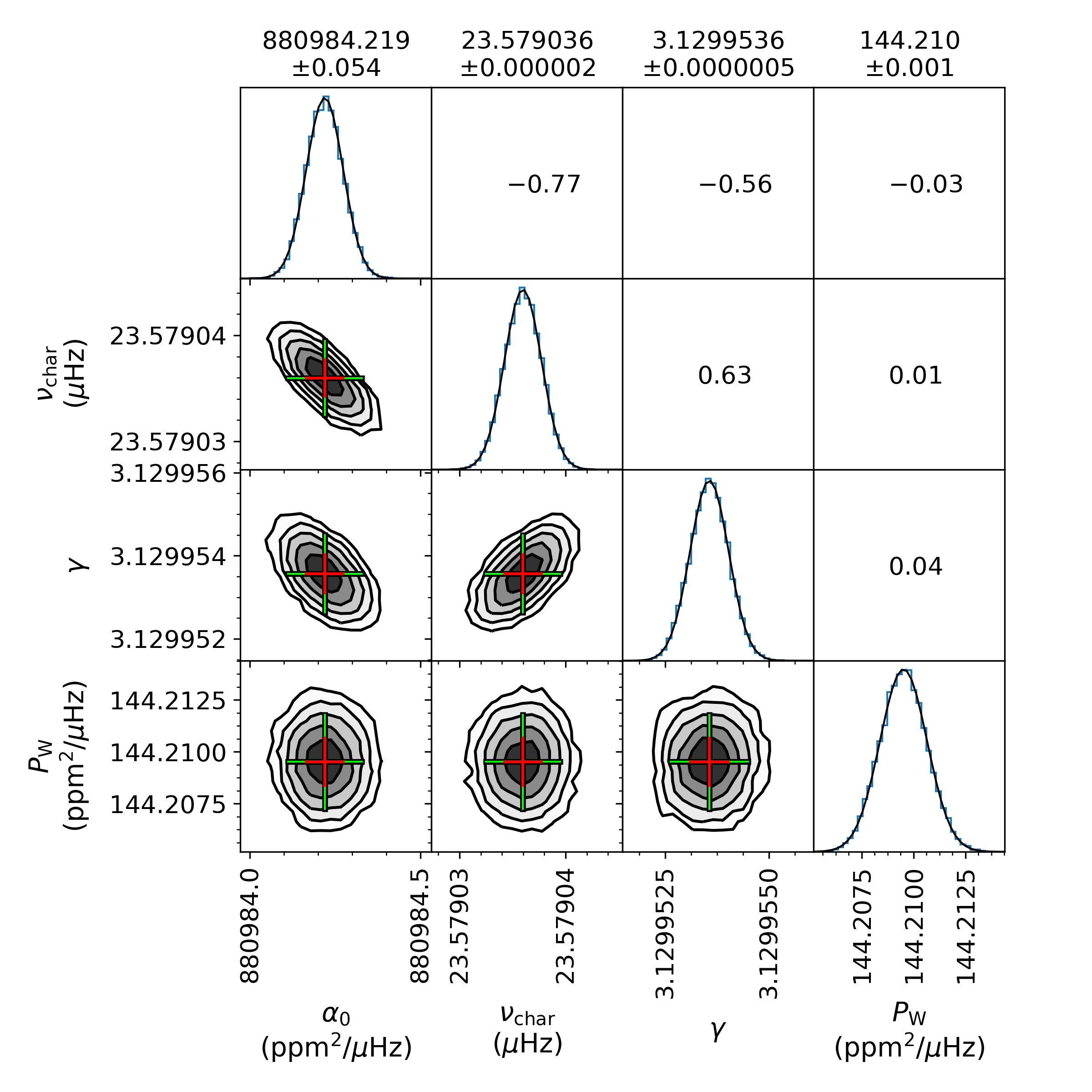}
\includegraphics[width=0.49\textwidth]{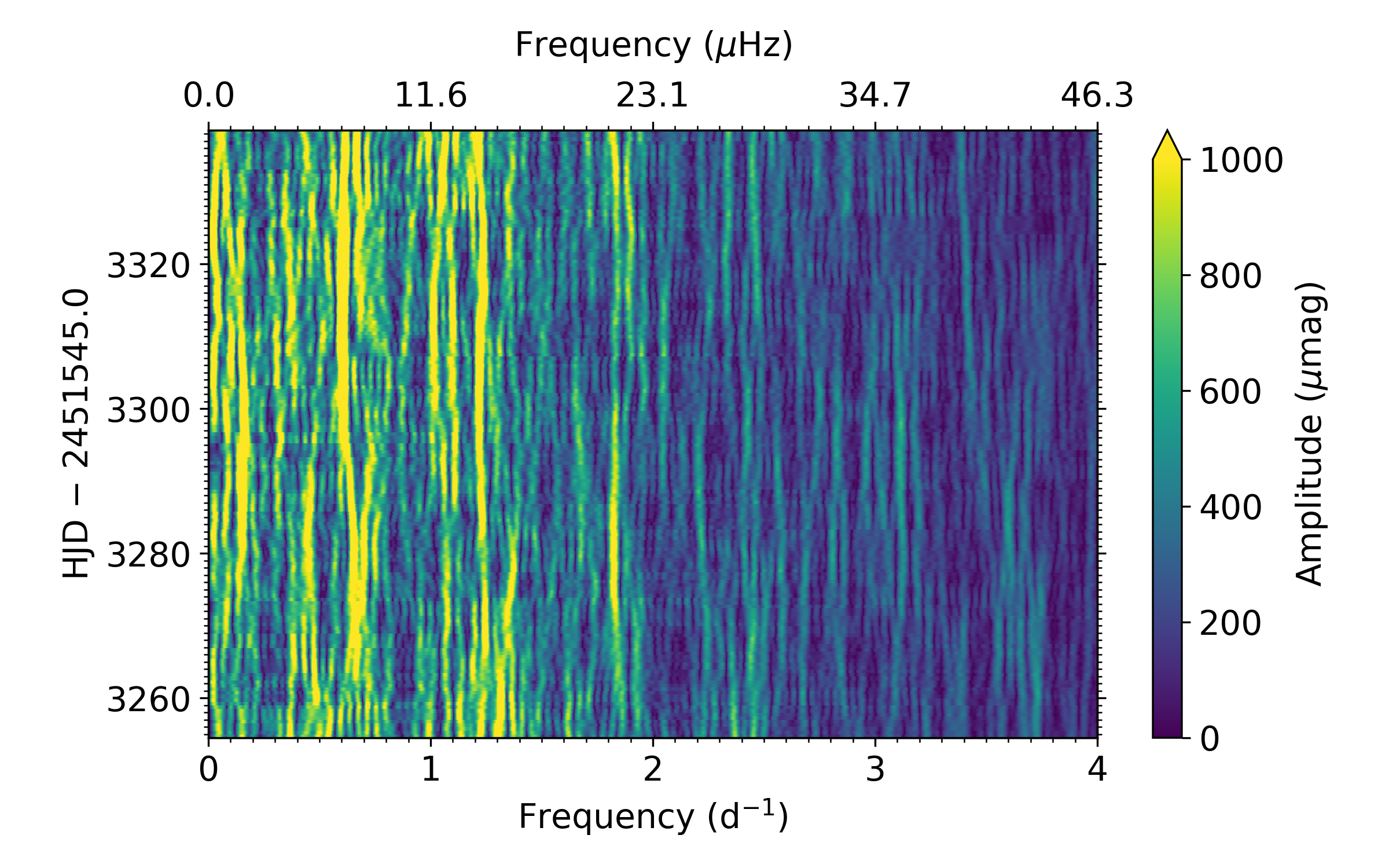}
\caption{Summary figure for the B star HD~51452, which has a similar layout as shown in Fig.~\ref{figure: HD46150}.}
\label{figure: HD51452}
\end{figure}		
	
\clearpage 

\begin{figure}
\centering
\includegraphics[width=0.49\textwidth]{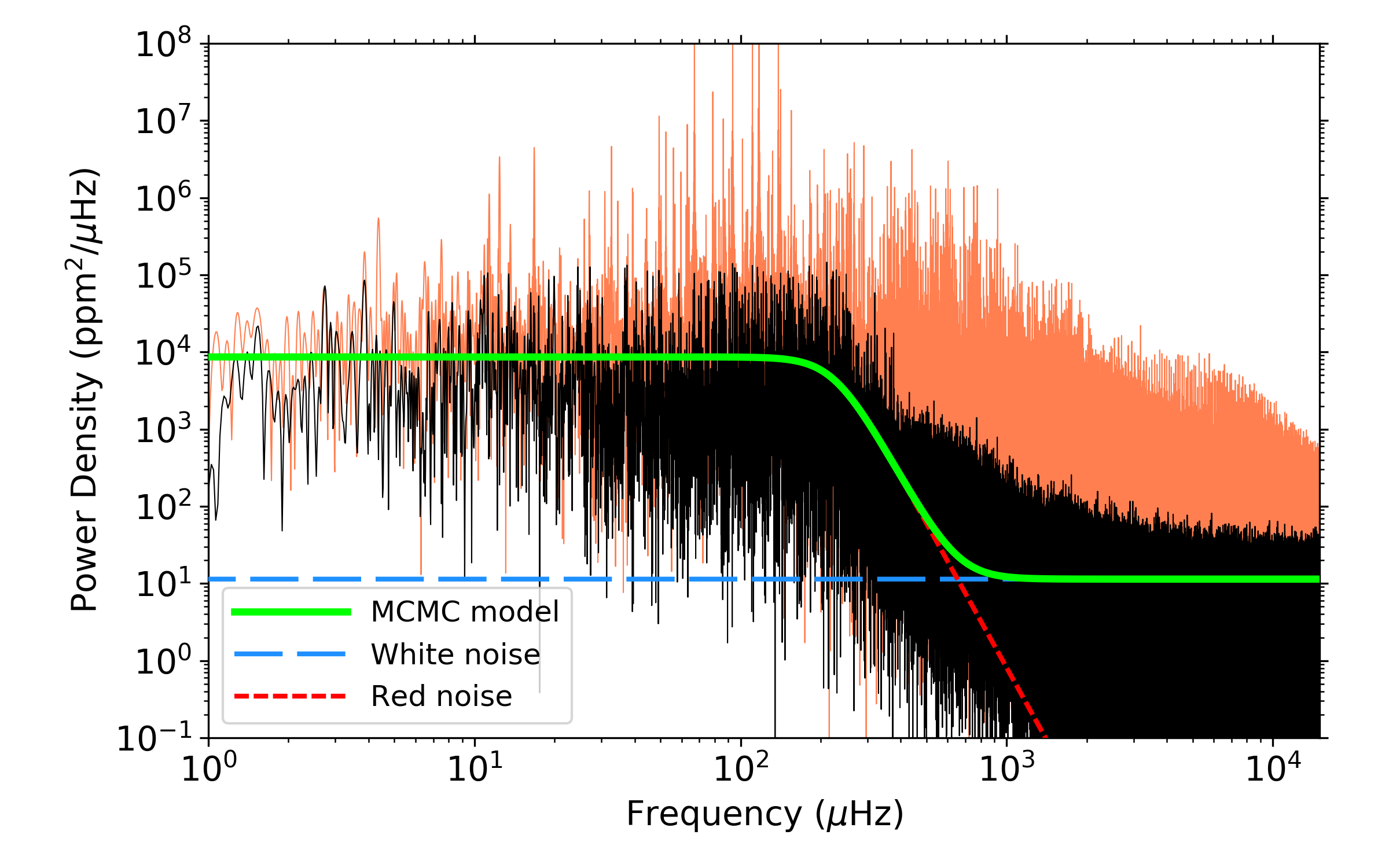}
\includegraphics[width=0.49\textwidth]{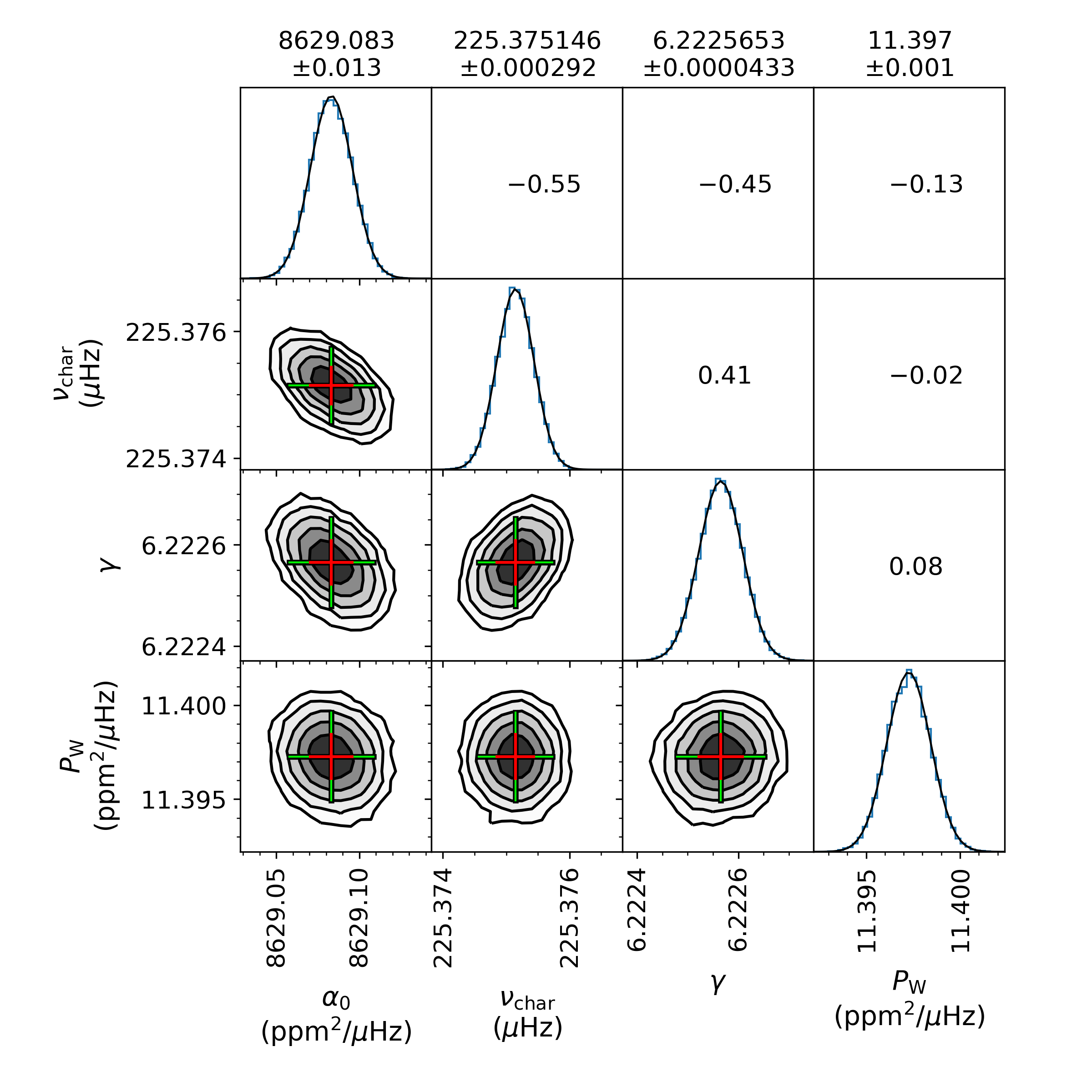}
\includegraphics[width=0.49\textwidth]{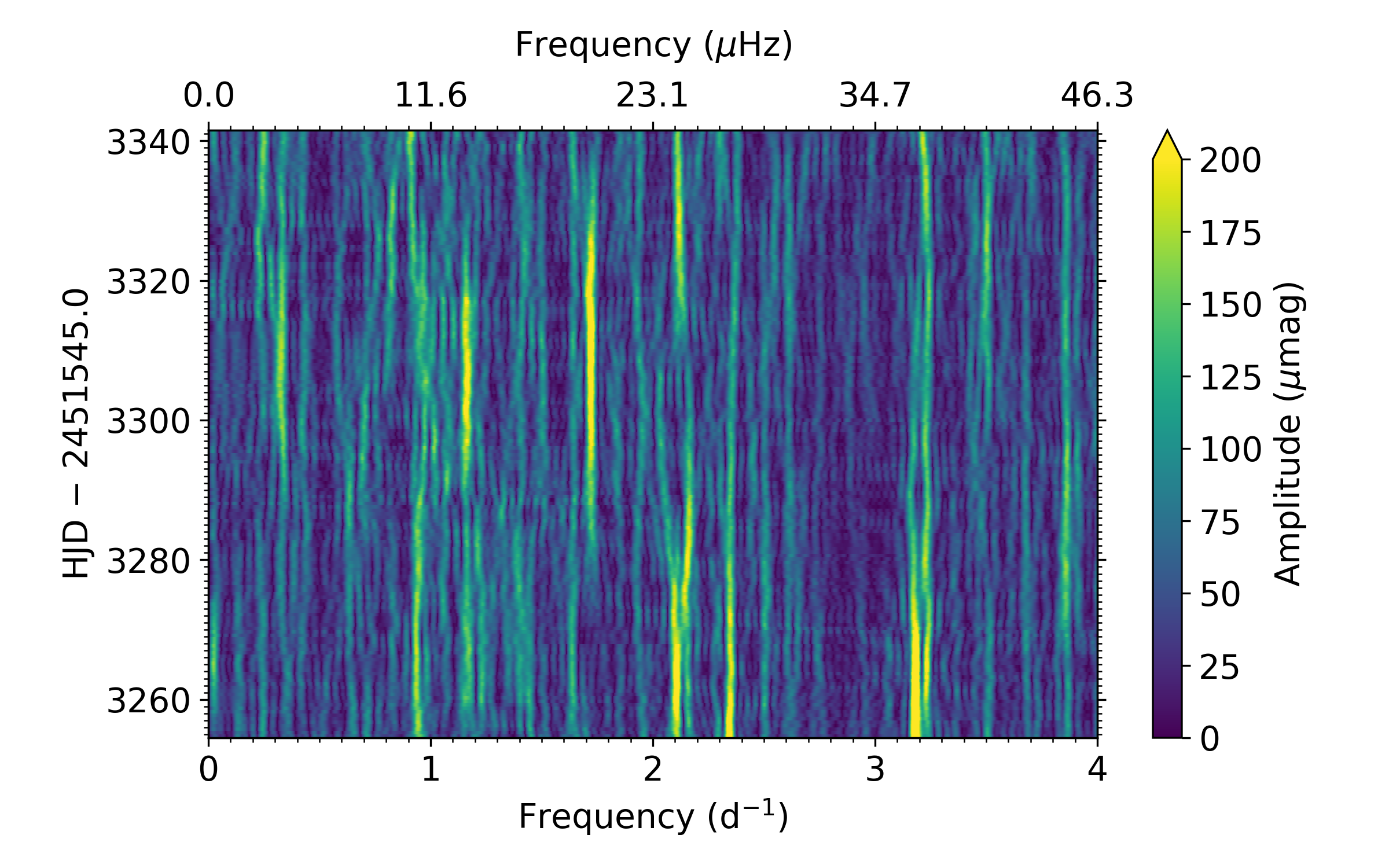}
\caption{Summary figure for the A star HD~51722, which has a similar layout as shown in Fig.~\ref{figure: HD46150}.}
\label{figure: HD51722}
\end{figure}


\begin{figure}
\centering
\includegraphics[width=0.49\textwidth]{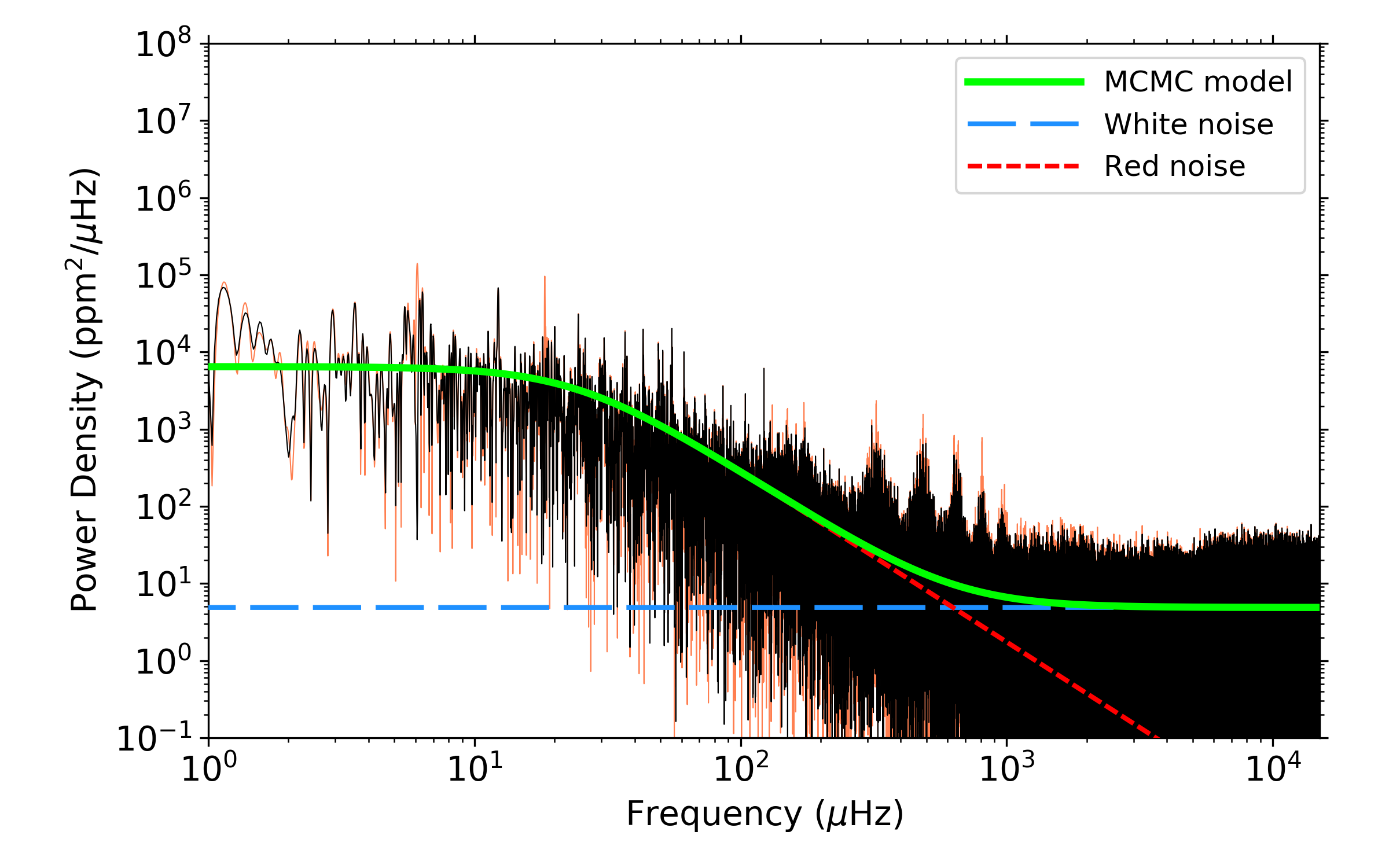}
\includegraphics[width=0.49\textwidth]{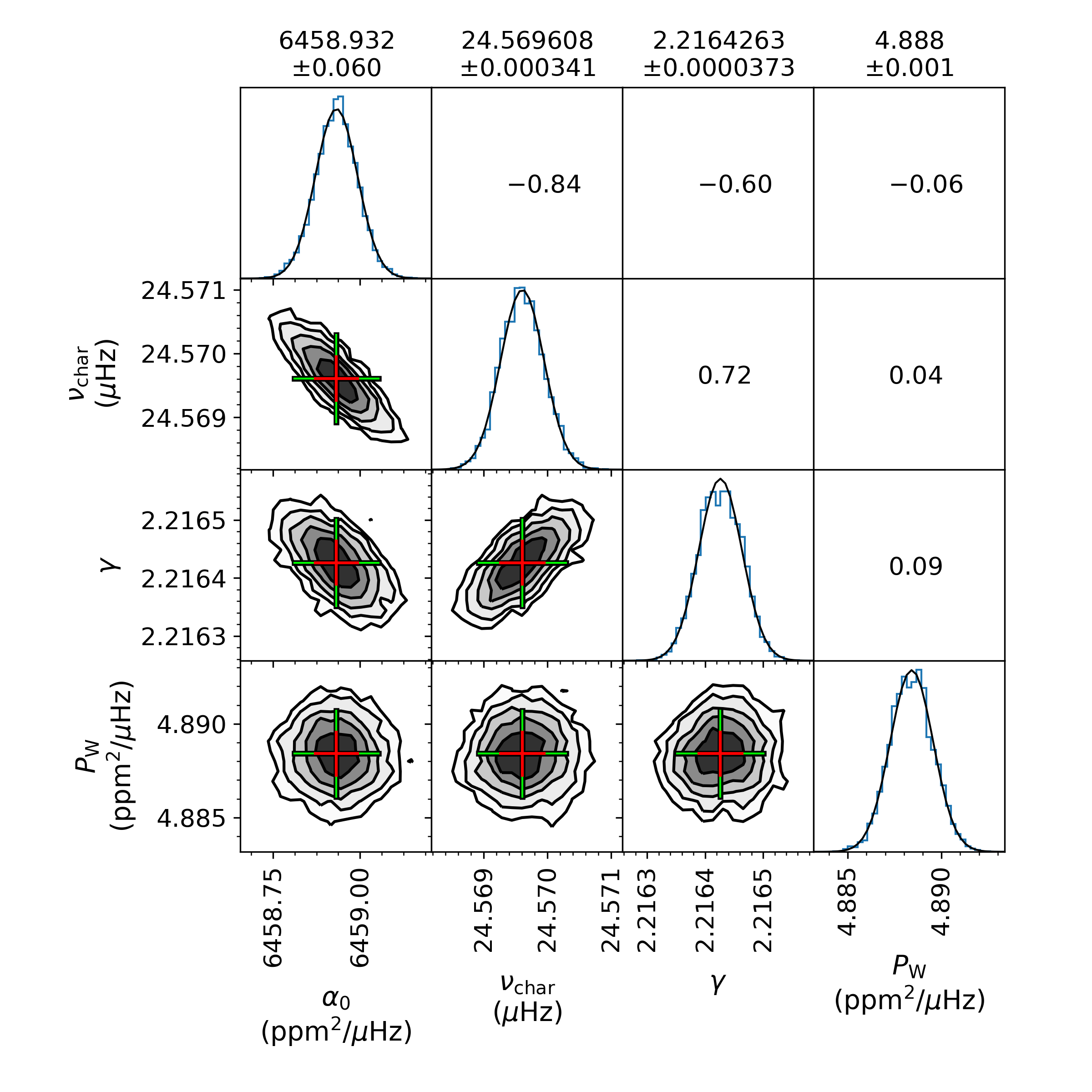}
\includegraphics[width=0.49\textwidth]{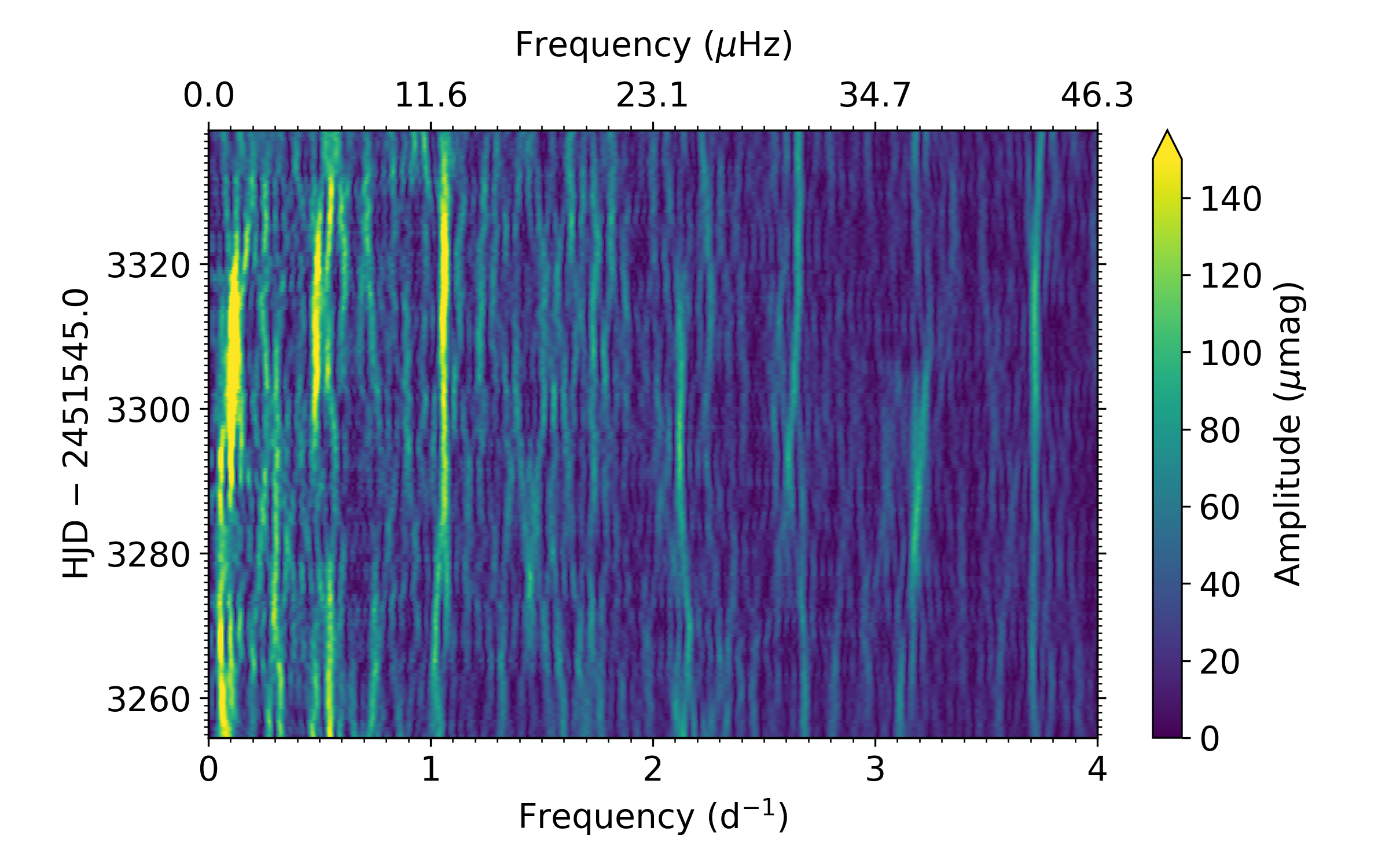}
\caption{Summary figure for the B star HD~51756, which has a similar layout as shown in Fig.~\ref{figure: HD46150}.}
\label{figure: HD51756}
\end{figure}	
	
\clearpage 

\begin{figure}
\centering
\includegraphics[width=0.49\textwidth]{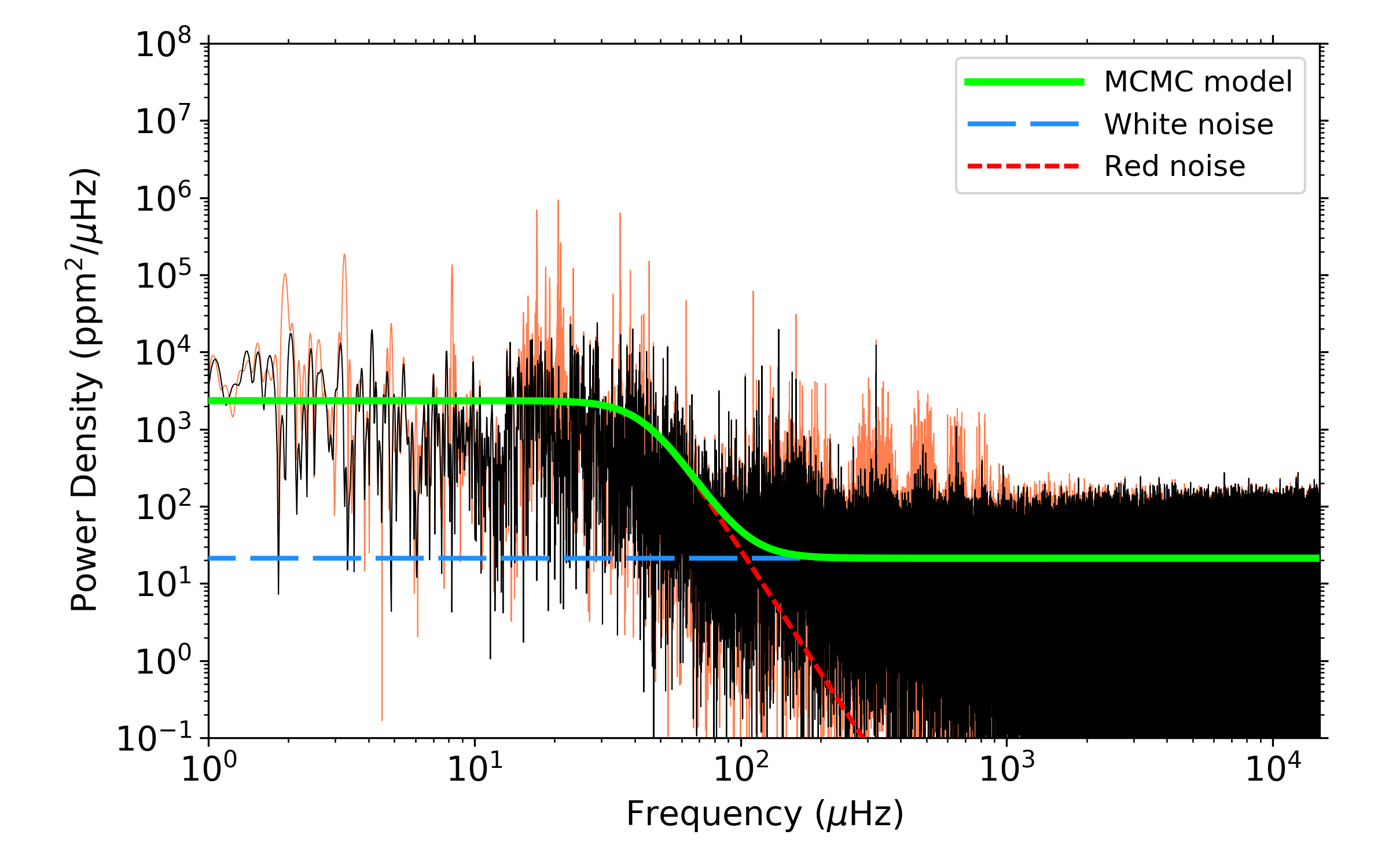}
\includegraphics[width=0.49\textwidth]{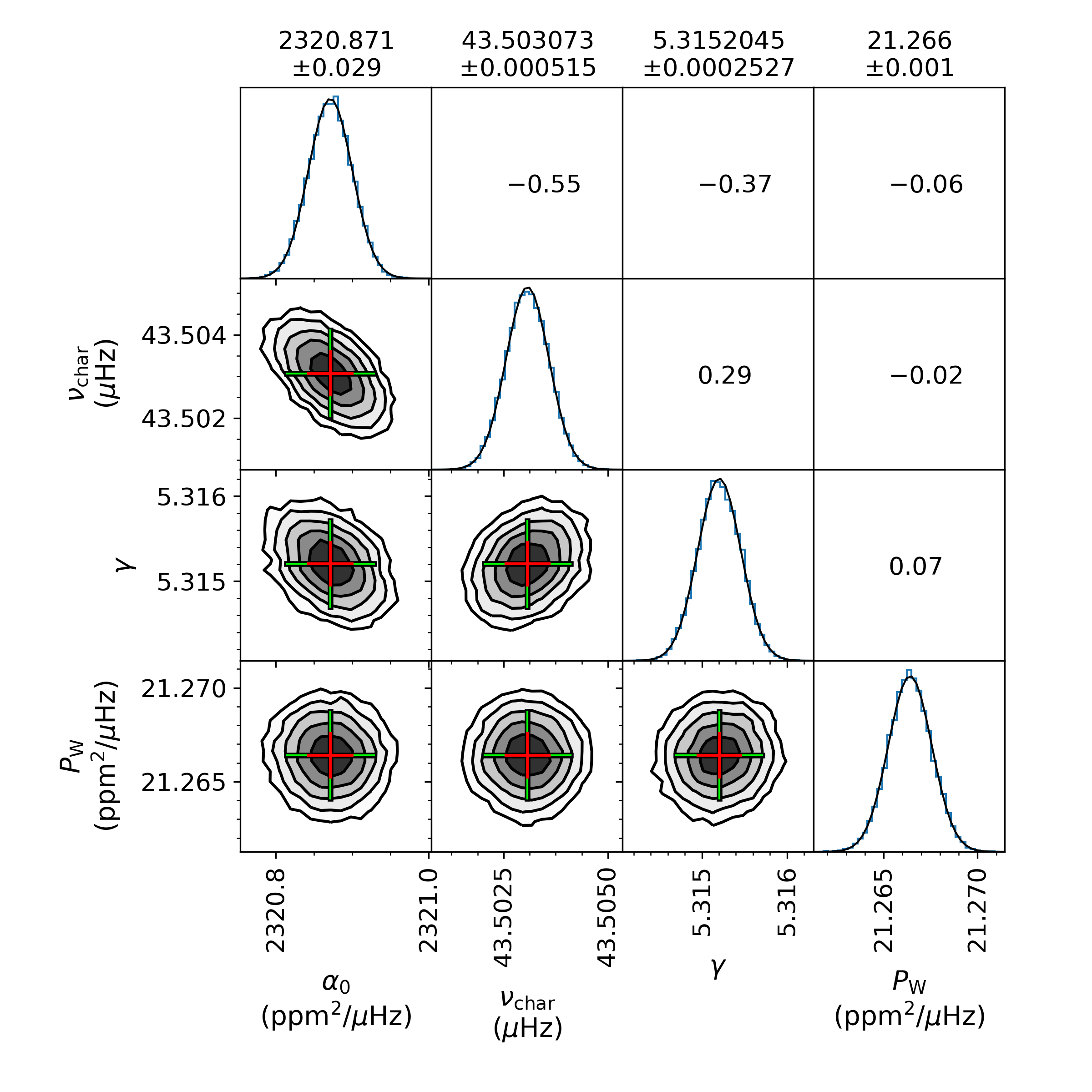}
\includegraphics[width=0.49\textwidth]{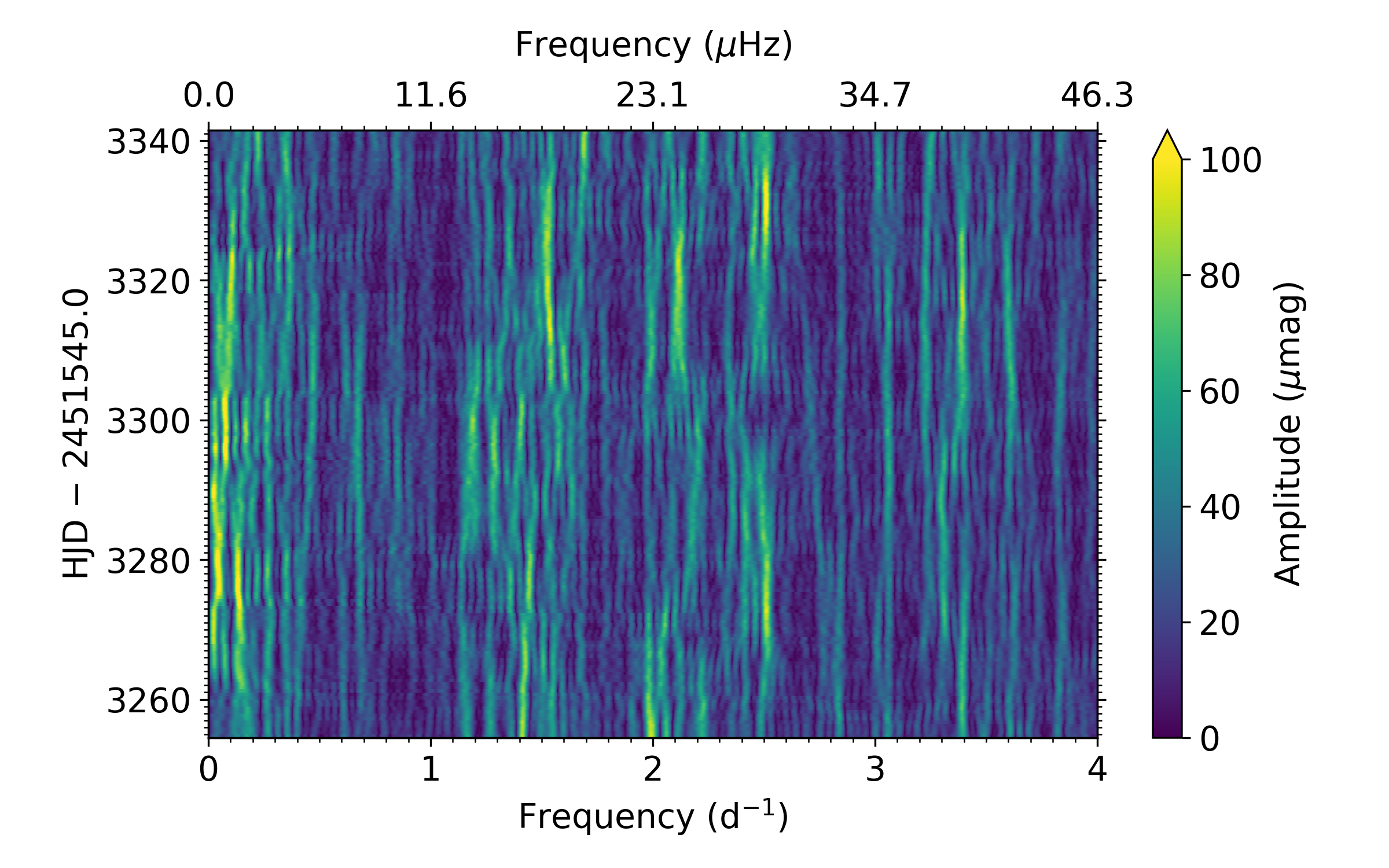}
\caption{Summary figure for the A star HD~52130, which has a similar layout as shown in Fig.~\ref{figure: HD46150}.}
\label{figure: HD52130}
\end{figure}


\begin{figure}
\centering
\includegraphics[width=0.49\textwidth]{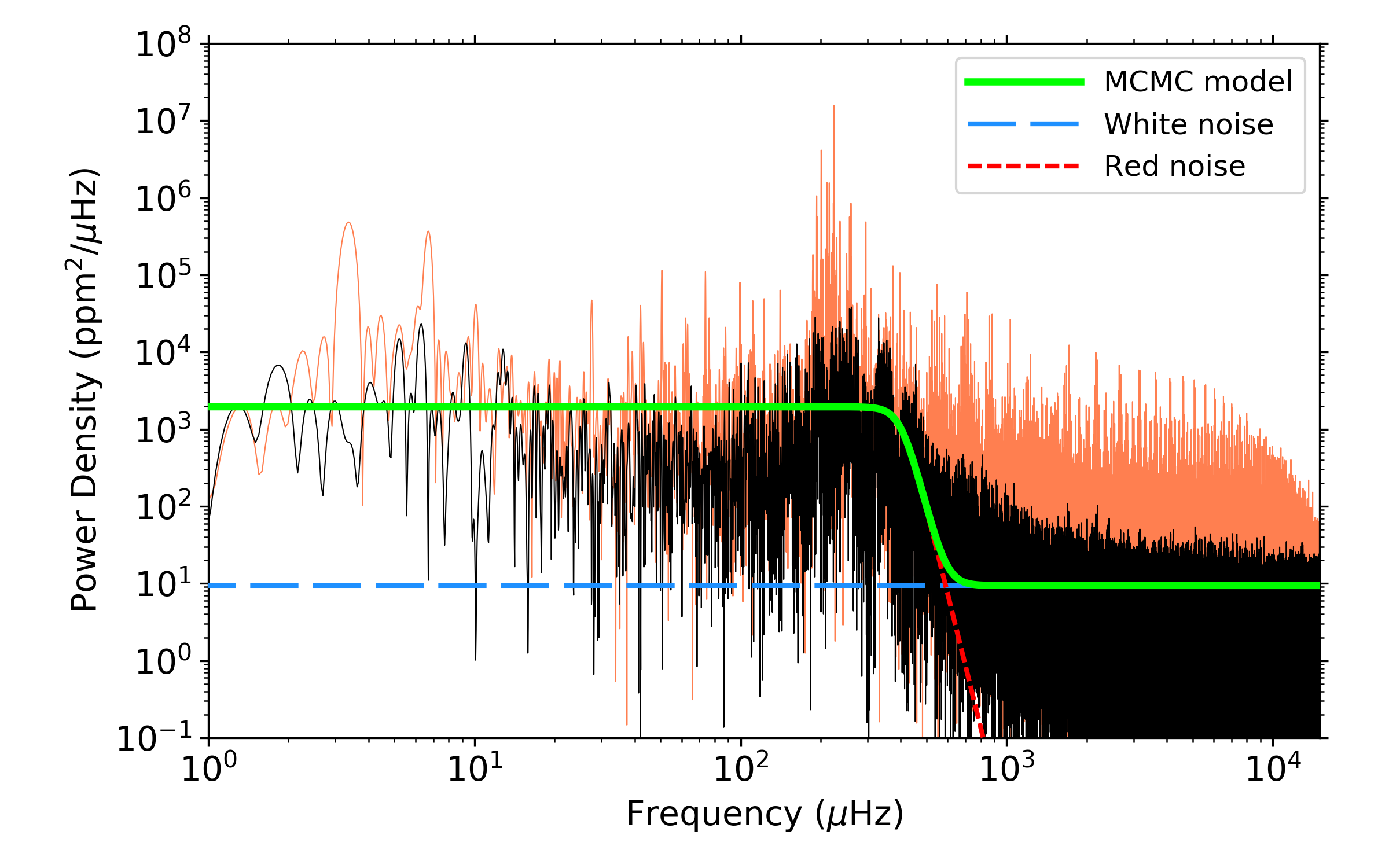}
\includegraphics[width=0.49\textwidth]{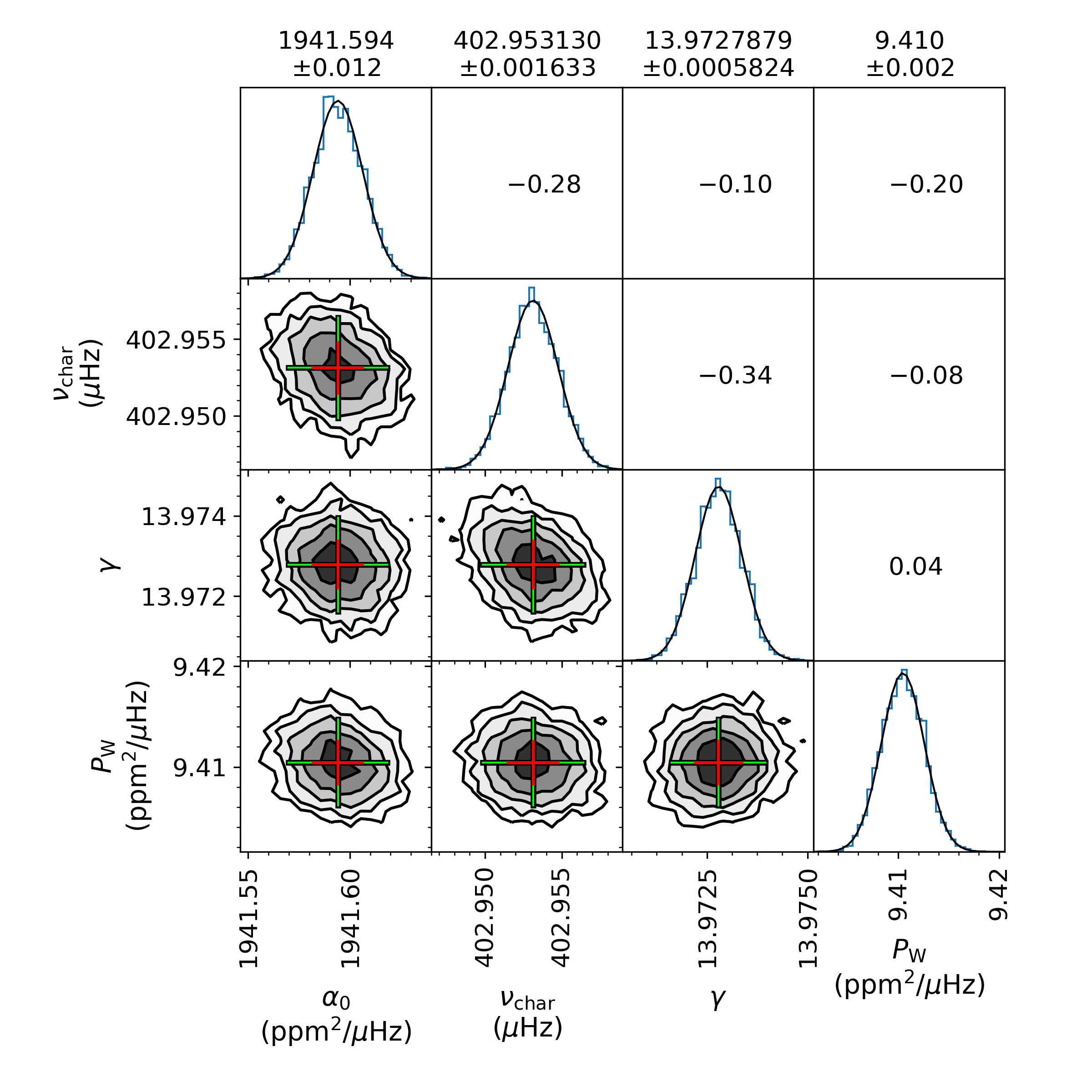}
\includegraphics[width=0.49\textwidth]{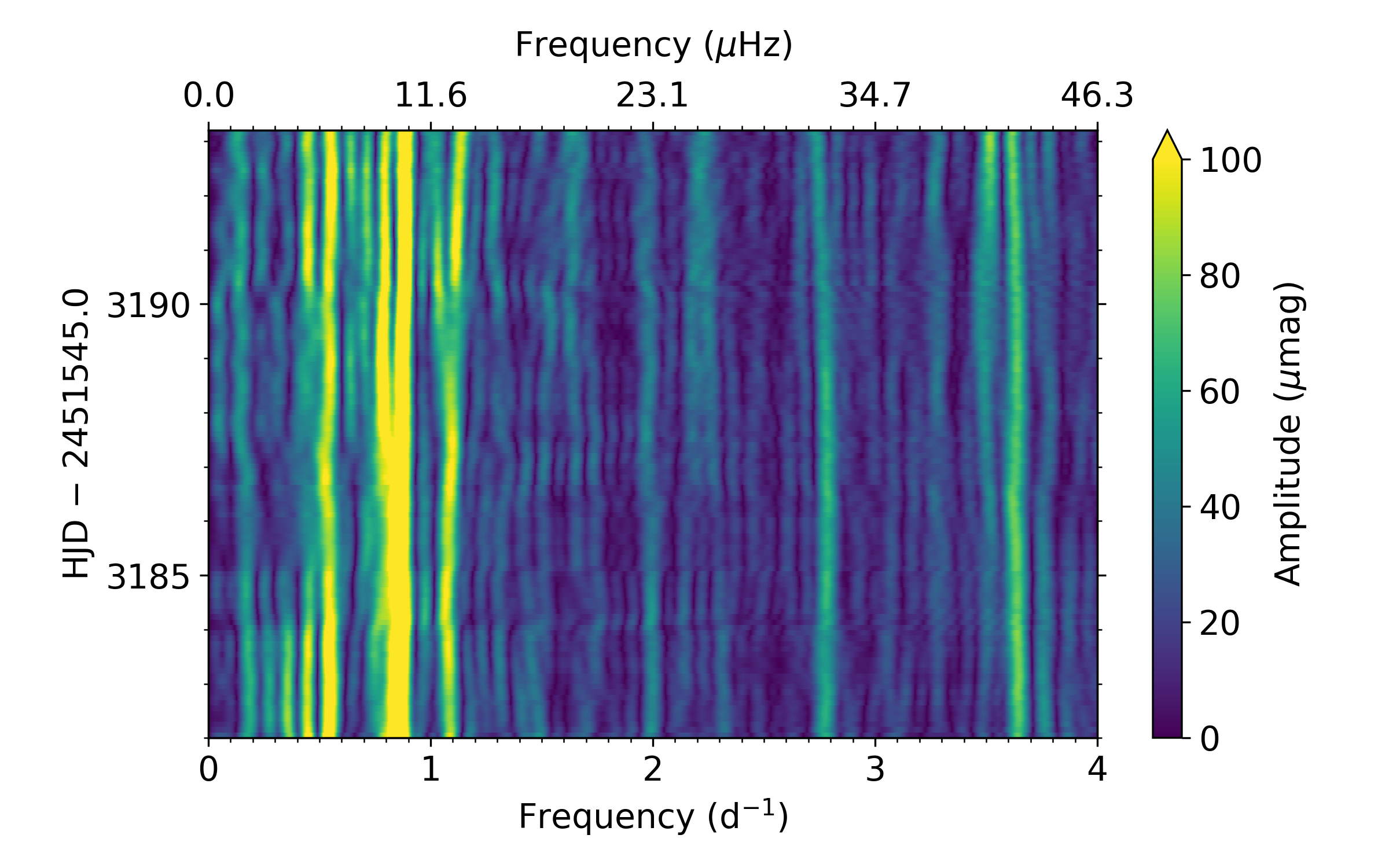}
\caption{Summary figure for the A star HD~174532, which has a similar layout as shown in Fig.~\ref{figure: HD46150}.}
\label{figure: HD174532}
\end{figure}

\clearpage 

\begin{figure}
\centering
\includegraphics[width=0.49\textwidth]{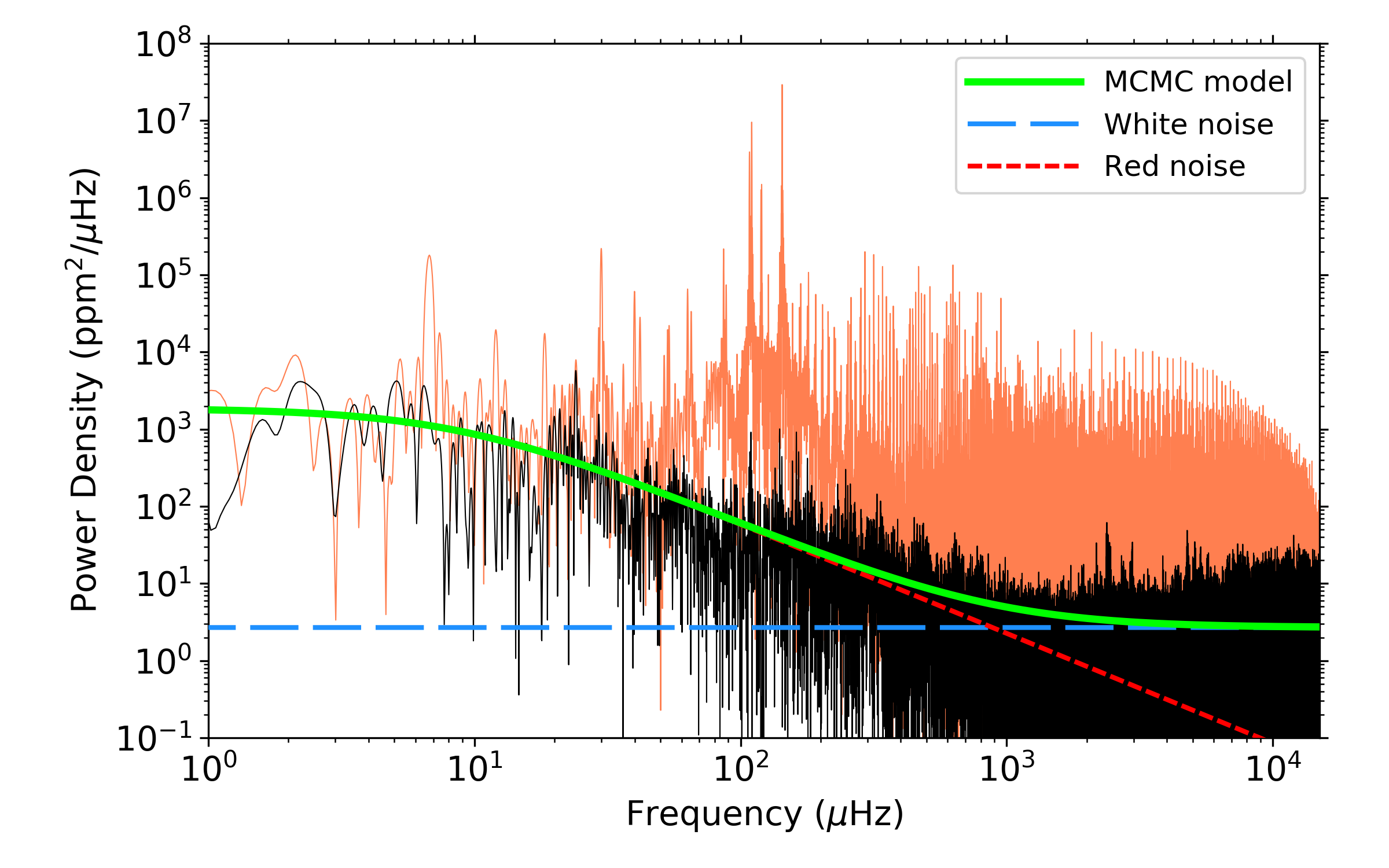}
\includegraphics[width=0.49\textwidth]{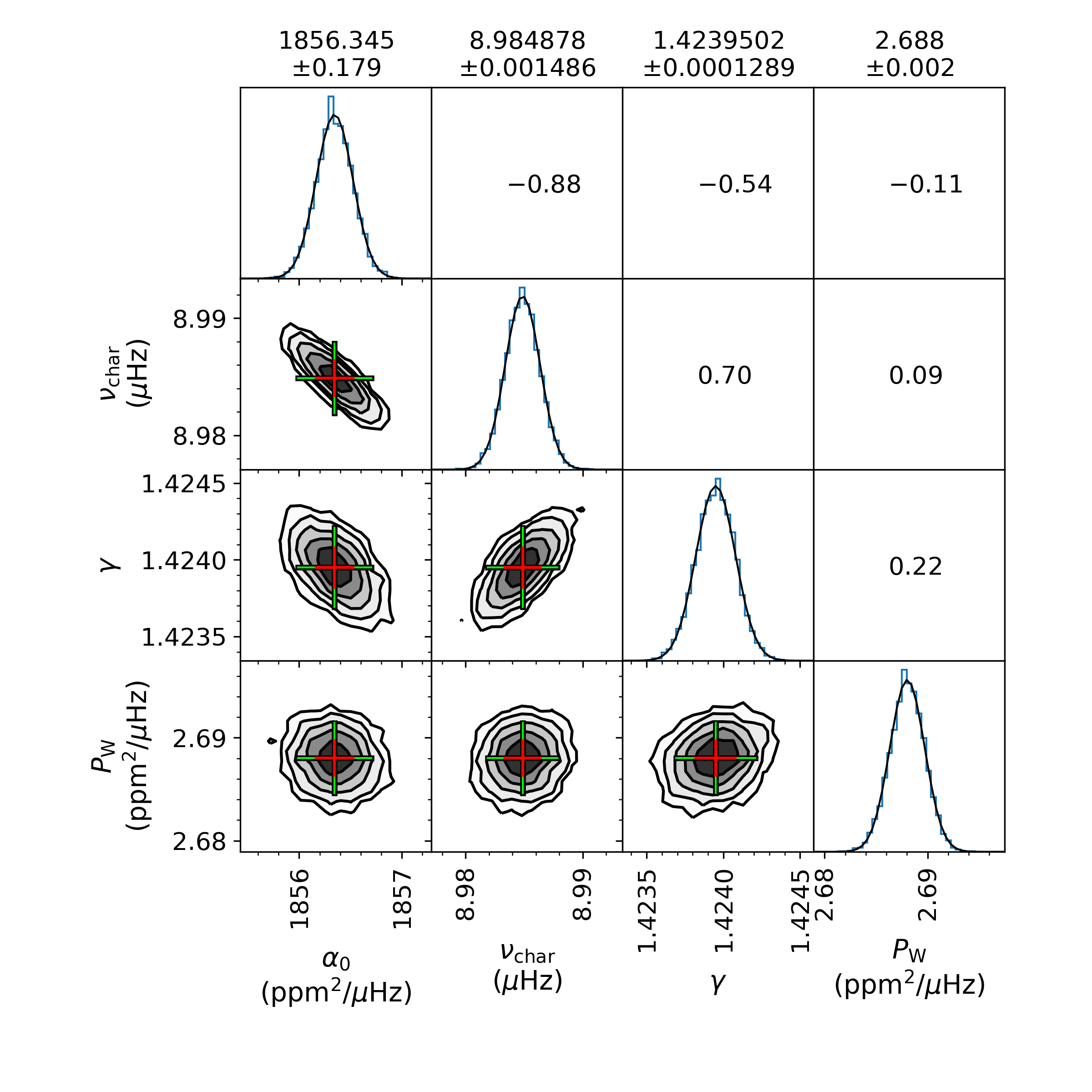}
\includegraphics[width=0.49\textwidth]{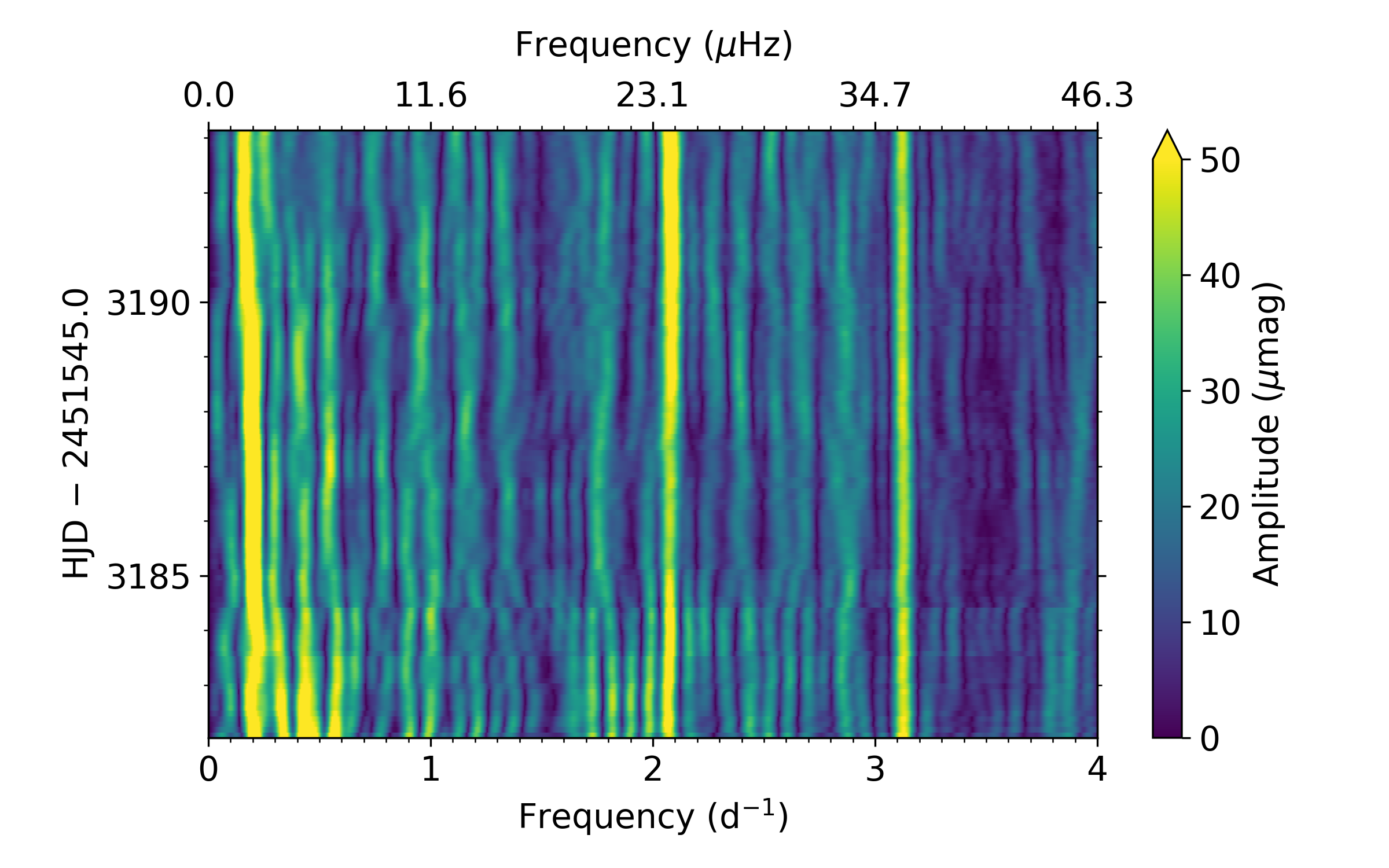}
\caption{Summary figure for the F star HD~174589, which has a similar layout as shown in Fig.~\ref{figure: HD46150}.}
\label{figure: HD174589}
\end{figure}


\begin{figure}
\centering
\includegraphics[width=0.49\textwidth]{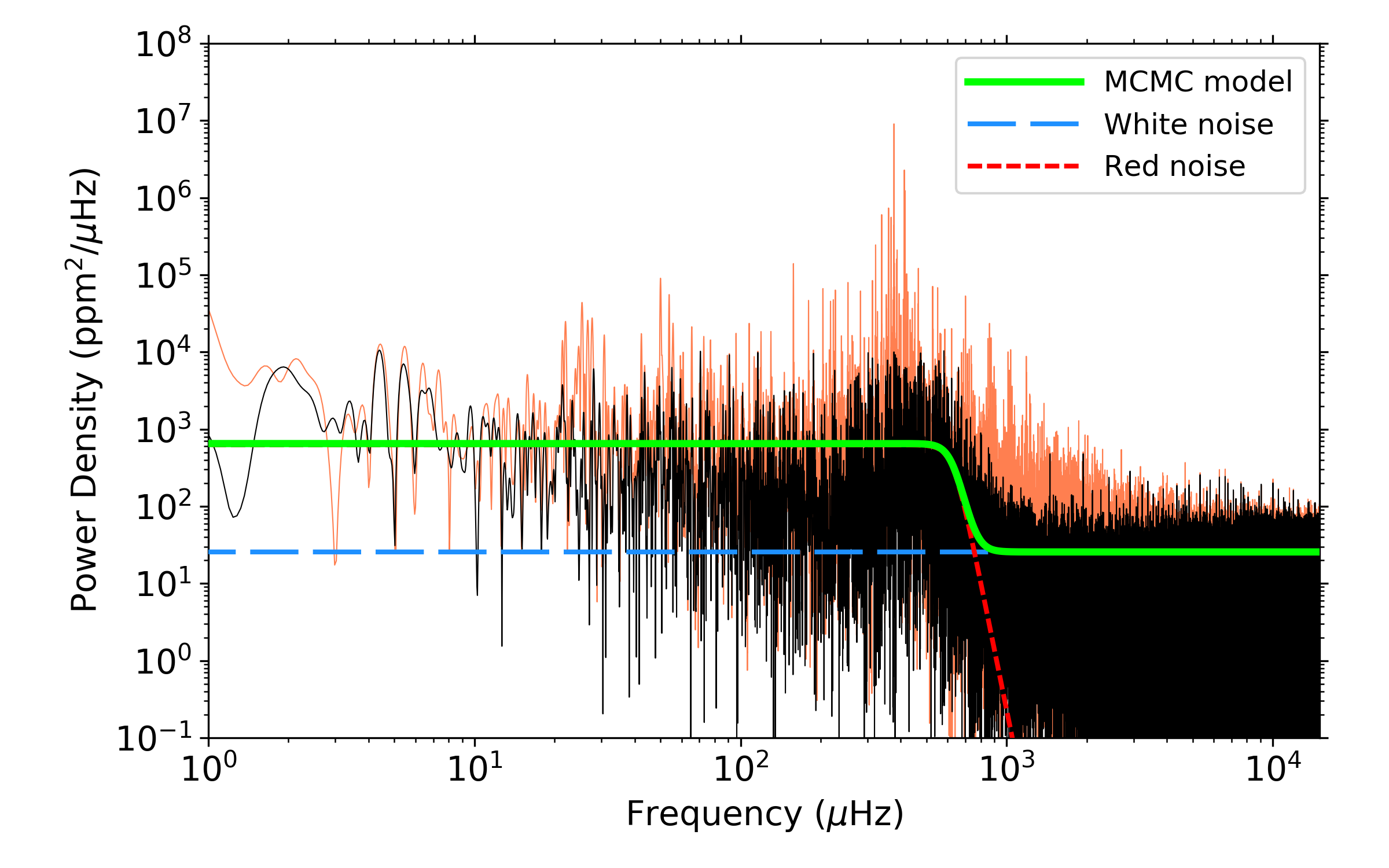}
\includegraphics[width=0.49\textwidth]{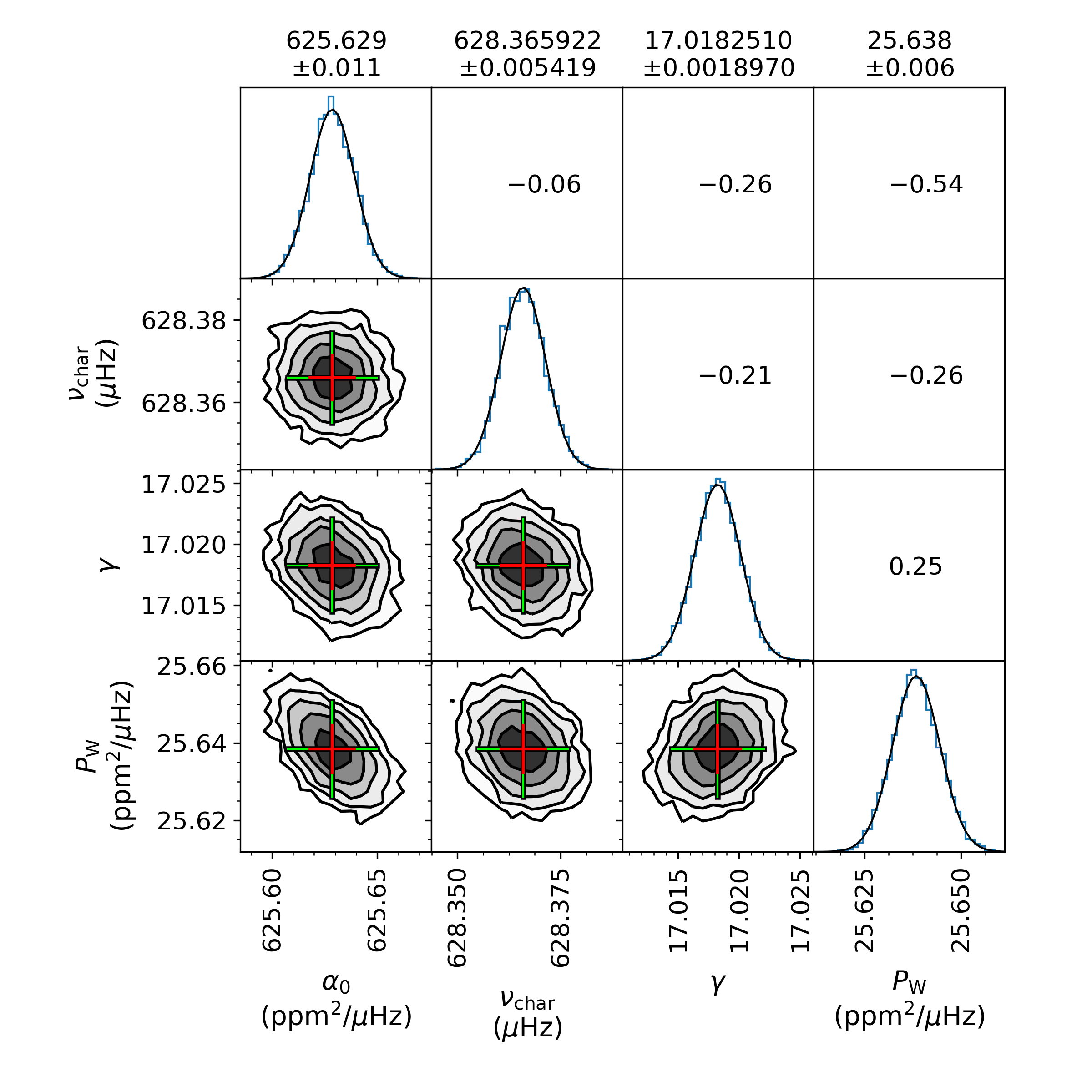}
\includegraphics[width=0.49\textwidth]{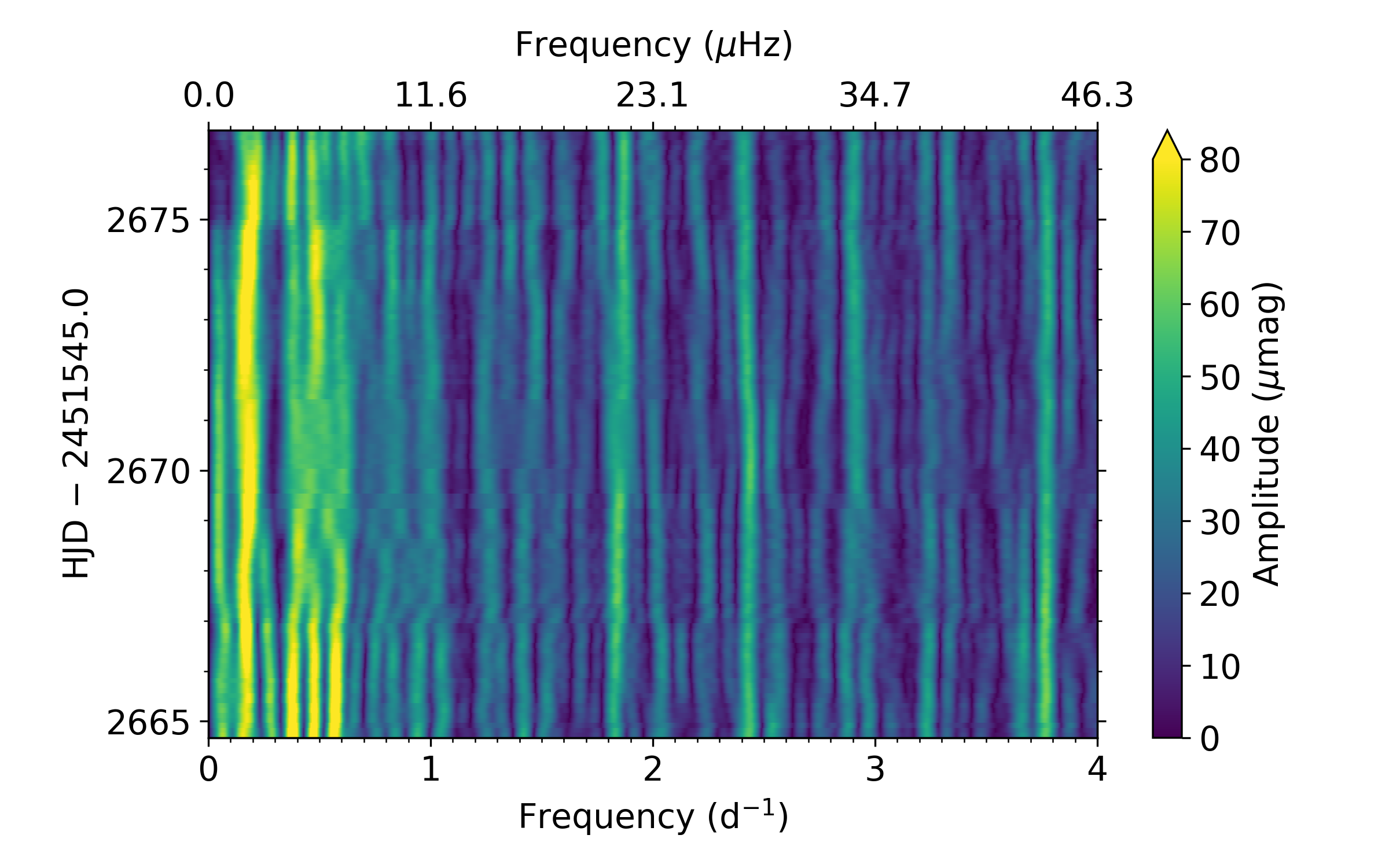}
\caption{Summary figure for the A star HD~174936, which has a similar layout as shown in Fig.~\ref{figure: HD46150}.}
\label{figure: HD174936}
\end{figure}

\clearpage 

\begin{figure}
\centering
\includegraphics[width=0.49\textwidth]{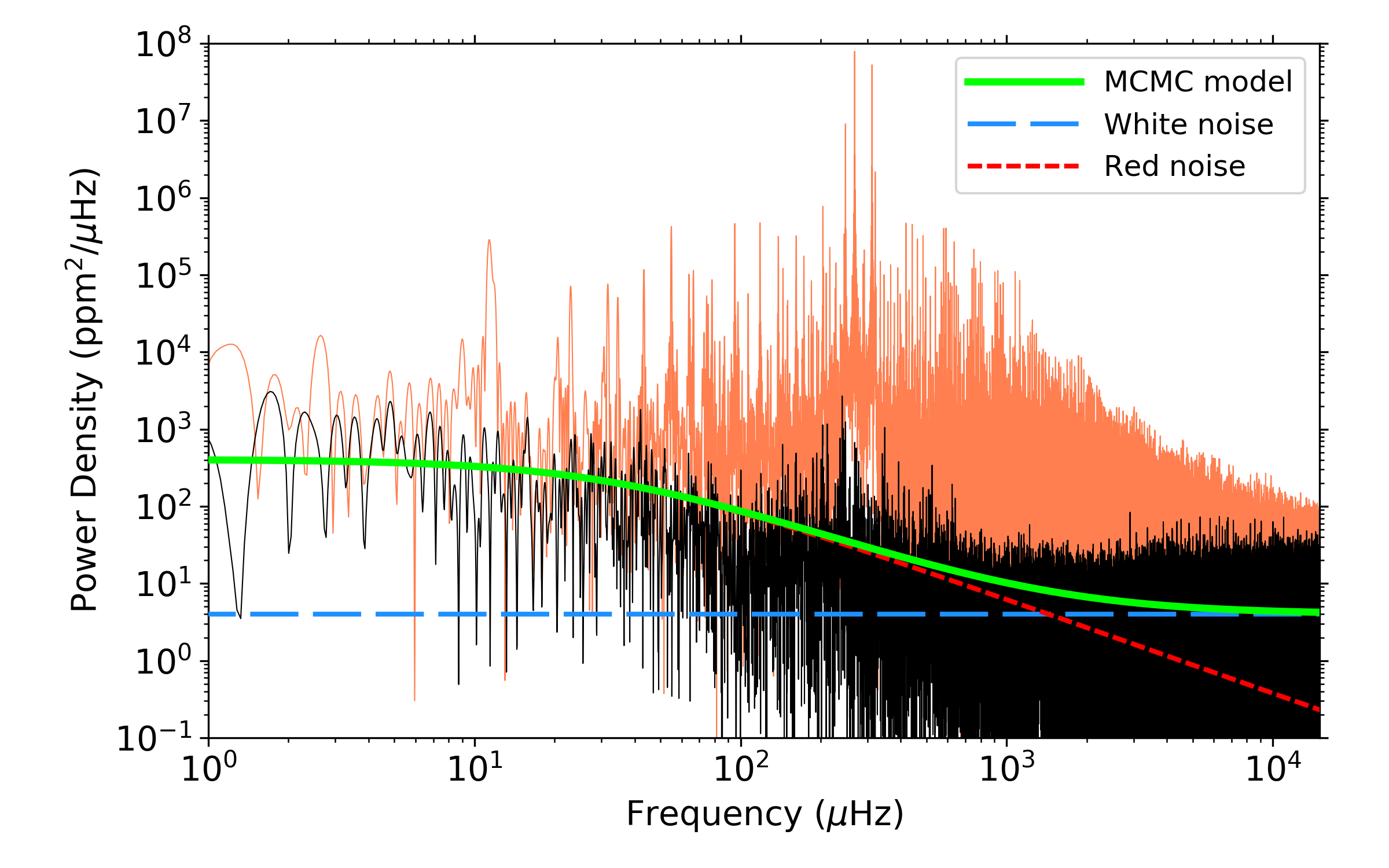}
\includegraphics[width=0.49\textwidth]{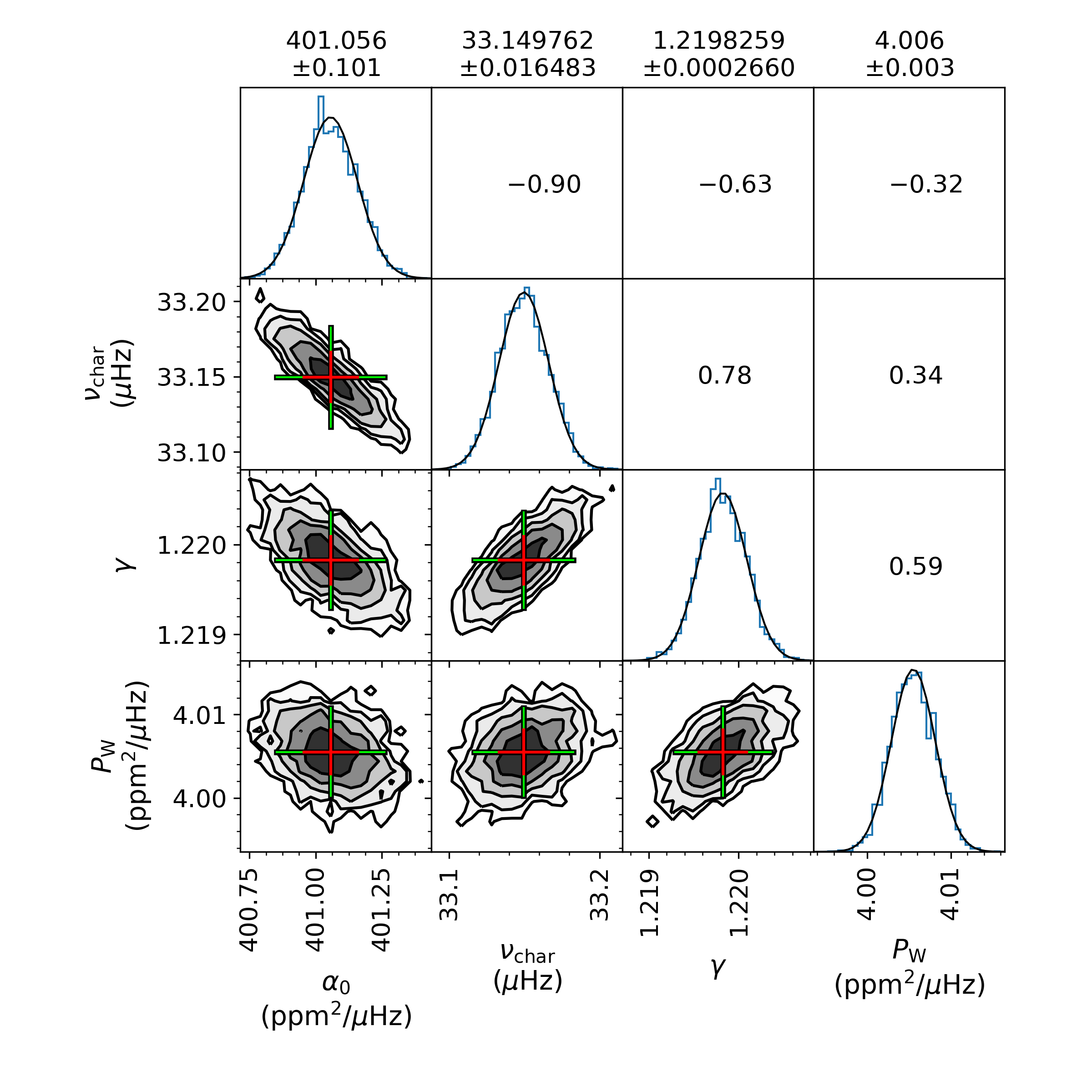}
\includegraphics[width=0.49\textwidth]{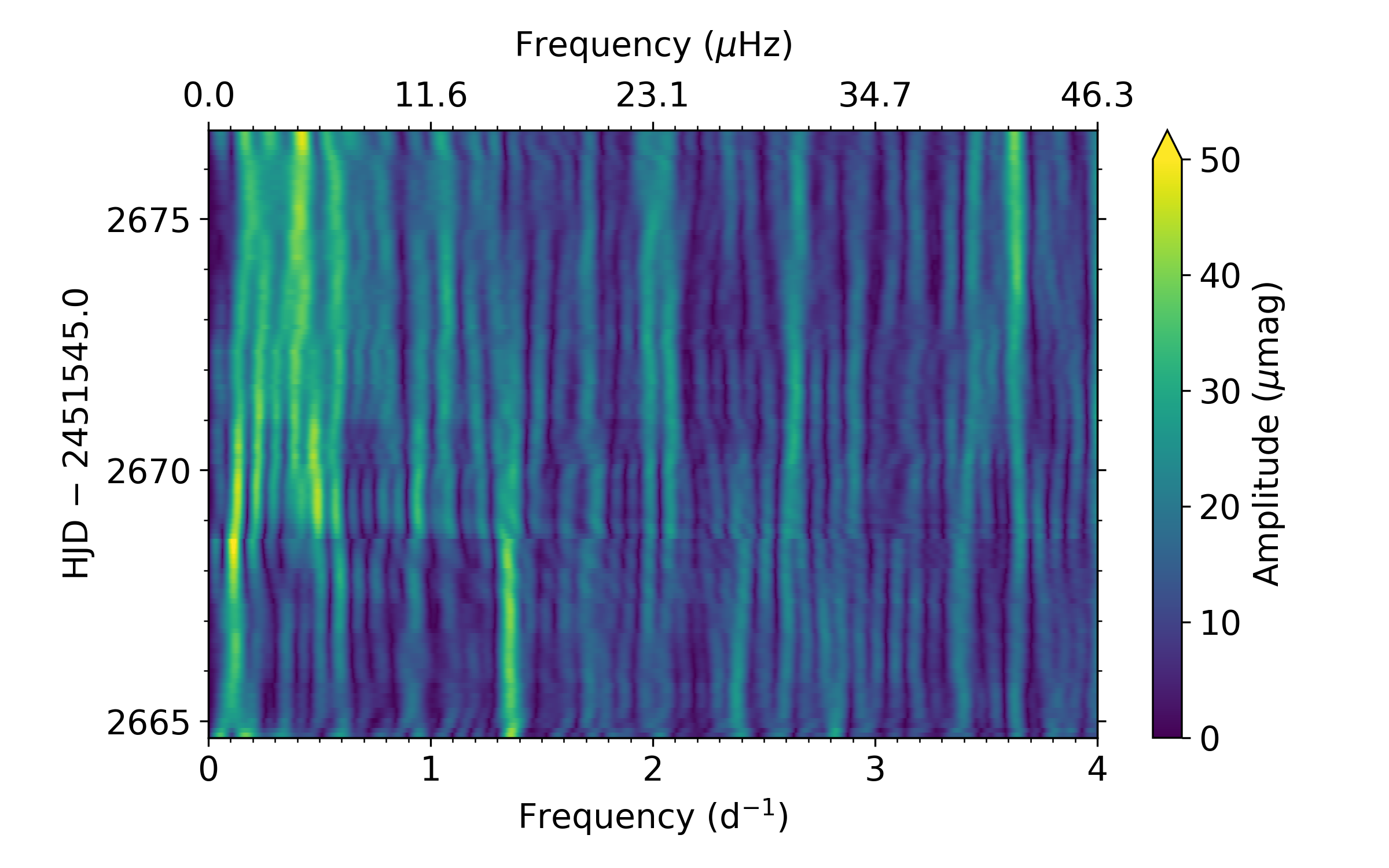}
\caption{Summary figure for the A star HD~174966, which has a similar layout as shown in Fig.~\ref{figure: HD46150}.}
\label{figure: HD174966}
\end{figure}


\begin{figure}
\centering
\includegraphics[width=0.49\textwidth]{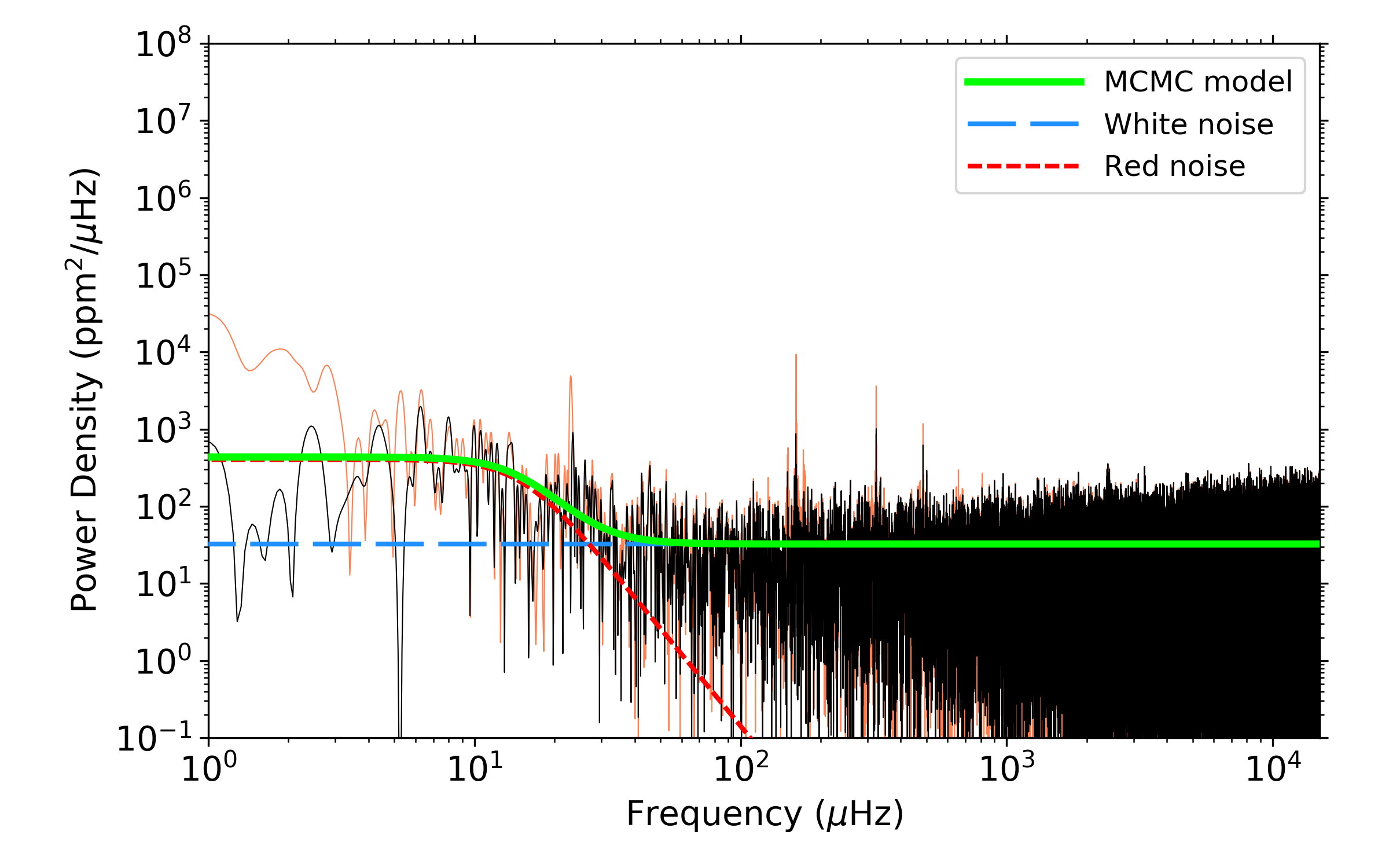}
\includegraphics[width=0.49\textwidth]{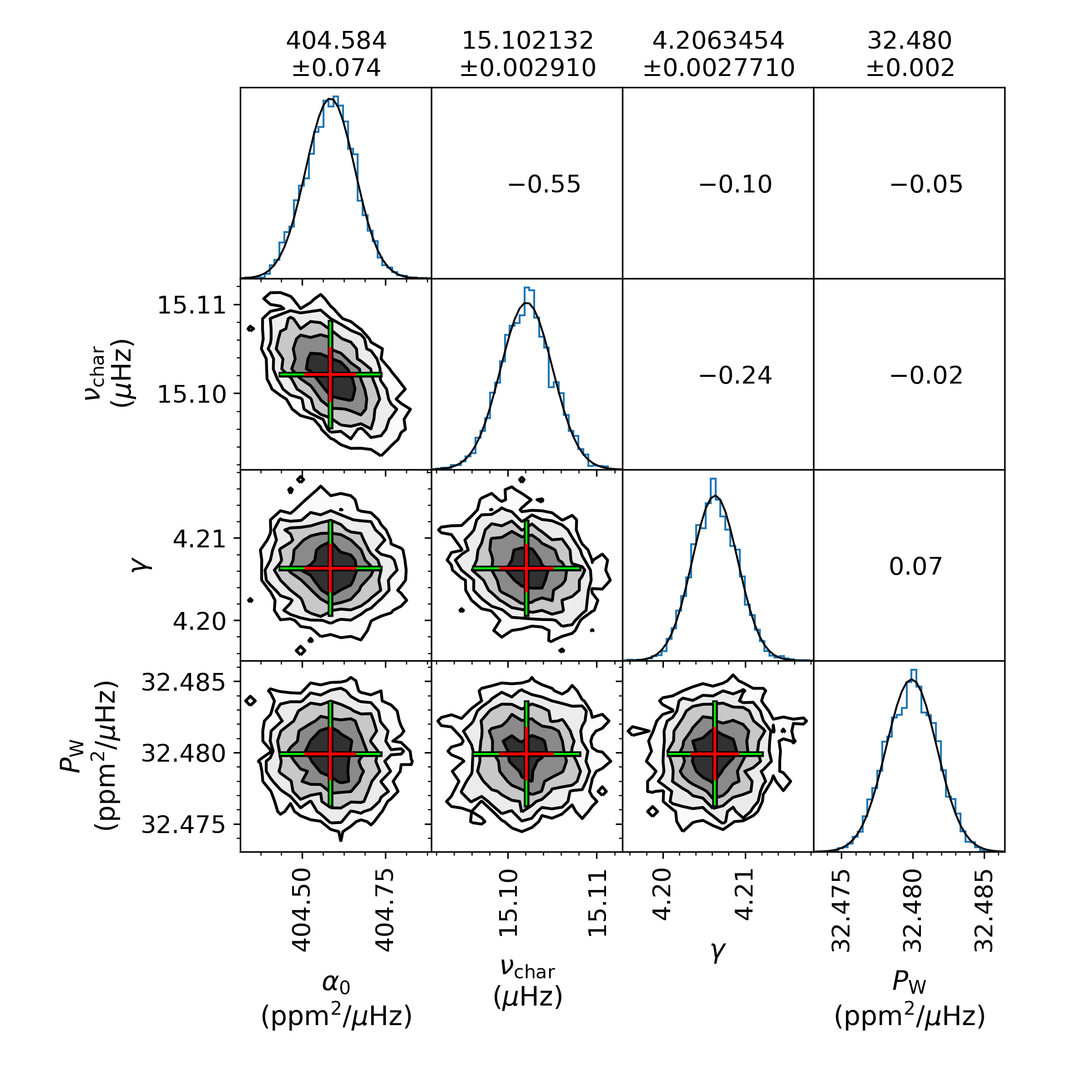}
\includegraphics[width=0.49\textwidth]{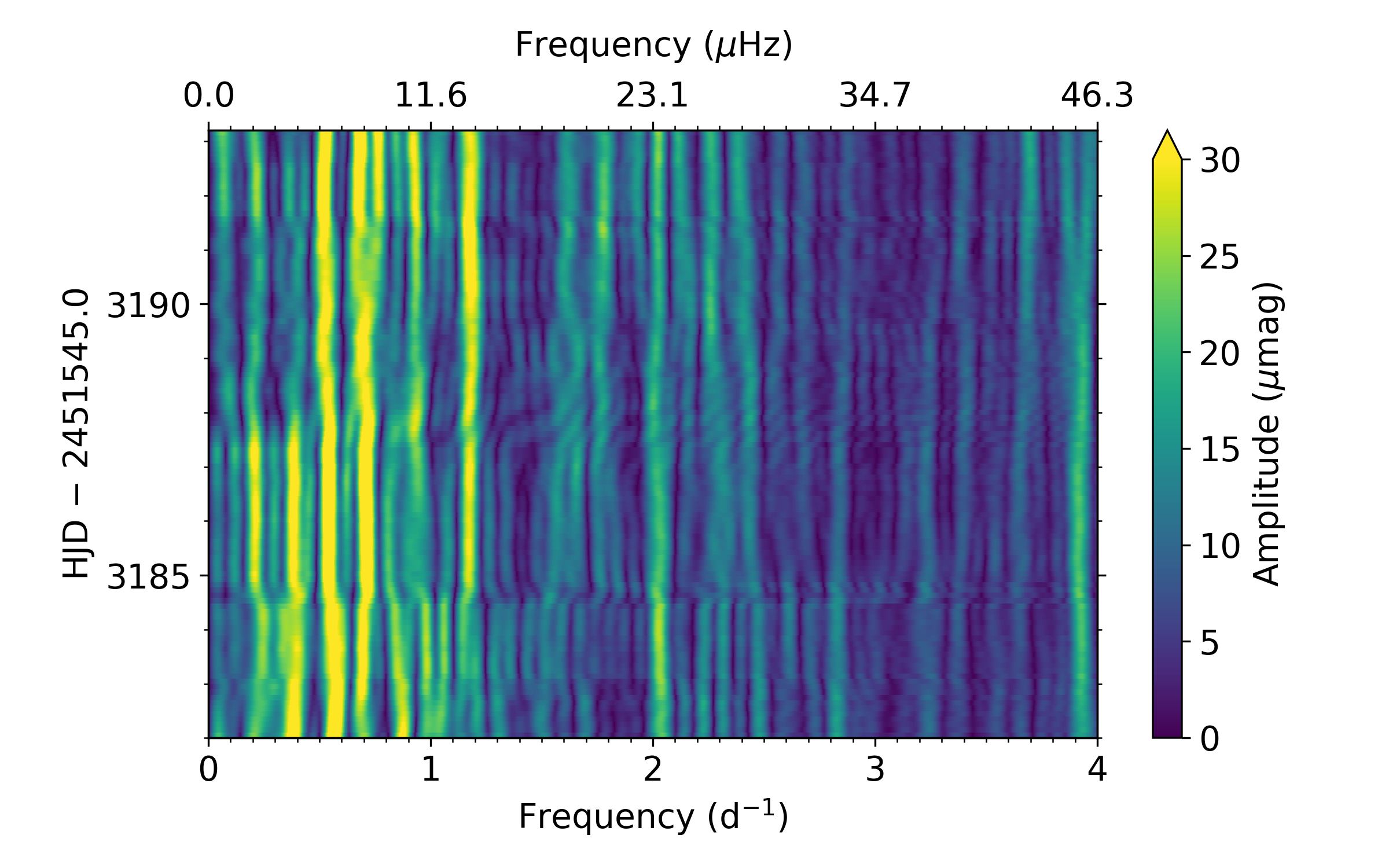}
\caption{Summary figure for the B star HD~174967, which has a similar layout as shown in Fig.~\ref{figure: HD46150}.}
\label{figure: HD174967}
\end{figure}

\clearpage 

\begin{figure}
\centering
\includegraphics[width=0.49\textwidth]{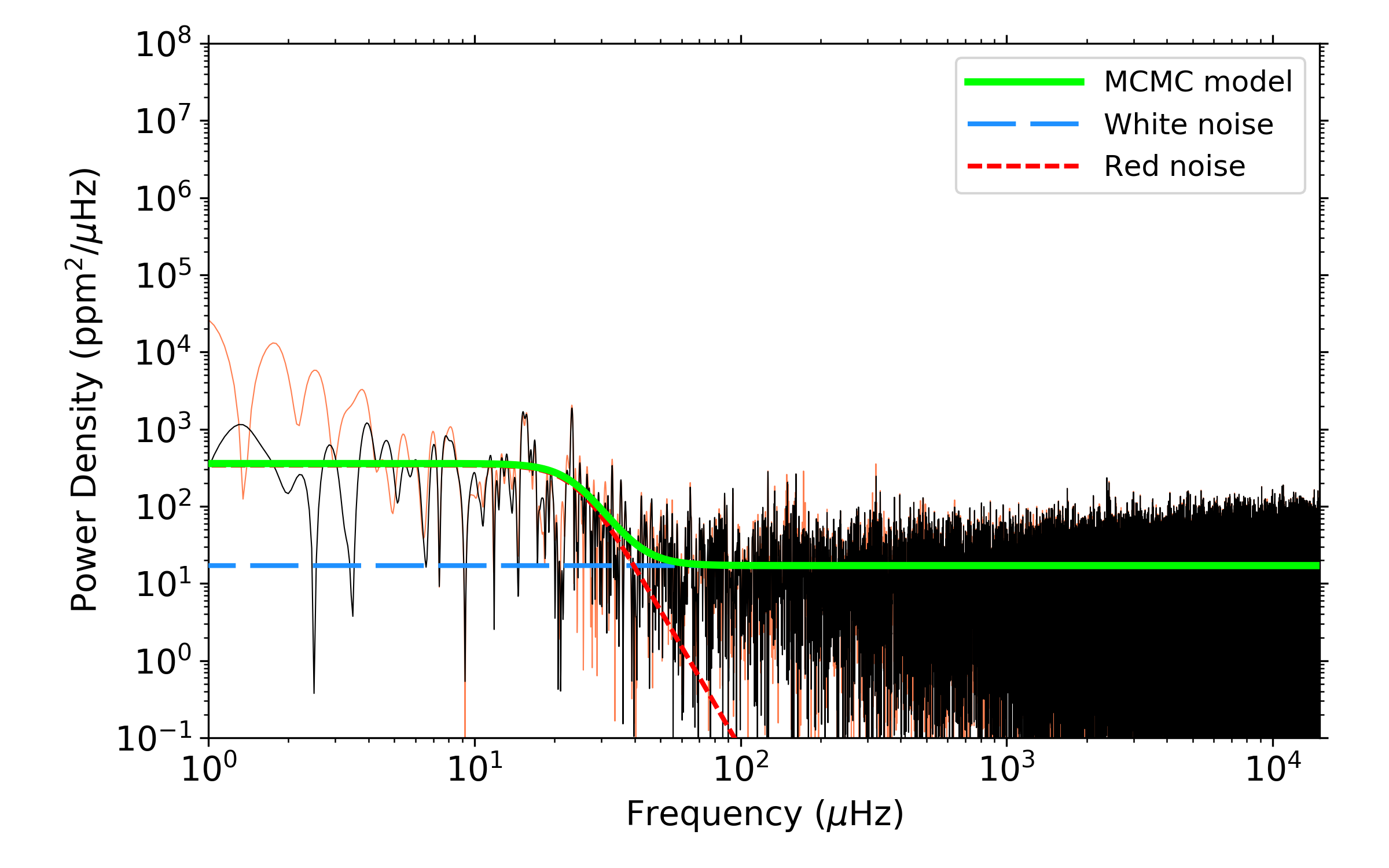}
\includegraphics[width=0.49\textwidth]{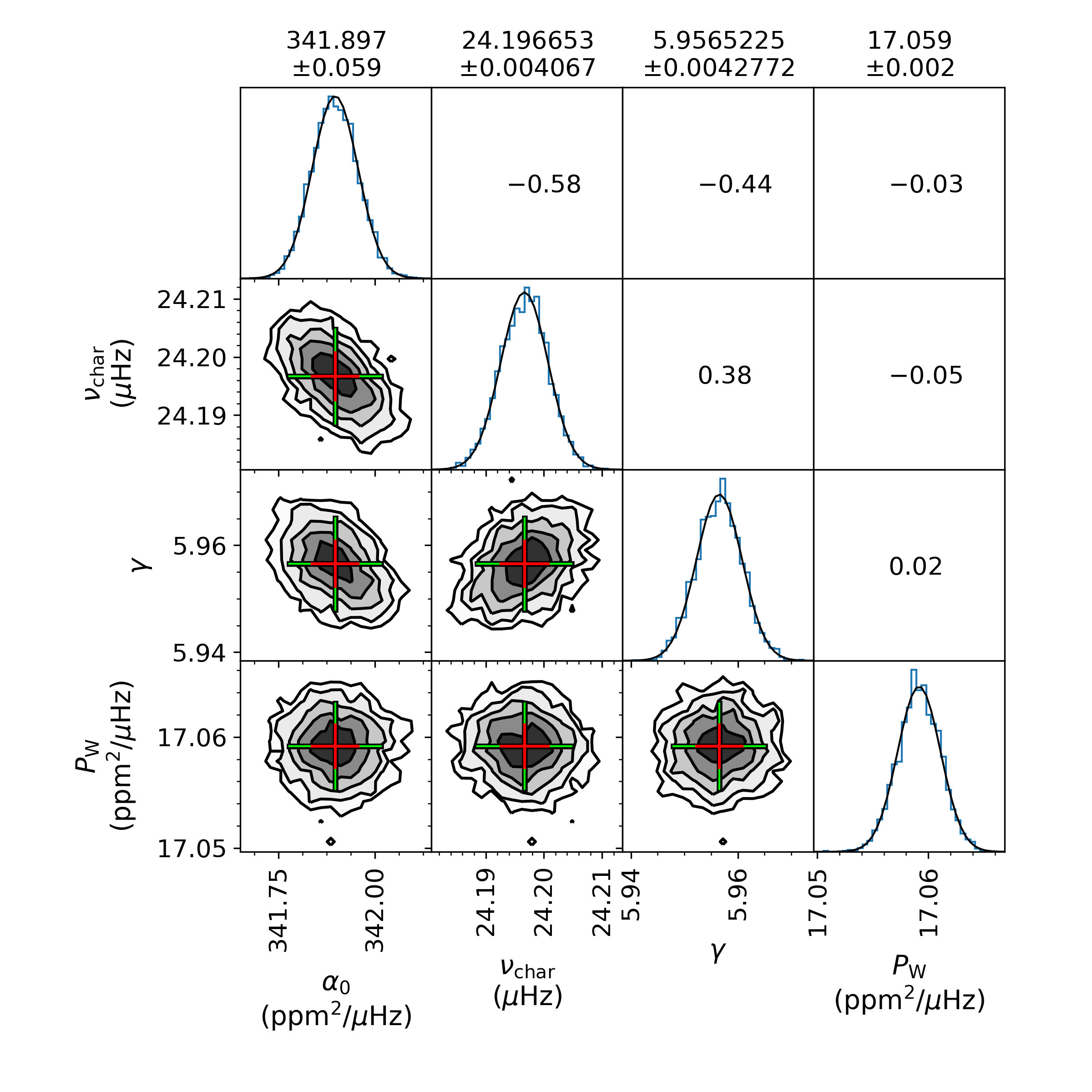}
\includegraphics[width=0.49\textwidth]{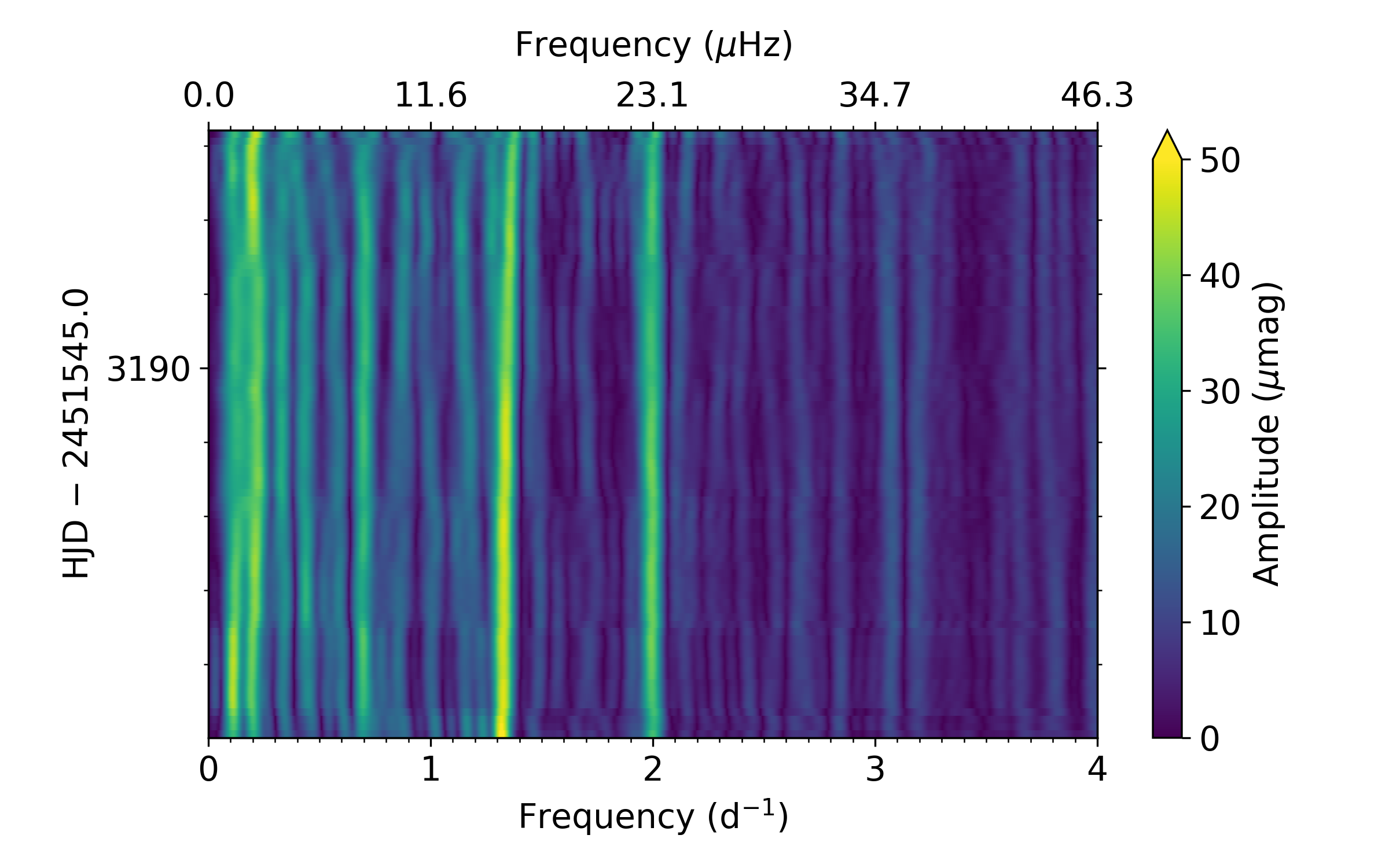}
\caption{Summary figure for the A star HD~174990, which has a similar layout as shown in Fig.~\ref{figure: HD46150}.}
\label{figure: HD174990}
\end{figure}


\begin{figure}
\centering
\includegraphics[width=0.49\textwidth]{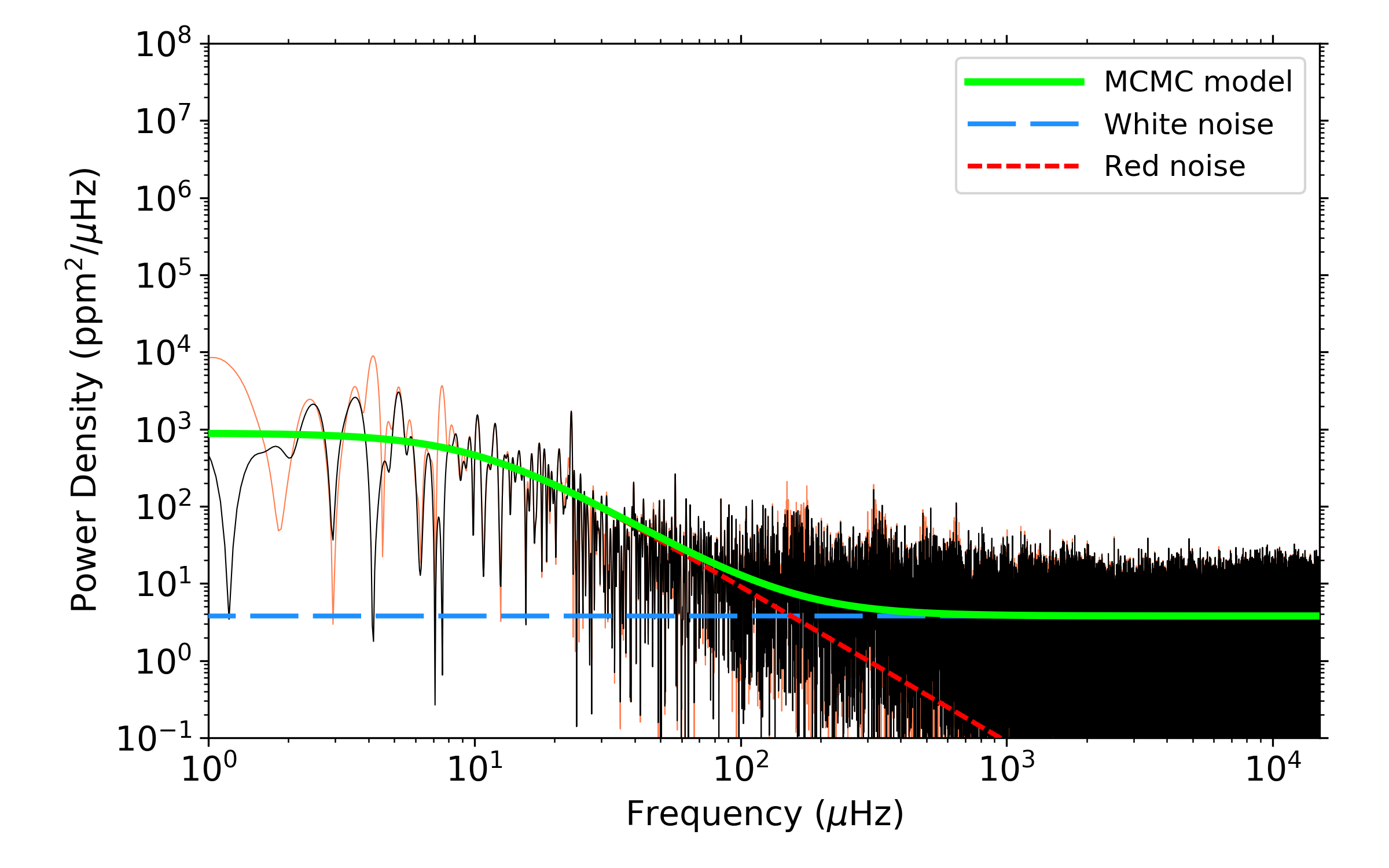}
\includegraphics[width=0.49\textwidth]{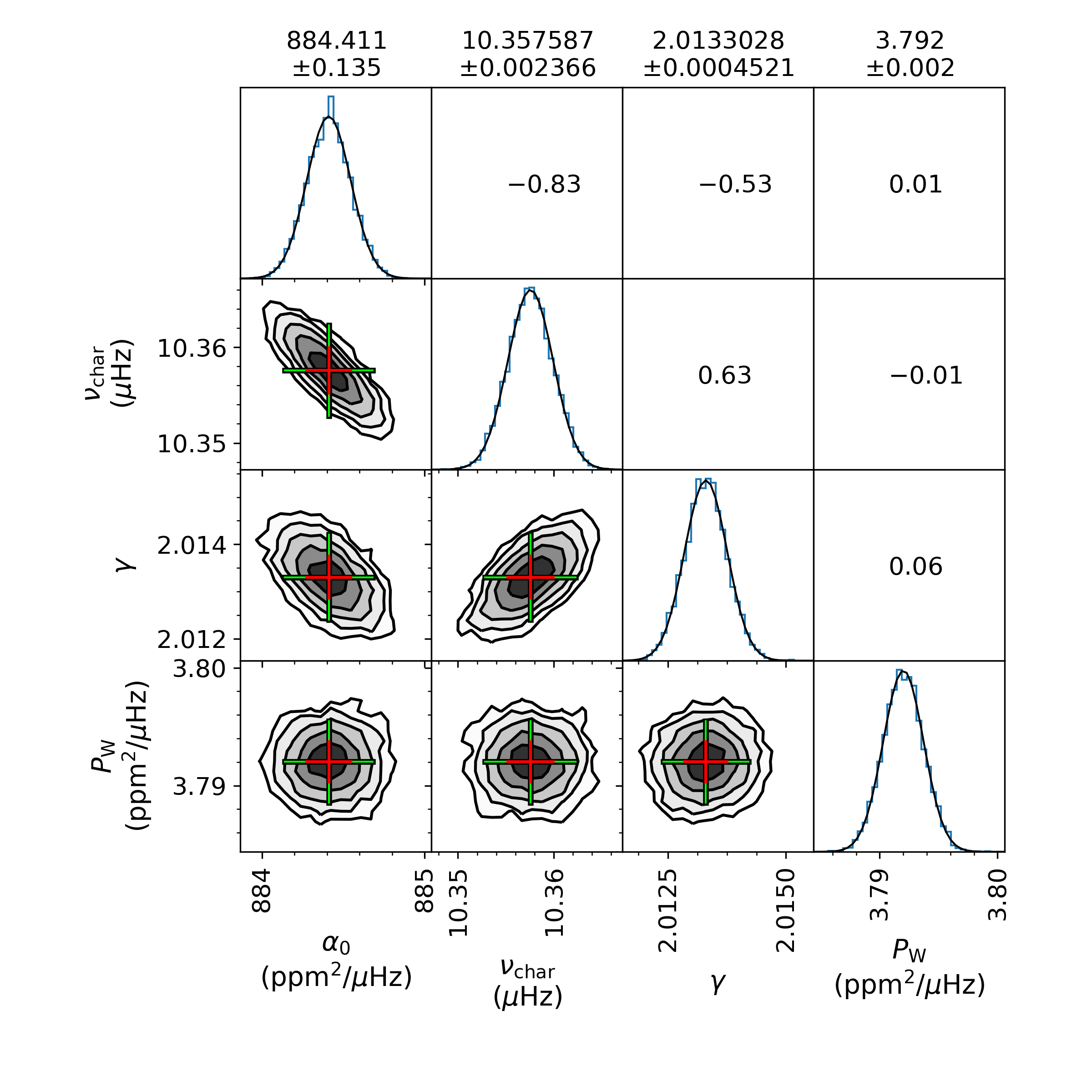}
\includegraphics[width=0.49\textwidth]{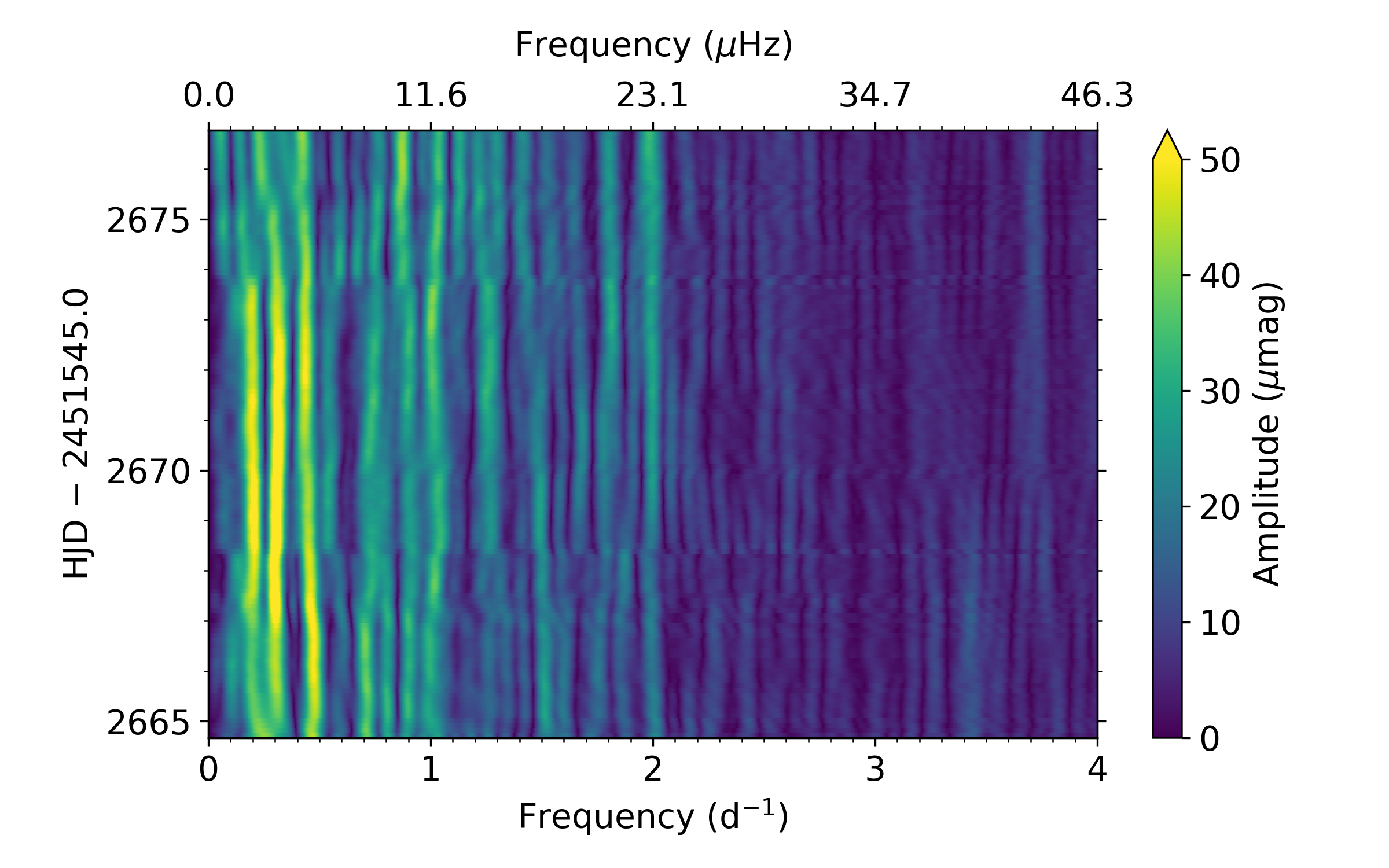}
\caption{Summary figure for the F star HD~175272, which has a similar layout as shown in Fig.~\ref{figure: HD46150}.}
\label{figure: HD175272}
\end{figure}

\clearpage 

\begin{figure}
\centering
\includegraphics[width=0.49\textwidth]{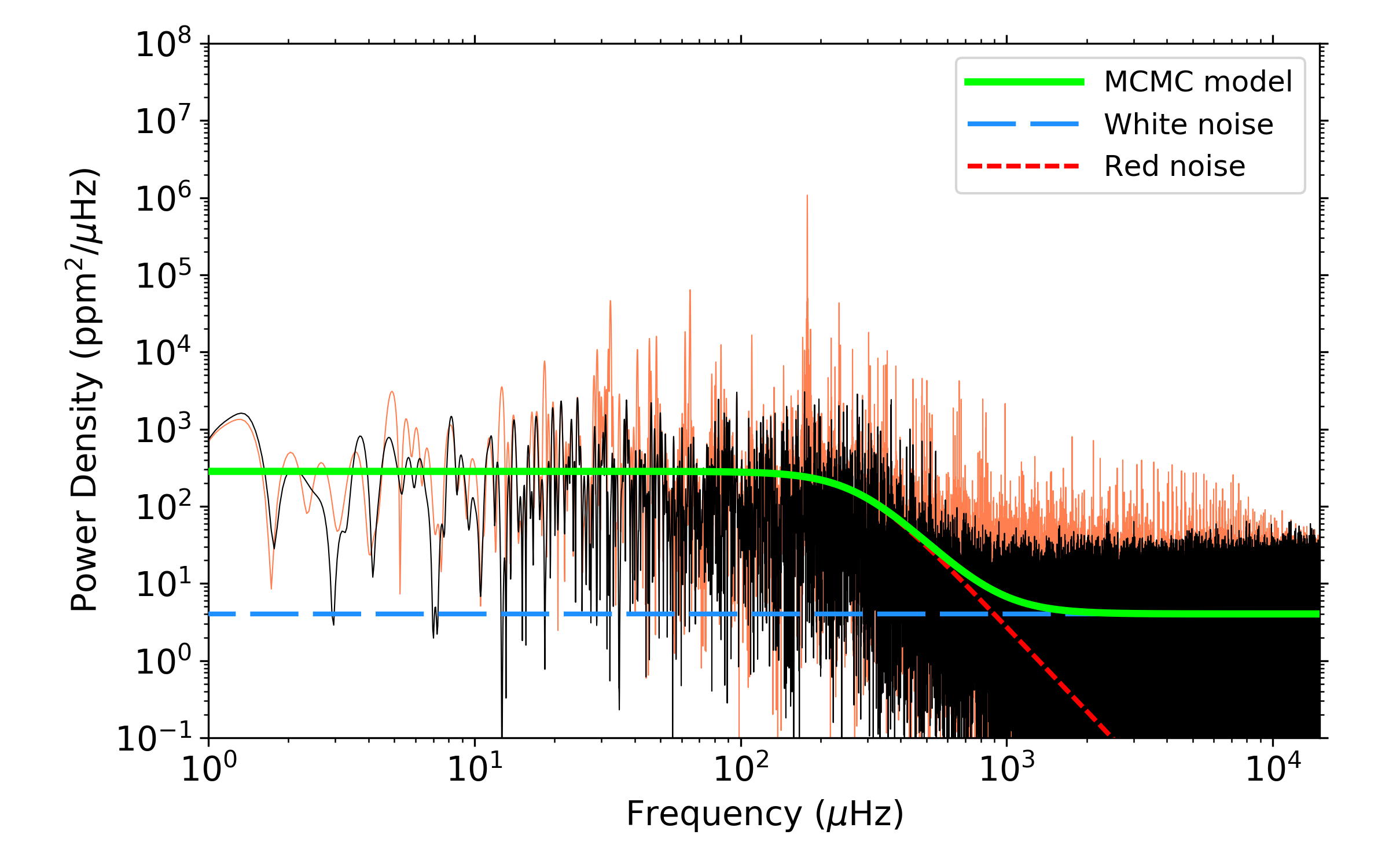}
\includegraphics[width=0.49\textwidth]{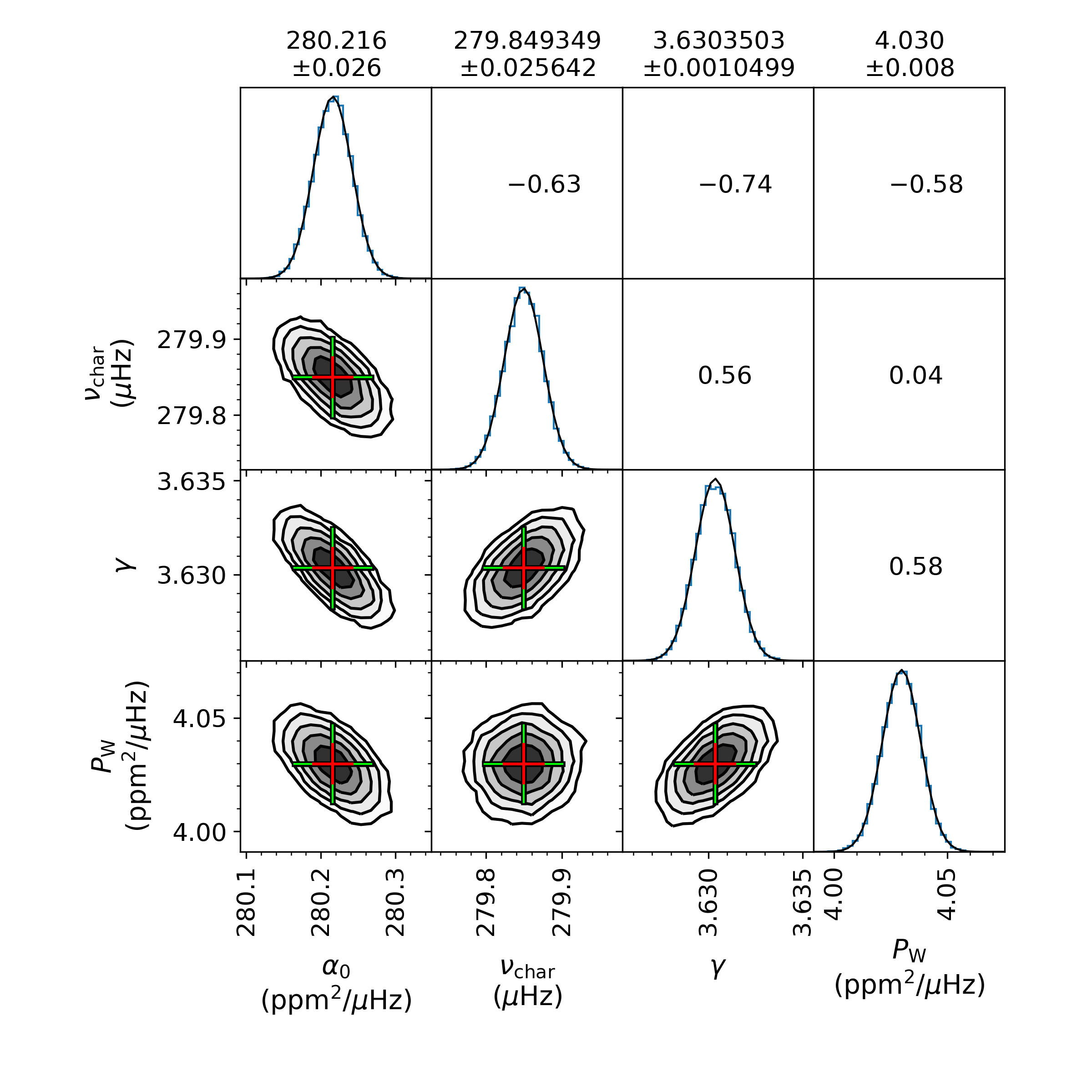}
\includegraphics[width=0.49\textwidth]{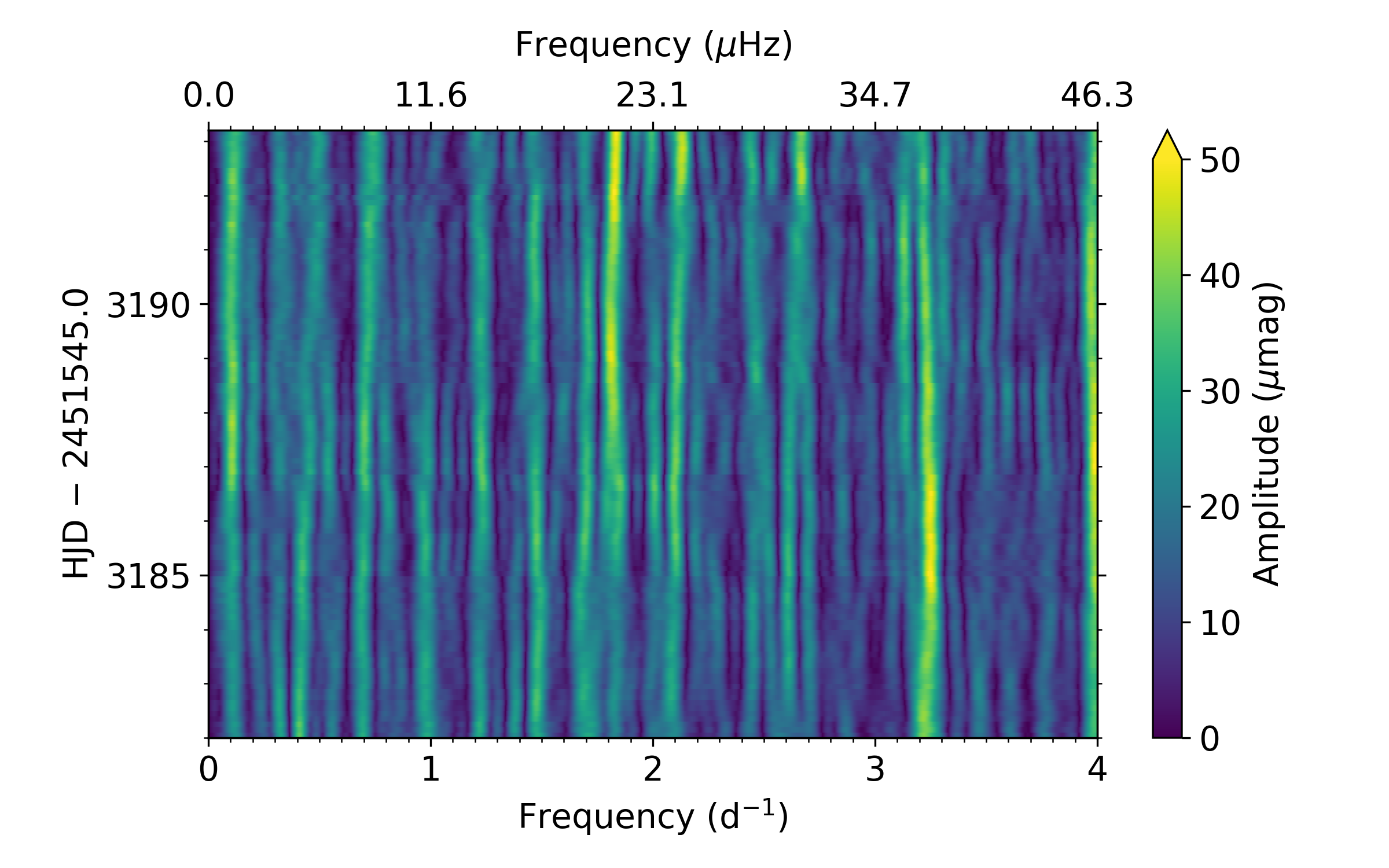}
\caption{Summary figure for the A star HD~175445, which has a similar layout as shown in Fig.~\ref{figure: HD46150}.}
\label{figure: HD175445}
\end{figure}


\begin{figure}
\centering
\includegraphics[width=0.49\textwidth]{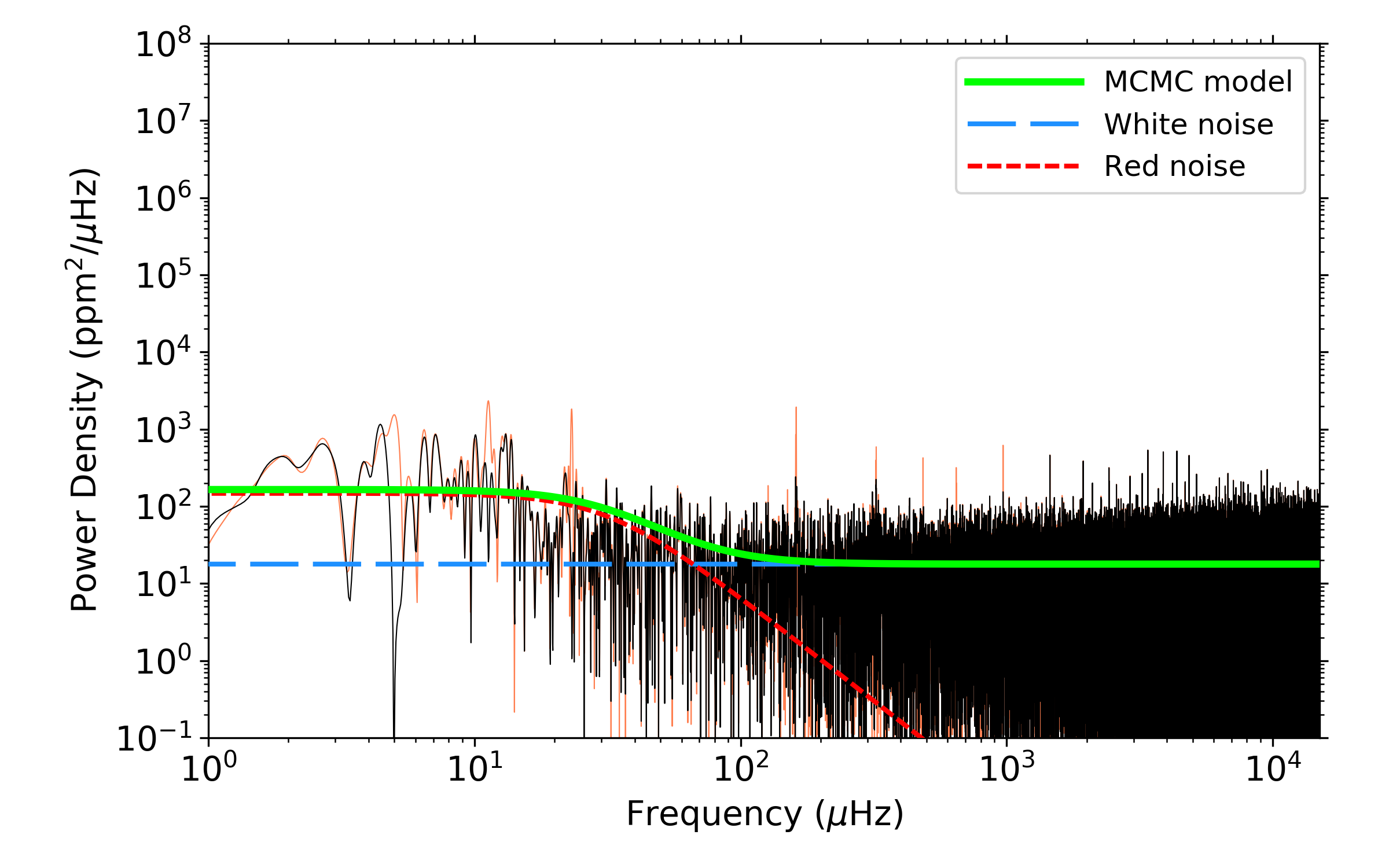}
\includegraphics[width=0.49\textwidth]{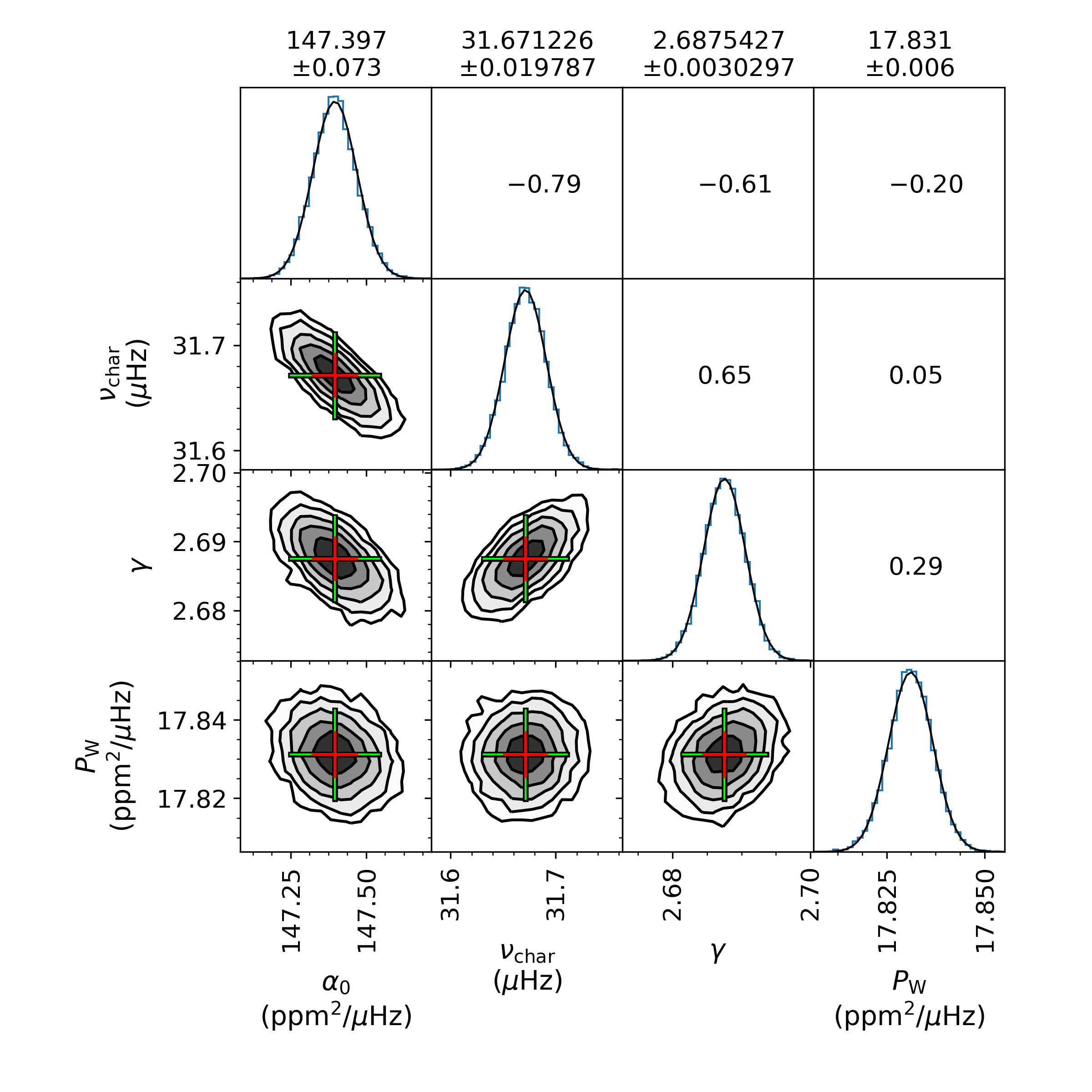}
\includegraphics[width=0.49\textwidth]{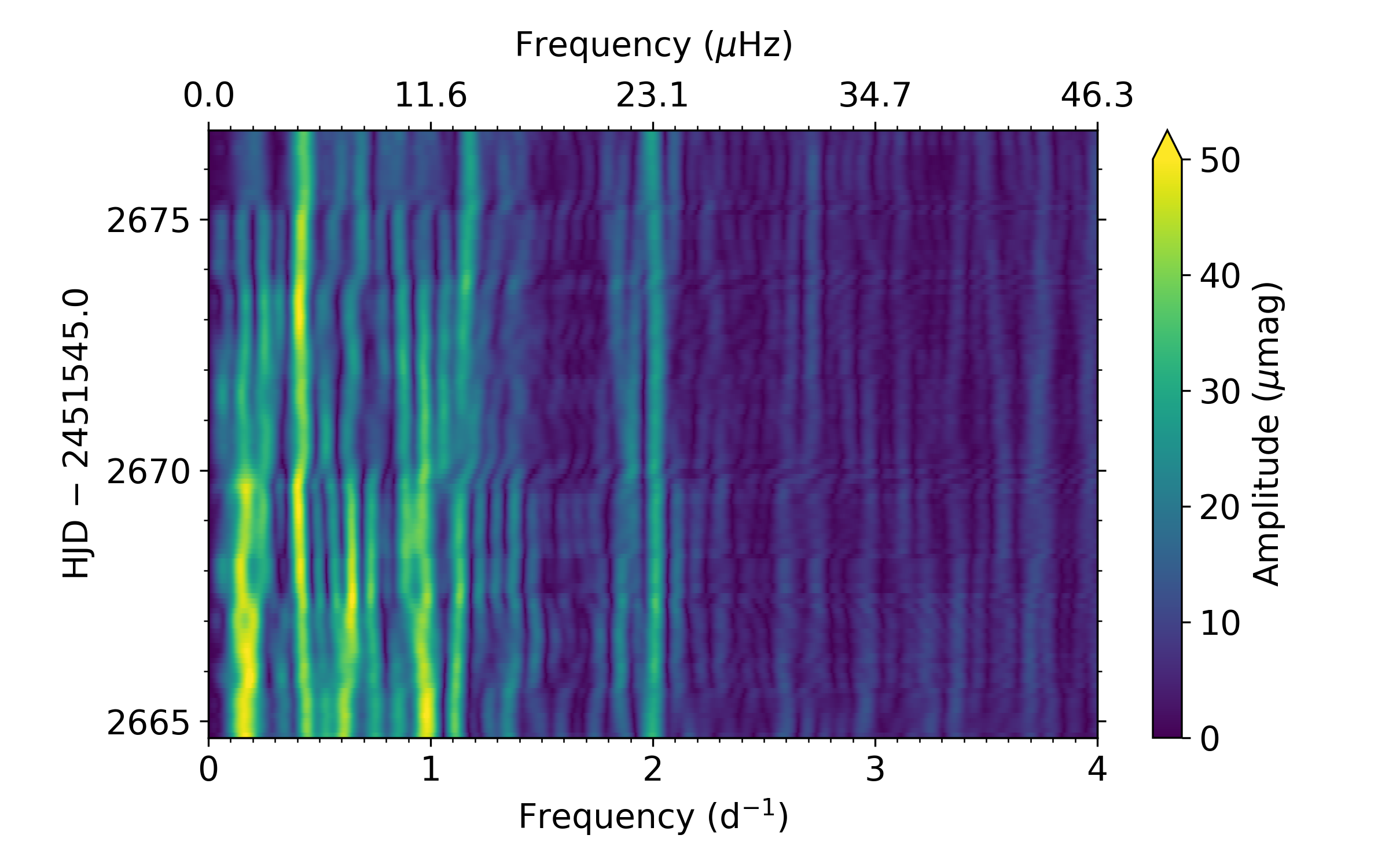}
\caption{Summary figure for the A star HD~175542, which has a similar layout as shown in Fig.~\ref{figure: HD46150}.}
\label{figure: HD175542}
\end{figure}

\clearpage 

\begin{figure}
\centering
\includegraphics[width=0.49\textwidth]{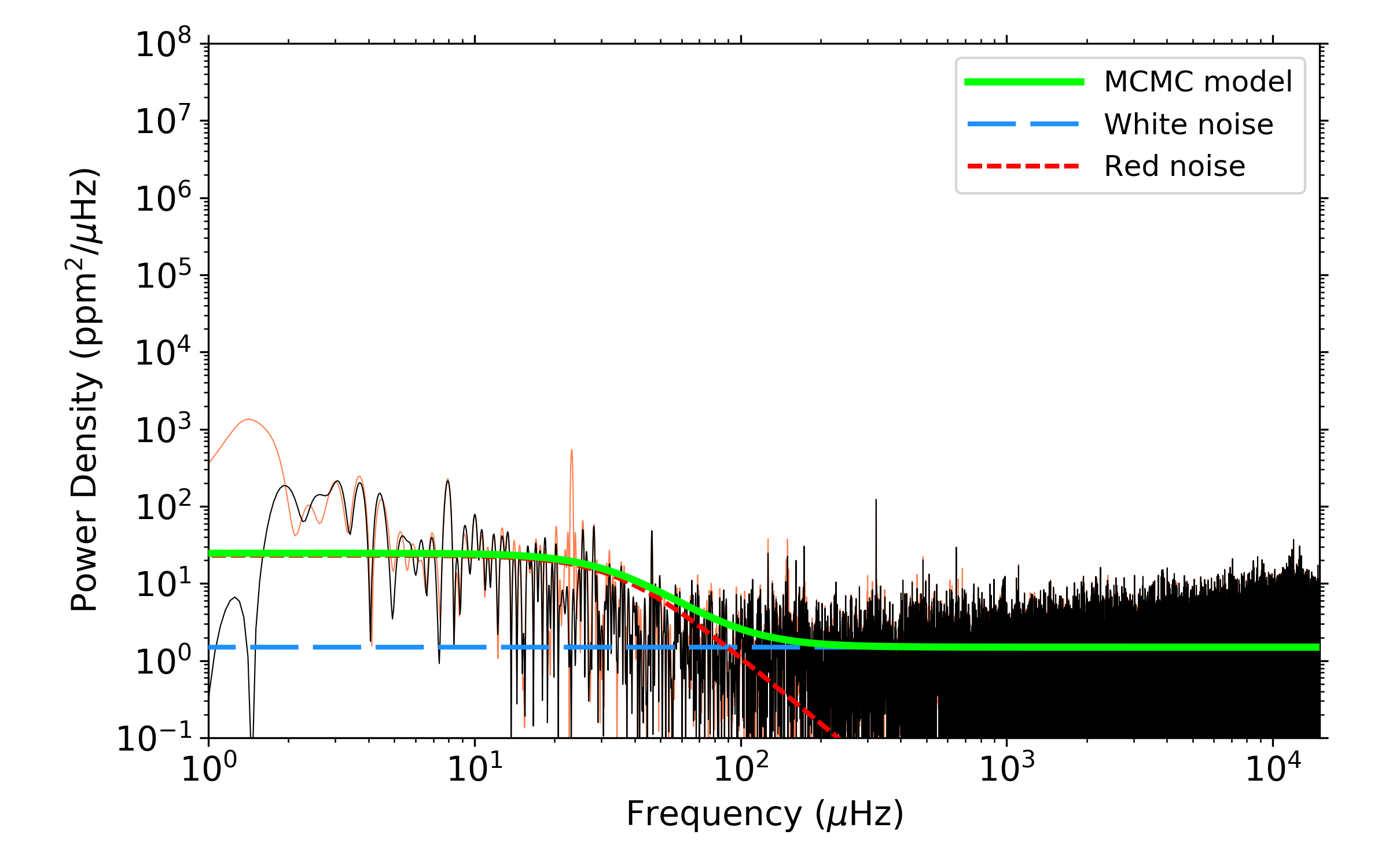}
\includegraphics[width=0.49\textwidth]{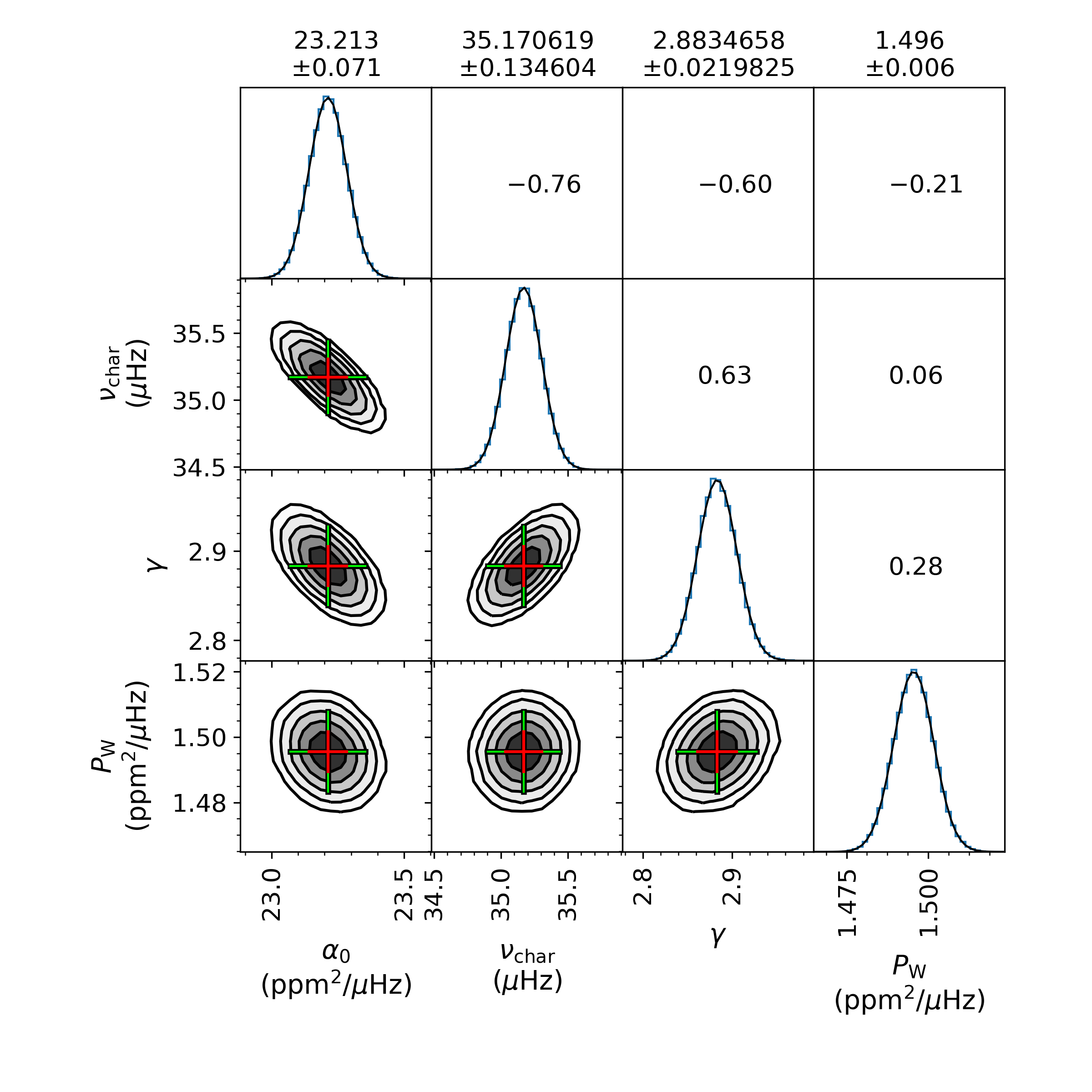}
\includegraphics[width=0.49\textwidth]{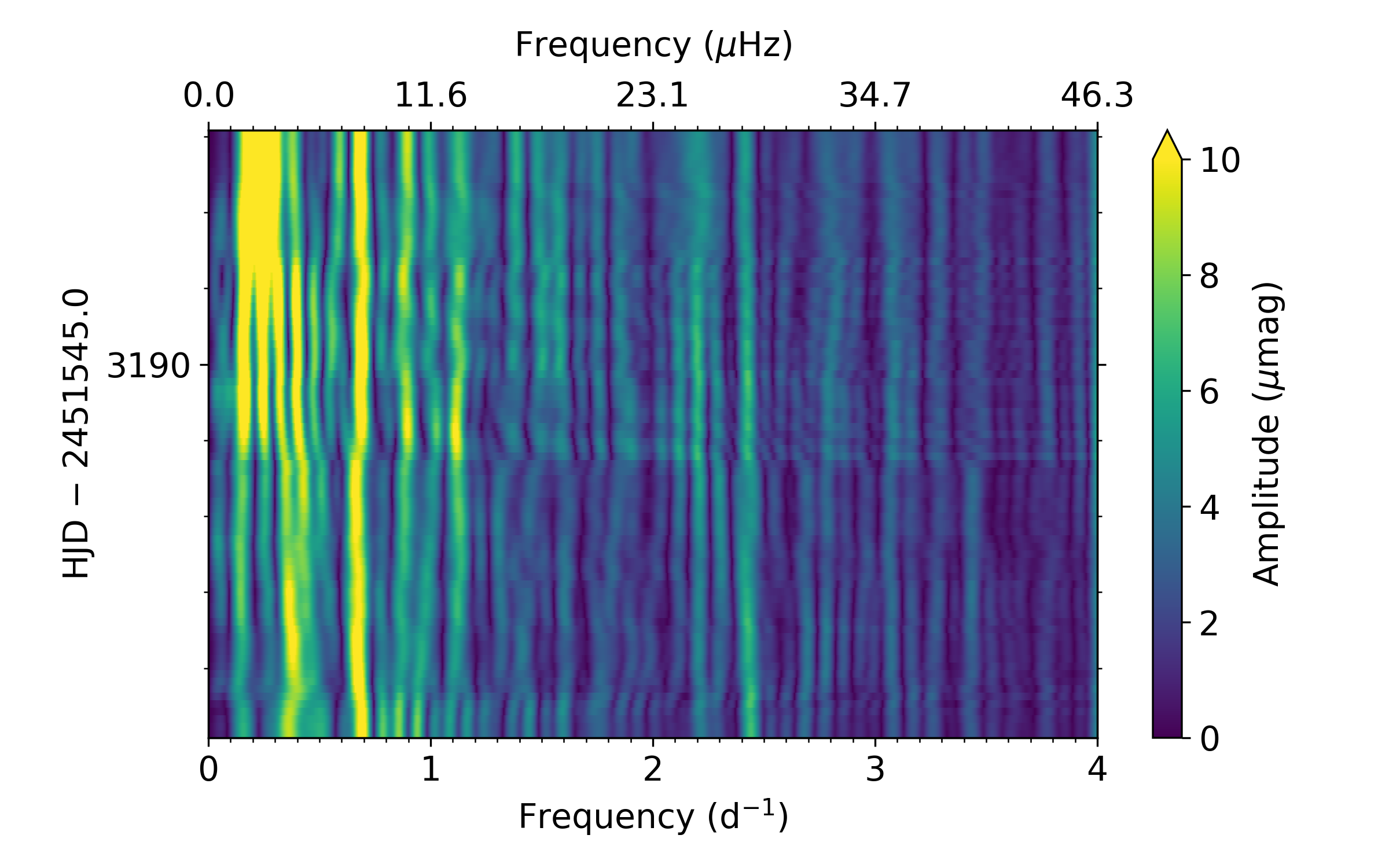}
\caption{Summary figure for the B star HD~175640, which has a similar layout as shown in Fig.~\ref{figure: HD46150}.}
\label{figure: HD175640}
\end{figure}


\begin{figure}
\centering
\includegraphics[width=0.49\textwidth]{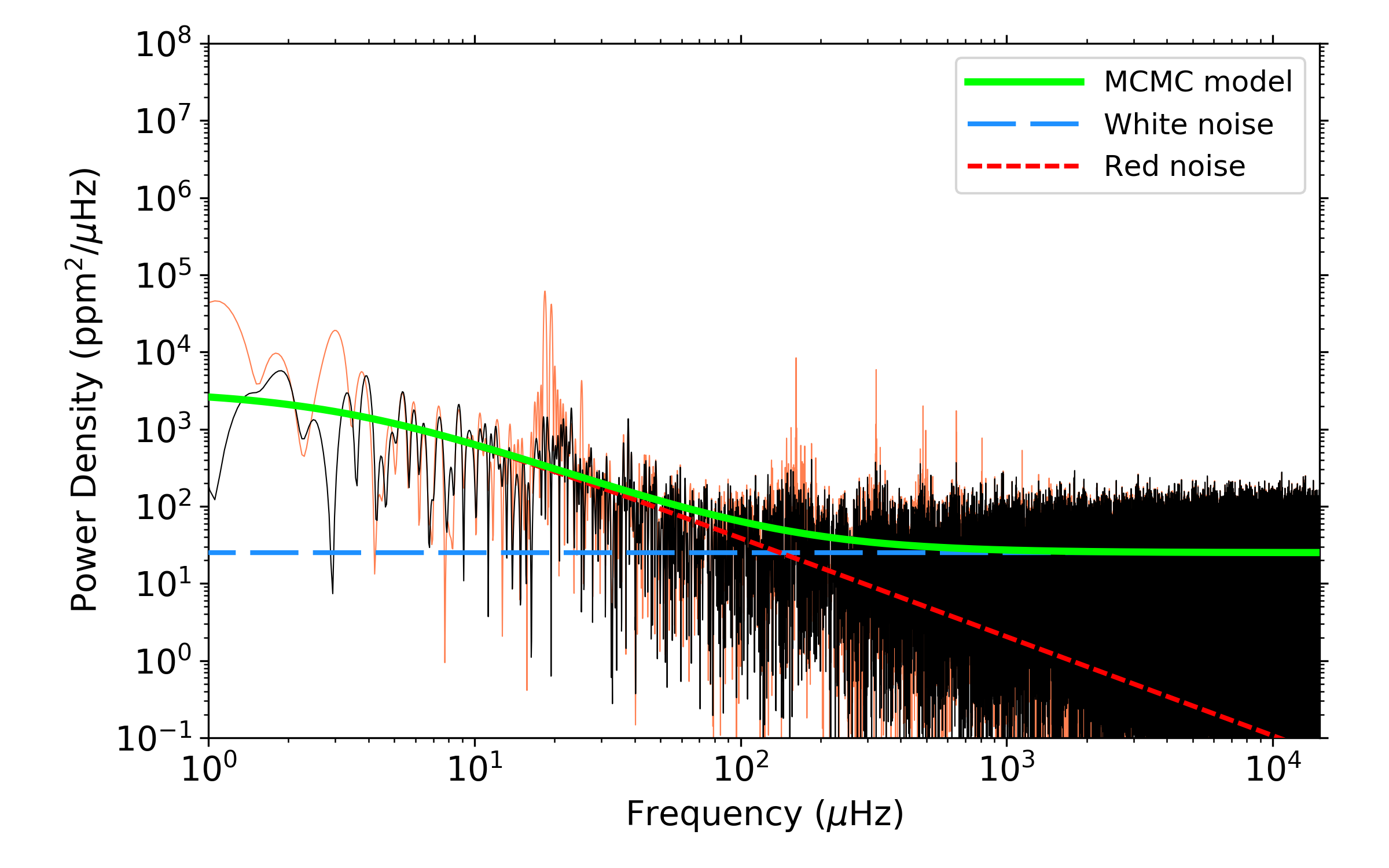}
\includegraphics[width=0.49\textwidth]{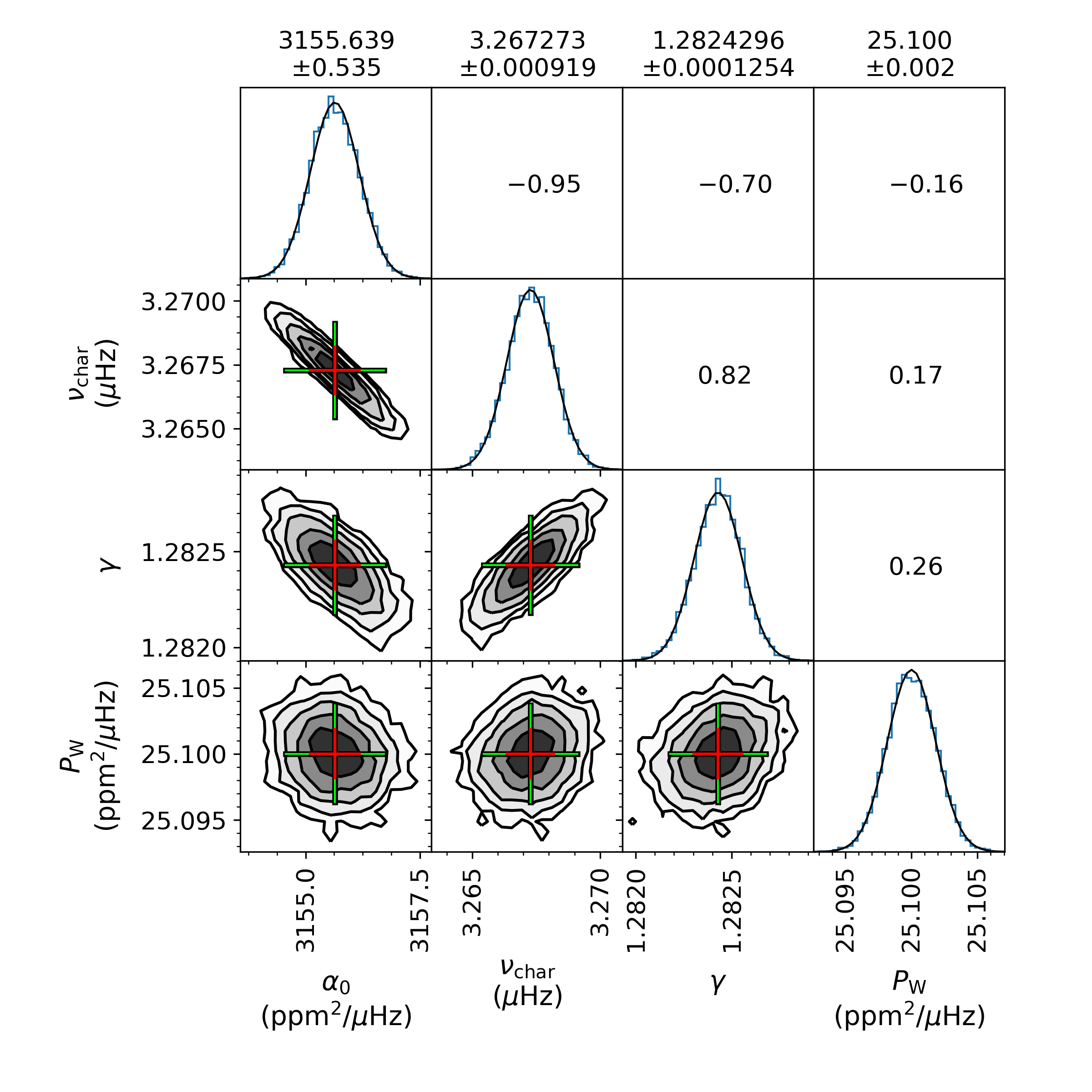}
\includegraphics[width=0.49\textwidth]{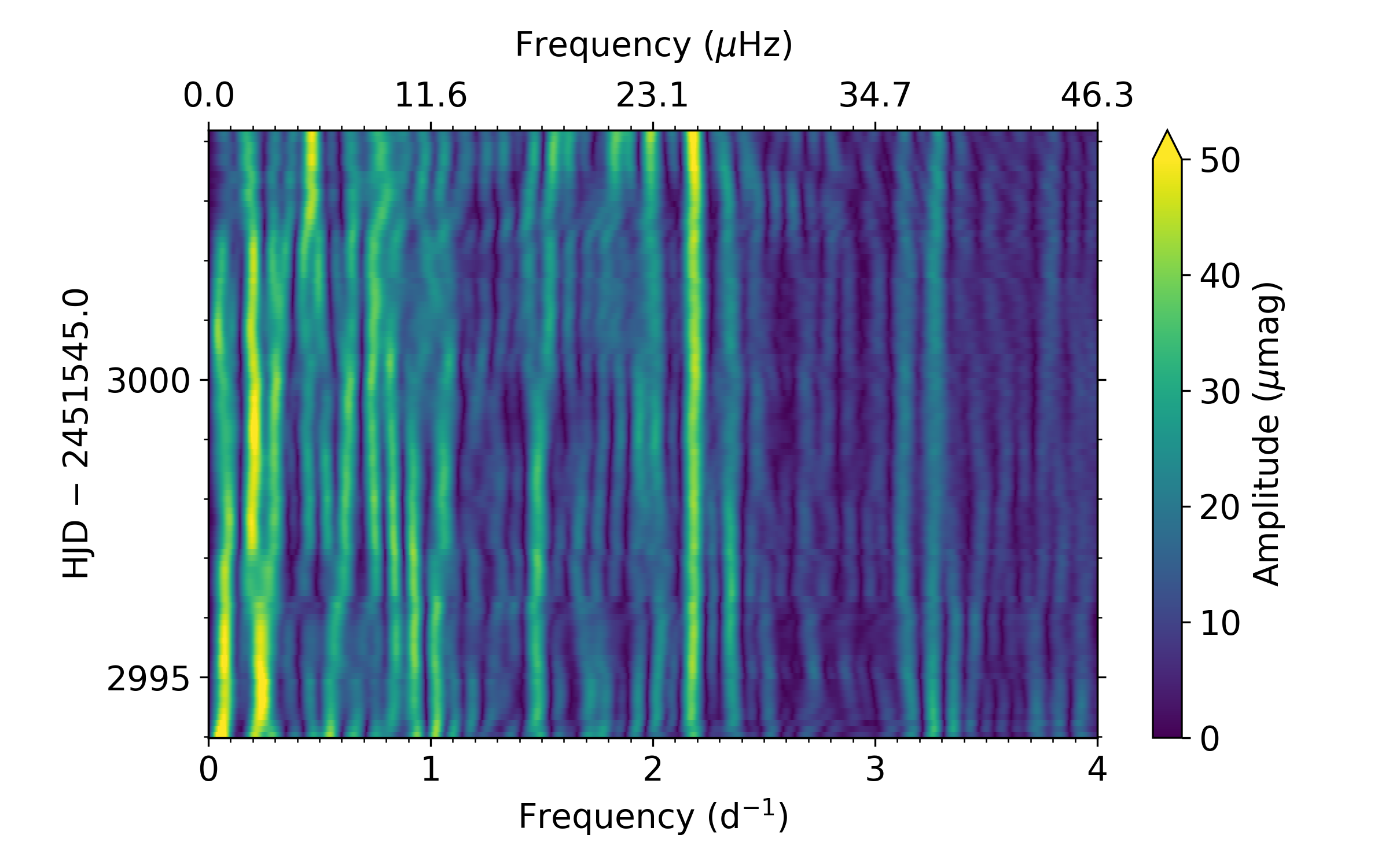}
\caption{Summary figure for the A star HD~263425, which has a similar layout as shown in Fig.~\ref{figure: HD46150}.}
\label{figure: HD263425}
\end{figure}


\end{appendix}


\end{document}